# Interactie tussen licht en geluid in nanoschaal siliciumgolfgeleiders

# Light-Sound Interaction in Nanoscale Silicon Waveguides

Raphaël Van Laer

UNIVERSITEIT GENT

Promotoren:

prof. dr. ir. Roel Baets
prof. dr. ir. Dries Van Thourhout

Examencommissie:

| | |
|---|---|
| prof. dr. ir. Luc Taerwe (voorzitter) | Universiteit Gent |
| prof. dr. ir. Roel Baets (promotor) | Universiteit Gent |
| prof. dr. ir. Dries Van Thourhout (promotor) | Universiteit Gent |
| prof. dr. ir. Günther Roelkens (secretaris) | Universiteit Gent |
| prof. dr. ir. Bart Kuyken | Universiteit Gent |
| prof. dr. ir. Filip Beunis | Universiteit Gent |
| dr. Ewold Verhagen | FOM Institute AMOLF |
| prof. dr. Amir Safavi-Naeini | Stanford University |

Universiteit Gent
Faculteit Ingenieurswetenschappen en Architectuur

Vakgroep Informatietechnologie
Technologiepark-Zwijnaarde 15 iGent, 9052 Gent, België

Dit werk kwam tot stand in het kader van een IWT doctoraatsbeurs voor strategisch basisonderzoek. Het werd gesteund door het ITN-netwerk cQOM en door het ERC InSpectra project.

# Dankwoord

*Let us step into the night and pursue that flighty temptress, adventure.*
J.K. Rowling

DOCTOREREN is op zoek gaan naar nieuwe continenten. Naar het schijnt vertrekken ontdekkingsreizigers met veel moed. Velen sterven onderweg aan ondervoeding of in een storm. Af en toe wordt er gemuit. Slechts enkelingen geraken ergens, meestal jaren later en duizenden kilometers verwijderd van hun geplande bestemming. Dit werk beschrijft de schatten van vier jaren avonturentocht, maar zonder de verhalen van verhongering en kannibalisme. Waarom zou je in hemelsnaam aan zo'n reis beginnen? Zoals in de middeleeuwen, begint dit met de Koning die een opdracht geeft.

Roel Baets hielp me al op een eerste tocht nog voor dit verhaal begint. Ik had een BAEF beurs binnengesleept om een jaar naar Amerika te gaan, maar Amerika zag dat minder zitten dan ik. Roel Baets gijpte de zeilen en wakkerde de Westenwinden aan. In Amerika kwam ik voor de eerste keer in contact met de snufjes die je als fotonicus nodig hebt. Lasers, EDFAs, fibroscopen en graag ook het vermijden van blindheid, Stéphane Clemmen legde me dit allemaal uit alsof het niks was. Hoe boeiend dit ook was, ik moet toegeven dat ik na een halfjaar verdwaald was. Er was geen duidelijk pad richting een schat en mijn hart lag nog in Zweden, bij uitstek het land van avontuurlijke (plunder)tochten. Op dit moment kwam Roel Baets weer in beeld met, zoals we nu gewoon zijn van hem, een heel straf en op het eerste zicht absurd idee. Ik begreep er allemaal niet veel van, maar blijkbaar kon "nanogras waaien in de fotonwind" en nog niemand had dat ooit gezien, dixit Roel Baets. Als dat zou lukken, amai, dat zou vonken geven en (letterlijk) regenbogen genereren. Tegelijk wist ik mijn eerdere Zweedse veroveringen te overtuigen om naar België verhuizen en we waren vertrokken! Het is wel duidelijk dat Roel Baets een hoofdrol speelde in al wat volgt, als mentor en ongeëvenaard kritisch vragensteller.

Ik kreeg een bureau toegewezen met het beste collegateam dat men zich kan wensen: zowel neuromorficus en diepdenker Thomas Van Vaerenbergh als co-Antwerpenaar en zeilexpert Peter De Heyn aan stuurboord, snelfietser en vaderfiguur Thijs Spuesens op de boeg en sprintster en vrolijkste collega Eva Ryckeboer aan bakboord. Zowel Thomas van Vaerenbergh als Peter De Heyn waren overigens echte specialisten in de studie van ringen (op een chip, niks met trouwen vandoen), waar ik uitermate veel van geleerd heb. Het vijandige schip aan de andere kant van de gang werd geleid door topatleet Yannick De Koninck en babbelaar Bart, intussen Prof. Kuyken. Wel, zo vijandig was het nu ook weer niet, maar er was toch sprake van enige frictie telkens Yannick De Koninck onze oase van rust kwam verstoren met zijn basketfanatisme (Achterwaarts! Met de ogen toe!) of indien er een Age of Empires/pizza veldslag plaatsvond (lang nadat we ons werk gedaan hadden en zeker nooit tijdens de middag – trouwens, Daan Martens, merci om de trebuchets van Pieter Wuytens naast mijn towncenter te vernietigen). Elk van deze mensen speelde een belangrijke rol bij het vermijden dat mijn boot toen strandde. En specifiek Bart Kuyken, die niet alleen een beroep als humorist gemist heeft maar ook nog eens sneller kan spreken (en denken) dan zijn schaduw. Zonder hem was ik wellicht nog steeds ergens aan het dobberen langs de kust. Neem nota, dit is iemand die weet hoe je fotonen met elkaar doet botsen, en met wat een schwung!

Toen leerde ik ook Dries Van Thourhout kennen. Hij is de echte expert in het doen trillen van kleine objecten en wist op kantelmomenten sleuteladvies aan te reiken. Samen met Bart Kuyken en Roel Baets vormde hij het triumviraat dat mij toen vakkundig op pad stuurde om dat nanogras te doen waaien in de fotonwind. Eerlijk gezegd weet ik niet meer wat ik dat eerste jaar juist uitspookte. Simulaties werden gedraaid, fabricage werd gepoogd en meetings gehouden, maar het leek allemaal vergeefs. Tegelijk kwam de competitie op dreef. Needless to say bereikten de stressniveau's een hoogtepunt. Ik zou graag zeggen dat ik daar als een echte avonturier kalm onder bleef, maar dat zou niet kloppen. Er lag daar immers een wetenschappelijke schat te wachten en we moesten die gewoon maar even opgraven!

Wat ik me iets beter herinner is dat we toen plots op de Olympische spelen (ja ja, voor Europese universiteiten en bedrijven) in Praag stonden. Medailles dat we daar binnenhaalden! Bart Kuyken verbeterde zijn persoonlijke minigolfrecord, Yannick De Koninck fietste sneller dan zijn banden liefhadden en Peter De Heyn haalde zelfs twee medailles denk ik, niet eens in het zeilen. Vriend en bedrijfskundige Derek Verleye versloeg toen (of beeld ik me dit nu in?) dramatisch het "gelleke", een strakke Duitse loper die tot dat moment door geen levend wezen in de wind gezet was. En ikzelf liep een respectabele 41 minuten op 10 kilometer, niet slecht voor iemand die ongeveer de beweging van een kamerplant krijgt, toch?

Alleszins, ik besloot toen dat dat nanogras nooit zou gaan waaien en begon op andere structuren te werken. Na een jaar werd op een of andere manier mijn eerste talk aanvaard en vloog ik met Bart Kuyken en Frédéric Peyskens naar San Jose. Buiten een epische fietstocht door San Francisco en over de Golden Gate bridge staat me daar vooral een bezoek aan een Mexicaanse bar in een dodelijke wijk van San Jose van voor. Pieter Geiregat, het is mij nog altijd niet duidelijk waarom jij een hotel aan de andere kant van die brug nam. We zijn daar nog een paar keer teruggeweest, maar even plezant als die eerste keer werd het toch nooit he? Het was vooral dankzij de heldendurf van Frédéric Peyskens dat we toen onvergetelijk getrakteerd werden. Ik ga dat hier nu niet allemaal vertellen om niemand te beschamen, misschien straks op de receptie. Laat me toch nog snel zeggen dat een lokale appel-etende pimp ons onderwierp aan een psychologische beoordeling. Zijn zorgvuldige conclusie: Bart Kuyken was "the cool nerd", Frédéric Peyskens "the actual nerd" en ik was "Harry Potter" als ik me het goed herinner. Eveneens in de context van San Jose maakte ik met Thomas Van Vaerenbergh enkele spannende momenten mee. Sneeuwval in mei, twee dagen rijden, spooksteden en een kapotte motor, wees gewaarschuwd voor Thomas Van Vaerenbergh en zijn auto.

Mijn onderzoek kwam na die conferentie in een stroomversnelling, indien niet door de vereenzelviging met de held Harry Potter dan wel door het opsnuiven van de fotonica glamour in San Jose. Wat later begon de fabricage eindelijk te werken en sloot ik mezelf op in de meetkamer om die dingen te doen trillen, wat enkele maanden later ook lukte. Het triumviraat was net als ik erg opgezet met de resultaten, dus werd ik weer op pad gestuurd naar een hele reeks conferenties. Ik vloog toen zelfs een keer met tweederde van het triumviraat naar San Diego en New Haven! Dat hadden we niet kunnen navertellen zonder enkele zeer doeltreffende slaappillen van Roel Baets. Maar goed, vooral van tel is dat ik mijn werk nooit had kunnen doen zonder de technische, morele en wetenschappelijke steun van vele mensen die elks een medaille verdienen. Zonder volgorde:

- Liesbet Van Landschoot, bedankt voor het opvrolijken van vele uren SEMmen, het verlenen van morele steun bij een hele reeks mislukkingen en voor enkele van de knapste SEM foto's uit de recente fotonica geschiedenis (waarvan één voor eeuwig op de cover van een bekend fotonica tijdschrift zal prijken)!

- Kristien De Meulder, bedankt voor die tientallen leveringen van USB kabels onder tijdsdruk, het fiksen van menig printerconnectie, herstellen van laptops en voor boeiende gesprekken over de alweer afgebrande Gävlebocken!

- Michael Vanslembrouck, zonder jou was ik nog steeds manueel data aan het nemen. Bedankt voor de talloze keren dat je een GPIB probleem oploste en voor diepe gesprekken in de Brug over buddha's, vrije wil en bewustzijn.

- Steven Verstuyft, de rots van de cleanroom. Ook wel de mysterieuze man die schaakgrootmeester is in zijn vrije tijd en elke machine als zijn broekzak kent.

- Ilse Meersman, Ilse Van Royen en Bert Coryn, bedankt voor het helpen bij een eindeloze serie administratieve taken.
- Günther Roelkens, voor het suggereren en helpen vormen van onze eerste plannen rond trillende objecten en fotonische winden.
- Alex Bazin, thanks a lot for fabricating the first suspended nanowires! These were really exciting times, I had a lot of fun doing this project.
- Bedankt Roel en Dries, voor de quasi ongelimiteerde vrijheid en steun in het achtervolgen van schijnbaar knotsgekke plannen. Deze schuld wordt wellicht best voorwaarts ingelost, door misschien ooit jonge onderzoekers in hun avonturen te begeleiden.
- Bedankt Joris Roels voor het uitvinden van hele kleine trillende snaren en het delen van jouw ervaring bij de start van mijn werk.
- Thanks Amin for bringing those keys, taking that awesome SEM with Liesbet and sharing your Iranian specialties! Hope you'll break your speed record yet again.
- Bedankt aan alle vorige generaties fotonici om de meetkamers en cleanroom zo ver te brengen nog voor ik wist wat een foton was.
- Paul and Jesper, I greatly enjoyed our many hours of dirty talk about photons, phonons and what they would do to one another, as well as our trips to and (not) skiing in Switzerland.
- Andreas en Sarah, bedankt voor dat heuglijke hoevefeest, ik wens jullie een hele mooie toekomst! Ik raad jullie wel aan jullie pannekoeken in het vervolg niet aan de gemeenschappelijke tafel te gaan opeten.
- Sam en Pauline, nog zo'n fotonica stronghold koppel, merci voor alle Japantips en de hulp bij menig ring- en highspeedprobleem.
- Bedankt Martijn Tassaert voor jouw inzichten in tapers en het voorbeeldgedrag wat efficiënte planning betreft.
- François, I will not forget that one particular event on a seemingly ordinary Thursday noon in Barcelona. Hope to meet you in Auckland.
- Haolan, your dedication to getting a PhD in photonics and becoming the next Schwarzenegger simultaneously continues to amaze me.
- Thank you Amir Safavi-Naeini and Ewold Verhagen for traveling far to be on my PhD committee.

- Martin Fiers, bedankt voor dat gezellige Halloweenmoment en de Limburgse referenties. Luceda for the win.

- Pieter Wuytens, nog zo'n snelfietser en globaal straffe kerel, wij hebben toch toffe momenten beleefd in die IEEE lounge he. Je hebt wellicht het lastigste onderzoeksproject ooit gekozen maar geen seconde twijfel ik dat dat goed komt.

Nu denk je wellicht dat die fotonica groep eerder een veredelde fietsclub is. Dat is maar een stuk van de waarheid, er wordt hier immers ook op hoog niveau gepingpongd. Ananth, when will we finally finish this tournament? Weiqiang, it was an honour being on your 'team'. Behalve tientallen reizende sportsterren loopt hier overigens ook een garde jonge wetenschapstalenten rond: Artur, Koen, Suzanne, Herbert, Kasper, Floris, Ashwyn, Yufei, Ang en Anton, volgens mij zijn jullie sterk bezig, niet te veel zorgen maken en gas blijven geven. You never fail until you stop trying, dixit Einstein.

Er zijn gelukkig ook nog mensen die minder geobsedeerd zijn door fotonen: de onmisbare familie- en vriendengroep. Ik denk dat jullie wel weten wie jullie zijn. Laat me in de context van dit werk snel Koen, Arno en Sofie bedanken voor de gezellige Trattoria escapades tijdens de week. En Jeroen, Carolien, Dimi, Elien, Luc, Sarah, Derek, Philippe, Eliane, Stefanie, Gert, Stijn en Julia voor allerlei etentjes, feestjes, skireizen, road- en citytrips, wandel- en vogelkijktochten, golfuitstapjes, vrijgezellenweekends en andere zaken die het leven waardevol maken. Bedankt mama, papa, zus, mami, bompa, Michel, Marc, Veerle voor de broodnodige thuismomenten en toffe uitstapjes allerlei (zoals Stockholm he zus!). Papa, dit boek is een rechtstreeks gevolg van onze jarenlange babbels over natuurfenomenen allerlei. Och tack så mycket Klas, Annika och Björn för den bästa tiden i Lidköping.

Ten slotte zou ik ook heel graag Emma bedanken. Det går jättebra med oss tror jag! Als iemand weet hoe veel zware momenten er achter dit doctoraat zitten ben jij het wel. Inte säker vad jag gjort för att förtjäna dig. Jag är helt redo för vår nästa äventyr. Tack, älskar dig och heja heja sötling!

Ziezo, en dit was nog maar het begin.

19 mei 2016
Raphaël Van Laer

# Samenvatting

DE WETENSCHAP van het licht tracht fotonen te manipuleren. Dit is met groot succes gelukt de voorbije decennia: fotonen zijn ideale informatiedragers gegeven hun laag propagatieverlies en hoge bandbreedte [1–3]. Daarom zijn ze nu al dominant wat lange-afstand communicatie betreft; ze versturen data aan steeds toenemende snelheden onder de oceaan. Sinds kort neemt fotonica ook chip-tot-chip communicatie over [4], waarbij het enkele belangrijke problemen – zoals warmtegeneratie en signaalvervorming [1, 2] – geassocieerd met elektronica oplost. Vandaar ontstond er een grote motivatie voor de ontwikkeling van fotonische circuits op de nanoschaal, de wet van Moore achterna [5]. Moderne optische golfgeleiders sluiten licht op in een gebied kleiner dan 0.1 $\mu m^2$ en kunnen gemaakt worden van het alomtegenwoordige materiaal silicium, wat massaal gebruikt wordt in de bestaande halfgeleiderindustrie.

Dit leidt tot de natuurlijke vraag of optica ook bepaalde bewerkingen kan uitvoeren [6, 7], wat vereist dat sommige fotonen de stroom van andere kunnen controleren [8]. De grote uitdaging bestaat uit het versterken van de typisch zwakke interacties tussen fotonen. Specifiek, foton-foton koppeling is verwaarloosbaar in vacuüm [9, 10], zodat het gebruik van intermediaire materiaalexcitaties de enige realistische route is om dit te doen. Aldus ontstond er een grote beweging richting het versterken van interactie tussen licht en materie, van het gebruik van het Kerr [11–14], Raman [15–17], vrije ladingsdrager [18, 19] en thermisch [20, 21] effect tot cavity QED [22, 23].

Hoewel ze nu reeds radiosignalen filteren in elke smartphone en laptop [24, 25], verschenen mechanische systemen pas in de fotonica wereld een decennium geleden. Initieel werden vooral megahertz vibraties opgewekt en uitgemeten in microtoroïdes [26], siliciumbalken [27–29] en nitride disks [30]. Zulke laagfrequente oscillatoren genereren niet-lineariteiten die grootteordes sterker zijn dan intrinsieke materiaaleffecten [31, 32]. Maar het is wenselijk om de mechanische frequenties te verhogen tot in het gigahertz gebied. Dit geeft hen toegang tot toepassingen in de bewerking van microgolven [33, 34] en laat hen sneller data behandelen.

In dit werk realiseren we een efficiënte en sterk maakbare optische nietlineariteit via gigahertz fononen. De niet-lineariteit wordt vaak *gestimuleerde Brillouin verstrooiing* (SBS) genoemd. Verder leggen we de focus op golfgelei-ders die zowel licht als geluid opsluiten, zodat hun interactie kan opbouwen over vele diffractielengtes. Deze golfgeleiders zijn optisch breedbanding in de zin dat de frequentie van de pomplaser vrij is binnen een bereik van circa 10 THz. We worden echter geconfronteerd met stijve gigahertz mechanica, waarbij verplaatsingen onder een picometer liggen. Daarnaast verliezen we ook op vlak van de efficiëntie geassocieerd met optische caviteiten [35].

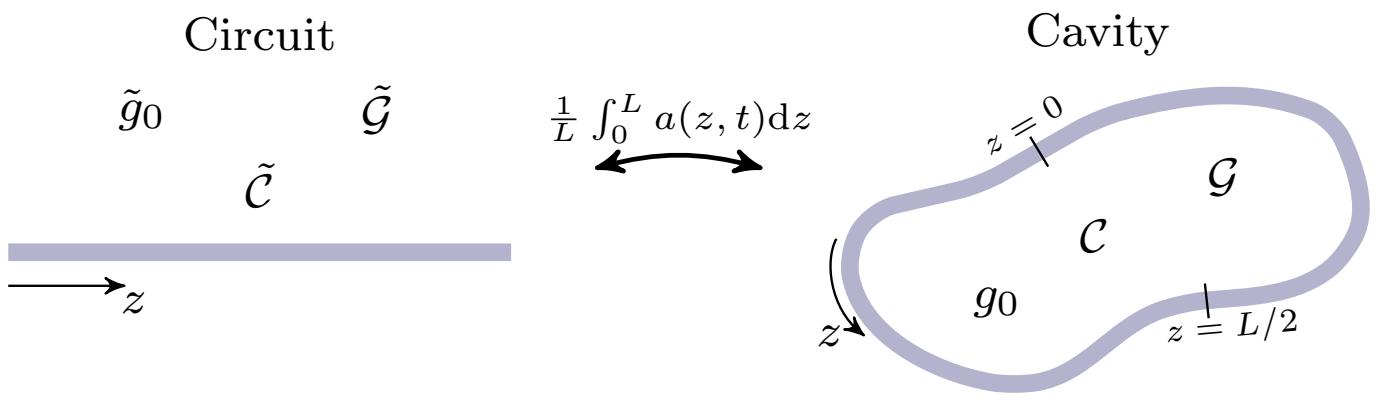

**Figure 1: Van optomechanica in circuits naar caviteiten.** We leiden expliciet de fysica van optomechanische caviteiten (rechts) af van die van Brillouin-actieve golfgeleiders (links). Vandaar kunnen zowel golfgeleider- als caviteitgebaseerde optomechanica geformuleerd worden in termen van vacuüm kopplingssterkes ($\tilde{g}_0$ and $g_0$), coöperativiteiten ($\tilde{\mathcal{C}}$ and $\mathcal{C}$) en gain parameters ($\tilde{\mathcal{G}}$ and $\mathcal{G}$).

Vooreerst onderzoeken we theoretisch de fysica van foton-fonon interacties in zowel golfgeleiders als caviteiten (fig.1). Het blijkt dat conventionele veronderstellingen niet meer gerechtvaardigd zijn in bestaande systemen. Traditioneel propageren fotonen verder dan fononen en wordt hun koppeling overspoeld door de propagatieverliezen. Dit geval leidt tot versterking van een optische probe die roodverschoven is ten opzichte van een sterke optische pomp; de klassieke *Brillouin versterking* [36, 37].

Als de interactie daarentegen sterker is dan de propagatieverliezen, kunnen fotonen worden geconverteerd in fononen, terug in fotonen, terug in fononen, etc. terwijl ze langsheen de golfgeleider vliegen. Dit noemen we spatiale sterke koppeling, naar analogie met hoofdthema's in andere takken van de fysica zoals cavity QED. Vervolgens produceert een gemiddeldeveld overgang de dynamica van hogefinesse caviteiten vertrekkende van die van golfgeleiders. Dit bewijst een verband tussen twee bekende parameters: de Brillouin versterkingsparameter [36, 37] en de vacuüm optomechanische koppelingssterkte [35]. Het verband verduidelijkt de samenhang tussen effecten zoals Brillouin versterking, koeling naar de grondtoestand [38], geïnduceerde transparantie [39], het optische veer effect [40] en traag licht geïnduceerd door geluid [41]. Tegelijk plaatst het diverse systemen – zoals Brillouin vezel lasers, microtoroïdes, plasmonische Ramancaviteiten en silicium golfgeleiders – in een bredere theorie van foton-fonon koppeling. Deze ideeën, geïnspireerd door symmetrie, kunnen resulteren in optische controle over geluid en warmte [42].

Vervolgens realiseren we experimenten op nanoschaal silicium-op-isolator golfgeleiders. Deze hoge-index-contrast golfgeleiders sluiten 193 THz licht sterk op via totale interne reflectie. Geluid beweegt daarentegen sneller in de siliciumkern dan in het siliciumdioxide substraat, wat akoestische opsluiting door middel van interne reflectie verbiedt. We vangen daarom fononen door het oxide substraat zo veel mogelijk te verwijderen. Daarbij gebruiken we het grote verschil in akoestische impedantie tussen de siliciumkern en de omringende lucht. Daarbovenop verenigen we de akoestische opsluiting met centimeterschaal interactielengtes door nog steeds een kleine oxide pilaar achter te laten (fig.2). Dit compromis tussen akoestische opsluiting en interactielengte leidt tot de eerste waarneming van Brillouin verstrooiing in silicium nanodraden.

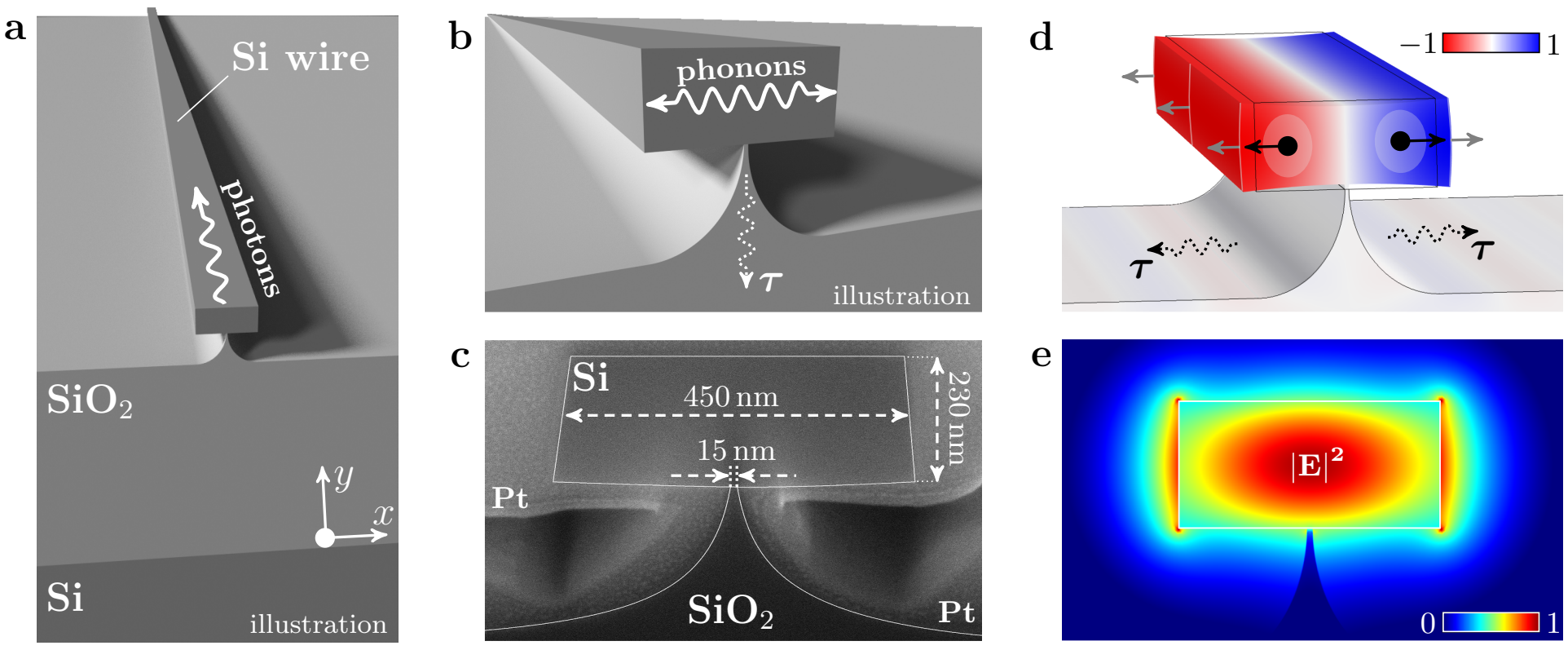

**Figure 2: Een siliciumdraad op een pilaar als een caviteit voor akoestische fononen.** **a**, Indruk van de siliciumdraad. Licht propageert langsheen de draad. Die sluit fotonen op dankzij het hoge optische contrast met het siliciumdioxide substraat en de lucht. **b**, In tegenstelling tot de fotonen, worden de fononen transversaal gevangen. Ze lekken nog gedeeltelijk door de pilaar, wat hun levensduur $\tau \approx 5\,\text{ns}$ bepaalt. **c**, Een scanning elektron beeld van de $450 \times 230\,\text{nm}$ cross-sectie. We fabriceren betrouwbaar pilaren zo nauw als 15 nm. **d**, De horizontale component van de waargenomen akoestische mode **u** (rood: $-$, blauw: $+$) aligneert met de bulk electrostrictieve krachten (zwarte pijlen) en de stralingsdruk op de randen (grijze pijlen). **e**, Elektrische veldnorm van de quasi-TE optische mode.

Verder tonen we in een reeks experimenten (fig.3) aan dat het Brillouin effect nu de sterkste derde-orde niet-lineariteit is van deze golfgeleiders: het samendrukken van zowel licht als geluid tot de $0.1\,\mu\text{m}^2$ kern leidt tot een uitermate efficiënt proces. Specifiek observeren we een Fabry-Pérot-achtige akoestische mode bij 9.2 GHz die goed overlapt met de fundamentele quasi-TE optische mode bij 193 THz. Het is opvallend dat zowel licht als geluid een golflengte van ongeveer $1\,\mu\text{m}$ hebben bij deze frequenties, wat gerelateerd is aan de goede overlap (fig.2d). Het versterkingsexperiment toont tot 175% aan/af Brillouin versterking aan, wat een eerder experiment [43] in silicium/nitride golfgeleiders met een factor 19 verbetert. Onze millimeter-lange golfgeleiders zijn in feite transparent in een band van 35 MHz. Het vierbundelmenging experiment ontdekt dat de Brillouin niet-lineariteit 2.5 keer sterker is dan het Kerr effect (fig.3c-d), in overeenstemming met het versterkingsexperiment. Deze golfgeleiders hebben akoestische kwaliteitsfactoren en gain parameters tot 306 en 3218 $\mathrm{W}^{-1}\mathrm{m}^{-1}$.

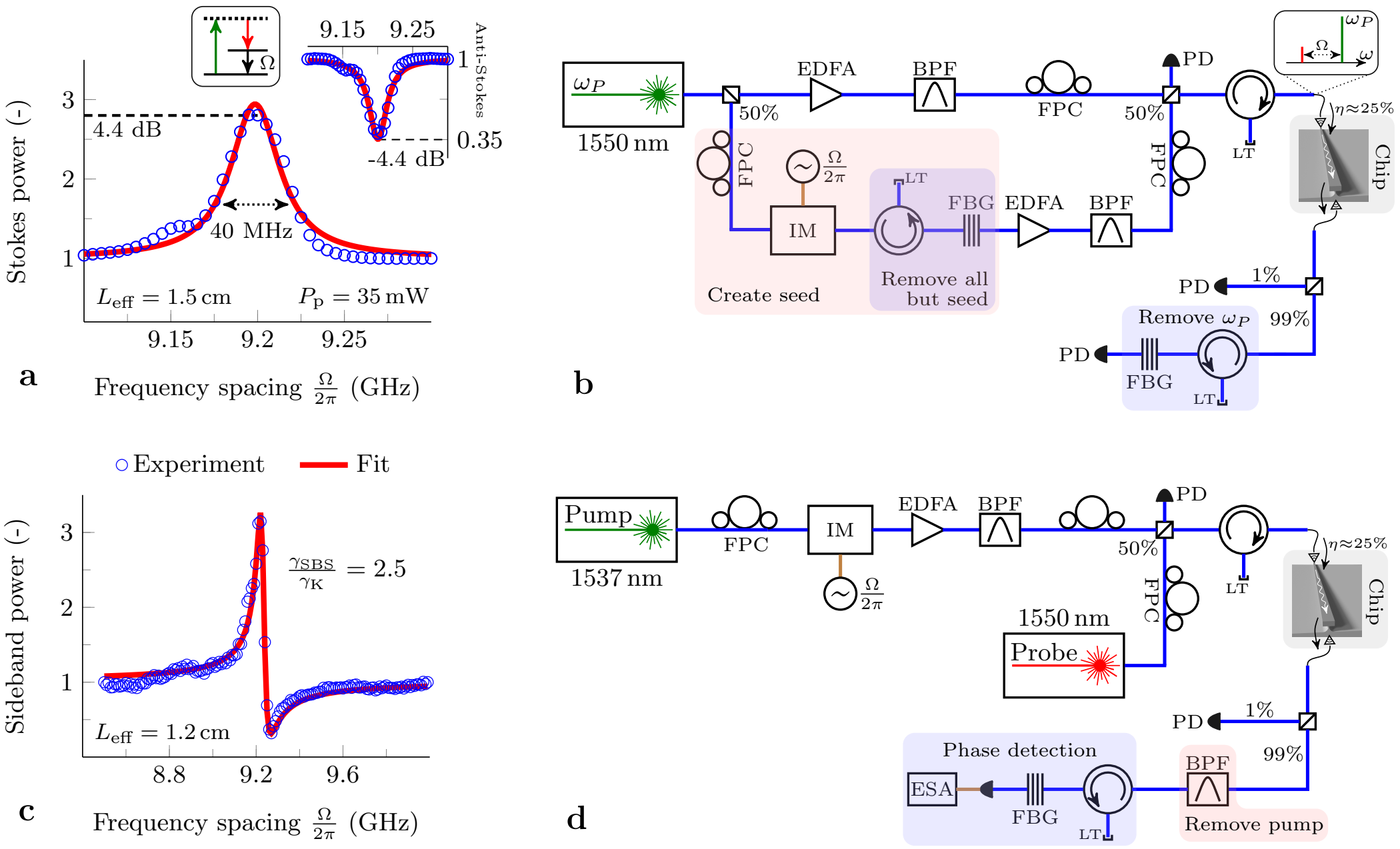

**Figure 3: Experimentele karakterisatie van de foton-fonon koppeling.** **a**, Een typische Lorentziaans versterkingsprofiel op een Stokes probe en (inset) een depletieprofiel op een anti-Stokes probe. In beide gevallen genereert de interactie akoestische fononen en roodverschuift ze fotonen (energiediagram). **b**, De vezelgebaseerde opstelling om voorwaartse SBS te bestuderen. **c**, Een typische Fano signatuur verkregen bij het cross fazemodulatie experiment, hetwelke we gebruiken om de Brillouin niet-lineariteit te calibreren ten opzichte van het Kerr effect ($\gamma_{SBS}/\gamma_K = 2.5$). **d**, De vezelgebaseerde opstelling om cross fazemodulatie te bestuderen.

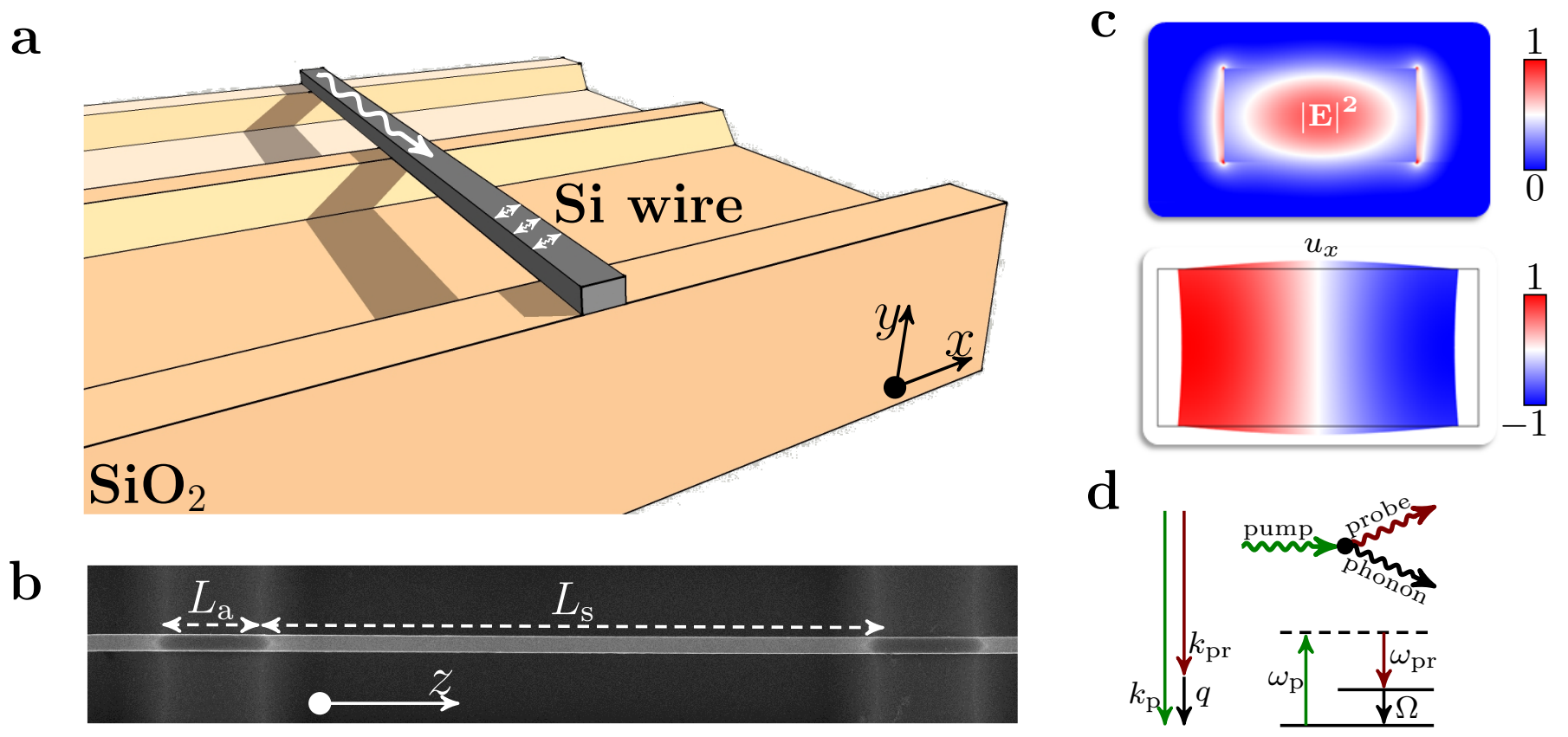

**Figure 4: Een serie vrijhangende silicium nanodraden.** **a**, Indruk van een silicium-op-isolator golfgeleider bestaande uit een serie van ophangingen en ankers. De fotonen propageren langsheen de draad terwijl de fononen gelokaliseerd zijn op hun *z*-punt van creatie. **b**, Scanning elektron beeld van een werkelijke ophanging met lengte $L_\mathrm{s} = 25.4\,\mu$m vastgehouden door $L_\mathrm{a} = 4.6\,\mu$m lange ankers. **c**, Fotonische (boven) en fononische (onder) modes. **d**, Het Brillouin proces converteert inkomende pomp fotonen met energie-momentum $(\hbar\omega_\mathrm{p}, \hbar k_\mathrm{p})$ naar roodverschoven probe (Stokes) fotonen $(\hbar\omega_\mathrm{pr}, \hbar k_\mathrm{pr})$ en fononen $(\hbar\Omega, \hbar q)$.

Vervolgens bestuderen we een reeks volledig vrijhangende siliciumdraden om de akoestische levensduur te vergroten (fig.4). Dit versterkt de kwaliteitsfactoren en gain parameters tot 1010 en $10^4\ \mathrm{W}^{-1}\mathrm{m}^{-1}$, de hoogste tot dusver in het gigahertzdomein. De Brillouin versterking overtreft nu de propagatieverliezen in de korte draden (fig.5). De netto versterking is gelimiteerd tot 0.5 dB door (1) het beschikbare pompvermogen (we zagen geen niet-lineaire absorptie), (2) de hogere propagatieverliezen na ophanging en (3) inhomogene verbreding van de akoestische resonantie. We observeren met name lijnverbreding van ongeveer 9.2 MHz voor 6 ophangingen tot meer dan 20 MHz voor 66 ophangingen. De verbreding wordt wellicht veroorzaakt door fluctuaties in de breedte van de golfgeleider, die de akoestische resonantiefrequentie moduleren langsheen de draden. Naast betere fabricage stellen we voor om dit effect uit te schakelen via de indirecte gevoeligheid van de mechanica aan de optische dispersie.

In een poging om nog sterkere licht-geluid koppeling op te zetten, simuleren we ten slotte nauwe silicium slotgolfgeleiders (fig.6). Een horizontale slot is aantrekkelijk, vermits die zowel (1) de fabricage van willekeurig kleine slots als (2) het efficiënt opwekken van de fundamentele buigmode toelaat. De simulatie schat dat gain parameters voorbij $10^5\,\mathrm{W^{-1}m^{-1}}$ beschikbaar zijn in 5 nm slots, meer dan een grootteorde boven de resultaten in alleenstaande siliciumdraden. Het valt nog af te wachten hoe groot de optische absorptie zou zijn in deze structuren.

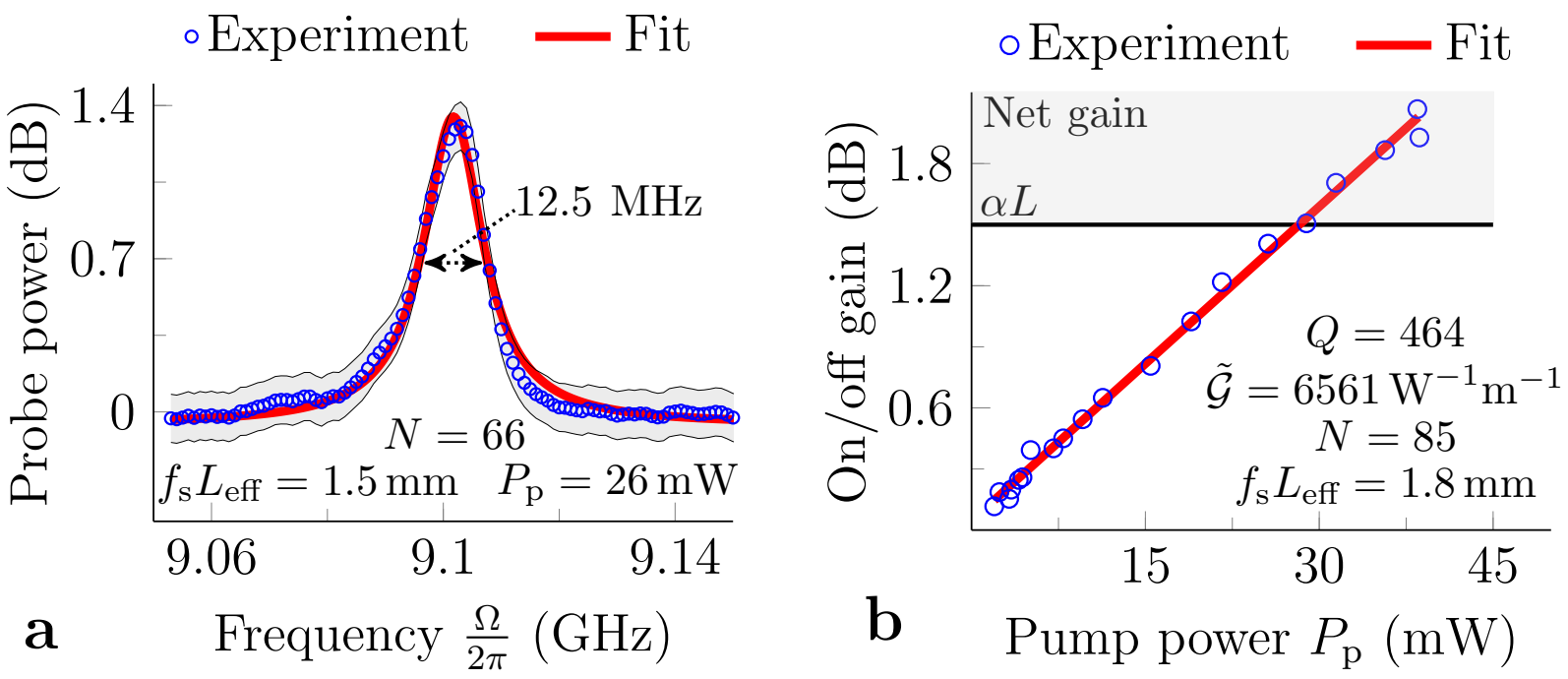

**Figure 5: Brillouin versterking boven de optische verliezen.** **a**, Voorbeeld van een Brillouin resonantie, in dit geval met een aan/af versterking van 1.4 dB, kwaliteitsfactor van $Q_m = 728$ en pompvermogen van 26 mW op de chip. Het grijze gebied wijst op de onzekerheid in het vermogen van de probe. **b**, Scan van de aan/af gain met pompvermogen. Transparantie wordt bereikt bij een vermogen van 30 mW. Bij $P_p > 30$ mW verlaten er meer probe fotonen de golfgeleider dan er binnenkomen. De helling geeft een Brillouin parameter van $\tilde{\mathcal{G}} = 6561\,\mathrm{W}^{-1}\mathrm{m}^{-1}$ met een kwaliteitsfactor van 464 in deze specifieke golfgeleider. Het is opmerkelijk dat de aan/af versterking lineair schaalt met pompvermogen in de hele sweep, wat duidt op de afwezigheid van vrije ladingsdragers in dit bereik.

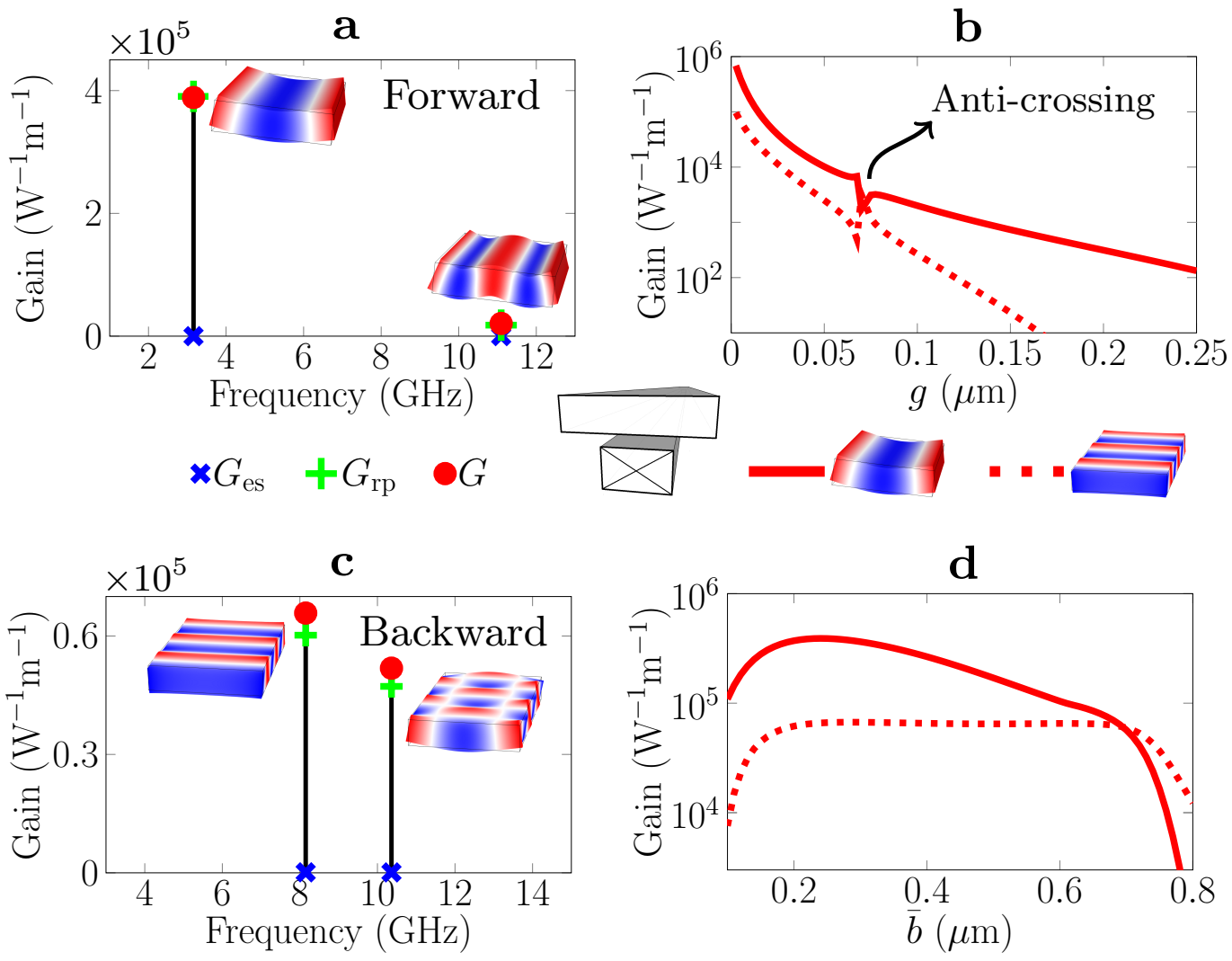

**Figure 6:** Zowel voorwaartse als achterwaartse SBS is zeer efficiënt in nauwe horizontale slots (**a**-**b**-**c**) en de buigmode is gevoelig aan $\bar{b}$ (**d**). De kleur van de modes duidt op het teken van $u_y$ (rood: +, blauw: −).

Dit gebied ligt wijd open. Op slechts enkele jaren was het getuige van grootteorde performantieverbeteringen en een opmerkelijke explosie aan invalshoeken. Naast hun duidelijke toepassingen in microgolf fotonica kunnen deze golfgeleiders andere dringende problemen helpen oplossen. In het bijzonder wordt het streven naar Moore's wet in de originele betekenis mogelijks binnenkort opgegeven [5]. De limieten van computers [44] motiveren reeds vandaag onderzoek naar steeds diversere technologieën, van millivolt schakelaars [45, 46] tot reversibele [47, 48] en kwantumcomputers [49, 50]. Fononen zijn misschien geen ideale informatiedragers, maar ze hebben zeker interessante eigenschappen als interfaces tussen fotonen, plasmonen [51], magnonen [52], excitonen [53] en andere deeltjes [54].

# Summary

THE SCIENCE of light seeks to manipulate photons. It has done so with great success in the past decades: photons are ideal messengers given their low propagation loss and high bandwidth [1–3]. Thus they are already the dominant long-haul information carriers, transmitting data at ever-increasing rates under the oceans. Photonics has now begun taking over chip-to-chip communication too [4], solving major issues – such as heat generation and signal distortion [1, 2] – associated with electronic interconnects. Hence there was a great push for photonic circuitry at the nanoscale, following the shrinking electronics on the path of Moore's law [5]. Modern optical waveguides trap light to cross-sections below 0.1 $\mu$m$^2$ and can be made from the ubiquitous material silicon, in line with mass-fabrication capabilities of existing semiconductor fabs.

This begs the question whether optics could perform certain computations too [6, 7], which requires some photons to control the flow of others [8]. The grand challenge lies in enhancing the weak interactions between photons. As photon-photon interactions are negligible in vacuum [9, 10], harnessing intermediate material excitations is the only viable route to do so. Thus there has been tremendous effort on improving light-matter coupling, from exploiting Kerr [11–14], Raman [15–17], free-carrier [18, 19] and thermal [20, 21] effects to cavity QED [22, 23].

Although they already filter radio-frequency signals in every smartphone and laptop [24, 25], mechanical systems did not arrive on the photonics scene until about a decade ago. Initially, mostly megahertz-range vibrations were excited and probed optically in e.g. microtoroids [26], silicon beams [27–29] and nitride disks [30]. These low-frequency oscillators generate nonlinearities orders of magnitude stronger than intrinsic material effects [31, 32]. It is desirable, however, to scale up the mechanical frequencies into the gigahertz range. This lets them enter the world of microwave photonics [33, 34] and handle higher data rates.

In this work, we realize an efficient and highly tailorable optical nonlinearity interfaced by gigahertz phonons. The nonlinearity is often called *stimulated Brillouin scattering* (SBS). In addition, we focus on waveguides that confine both light and sound, so that their interaction can build up over many diffraction lengths. These waveguides are optically broadband in the sense that the driving laser's frequency is free within a range of about 10 THz. However, we are confronted with the difficulty of stiff gigahertz mechanics yielding displacements below a picometer. In addition, we lose on the power-efficiency associated with optical cavities [35].

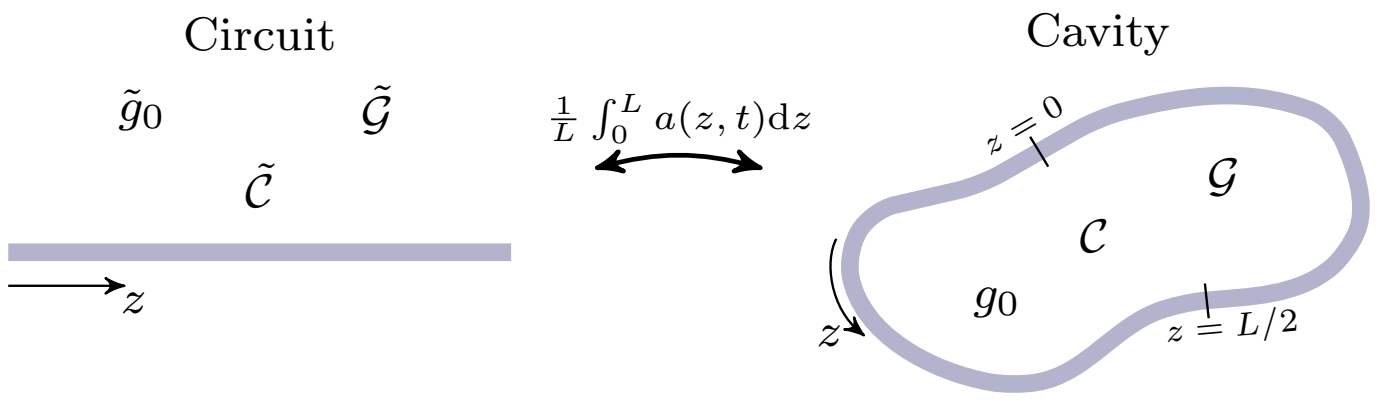

**Figure 1: From circuit to cavity optomechanics.** We explicitly derive the physics of optomechanical cavities (right) from that of Brillouin-active waveguides (left). Therefore, both traveling-wave and cavity-based optomechanics can be cast in terms of vacuum coupling rates ($\tilde{g}_0$ and $g_0$), cooperativities ($\tilde{\mathcal{C}}$ and $\mathcal{C}$) and gain coefficients ($\tilde{\mathcal{G}}$ and $\mathcal{G}$).

To begin with, we theoretically explore the physics of photon-phonon interactions in both waveguides and cavities (fig.1). Conventional assumptions turn out to no longer be warranted in existing systems. For instance, photons traditionally travel farther than phonons and their interaction strength is normally swamped by the propagation losses. This case results in amplification of an optical seed slightly red-detuned from a strong optical pump; the classical *Brillouin gain* [36, 37].

If the interaction on the contrary exceeds the propagation losses, photons can be converted into phonons, back into photons, back into phonons, etc. as they fly along the waveguide. We call this spatial strong coupling, bringing Brillouin scattering in line with major themes in other areas of physics such as cavity QED. Further, a mean-field transition generates the dynamics of high-finesse cavities from that of waveguides. It proves a link between two well-known figures of merit: the Brillouin gain coefficient [36, 37] and the vacuum optomechanical coupling rate [35]. The link elucidates the connections between effects such as Brillouin gain, ground-state cooling [38], induced transparency [39], the optical spring effect [40] and sound-induced slow light [41]. Simultaneously, it places a diverse set of systems – such as Brillouin fiber lasers [55], microtoroids [56], plasmonic Raman cavities [51] and silicon waveguides – in a broader theory of photon-phonon coupling. These ideas, inspired by symmetry, may result in optical control over the flow of sound and heat [42].

Next, we move on to experiments on nanoscale silicon-on-insulator waveguides. These high-index-contrast waveguides strongly confine 193 THz light by total internal reflection. However, sound moves faster in the silicon-dioxide substrate than in the silicon core, forbidding acoustic confinement by inter-nal reflection. Therefore, we instead trap phonons by removing the oxide substrate as much as possible. We thus exploit the huge mismatch in acoustic impedance between the silicon core and the surrounding air. In addition, we reconcile the phononic confinement with centimeter-scale interaction lengths by still leaving a small oxide pillar (fig.2). This compromise between acoustic confinement and interaction length leads to the first observation of Brillouin scattering in silicon nanowires.

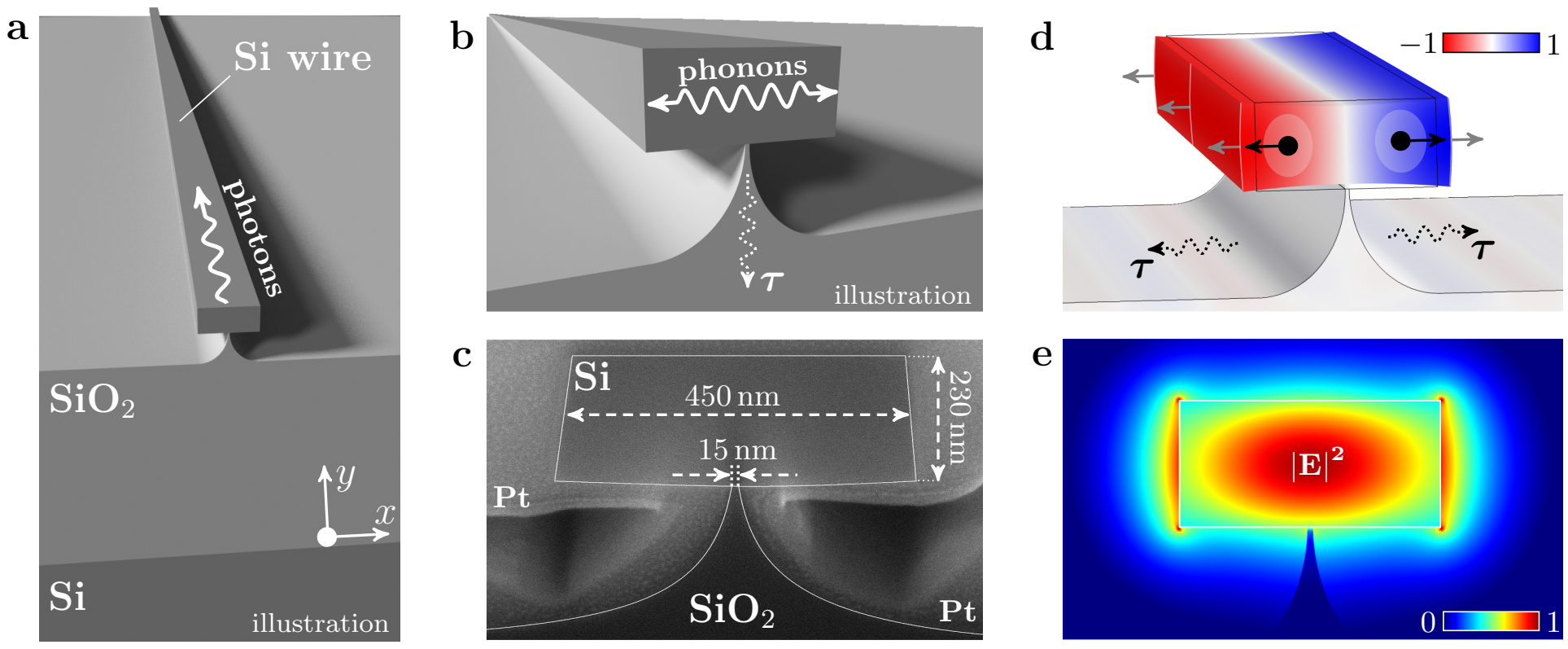

**Figure 2: A silicon wire on a pillar as an acoustic phonon cavity.** **a**, Top view of the silicon wire. Light propagates along the wire. It confines photons owing to the high optical contrast with the silicon dioxide substrate and the air. **b**, Unlike the photons, the phonons are trapped transversally. The leakage of phonons through the pillar determines their lifetime $\tau \approx 5\,\text{ns}$. **c**, A scanning electron micrograph of the $450 \times 230\,\text{nm}$ cross-section. We fabricate pillars as narrow as 15 nm reliably. **d**, The horizontal component of the observed acoustic mode **u** (red: $-$, blue: $+$) aligns with the bulk electrostrictive forces (black arrows) and the boundary radiation pressure (grey arrows). **e**, Electric field norm of the quasi-TE optical mode.

In a series of experiments (fig.3), we show that the Brillouin effect is now the strongest third-order nonlinearity of these waveguides. Indeed, compressing both light and sound to the $0.1\,\mu\text{m}^2$ core results in an exquisitely efficient process. In particular, we observe a Fabry-Pérot-like acoustic mode at 9.2 GHz that has a good overlap with the fundamental quasi-TE optical mode at 193 THz. Notably, the wavelengths of both light and sound are about $1\,\mu\text{m}$ at these frequencies – which is related to the good overlap (fig.2d). The gain experiment (fig.3a-b) shows up to 175% on/off Brillouin gain in a 4 cm-long spiral, improving on a previous result [43] in silicon/nitride waveguides by a factor 19. The shorter wires are essentially transparent in a 35 MHz-wide band. The four-wave mixing experiment finds the Brillouin nonlinearity to be 2.5 times stronger than the Kerr effect (fig.3c-d), in agreement with the gain experiment. These devices enable acoustic quality factors and gain coefficients up to 306 and $3218\,\text{W}^{-1}\text{m}^{-1}$.

To eliminate the acoustic leakage, we next study a cascade of fully sus-pended silicon nanowires (fig.4). This enhances the quality factors and gain coefficients up to 1010 and $10^4\,\mathrm{W}^{-1}\mathrm{m}^{-1}$, the highest so far among gigahertz-class devices. The Brillouin gain now exceeds the propagation loss in the shorter wires (fig.5). The net gain remains limited to 0.5 dB by (1) the available pump power (no free-carrier effects were seen), (2) the higher propagation losses after suspension and (3) inhomogeneous broadening of the acoustic resonance. In particular, we observe line broadening from about 9.2 MHz for 6 suspensions to more than 20 MHz for 66 suspensions. The broadening is likely caused by static fluctuations in the waveguide's width, which modulate the acoustic resonance frequency. Better fabrication aside, we suggest to tackle this effect via the indirect sensitivity of the mechanics to the optical dispersion.

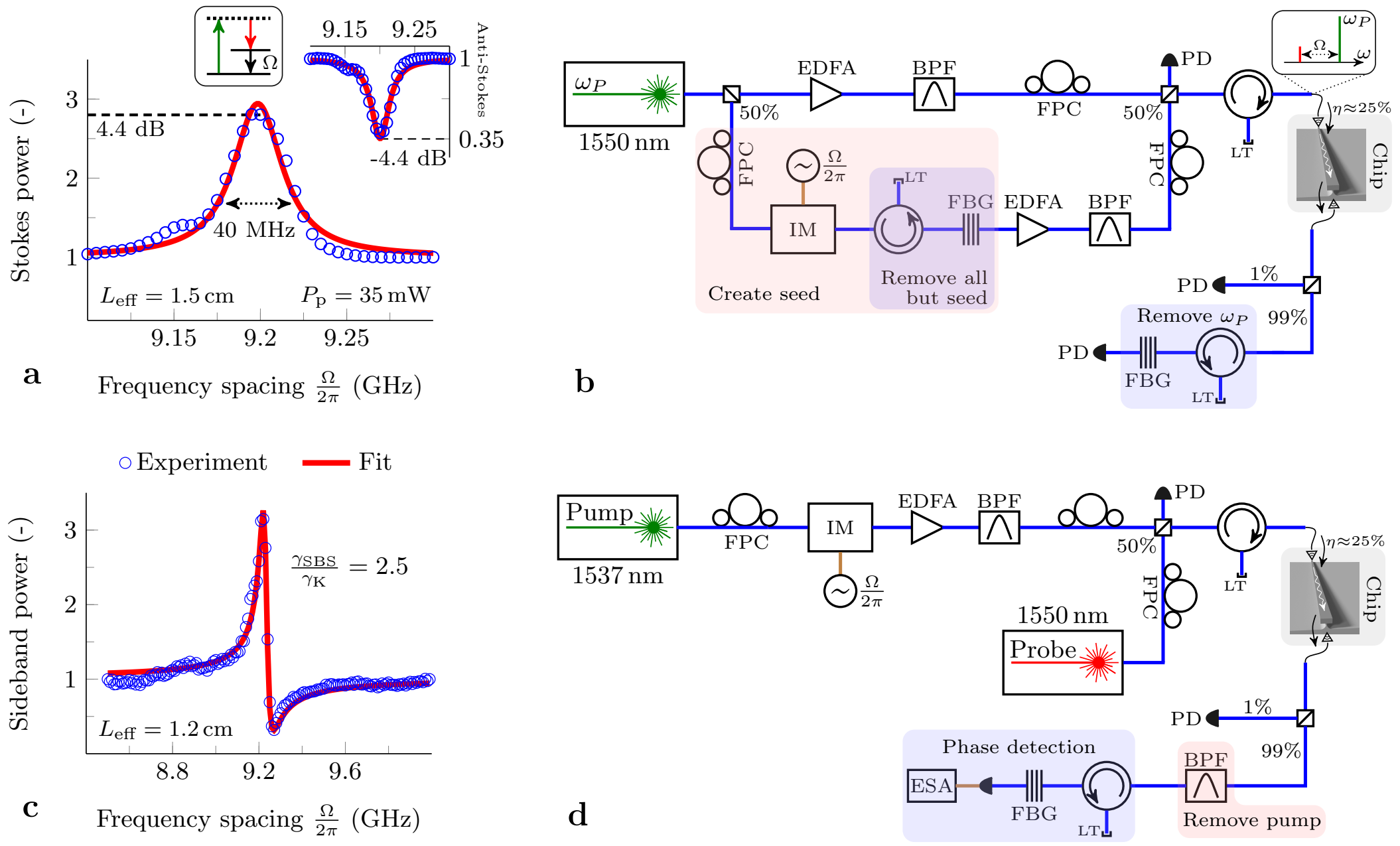

**Figure 3: Experimental characterization of the photon-phonon coupling. a**, A typical Lorentzian gain profile on a Stokes seed and (inset) a depletion profile on an anti-Stokes seed. In both cases, the interaction generates acoustic phonons and red-shifted photons (energy diagram). **b**, The fiber-based set-up used to study forward SBS. **c**, A typical Fano signature obtained from the cross-phase modulation experiment, which we use to calibrate the Brillouin with respect to the Kerr nonlinearity ($\gamma_{\text{SBS}}/\gamma_{\text{K}} = 2.5$). **d**, The fiber-based set-up used to study cross-phase modulation.

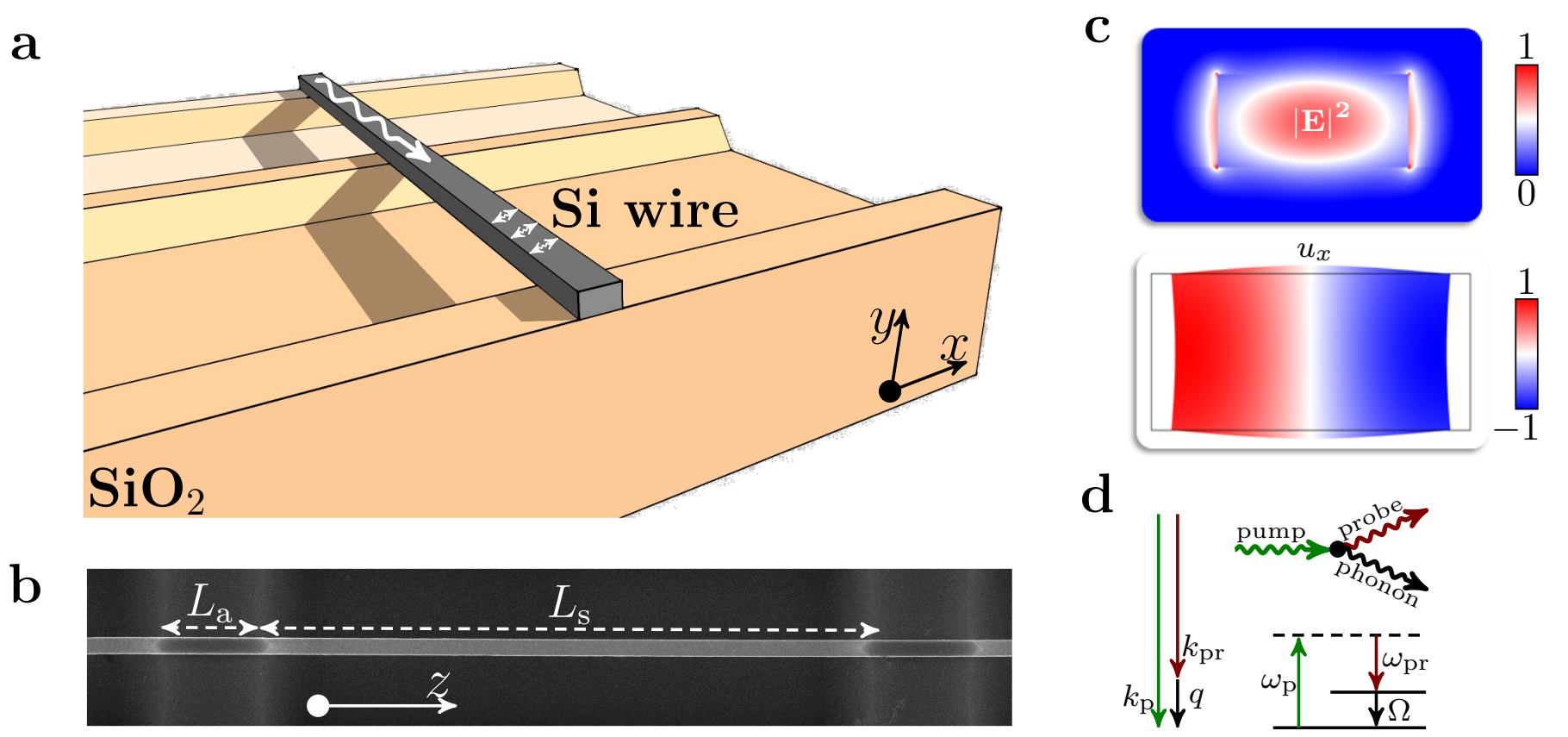

**Figure 4: A series of suspended silicon nanowires.** **a**, Impression of a silicon-on-insulator waveguide that consists of a series of suspensions and anchors. The photons propagate along the wire while the phonons are localized at their $z$-point of creation. **b**, Scanning electron micrograph of an actual suspension of length $L_s = 25.4\,\mu$m held by $L_a = 4.6\,\mu$m long anchors. **c**, Photonic (top) and phononic (bottom) modes. **d**, The Brillouin process converts incoming pump photons with energy-momentum $(\hbar\omega_p, \hbar k_p)$ into redshifted probe (Stokes) photons $(\hbar\omega_{pr}, \hbar k_{pr})$ and phonons $(\hbar\Omega, \hbar q)$.

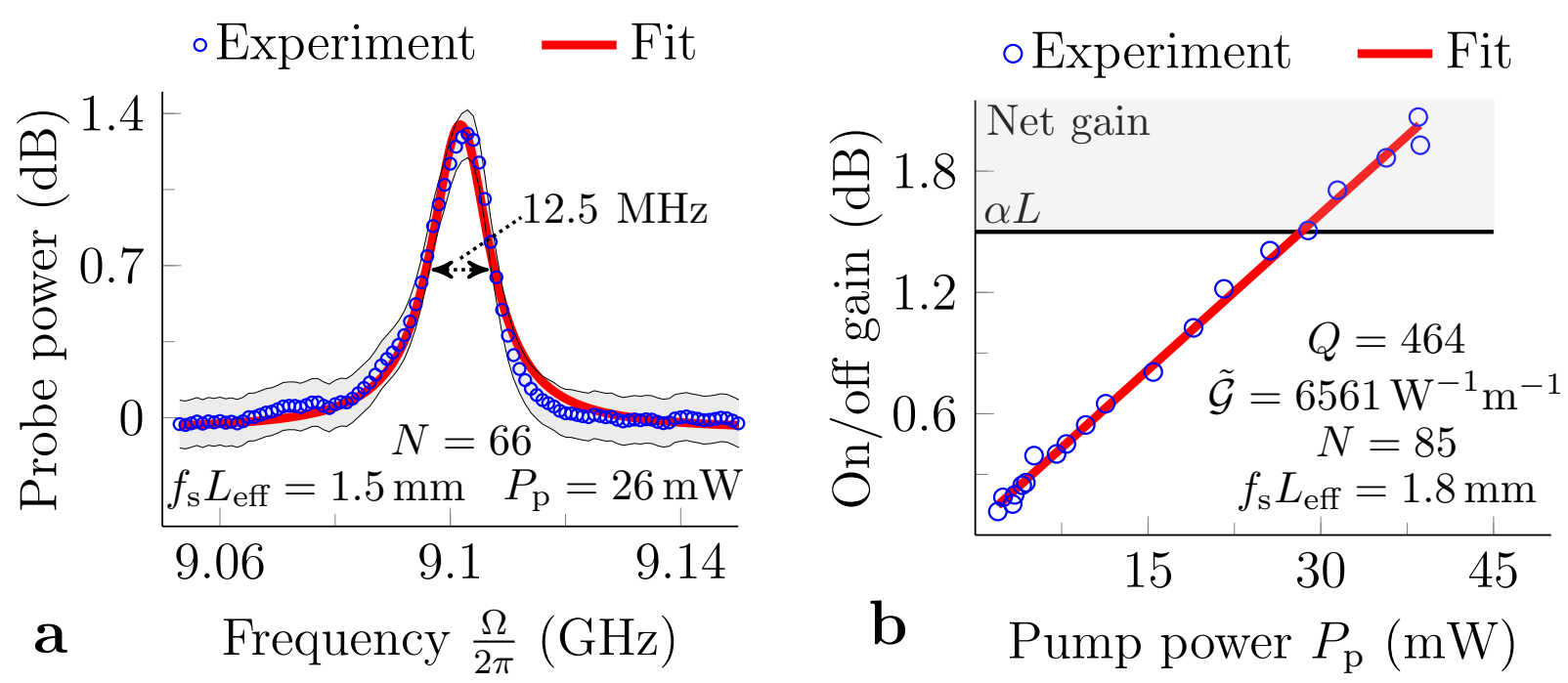

**Figure 5: Brillouin gain exceeding the optical losses.** **a**, An example of a Brillouin gain resonance, in this case with an on/off gain of 1.4 dB, quality factor of $Q_m = 728$ and an on-chip input pump power of 26 mW. The shaded grey area indicates uncertainty in the probe power. **b**, Scan of the on/off gain with pump power. At a pump power of 30 mW the transparency point is reached. For $P_p > 30$ mW, more probe photons leave than enter the waveguide. The slope yields the Brillouin gain coefficient $\tilde{\mathcal{G}} = 6561\,\mathrm{W}^{-1}\mathrm{m}^{-1}$ with a quality factor of 464 in this particular waveguide. Notably, the on/off gain scales linearly with pump power across the entire sweep – indicating the absence of free-carrier absorption in this range.

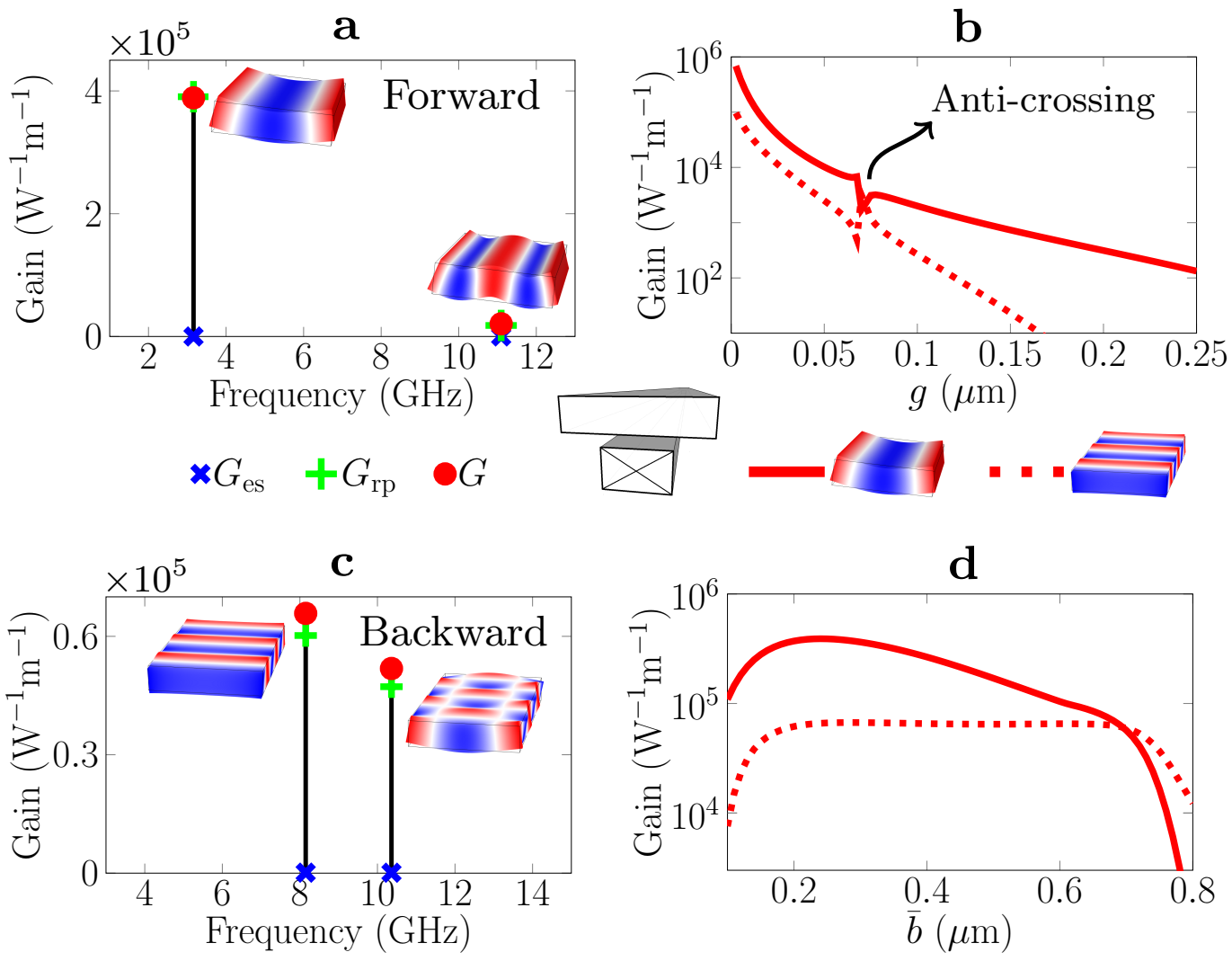

**Figure 6:** Both forward and backward SBS is very efficient in narrow horizontal slots (**a**-**b**-**c**) and the flexural mode is sensitive to $\bar{b}$ (**d**). The color of the modes indicates the sign of $u_y$ (red: $+$, blue: $-$).

Finally, we simulate narrow-gap silicon slot waveguides (fig.6) in an effort to engineer yet stronger light-sound interactions. A horizontal-slot geometry is attractive as it enables (1) the fabrication of arbitrarily small gaps and (2) the fundamental flexural mode to be excited efficiently. We simulate gain coefficients beyond $10^5\,\mathrm{W}^{-1}\mathrm{m}^{-1}$ for 5 nm-gap slots, an order of magnitude above those observed in stand-alone silicon nanowires. We actually fabricated such slots but it remains to be seen how the device fares in terms of optical absorption.

This field is wide open. In only a few years, it witnessed order-of-magnitude performance improvements and a noteworthy explosion of approaches. Besides their clear applications in microwave processing, these waveguides may help solve other pressing issues. Indeed, the pursuit of Moore's law in its original sense may soon be abandoned [5]. The limits on computation [44] are driving investigations into exceedingly diverse technologies, from millivolt switches [45, 46] to reversible [47, 48] and quantum [49, 50] computing. Phonons may not be ideal messengers, but they certainly have interesting properties as interfaces beween photons, plasmons [51], magnons [52], excitons [53] and others [54].

# Contents

# 1

# Introduction to photons and phonons at the nanoscale

*There is plenty of room at the bottom.*
Richard Feynman

**Contents**

LIGHT AND SOUND *are everywhere. Many animals have small holes in their heads, sometimes called "eyes" and "ears", through which they register light and sound to figure out what is going on around them. Some of these animals have even evolved ways to communicate via light and sound. But only humans, as far as we currently know, understand pretty well what light and sound actually is. Both are vibrations, light in the electromagnetic field; sound in the positions of atoms. This work deals with light we cannot see and sound we cannot hear – and in particular the interaction between them. In fact, intense light creates sound. And in reverse, intense sound changes the direction and frequency of light. By trapping both* 200 THz *electromagnetic and* 10 GHz *mechanical vibrations in nanoscale waveguides, we boost the strength of their interaction tremendously. We also propose novel structures in which sound takes the spotlight. All in all, these efforts may help solve long-standing challenges in our control over the flow of information and serve as a platform for quantum information and studies of quantum decoherence.*

## 1.1 Background and challenges

Sensitive detection of light shows that it comes in discrete lumps called photons. The science of light seeks to understand, manipulate and encode information onto these photons. It has done so with great success in the past decades. In particular, photons are ideal messengers given their low propagation loss and high bandwidth [1–3]. Thus they are already the dominant long-distance information carriers, transmitting data at ever-increasing rates under the oceans. However, their application to short-distance communication remained unexplored until the early 21st century. Since then there has been massive interest in replacing electronic by optical interconnects [4, 8, 57, 58]. Hence there was a great push for photonic circuitry at the nanoscale, following the shrinking electronics on the path of Moore's law [5]. Modern optical waveguides trap light to cross-sections below 0.1 $\mu m^2$ and can be made from the ubiquitous material silicon, in line with mass-fabrication capabilities of existing semiconductor fabs.

Photonics is thus taking over chip-to-chip and perhaps even on-chip communication [4], solving major issues – such as heat generation, signal distortion and cross-talk [1, 2] – associated with electronic interconnects. This begs the question whether optics could perform certain computations too [6, 7], avoiding optics-to-electronics conversions and dramatically speeding up acquisition [7]. This requires some photons to control the flow of others [8]. The grand challenge lies in enhancing the weak interactions between photons, essentially due to the absence of a photonic Coulomb force. As photon-photon interactions are negligible in vacuum [9, 10], harnessing intermediate material excitations is the only viable route to do so. Thus there has been tremendous effort on improving light-matter coupling, from exploiting Kerr [11–14], Raman [15–17], free-carrier [18, 19] and thermal [20, 21] effects to cavity QED [22, 23]. These are all nonlinear or time-variant systems [36, 37]: the system's output should no longer be simply proportional to its input, violating the superposition principle.

Piezoelectric mechanical oscillators already filter radio-frequency signals in every smartphone and laptop [24, 25]. Phonons by themselves also receive attention in the context of controlling heat and sound flow [42, 59]. However, they are mainly seen as ideal transducers between other particle species, such as photons [60], plasmons [51], magnons [52], excitons [53] and others [54]. Thoroughly studied in ion traps [61] and gravitational wave detectors [62], mechanical systems did not arrive on the photonics scene until about a decade ago. Initially, mostly megahertz-range vibrations were excited and probed optically in e.g. microtoroids [26], silicon beams [27–29] and nitride disks [30]. These low-frequency oscillators generate nonlinearities orders of magnitude stronger than intrinsic material effects [31, 32]. It is desirable, however, to scale up the mechanical frequencies into the gigahertz range.

This lets them enter the world of microwave photonics [33, 34] and handle higher data rates.

In this work, we realize an efficient and highly tailorable optical nonlinearity interfaced by gigahertz phonons. As we will see, the nonlinearity is often called *Brillouin scattering*. In addition, we focus on waveguides that confine both light and sound, so that their interaction can build up over many diffraction lengths. These waveguides are optically broadband in the sense that the driving laser's frequency is free within a range of about 10 THz. However, we are confronted with the difficulty of stiff gigahertz mechanics yielding displacements below a picometer. In addition, we lose on the power-efficiency associated with optical cavities [35].

In short, this work aims to address the following challenges:

- to grasp the physics of photon-phonon interactions across the widest range of structures and materials,
- to create strong, engineerable phonon-mediated photon-photon control in small-core silicon waveguides,
- to scale up the resonance frequency of these phonons into the gigahertz range,
- and to perform optical signal processing tasks, such as amplification, using the newly developed structures.

Finally, the pursuit of Moore's law in its original sense may soon be abandoned [5]. The limits on computation [44] are driving investigations into exceedingly diverse technologies, from millivolt switches [45, 46] to reversible [47, 48] and quantum [49, 50] computing. We expect that efficient photon-phonon coupling may play a role on many of these fronts. Besides exploiting phonons as interfaces, we will also propose devices in which they take center stage.

## 1.2 Overview of results

This work consists of seven chapters, whose content is illustrated in fig.1.1.

Chapter 2 introduces photon-phonon interaction and its theoretical description. It starts off by describing the basic mechanisms. Next, it develops a quantum field theory for the spatiotemporal dynamics of the process.

Chapter 3 builds on the set of dynamical equations derived in chapter 2. It explores the range of effects that are contained within them, most of which are yet to be observed. Depending on their relative damping, phonons can amplify traveling photons or vice versa. If the coupling strength exceeds the propagation losses, the interaction produces Rabi oscillations along the waveguide. It also considers the link between waveguide- and cavity-based optomechanics. For instance, it proves a connection between the Brillouin gain coefficient and the vacuum optomechanical coupling rate. This chapter is based on [63].

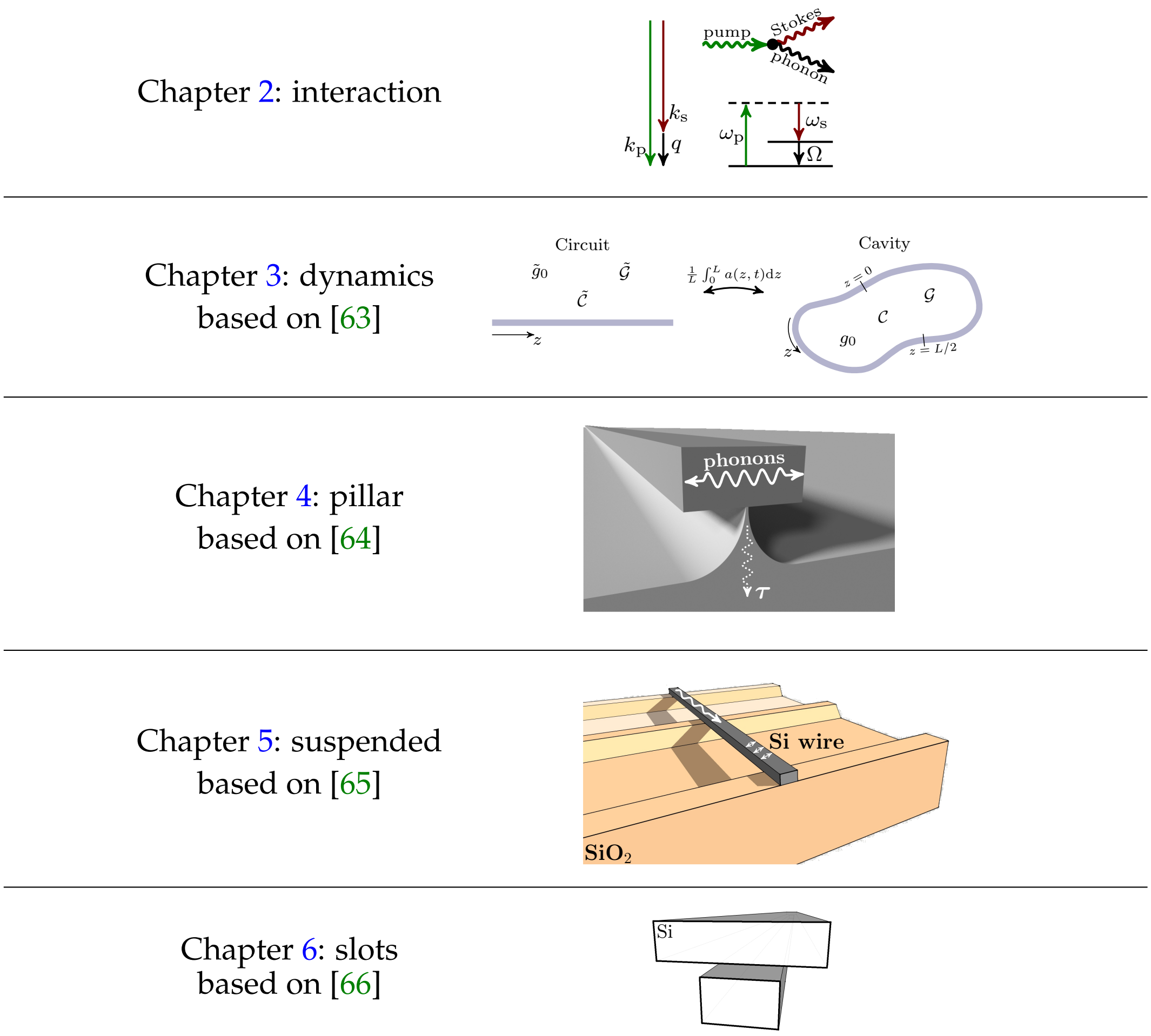

**Figure 1.1: Outline of this work.** Chapters 2, 3, 4, 5 and 6 are based on papers [63–66]. Papers [67, 68] extend chapter 3, but are not treated in detail here.

Chapter 4 deals with the first observation of Brillouin scattering in nanoscale silicon waveguides. The waveguides confine 193 THz light by total internal reflection and 10 GHz acoustic vibrations by impedance mismatch. The acoustic quality factor remains limited to about 300 because of leakage into silica substrate. These waveguides are optically transparent in a narrow band of frequencies at a pump power of 25 mW. We demonstrate Brillouin gain above a factor 2. Besides this amplification, we observe optomechanical wavelength conversion – translating a 10 GHz microwave signal across 1 THz. This chapter is based on [64].

Chapter 5 extends the experimental work of chapter 4. We fabricate a cascade of fully suspended nanowires held in between by silica anchors. Acoustic leakage is thus eliminated in the suspended sections, boosting the mechanical quality factor from 300 to 1000. This enables the first observation of Brillouin amplification exceeding the propagation losses in silicon. The amount of amplification is mostly limited by a rapid drop in acoustic quality as the number of suspensions increases. We propose a novel mechanism to cancel this broadening. This chapter is based on [65].

Chapter 6 looks at the potential of silicon slot waveguides to enhance the optomechanical coupling. Specifically, we look into narrow-gap horizontal slot waveguides. For certain dimensions, these waveguides support opto-acoustic modes with an interaction efficiency simulated an order of magnitude above those of single-nanobeam systems. This chapter is based on [66].

Finally, chapter 7 considers the prospects for phonon-based integrated circuits.

## 1.3 Output

This thesis is based on the following peer-reviewed articles:

1. R. Van Laer, R. Baets and D. Van Thourhout, "Unifying Brillouin scattering and cavity optomechanics," *Physical Review A*, pp. 1–16, Apr. 2016

2. R. Van Laer, B. Kuyken, D. Van Thourhout and R. Baets, "Interaction between light and highly confined hypersound in a silicon photonic nanowire," *Nature Photonics*, vol. 9, pp. 199–203, Mar. 2015

3. R. Van Laer*, A. Bazin*, B. Kuyken, R. Baets and D. Van Thourhout, "Net on-chip Brillouin gain based on suspended silicon nanowires," *New Journal of Physics*, vol. 17, no. 115005, 2015

4. R. Van Laer, B. Kuyken, D. Van Thourhout and R. Baets, "Analysis of enhanced stimulated Brillouin scattering in silicon slot waveguides," *Optics Letters*, vol. 39, pp. 1242–5, Mar. 2014

5. R. Van Laer et al., "Wideband traveling-wave optomechanics below the phonon resonance," *(in preparation)*, May 2016

6. C. Wolff, R. Van Laer, M. J. Steel, B. J. Eggleton and C. G. Poulton, "Brillouin resonance broadening due to structural variations in nanoscale waveguides," *New Journal of Physics*, vol. 18, no. 025006, 2016

The IEEE Graduate Student Fellowship and the Optics & Photonics "Optics in 2015" recognition [69] were awarded in the context of chapter 4. The author presented this work internationally:

1. "Circuit optomechanics for basic physics and novel devices"
*Seminar in the Kippenberg lab at EPFL (Lausanne, invited, 2015)*

2. "Harnessing photon-phonon coupling in integrated optical circuits"
*Frontiers in Optics/Laser Science (San Jose, invited, 2015)*

3. "Traveling-wave optomechanics for basic physics and novel devices"
*Seminar in the Ginzton Laboratory at Stanford University (Palo Alto, invited, 2015)*

4. "Brillouin scattering in integrated circuits"
*Presentation in the Vahala group at Caltech (Pasadena, 2015)*

5. "Towards silicon phononics: exploiting Brillouin scattering and cavity optomechanics in integrated circuits"
*Workshop on Optomechanics and Brillouin scattering (Sydney, invited, 2015)*

6. "Unifying Brillouin scattering and cavity optomechanics in silicon photonic wires"
*Conference on Lasers and Electro-Optics (San Jose, 2015)*

7. "Brillouin scattering and optomechanics in silicon photonic nanowires"
*Marie Curie ITN Cavity Quantum Optomechanics Workshop (Diavolezza, 2015)*

8. "Highly efficient light-sound interaction in silicon photonic nanowires"
*IEEE Photonics Society Benelux Symposium (Enschede, 2014)*

9. "Photon-phonon coupling in silicon nanowires"
*Seminar in the Rakich group at Yale University (New Haven, 2014)*

10. "Demonstration of 4.4 dB Brillouin gain in a silicon photonic wire"
*IEEE Photonics Conference (San Diego, 2014)*

11. "Observation of 4.4 dB Brillouin gain in a silicon photonic wire"
    *OSA Nonlinear Optics Meeting (Barcelona, 2014)*

12. "Strong stimulated Brillouin scattering in an on-chip silicon slot waveguide"
    *Conference on Lasers and Electro-Optics (San Jose, 2013)*

13. "Brillouin scattering in silicon slot waveguides"
    *Marie Curie ITN Cavity Quantum Optomechanics Workshop (Diavolezza, 2013)*

14. "An ultra-high frequency optomechanical oscillator"
    *IEEE Photonics Society Benelux Symposium (Mons, 2012)*

# 2

# Interaction between photons and phonons

*...makes light makes sound makes light makes sound makes...*
Léon Brillouin and Leonid Mandelstam

## Contents

WHEN LIGHT *bounces off a mirror, it transfers momentum to it. Therefore, the mirror moves away from the incident light. Some of the incident optical energy was converted into the mirror's kinetic energy. This implies that the light was slightly Doppler red-shifted. The mirror thus changed the frequency and direction of the incident light. In other words, incoming pump light was converted into reflected and red-shifted Stokes light and the mirror's motion. The energy and momentum of the pump light matches that of the Stokes light and that of the mirror combined. From the perspective of a light beam, sound is nothing but a series of moving mirrors. Light*

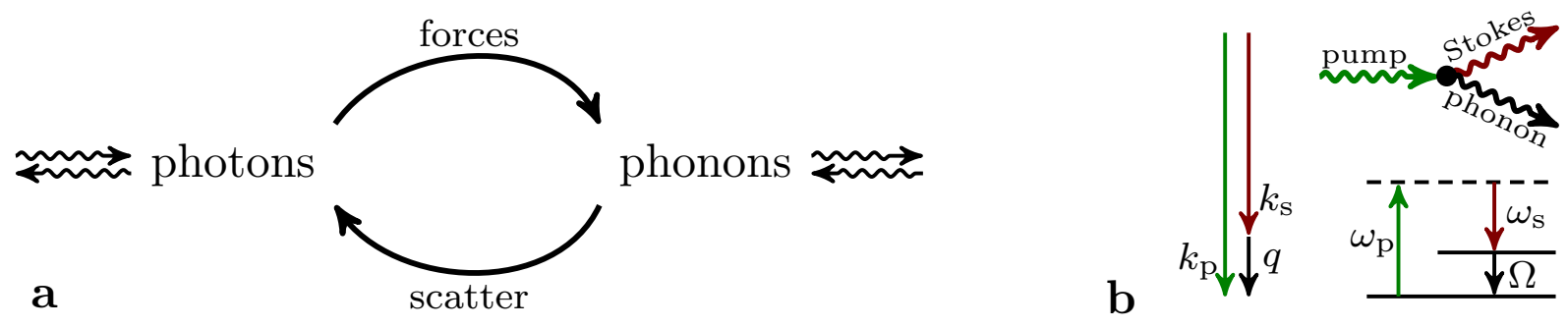

**Figure 2.1: The photon-phonon feedback loop.** **a**, The photon-phonon coupling results in a feedback loop: photons generate phonons via optical forces, whereas phonons scatter photons via changes in refractive index. Thereby, they imprint frequency-shifted sidebands onto the light. Both photons and phonons are coupled to the external world (⇜⇝). This leads to irreversible damping in their dynamics. Simultaneously, the external world yields noisy inputs. In this work, usually the process is initiated by injecting coherent, intensity-modulated light. **b**, When light resonantly produces sound, it must lose some energy. Therefore, pump photons of energy-momentum $(\hbar\omega_p, \hbar k_p)$ are converted into red-shifted *Stokes* photons $(\hbar\omega_s, \hbar k_s)$ and phonons $(\hbar\Omega, \hbar q)$ – with the conservation laws $\omega_p = \omega_s + \Omega$ and $k_p = k_s + q$. Therefore, this is usually an inelastic process: energy flows from light into the material excitation. The Stokes light can also counter-propagate with respect to the pump light ($k_s < 0$), then the phonons carry a large momentum $\hbar q$. This is called *backward* scattering. Usually the phonons are heavily damped, which sets up an endless source of gain for the Stokes wave. However, the process can also run in reverse when the damping is small. We expand on these effects in section 2.1 and chapter 3.

*whose intensity is periodically turned on and off therefore creates sound, which then red-shifts the incident light. This implies that the light's intensity oscillates more violently, which therefore creates more sound, and next, more red-shifted light. To make this feedback loop efficient, we trap both light and sound to some of the smallest known waveguides. This chapter elaborates on this process in increasingly refined models, culminating in a field theory that captures interactions between even very weak light and sound.*

## 2.1 Mechanisms

In this section, we expand on the basic mechanisms underlying photon-phonon coupling. These mechanisms are at the core of cavity optomechanics, Raman and Brillouin scattering. Similar effects are found across many areas of physics, such as in the coupling between atoms and an optical cavity field (cavity QED) [23, 70, 71], in parametric amplifiers [11, 12, 72] and in plasmon-enhanced Raman scattering [51, 73]. In fact, Brillouin scattering is sometimes defined to include coupling between photons and magnons [74] or plasmons [75]. Therefore, a thorough understanding of photon-phonon coupling offers a broad window on contemporary physics and nonlinear optics [36, 37, 76] specifically. This work focuses on waveguides: structures that confine both photons and phonons in two dimensions, so there is only one dimension $z$ along which they can propagate. In addition, the emphasis lies on *acoustic* phonons [77]. The structure is assumed to have translational symmetry, such that both photons $\omega_k$ and phonons $\Omega_q$ are characterized by their respective dispersion relations [78–80]. However, much of the physics is equally well applicable to bulk media [81] and cavities [35]. We will consider systems besides waveguides along the way. The link to cavity optomechanics [35] is thoroughly investigated in chapter 3.

### 2.1.1 The feedback loop

The essence of photon-phonon coupling is a feedback loop (fig.2.1a). The loop may initiate at any point. In this work, we inject intense and coherent light into the structure. We make sure that the intensity of this light field oscillates sinusoidally in time at frequency $\Omega$ and in space at wavevector $q$. Equivalently, the injected light field consists of two waves: the so-called *pump* at frequency $\omega_{\mathrm{p}}$ and wavevector $k_{\mathrm{p}}$, which is assumed to be strong, and that of the *Stokes seed* or *red-detuned probe* at frequency $\omega_{\mathrm{s}} = \omega_{\mathrm{p}} - \Omega$ and wavevector $k_{\mathrm{s}} = k_{\mathrm{p}} - q$, which is assumed to be weak. Thus, the total optical power contains a temporal beat note at frequency $\Omega$ and a spatial beat note at wavevector $q$. In this work, we typically have $\frac{\omega_{\mathrm{p,s}}}{2\pi} \approx 190\,\mathrm{THz}$ and $\frac{\Omega}{2\pi} \approx 10\,\mathrm{GHz}$. These parameters can vary by many orders of magnitude without affecting the physics.

As in the mirror analogy, this optical beat note generates motion of the atoms – and thus phonons – inside the material at frequency $\Omega$ and wavevector $q$. Once these phonons are present, they modulate the refractive index distribution of the structure. The refractive index fluctuations subsequently Doppler red-shift pump photons, thereby creating more probe photons. This makes the beat note even stronger, completing the feedback loop. All in all, incident pump photons were converted into outgoing Stokes photons and phonons (fig.2.1a).

To observe the strongest effect, the beat note must coincide with one of the waveguide's phononic resonances. In practice, one typically scans the probe's frequency $\omega_{\mathrm{s}}$ at fixed pump frequency $\omega_{\mathrm{p}}$. This indirectly sweeps the beat frequency $\Omega = \omega_{\mathrm{p}} - \omega_{\mathrm{s}}$ and wavevector $q = k_{\mathrm{p}} - k_{\mathrm{s}}$. Resonance or *phase-matching* occurs when this beat crosses the phonons' dispersion relation $\Omega_q$ (fig.2.2 and 2.3a). This is another way of stating energy-momentum conservation [36]. The intersection point depends strongly on the dispersion and propagation direction of the interacting light fields. The pump and probe can co- or counter-propagate along the waveguide and they may reside in the same or a different spatial mode. So in general, there are four potential configurations of pump-probe interaction, called *forward intra-*, *forward inter-*

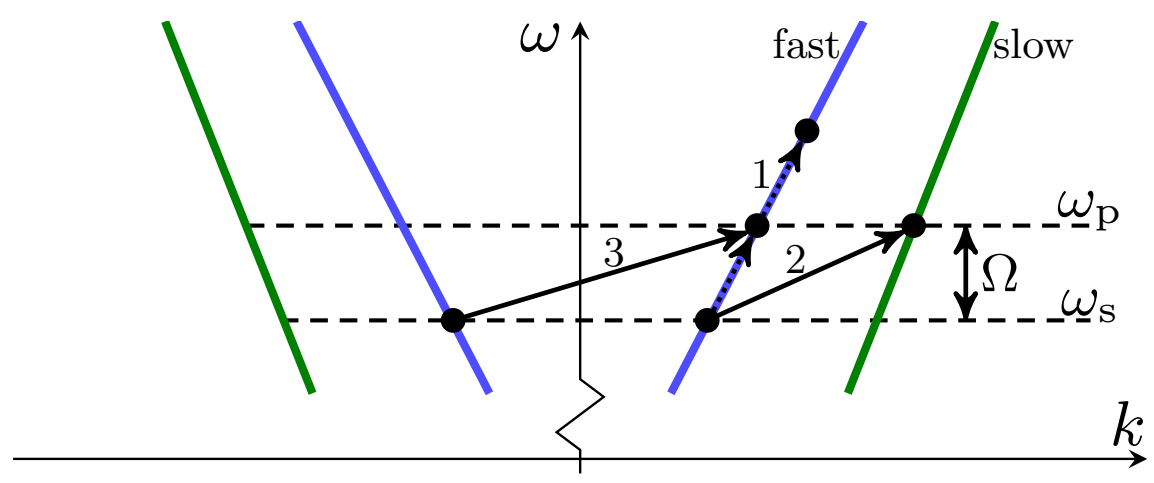

**Figure 2.2: Phase-matching diagrams.** The optical dispersion relation $\omega_k$ shows phonon-mediated coupling between co- (1 and 2) or counter-propagating (3) photons and between two identical (intra-, 1) or two different (inter-modal, 2) optical modes. Therefore, there are generally four types of photon-phonon coupling, of which three are indicated in this diagram. The fourth is counter-coupling between two different optical modes; for instance, a backward fast mode and forward slow mode. The optical dispersion relations are approximately straight lines on the relevant scale. The fast and slow modes could be, for instance, the quasi-TE (slow) and quasi-TM (fast) optical modes of a single-mode silicon waveguide.

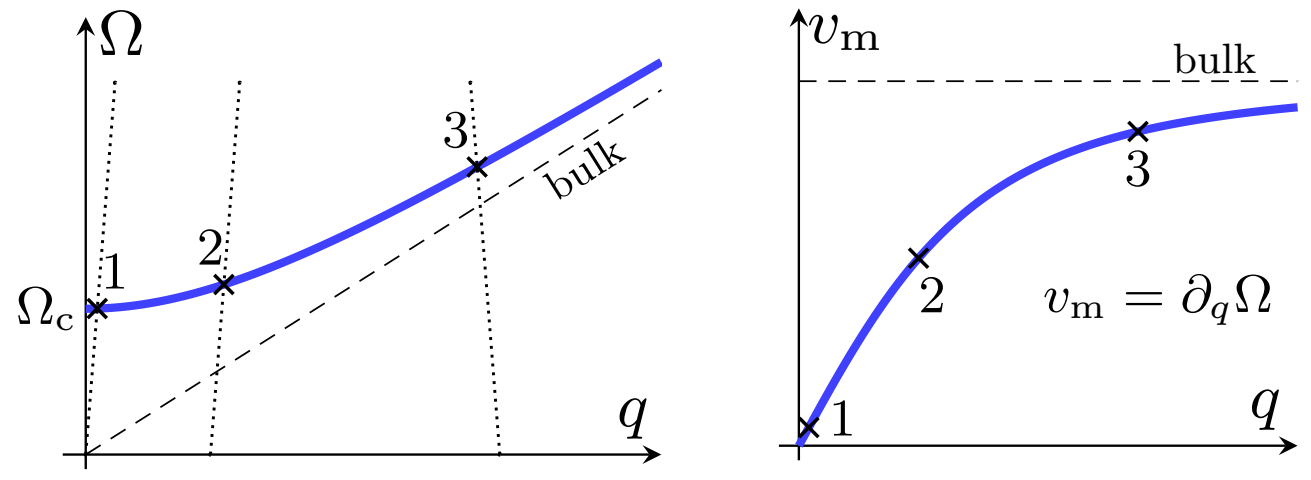

**Figure 2.3: Example of phonon dispersion relation.** **a**, The frequency $\Omega_q$ of transversally trapped acoustic phonons typically has a cut-off $\Omega_c$ for low $q$ and approaches the bulk relation for large $q$. To excite the acoustic mode optically, one scans the detuning $\Omega = \omega_p - \omega_s$ by sweeping $\omega_s$. The sweep yields the dashed lines (compare to fig.2.2), which generate the phase-matched points 1, 2 and 3 depending on the optical dispersion relations. Thus, one observes resonances when the frequency $\Omega$ and wavevector $q$ of the optical beat note coincide with the phonon dispersion relation. **b**, The phonon group velocity $v_m$ vanishes for low $q$ and becomes the bulk speed for large $q$. Therefore, forward intra-modal scattering (1) is often called *Raman-like* scattering, as each cross-section of the waveguide can be considered an independent Raman-active molecule.

, *backward intra-* and *backward inter*-modal coupling. As we will see, the feedback loop (fig.2.1a) may have different consequences for each of these configurations. In chapters 4 and 5, we focus on forward intra-modal scattering.

In the above picture, the process was triggered or *stimulated* by the presence of Stokes seed photons. However, the damping is critical to the precise behavior of the feedback loop. In particular, it dictates whether probe photons or phonons stimulate the conversion. Indeed, the optical beat note excites phonons. If the phononic mode is highly damped compared to the probe photons, it rapidly decays back to the ground state. The next conversion event is triggered by the probe photons created in previous events, not by the phononic mode since it has decayed already. The phononic mode is simply carried along, slaved to the optical waves. This sets up an endless source of amplification for the probe photons, at least as long as the pump remains undepleted. In waveguides, the probe light then effectively sees lower propagation loss. In cavities, the amplification results in line narrowing: the probe effectively sees a longer cavity lifetime. Many cavities, however, feature smaller phononic than photonic damping. In those cases the phonons actually trigger the process: the phononic linewidth narrows so the phonons' decay slows down.

The nature of this amplification and line narrowing can also be understood in terms of a mass-on-a-spring model [82]. The mass has a certain resonance frequency and its motion is damped by a viscous force. Suppose we measure the position of the freely moving mass. Next, we feed back this information by driving the mass with a force that is in phase with the velocity. This results in an effectively lower linewidth for the oscillator, since this force cannot be distinguished from the instrinsic viscous damping. On the other hand, a feedback force in anti-phase with the velocity yields an effectively larger linewidth, analogous to the Purcell drop of an emitter's lifetime close to a cavity [83–85]. Similarly, a force in phase with the position generates a shift of the spring constant. Variations on this theme are responsible for most effects studied here. This is precisely how an optomechanical cavity operates, in this context the feedback is often termed *dynamical back-action* [35].

If the interaction strength exceeds the total damping, both probe photons and phonons trigger the loop simultaneously. To achieve this *strong coupling* was a major goal of cavity QED [23, 86–89], serving as a testbed for studies of the boundary between quantum and classical behavior. It has been realized in certain opto- and electromechanical cavities as well [56, 90–92], but not yet in waveguides (see chapter 3). A blue-detuned probe – often called the *anti-Stokes seed* – experiences Rabi oscillations in the strong-coupling case: the probe photons convert into phonons, then back into probe photons and so on.

The loop can also initiate from noise, and in particular from thermal, incoherent phonons. Such spontaneous scattering at first builds up slowly.

However, each spontaneous scattering event red- or blue-shifts an incident pump photon. The shifted and pump photons interfere, so they produce a beat note that excites phonons. After a while, the optically generated phonon density may exceed the thermal phonon density. Above this threshold, one may see a transition to stimulated scattering, which could result in endless amplification. Whether this actually happens depends on the pump-probe configuration and the phonon frequency. For instance, spontaneous forward intra-modal coupling does not result in exponential amplification unless the phonons' frequency is very high. Instead, one observes cascading into higher-order sidebands (fig.2.2) and algebraic growth [93–95]. In contrast, spontaneous backward scattering does yield exponential amplification above a certain power threshold [96] – although it can also produce combs despite the lack of higher-order phase-matching [97].

Clearly, this deceptively simple feedback loop in fact generates a wealth of interesting effects. We will investigate the dynamics of waveguides and cavities in greater detail in chapter 3. The range of effects has by no means been explored fully. The experiments presented in chapters 4 and 5 are located in the weak-coupling regime and exhibit much higher phononic than photonic damping. Therefore, they concern the amplification of red-detuned probe photons, a situation that is often called *Brillouin gain* when the interaction involves acoustic phonons. Before developing a full model in section 2.2, we first zoom in on two components of the feedback loop: scattering and optical forces.

### 2.1.2 Phonons scatter photons

We first look at the scattering or so-called *forward-action*. We treat the optical forces or *back-action* in the next section. The presence of phonons modulates the refractive index distribution of the waveguide, generating motional sidebands in the optical spectrum. This occurs via two routes:

- **The photoelastic effect.** Consider a volume of matter consisting of polarizable particles (fig.2.4). An external electric field induces dipoles on each of these particles, yielding some polarization. The particles' dipoles combine into a macroscopic polarization that is proportional to the external electric field. The proportionality constant is the permittivity. Next, the matter is compressed by an acoustic wave. Thus there are more particles in the same volume. Each of the particles' polarizations goes up as well, since the local electric field is larger (i.e. Clausius-Mossotti [99]). Hence compression increases the total polarization, and therefore the permittivity, for two reasons: (1) higher particle density and (2) larger local electric field. This effect is described by the Pockels tensor, which connects strain to permittivity changes. It is a small effect, but it occurs in the entire bulk of a material. This simplified picture does not account for negative photoelastic coefficients. The precise size, origin and sign of a material's photoelastic coefficients depends strongly on its composition and the type of inter-atomic forces. In particular, compression may result in charge redistributions that reduce the polarization more than the dipole density increases [100, 101]. The photoelastic coefficients can be predicted by density functional theory [102, 103].

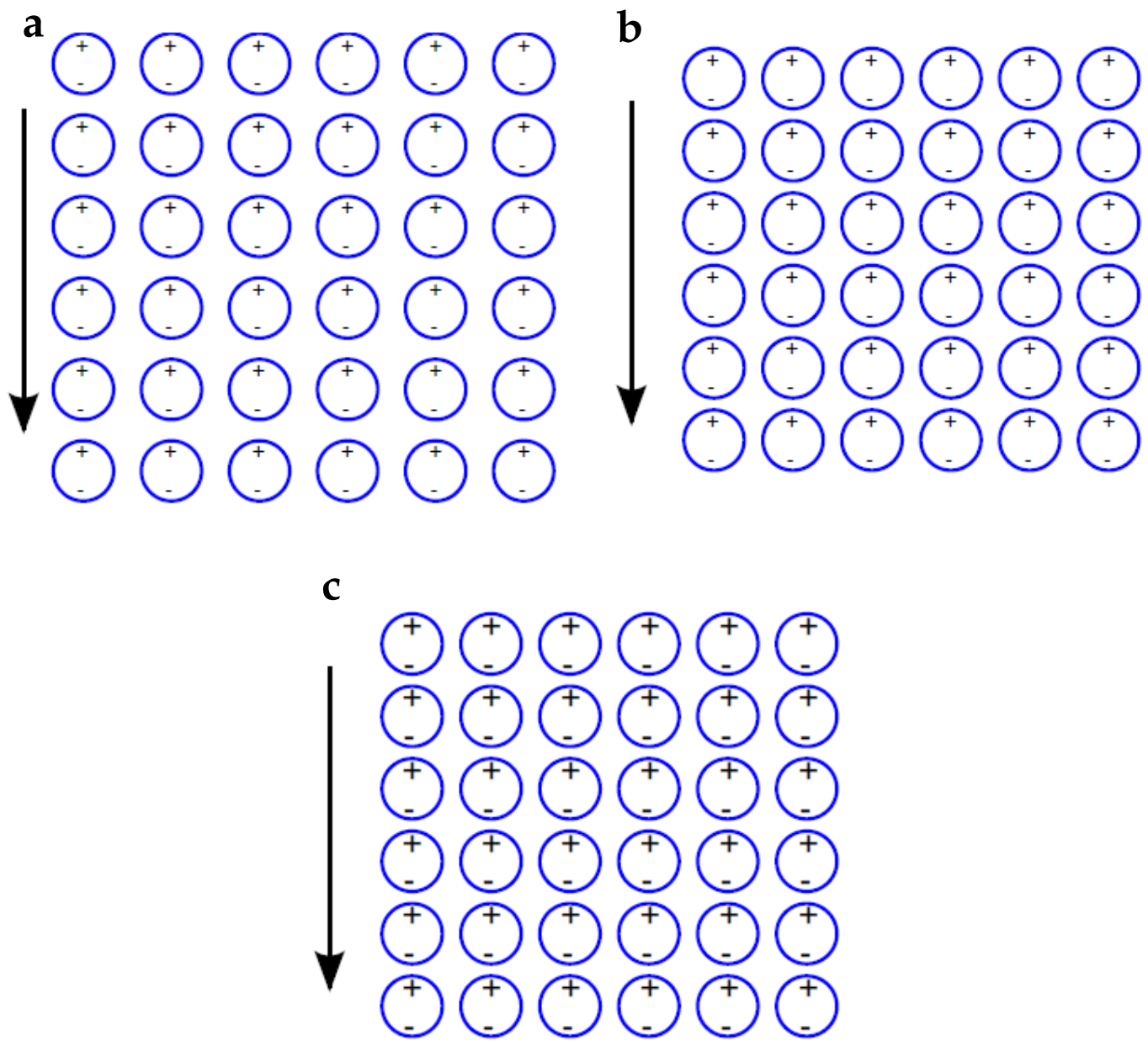

**Figure 2.4: The photoelastic effect has two causes.** **a**, Some volume of dielectric is polarized by an external electric field, producing a polarization density proportional to the field. **b**, Compression enhances the polarization density. **c**, At the same time, the polarization on each particle increases because of dipole-dipole coupling, yielding an even larger polarization density. Figure credit: C. Wolff [98].

- **Boundary motion.** The boundaries between materials move in presence of an acoustic wave. This sets up large changes in the electric field in the boundary region (fig.2.5). Indeed, a moving boundary drags the electromagnetic field along with it because of Maxwell's continuity conditions. For a boundary between a dielectric and air, this results in large changes in polarizability: from negligible polarizability in air to the significant polarizability of the dielectric. The change in total polarization thus scales with the difference in permittivities of the media. This boundary effect closely corresponds to the mirror analogy at the beginning of this chapter. The boundary effect can be seen as a limiting case of photoelasticity. It is a large effect, but it occurs only in a small region.

In this work, the strain or boundary motion is generated by the optical field intensity (see next section), such that the macroscopic polarization scales with the cube of the electric field. It is therefore called a *third-order nonlinearity* [36, 37] along with the Kerr and Raman effect. The boundary effect is a purely geometric nonlinearity: it disappears in bulk media. Crucially, the photoelastic and boundary-induced changes in the total polarization need not have the same sign. For instance, compressing an object may increase the bulk refractive index, while reducing the polarization at the boundaries. In large objects, the bulk far exceeds the boundary contribution. However, in micro- and nanoscale devices, these effects can be of similar size [104]. Therefore, one has to make sure that they interfere constructively – as we show experimentally in chapter 4. In principle, scattering can occur on fluctuations in absorption as well [36]. However, we purposely avoid this effect to minimize damping. Finally, we measure the scattering optically, so the optical field needs to be strong where the index changes. In this sense, the optical field acts as a weighting function for the permittivity fluctuations.

### 2.1.3 Photons create phonons

The previous section dealt with half of the feedback loop. This is sufficient to understand the entire loop because these processes are reversible. More precisely, as the interaction globally conserves energy, any change in the optical energy must be compensated by the same amount of mechanical work. For infinitesimal displacements, the mechanical work is the product of the displacement and the optical force. Therefore, the force is just the derivative or gradient of the optical energy with respect to motion. Thus one speaks of *gradient forces* [36, 105]. Such forces were first used to trap and manipulate particles in the focus of an intense laser [106–108]. The optical energy can change rapidly from place to place in small objects, such that one may achieve large gradient forces. The reverse processes are:

- **Electrostriction: reversed photoelasticity.** Compression causes a permittivity shift, and thus a change in the optical energy. Since energy is conserved, this change in optical energy equals the mechanical work done by some force – which is called the *electrostrictive* force in this case. It is proportional to the gradient of the local optical energy, thereby pulling polarizable particles into regions of high intensity. Its strength is also determined by the photoelastic Pockels tensor. This force attempts to deform the bulk of a material (fig.2.6a).

- **Radiation pressure: reversed boundary motion.** Radiation pressure results from the reflections of photons off material interfaces (fig.2.6b). In case the photons are confined to a waveguide, the totally reflecting photons drive a radiation pressure perpendicular to the waveguide's edges (fig.2.6c). This radiation pressure can be seen as a limiting case of electrostriction, since the boundaries are pulled into regions of high optical energy just like a polarizable particle is into a laser focus (fig.2.6d). The strength of this force scales with the difference in photon momenta, and thus permittivities, of the media. It attempts to move the interfaces between materials.

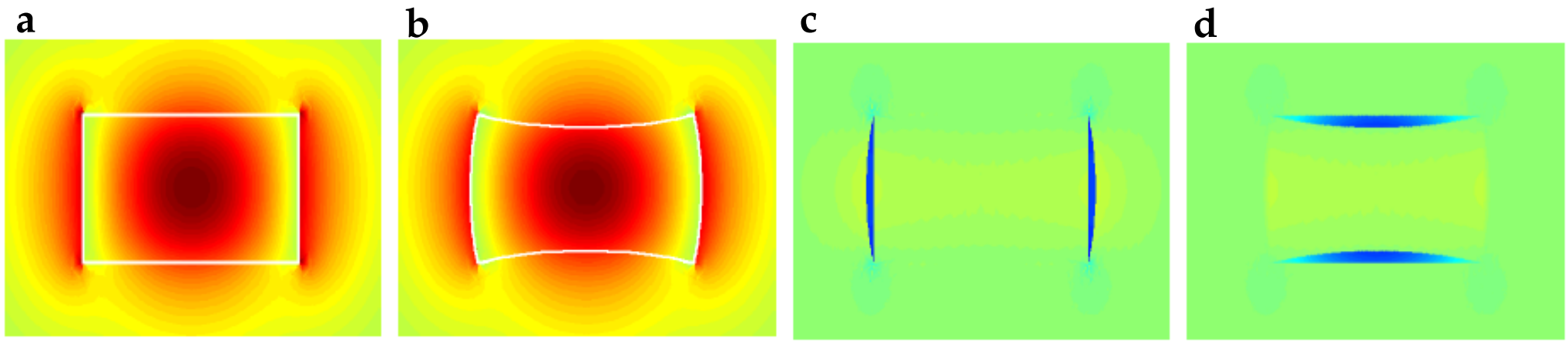

**Figure 2.5: Boundary motion produces large optical field changes at boundaries. a**, Horizontal electric field component in a nanoscale silicon waveguide surrounded by air. This is a cross-section perpendicular to the waveguide's propagation axis. **b**, The boundaries (white) of the waveguide move, dragging along the electromagnetic field. This results in large changes in the horizontal electric (**c**) and induction (**d**) field near the boundaries. Figure credit: C. Wolff [98].

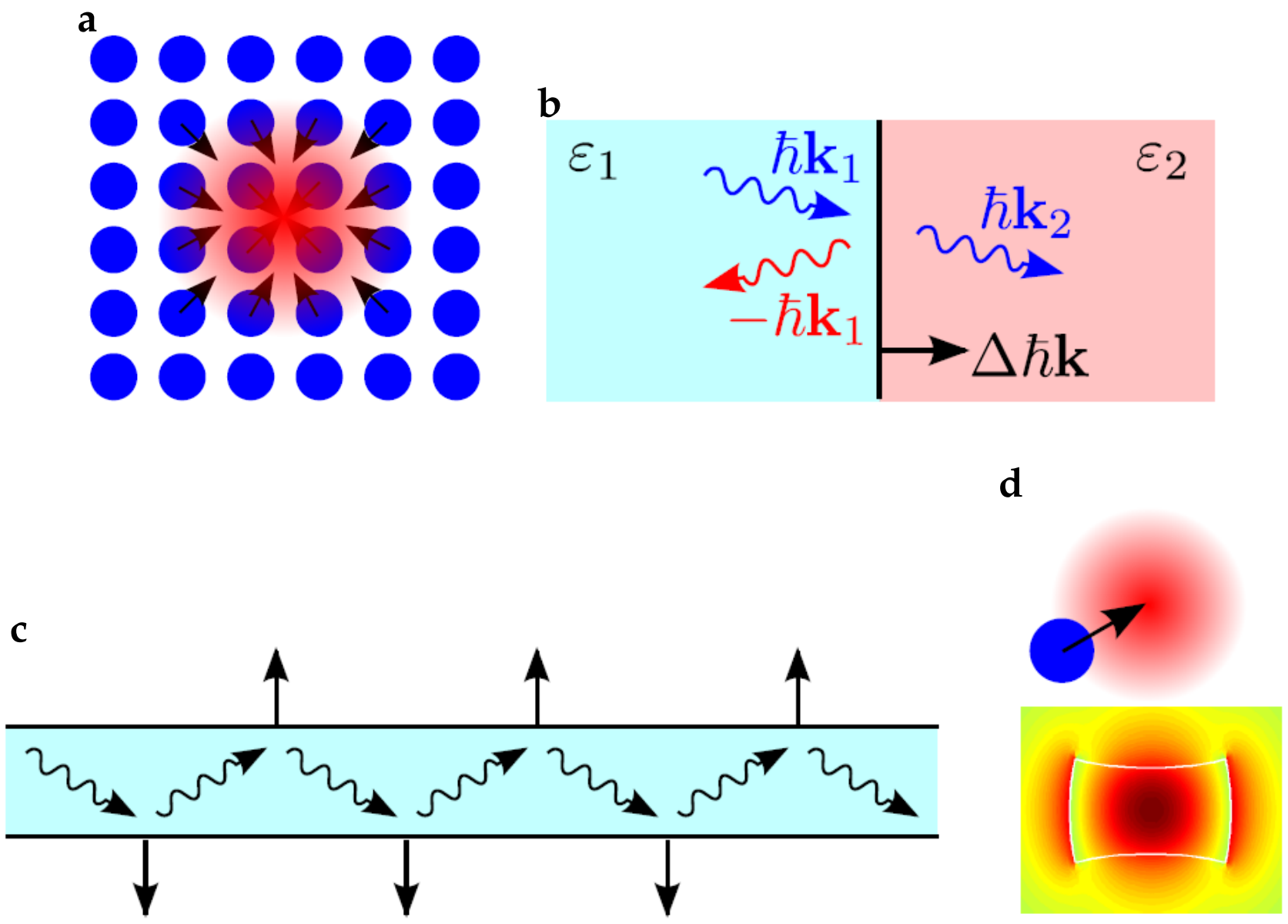

**Figure 2.6: Optical forces in bulk and at boundaries. a**, The bulk electrostrictive force compresses the material, pulling polarizable particles into regions of high optical energy. **b**, An incoming photon of momentum $\hbar\mathbf{k}_1$ reflects on an interface between materials with permittivities $\epsilon_1$ and $\epsilon_2$. Part of the incident momentum is carried away by a photon $\hbar\mathbf{k}_2$, another part is transferred to the material interface. **c**, A photon ($\rightsquigarrow$) is trapped in a waveguide by total internal reflection. Each reflection transfers some momentum to the waveguide's boundary. The radiation pressure is perpendicular to the boundary. **d**, The momentum transfer to boundaries can also be understood from an energetic perspective: the boundaries are pulled towards regions of high optical energy, precisely like a polarizable particle is trapped in an optical tweezer. Figure credit: C. Wolff [98].

Optical forces are slightly controversial. First, there is the century-old Abraham-Minkowski dilemma about the momentum of light in a medium [109–121]: is it the Minkowski momentum $\hbar \mathbf{k}_0 n$ or the Abraham momentum $\hbar \frac{\mathbf{k}_0}{n}$ with $\mathbf{k}_0$ the vacuum wavevector and $n$ the refractive index? The crux of the matter is that a photon in a medium partly consists of polarized dipoles, so one can divide the momentum between the light and the material excitation in several ways. Nelson [111] offers an interesting perspective. He notes that momentum is conserved by virtue of the homogeneity of space via Noether's theorem [122, 123]. In addition, there is another conserved quantity called *pseudomomentum* corresponding to the homogeneity of the material. This pseudomomentum is carried by the material's response fields, such as the polarization. The sum of the momentum and pseudomomentum is also conserved and is termed the *wave-momentum*. Light can travel both in vacuum and in a material, so it posseses both momentum and pseudo-momentum. Sound can travel only in a material, so its pseudomomentum equals its wave-momentum – sometimes called the phonons' *crystal momentum* [77]. Phonons do not carry momentum as they involve only relative motion of atoms. Nelson shows that it is the wave-momentum that shows up in the phase-matching conditions. This is nearly equivalent to applying the Minkowski momentum $\hbar \mathbf{k}_0 n$ for light [111]. The fields of nonlinear optics and optomechanics assume this implicitly. It is consistent with calculating optical forces as the gradient of the optical energy. We continue this tradition and find good agreement with experiments in chapters 4 & 5.

Second, the terminology regarding optical forces is confounding. One encounters terms such as electrostrictive surface stress [36, 104, 124, 125], radiation pressure [35, 64, 104], gradient force [29, 105], scattering force [106], photothermal force [126], ponderomotive force [127], bolometric force [128, 129], etc. Most of these concepts overlap or are limiting cases of one another. Some have slight differences in their detailed manifestation. Adding to the confusion, the terms are applied inconsistently. All lossless forces, however, can be traced back to their time-reversed scattering mechanisms. These provide a safe route to conceptualize and calculate the optical forces in nanostructures. In this work, we apply the terms electrostriction and radiation pressure as defined above. The former is reversed photoelasticity and operates in the bulk, the latter is reversed boundary motion and operates on interfaces. This implies that large optical forces are found in structures that are very sensitive to motion. In waveguides, this requires that the optical dispersion relation $\omega_k$ must strongly depend on dimensional fluctuations.

## 2.2 Quantum field treatment

In this section, we move from intuition to a quantum field theory. Traditionally, the dynamics of Brillouin scattering was derived from Maxwell's and the elasticity [36, 37, 76, 125] equations directly, see [125] for a treatment of nanoscale waveguides. Such approaches make heavy use of the nonlinear polarization, but do not hold for weak light or sound. In addition, these treatments assume strong acoustic damping, see chapter 3. Here, we take a modern standpoint in analogy to cavity optomechanics [35]. We start from the Hamiltonian of the system. Next, we obtain the spatiotemporal evolution of the opto-acoustic field operators from the Heisenberg equations of motion. Following [130], this approach describes opto-acoustic fields at the quantum level. Thus it explicitly contains creation and annihilation events.

### 2.2.1 Optics and mechanics without interaction

Specifically, we treat photon-phonon coupling in a waveguide along the $z$-axis. The opto-acoustic fields are thus confined along the $x$- and $y$-axes. This quantizes their wavevectors along these axes, resulting in a series of optical modes $\gamma$ and acoustic modes $\eta$. The optical modes $\gamma$ have dispersion relations $\omega_{\gamma k}$, whereas the acoustic modes $\eta$ have dispersion relations $\Omega_{\eta q}$. The interaction-free Hamiltonian $\mathcal{H}_0$ of the waveguide is

$$\mathcal{H}_0 = \sum_{\gamma} \int \mathrm{d}k \, \hbar\omega_{\gamma k} a^\dagger_{\gamma k} a_{\gamma k} + \sum_{\eta} \int \mathrm{d}q \, \hbar\Omega_{\eta q} b^\dagger_{\eta q} b_{\eta q} \tag{2.1}$$

with $a^\dagger_{\gamma k}$ and $a_{\gamma k}$ the creation and annihilation operators of optical mode $\gamma$ at wavevector $k$, $b^\dagger_{\eta q}$ and $b_{\eta q}$ the creation and annihilation operators of acoustic mode $\eta$ at wavevector $q$, $a^\dagger_{\gamma k} a_{\gamma k}$ the photon density operator of mode $\gamma$ at wavevector $k$ and $b^\dagger_{\eta q} b_{\eta q}$ the phonon density operator of mode $\eta$ at wavevector $q$. Integrals are assumed to go from $-\infty$ to $\infty$ although modal cut-offs in fact restrict this range. This is a good approximation in case of narrow-band envelopes far from cut-offs [130]. Crucially, (2.1) is accompanied by the commutators

$$\begin{aligned} \left[a_{\gamma k}, a^\dagger_{\gamma' k'}\right] &= \delta_{\gamma\gamma'} \delta(k - k') \\ \left[b_{\eta q}, b^\dagger_{\eta' q'}\right] &= \delta_{\eta\eta'} \delta(q - q') \end{aligned} \tag{2.2}$$

with $\delta_{\gamma\gamma'}$ the Kronecker delta, $\delta(k - k')$ the Dirac delta distribution and similar for $b_{\eta q}$. All other commutators vanish.

The free Hamiltonian (2.1) is a reformulation of

$$\begin{aligned} \mathcal{H}_0 = &\frac{1}{2\mu_0} \int B^i(\mathbf{r}) B^i(\mathbf{r}) \mathrm{d}\mathbf{r} + \frac{1}{2\epsilon_0} \int D^i(\mathbf{r}) \frac{1}{\epsilon_\mathrm{r}^{ij}(\mathbf{r})} D^j(\mathbf{r}) \mathrm{d}\mathbf{r} \\ &+ \int \frac{\pi^i(\mathbf{r}) \pi^i(\mathbf{r})}{2\rho(\mathbf{r})} \mathrm{d}\mathbf{r} + \frac{1}{2} \int S^{ij}(\mathbf{r}) c^{ijkl}(\mathbf{r}) S^{kl}(\mathbf{r}) \mathrm{d}\mathbf{r} \end{aligned} \tag{2.3}$$

with $\mathbf{B}(\mathbf{r})$ the magnetic field, $\mathbf{D}(\mathbf{r})$ the electric induction field, $\epsilon_\mathrm{r}^{ij}$ the relative permittivity, $\boldsymbol{\pi}(\mathbf{r})$ the acoustic momentum density, $S^{ij}(\mathbf{r}) = \frac{1}{2}\left(\partial_i u^j(\mathbf{r}) + \partial_j u^i(\mathbf{r})\right)$ the strain tensor, $\mathbf{u}(\mathbf{r})$ the acoustic displacement field and $c^{ijkl}(\mathbf{r})$ the stiffness tensor. One can express (2.3) as (2.1) via

$$\mathbf{D}(\mathbf{r}) = \sum_{\gamma} \int \frac{\mathrm{d}k}{\sqrt{2\pi}} \sqrt{\frac{\hbar\omega_{\gamma k}}{2}} a_{\gamma k} \mathbf{d}_{\gamma k}(x,y) e^{ikz} + \text{h.c.} \tag{2.4}$$

$$\mathbf{u}(\mathbf{r}) = \sum_{\eta} \int \frac{\mathrm{d}q}{\sqrt{2\pi}} \sqrt{\frac{\hbar\Omega_{\eta q}}{2}} b_{\eta q} \mathbf{u}_{\eta q}(x,y) e^{iqz} + \text{h.c.} \tag{2.5}$$

$$\left[D^i(\mathbf{r}), B^j(\mathbf{r}')\right] = i\hbar\epsilon^{ilj}\partial_l \delta(\mathbf{r}-\mathbf{r}') \tag{2.6}$$

$$\left[u^n(\mathbf{r}), \pi^m(\mathbf{r}')\right] = i\hbar\delta^{nm}\delta(\mathbf{r}-\mathbf{r}') \tag{2.7}$$

where h.c. stands for Hermitian conjugate and all other commutators vanish [130]. Vacuum fluctuations are neglected in (2.1) as they disappear in the eventual dynamics.

Next, we derive the interaction-free dynamics from (2.1). We take

$$\begin{aligned} a_{\gamma k} &= \int \frac{\mathrm{d}z}{\sqrt{2\pi}} a_{\gamma}(z) e^{-i(k-k_\gamma)z} \\ b_{\eta q} &= \int \frac{\mathrm{d}z}{\sqrt{2\pi}} b_{\eta}(z) e^{-i(q-q_\eta)z} \end{aligned} \tag{2.8}$$

where the excitation of the modes $\gamma$ and $\eta$ is centered around wavevectors $k_\gamma$ and $q_\eta$. This leads from (2.2) to the equal-time commutators

$$\begin{aligned} \left[a_\gamma(z,t), a^\dagger_{\gamma'}(z',t)\right] &= \delta_{\gamma\gamma'}\delta(z-z') \\ \left[b_\eta(z,t), b^\dagger_{\eta'}(z',t)\right] &= \delta_{\eta\eta'}\delta(z-z') \end{aligned} \tag{2.9}$$

where we used $\delta(z-z') = \int \frac{\mathrm{d}k}{2\pi} e^{i(k-k_\gamma)(z-z')}$. The optics and mechanics can be treated identically. For the optics, we get

$$\int \mathrm{d}k\, \hbar\omega_{\gamma k} a^\dagger_{\gamma k} a_{\gamma k} = \int \mathrm{d}k \int \frac{\mathrm{d}z}{\sqrt{2\pi}} \int \frac{\mathrm{d}z'}{\sqrt{2\pi}} \hbar\omega_{\gamma k} a^\dagger_\gamma(z) a_\gamma(z') e^{i(k-k_\gamma)(z-z')} \tag{2.10}$$

To simplify this expression, we expand the dispersion relation $\omega_{\gamma k}$ around the excitation wavevector $k_\gamma$. In particular, we use

$$\begin{aligned} \omega_{\gamma k} &= \sum_{n=0}^{\infty} \frac{1}{n!} \partial_k^n \left.\omega_{\gamma k}\right|_{k=k_\gamma} (k-k_\gamma)^n \\ (k-k_\gamma)^n e^{i(k-k_\gamma)(z-z')} &= (i\partial_{z'})^n e^{i(k-k_\gamma)(z-z')} \\ \int \mathrm{d}z' a_\gamma(z') (i\partial_{z'})^n \delta(z-z') &= (-i\partial_z)^n a_\gamma(z) \end{aligned} \tag{2.11}$$

Substituting these into (2.10) yields

$$\int \mathrm{d}k\, \hbar\omega_{\gamma k} a^\dagger_{\gamma k} a_{\gamma k} = \int \mathrm{d}z\, a^\dagger_\gamma(z)\hbar\xi_{\gamma z} a_\gamma(z) \quad \text{with} \quad \xi_{\gamma z} = \sum_{n=0}^{\infty} \frac{1}{n!}\, \partial^n_k \omega_{\gamma k}\big|_{k=k_\gamma} (-i\partial_z)^n \tag{2.12}$$

$$\int \mathrm{d}q\, \hbar\Omega_{\eta q} b^\dagger_{\eta q} b_{\eta q} = \int \mathrm{d}z\, b^\dagger_\eta(z)\hbar\zeta_{\eta z} b_\eta(z) \quad \text{with} \quad \zeta_{\eta z} = \sum_{n=0}^{\infty} \frac{1}{n!}\, \partial^n_q \Omega_{\eta q}\Big|_{q=q_\eta} (-i\partial_z)^n \tag{2.13}$$

where (2.13) resulted from identical operations on the mechanics. The operators $\xi_{\gamma z}$ and $\zeta_{\eta z}$ are Hermitian since $\partial_z$ is anti-Hermitian. Therefore (2.1) becomes

$$\mathcal{H}_0 = \sum_\gamma \int \mathrm{d}z\, a^\dagger_\gamma(z)\hbar\xi_{\gamma z} a_\gamma(z) + \sum_\eta \int \mathrm{d}z\, b^\dagger_\eta(z)\hbar\zeta_{\eta z} b_\eta(z) \tag{2.14}$$

Using the Heisenberg equation of motion and (2.9), the dynamics is

$$\begin{aligned} \partial_t a_\gamma(z) &= -\frac{i}{\hbar}\left[a_\gamma(z), \mathcal{H}_0\right] \\ &= -i \int \mathrm{d}z' \left[a_\gamma(z), a^\dagger_\gamma(z')\right] \xi_{\gamma z'} a_\gamma(z') \\ &= -i\xi_{\gamma z} a_\gamma(z) \end{aligned} \tag{2.15}$$

If one restricts the Taylor-expansion of the dispersion to first-order, the dynamics is

$$\partial_t a_\gamma(z) + v_\gamma \partial_z a_\gamma(z) = -i\omega_\gamma a_\gamma(z) \tag{2.16}$$

$$\partial_t b_\eta(z) + v_\eta \partial_z b_\eta(z) = -i\Omega_\eta b_\eta(z) \tag{2.17}$$

with $v_\gamma = \partial_k \omega_{\gamma k}\big|_{k=k_\gamma}$ and $v_\eta = \partial_q \Omega_{\eta k}\big|_{q=q_\eta}$ the photonic and phononic group velocities and $\omega_\gamma$ and $\Omega_\eta$ the frequencies corresponding to the excitations' center wavevectors.

### 2.2.2 The interaction Hamiltonian

Here we treat the interaction between photons and phonons. Under common assumptions that will soon be clarified, the waveguide's interaction Hamiltonian $\mathcal{V}$ equals

$$\mathcal{V} = \sum_{\gamma\gamma'\eta} \frac{\hbar g_{\gamma\gamma'\eta}}{\sqrt{2\pi}} \int \mathrm{d}k \int \mathrm{d}q\, a^\dagger_{\gamma k} a_{\gamma'(k-q)} b_{\eta q} + \text{h.c.} \tag{2.18}$$

This is a reformulation of

$$\mathcal{V} = \frac{1}{2\epsilon_0} \int D^i(\mathbf{r})\delta\beta^{ij}(\mathbf{r};\mathbf{u}(\mathbf{r}))D^j(\mathbf{r})\mathrm{d}\mathbf{r} \tag{2.19}$$

with $\delta\beta^{ij}$ the shift in the inverse permittivity by the acoustic displacement $\mathbf{u}(\mathbf{r})$:

$$\delta\beta^{ij}(\mathbf{r};\mathbf{u}(\mathbf{r})) = p^{ijlm}(x,y)S^{lm}(\mathbf{r}) + \delta^{ij}\beta_{\text{ref}}(\mathbf{r}-\mathbf{u}(\mathbf{r})) \tag{2.20}$$

where $p^{ijlm}(x,y)$ is the photoelastic Pockels tensor and $\beta_{\text{ref}}(\mathbf{r}) = \frac{1}{\epsilon(\mathbf{r})}$ the inverse permittivity profile in absence of movement. The first term operates in the bulk of the waveguide, generating both photoelastic scattering and the electrostrictive gradient force (section 2.1.2). The second term operates at the boundaries between materials, generating both boundary-induced scattering and radiation pressure (section 2.1.3). To derive (2.18) from (2.19), for the sake of brevity we limit ourselves to the boundary term in the following. Assuming small displacements, we write the perturbation as

$$\delta\beta^{ij}(\mathbf{r};\mathbf{u}(\mathbf{r})) = \delta^{ij}\beta_{\text{ref}}(\mathbf{r}-\mathbf{u}(\mathbf{r})) \approx -\delta^{ij}u^l(\mathbf{r})\partial_l\beta_{\text{ref}}(\mathbf{r}) \tag{2.21}$$

Inserting (2.4) and (2.21) into (2.19) leads to

$$\begin{aligned}\mathcal{V} = -\frac{1}{2\epsilon_0}\sum_{\gamma\gamma'}\int \mathrm{d}\mathbf{r}\int\frac{\mathrm{d}k}{\sqrt{2\pi}}\int\frac{\mathrm{d}k'}{\sqrt{2\pi}}\sqrt{\frac{\hbar\omega_{\gamma k}}{2}}\sqrt{\frac{\hbar\omega_{\gamma' k'}}{2}} \\ \times\left(a^\dagger_{\gamma k}a_{\gamma' k'}d^{i\star}_{\gamma k}d^i_{\gamma' k'}e^{i(k'-k)z}u^l\partial_l\beta_{\text{ref}} + \text{h.c.}\right)\end{aligned} \tag{2.22}$$

where we neglected terms in $a_{\gamma k}a_{\gamma' k'}$ and $a^\dagger_{\gamma k}a^\dagger_{\gamma' k'}$ in the *rotating-wave approximation*: these terms cannot efficiently drive the mechanics. Here we used the commutator (2.2) but left out the resulting vacuum terms as these disappear in the eventual dynamics. Next, we insert (2.5) and split $\int \mathrm{d}\mathbf{r}$ into a longitudinal integral $\int \mathrm{d}z$ and a transverse overlap integral $\int \mathrm{d}x\mathrm{d}y$. This results in

$$\mathcal{V} = \sum_{\gamma\gamma'\eta}\int \mathrm{d}z\int\frac{\mathrm{d}k}{\sqrt{2\pi}}\int\frac{\mathrm{d}k'}{\sqrt{2\pi}}\int\frac{\mathrm{d}q}{\sqrt{2\pi}}\,\hbar g_{(\gamma k;\gamma' k';\eta q)}a^\dagger_{\gamma k}a_{\gamma' k'}b_{\eta q}e^{i(k'+q-k)z} + \text{h.c.} \tag{2.23}$$

with $g_{(\gamma k;\gamma' k';\eta q)}$ the transverse overlap integral [130]. This coupling coefficient can be taken real and positive without loss of generality (appendix A). Subsequently assuming an infinitely long waveguide enforces wave-momentum conservation (section 2.1.3) via $\int \frac{\mathrm{d}z}{2\pi}e^{i(k'+q-k)z} = \delta(k'+q-k)$. Eliminating the integral over $k'$ via wave-momentum conservation yields

$$\mathcal{V} = \sum_{\gamma\gamma'\eta}\int \mathrm{d}k\int \mathrm{d}q\,\frac{\hbar g_{(\gamma k;\gamma' k';\eta q)}}{\sqrt{2\pi}}a^\dagger_{\gamma k}a_{\gamma'(k-q)}b_{\eta q} + \text{h.c.} \tag{2.24}$$

$$= \sum_{\gamma\gamma'\eta}\frac{\hbar g_{\gamma\gamma'\eta}}{\sqrt{2\pi}}\int \mathrm{d}k\int \mathrm{d}q\, a^\dagger_{\gamma k}a_{\gamma'(k-q)}b_{\eta q} + \text{h.c.} \tag{2.25}$$

where we took $g_{(\gamma k;\gamma' k';\eta q)} = g_{\gamma\gamma'\eta}$ constant in the relevant range of wavevectors. The bulk term can be treated identically [130], thereby yielding another contribution to the overlap integral $g_{\gamma\gamma'\eta}$. So we derived (2.18) from (2.19). Next, we insert the Fourier transforms (2.8) to obtain

$$\mathcal{V} = \sum_{\gamma\gamma'\eta} \hbar g_{\gamma\gamma'\eta} \int \mathrm{d}z\, a_{\gamma}^{\dagger}(z) a_{\gamma'}(z) b_{\eta}(z) e^{i\left(k_{\gamma'}+q_{\eta}-k_{\gamma}\right)z} + \text{h.c.} \tag{2.26}$$

where we used $\int \frac{\mathrm{d}z}{2\pi} e^{ikz} = \delta(z)$ twice. We now look at the dynamics without the free Hamiltonian $\mathcal{H}_0$. On the photonic side, the Heisenberg equation of motion yields:

$$\partial_t a_{\gamma}(z) = -\frac{i}{\hbar}\left[a_{\gamma}(z), \mathcal{V}\right] \tag{2.27}$$

$$= -i \sum_{\gamma_1\gamma_2\eta} g_{\gamma_1\gamma_2\eta} \int \mathrm{d}z' \left[a_{\gamma}(z), a_{\gamma_1}^{\dagger}(z') a_{\gamma_2}(z') b_{\eta}(z') e^{i\left(k_{\gamma_2}+q_{\eta}-k_{\gamma_1}\right)z'} + \text{h.c.}\right] \tag{2.28}$$

$$= -i \sum_{\gamma'\eta} g_{\gamma\gamma'\eta} \left(a_{\gamma'}(z) b_{\eta}(z) e^{i\left(k_{\gamma'}+q_{\eta}-k_{\gamma}\right)} + b_{\eta}^{\dagger}(z) a_{\gamma'}(z) e^{i\left(k_{\gamma'}-q_{\eta}-k_{\gamma}\right)}\right) \tag{2.29}$$

where we applied (2.9). On the phononic side, we similarly arrive at

$$\partial_t b_{\eta}(z) = -i \sum_{\gamma\gamma'\eta'} g_{\gamma\gamma'\eta'} \int \mathrm{d}z' \left[b_{\eta}(z), a_{\gamma'}^{\dagger}(z') a_{\gamma}(z') b_{\eta'}^{\dagger}(z') e^{i\left(k_{\gamma}-k_{\gamma'}-q_{\eta'}\right)z'}\right] \tag{2.30}$$

$$= -i \sum_{\gamma\gamma'} g_{\gamma\gamma'\eta} a_{\gamma'}^{\dagger}(z) a_{\gamma}(z) e^{i\left(k_{\gamma}-q_{\eta}-k_{\gamma'}\right)z} \tag{2.31}$$

The phase-mismatch terms $e^{i\left(k_{\gamma}-k_{\gamma'}-q_{\eta}\right)z}$ are absent in most treatments [36, 37, 125]: it is usually assumed that the phonons do not propagate. This produces a large uncertainty in their momentum $\hbar q_{\eta}$, such that wave-momentum conservation need no longer be exactly satisfied. In absence of damping, however, one can externally inject optical and acoustic beams that suffer from a phase-mismatch. The phase relations between the waves, and thus the direction of the process, will then reverse with spatial period $\frac{\pi}{\Delta k}$ and $\Delta k$ the wavevecotor mismatch. A finite waveguide length produces a similar violation of momentum conservation: $\int \frac{\mathrm{d}z}{2\pi} e^{i(k'+q-k)z}$ no longer equals a delta distribution if the integration range is finite. Noether's theorem guarantees wave-momentum conservation only if space and the material are invariant with respect to translations (section 2.1.3).

### 2.2.3 The complete dynamics

Here we combine the results of sections 2.2.1 and 2.2.2 to derive the full dynamics. The full Hamiltonian is given by $\mathcal{H} = \mathcal{H}_0 + \mathcal{V}$:

$$\frac{\mathcal{H}}{\hbar} = \sum_{\gamma} \int \mathrm{d}k\, \omega_{\gamma k} a^{\dagger}_{\gamma k} a_{\gamma k} + \sum_{\eta} \int \mathrm{d}q\, \Omega_{\eta q} b^{\dagger}_{\eta q} b_{\eta q} + \left( \sum_{\gamma\gamma'\eta} \frac{g_{\gamma\gamma'\eta}}{\sqrt{2\pi}} \int \mathrm{d}k \int \mathrm{d}q\, a^{\dagger}_{\gamma k} a_{\gamma'(k-q)} b_{\eta q} + \text{h.c.} \right) \tag{2.32}$$

$$= \sum_{\gamma} \int \mathrm{d}z\, a^{\dagger}_{\gamma}(z) \xi_{\gamma z} a_{\gamma}(z) + \sum_{\eta} \int \mathrm{d}z\, b^{\dagger}_{\eta}(z) \zeta_{\eta z} b_{\eta}(z) + \left( \sum_{\gamma\gamma'\eta} g_{\gamma\gamma'\eta} \int \mathrm{d}z\, a^{\dagger}_{\gamma}(z) a_{\gamma'}(z) b_{\eta}(z) e^{i\left(k_{\gamma'}+q_{\eta}-k_{\gamma}\right)z} + \text{h.c.} \right)$$

This Hamiltonian sets up the dynamics

$$\partial_t a_{\gamma}(z) = -i\xi_{\gamma z} a_{\gamma}(z) - i \sum_{\gamma'\eta} g_{\gamma\gamma'\eta} \left( a_{\gamma'}(z) b_{\eta}(z) e^{i\left(k_{\gamma'}+q_{\eta}-k_{\gamma}\right)} + b^{\dagger}_{\eta}(z) a_{\gamma'}(z) e^{i\left(k_{\gamma'}-q_{\eta}-k_{\gamma}\right)} \right) \tag{2.33}$$

$$\partial_t b_{\eta}(z) = -i\zeta_{\eta z} b_{\eta}(z) - i \sum_{\gamma\gamma'} g_{\gamma\gamma'\eta} a^{\dagger}_{\gamma'}(z) a_{\gamma}(z) e^{i\left(k_{\gamma}-q_{\eta}-k_{\gamma'}\right)z}$$

These equations capture coupling between several photonic and phononic modes even for high dispersion. From here on, we take the group velocities $v_{\gamma}$ and $v_{\eta}$ as positive, introducing explicit minus signs for counter-propagating waves. We also flux-normalize the envelopes and remove the fast-rotating temporal terms through the rescaling

$$a_{\gamma}(z) \mapsto \sqrt{v_{\gamma}}\, a_{\gamma}(z) e^{i(\omega_{\gamma}+\Delta_{\gamma})t} \tag{2.34}$$
$$b_{\eta}(z) \mapsto \sqrt{v_{\eta}}\, b_{\eta}(z) e^{i\left(\Omega_{\eta}+\Delta_{\eta}\right)t}$$

with $\Delta_{\gamma}$ and $\Delta_{\eta}$ detunings between the injected frequencies and the dispersion relation. This transforms the commutators (2.9) into

$$\left[a_{\gamma}(z,t), a^{\dagger}_{\gamma'}(z',t)\right] = \sqrt{v_{\gamma} v_{\gamma'}}\, \delta_{\gamma\gamma'} \delta(z-z') \tag{2.35}$$
$$\left[b_{\eta}(z,t), b^{\dagger}_{\eta'}(z',t)\right] = \sqrt{v_{\eta} v_{\eta'}}\, \delta_{\eta\eta'} \delta(z-z')$$

Since the phonon frequencies are small in this work, we limit the operators $\xi_{\gamma z}$ and $\zeta_{\eta z}$ to first-order dispersion. The rescaling (2.34) then leads to

$$v_{\gamma}^{-1} \partial_t a_{\gamma} + \text{sign}[v_{\gamma}] \partial_z a_{\gamma} = -i \sum_{\gamma'\eta} \tilde{g}_{\gamma\gamma'\eta} \left( a_{\gamma'} b_{\eta} + b^{\dagger}_{\eta} a_{\gamma'} \right) - \tilde{\chi}_{\gamma}^{-1} a_{\gamma} \tag{2.36}$$
$$v_{\eta}^{-1} \partial_t b_{\eta} + \text{sign}[v_{\eta}] \partial_z b_{\eta} = -i \sum_{\gamma\gamma'} \tilde{g}_{\gamma\gamma'\eta} a^{\dagger}_{\gamma'} a_{\gamma} - \tilde{\chi}_{\eta}^{-1} b_{\eta}$$

where we defined the coupling strengths $\tilde{g}_{\gamma\gamma'\eta} = \frac{g_{\gamma\gamma'\eta}}{\sqrt{v_\gamma v_{\gamma'} v_\eta}}$ and spatial response functions $\tilde{\chi}_\gamma^{-1} = \frac{\alpha_\gamma}{2} - i\tilde{\Delta}_\gamma$ with $\alpha_\gamma$ the propagation losses and $\tilde{\Delta}_\gamma = v_\gamma^{-1}\Delta_\gamma$ the excitations' wavevector detuning from the dispersion relation. The sign-function captures counter-propagating fields. We assumed a phase-matched configuration where $k_\gamma = k_{\gamma'} \pm q_\eta$ and $\omega_\gamma + \Delta_\gamma = \omega_{\gamma'} + \Delta_{\gamma'} \pm (\Omega_\eta + \Delta_\eta)$ for all involved modes. Usually, however, the sums $\sum_{\gamma'\eta}$ and $\sum_{\gamma\gamma'}$ only produce one or two phase-matched terms. The losses $\alpha_\gamma$ were introduced on the assumption that they only weakly affect the opto-acoustic modes. They could be derived explicitly by including coupling to a bath in the Hamiltonian (2.32) and assuming the bath is memoryless [131]. Noise terms associated with the losses $\alpha_\gamma$ were neglected in (2.36).

The simplest and most relevant case is the interaction between two photonic modes, called the pump and Stokes seed, and one phononic mode. This yields the dynamics

$$v_{\mathrm{p}}^{-1}\partial_t a_{\mathrm{p}} + \partial_z a_{\mathrm{p}} = -i\tilde{g}_0 a_{\mathrm{s}} b - \tilde{\chi}_{\mathrm{p}}^{-1} a_{\mathrm{p}} \tag{2.37}$$

$$v_{\mathrm{s}}^{-1}\partial_t a_{\mathrm{s}} \pm \partial_z a_{\mathrm{s}} = -i\tilde{g}_0 b^\dagger a_{\mathrm{p}} - \tilde{\chi}_{\mathrm{s}}^{-1} a_{\mathrm{s}} \tag{2.38}$$

$$v_{\mathrm{m}}^{-1}\partial_t b + \partial_z b = -i\tilde{g}_0 a_{\mathrm{s}}^\dagger a_{\mathrm{p}} - \tilde{\chi}_{\mathrm{m}}^{-1} b \tag{2.39}$$

where we allowed for a counter-propagating Stokes field ($\pm$) and with $\tilde{g}_0$ the waveguide's *vacuum coupling rate*. The interpretation of $\tilde{g}_0$ will be discussed in the next chapter. Expressions for $\tilde{g}_0$ in terms of the opto-acoustic fields are given in [125, 130]. A counter-propagating acoustic mode could be described via another minus sign next to $\partial_z b$ in (2.39). These equations predict a wealth of effects, most of them currently unobserved. We explore the landscape of possibilities in the next chapter.

# 3
# Dynamics and link to cavity optomechanics

*When the solution is simple, God is answering.*
Albert Einstein

This chapter is based on [63].

**Contents**

HERE *we explore the landscape of dynamical effects in both optomechanical waveguides and cavities. So far, Brillouin scattering and cavity optomechanics were mostly disconnected branches of research – although both deal with photon-phonon coupling. This begs for the development of a broader theory that contains both fields. In this chapter, we derive the dynamics of optomechanical cavities from that of Brillouin-active waveguides. This explicit transition elucidates the link between phenomena such as Brillouin amplification and electromagnetically induced transparency. It proves that effects familiar from cavity optomechanics all have traveling-wave partners, but not vice versa. We reveal a close connection between two pa-* *rameters of central importance in these fields: the Brillouin gain coefficient and the zero-point optomechanical coupling rate. This enables comparisons between systems as diverse as ultracold atom clouds, plasmonic Raman cavities and nanoscale silicon waveguides. In addition, back-of-the-envelope calculations show that unobserved effects, such as photon-assisted amplification of traveling phonons, are now accessible in existing systems. Finally, we formulate both circuit- and cavity-oriented optomechanics in terms of vacuum coupling rates, cooperativities and gain coefficients, thus reflecting the similarities in the underlying physics.*

## 3.1 Introduction

Brillouin scattering [132] and cavity optomechanics [35] have been intensively studied in recent years. Both concern the interaction between light and sound, but they were part of separate traditions. Already in the early 1920s, diffraction of light by sound was studied by Léon Brillouin. Therefore, such inelastic scattering is called *Brillouin scattering* [36, 37]. The effect is known as *stimulated* Brillouin scattering [133–135] when a strong intensity-modulated light field generates the sound, often with classical applications such as spectral purification [136] and microwave signal processing [137] in mind. In contrast, cavity optomechanics arose from Braginsky's efforts to understand the limits of gravitational wave detectors in the 1970s – and greatly expanded since the demonstration of phonon lasing in microtoroids [26]. By and large, it aims to control both optical and mechanical quantum states [56, 138, 139].

Historically, a number of important differences hindered their merger. For instance, SBS generally dealt with high-group-velocity and cavity optomechanics with low-group-velocity acoustic phonons. In addition, bulk electrostrictive forces usually dominated phonon generation in SBS – while radiation pressure at the boundaries took this role in cavity optomechanics. Further, cavity optomechanics typically studied resonators with much lower phonon than photon dissipation – whereas Brillouin lasers [136, 140, 141] operate in the reversed regime [142]. Finally, SBS is often studied not in cavities but in optically broadband waveguides [132]. Thus, particular physical systems used to be firmly placed in either one or the other research paradigm.

Lately, the idea that these are mostly superficial classifications has been gaining traction. Indeed, in both cases light generates motion and the motion phase-modulates light. Next, this spatiotemporal phase-modulation creates motional sidebands – which interfere with those initially present (chapter 2). The research fields share this essential feedback loop. Some connections have already been made. For instance, electrostrictive forces were exploited for sideband cooling [143, 144] and induced transparency [145, 146] while radiation pressure contributed to SBS in dual-web fibers [147] and silicon waveguides [43, 64, 65, 104].

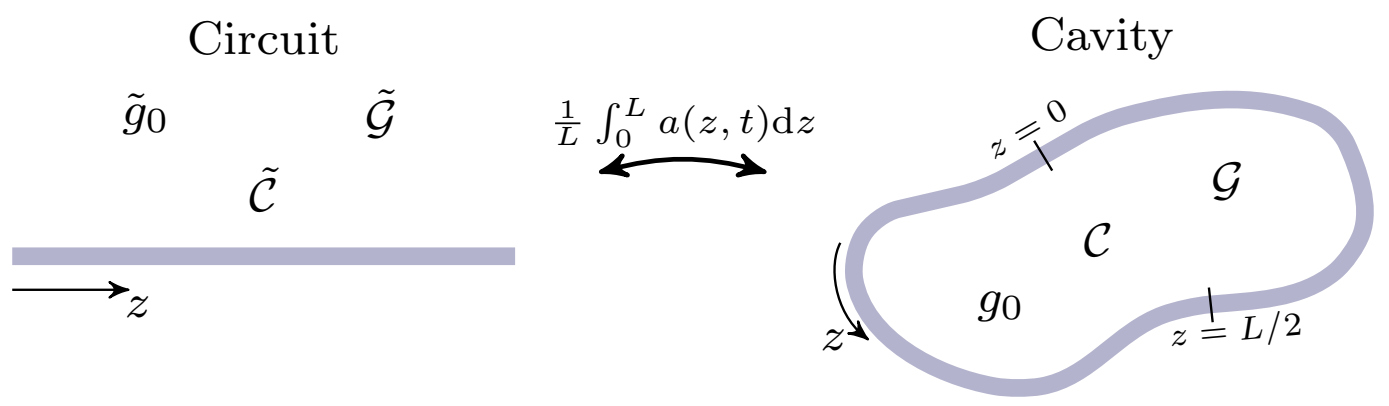

**Figure 3.1: From circuit to cavity optomechanics.** We explicitly derive the physics of optomechanical cavities (right) from that of Brillouin-active waveguides (left). Therefore, both traveling-wave and cavity-based optomechanics can be cast in terms of vacuum coupling rates ($\tilde{g}_0$ and $g_0$), cooperativities ($\tilde{\mathcal{C}}$ and $\mathcal{C}$) and gain coefficients ($\tilde{\mathcal{G}}$ and $\mathcal{G}$).

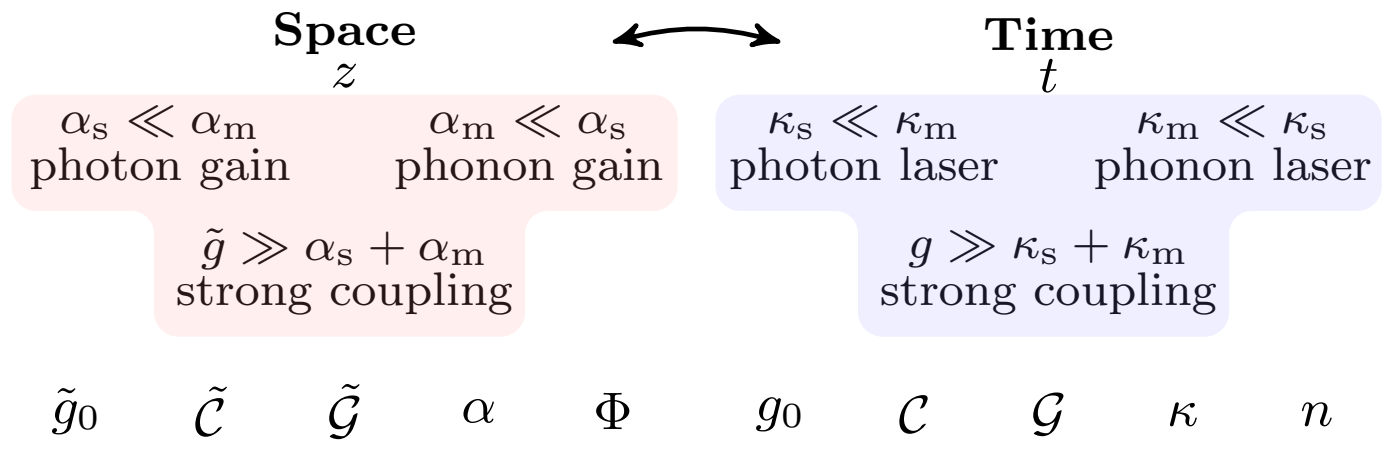

**Figure 3.2: Symmetry of circuit and cavity optomechanics.** Each temporal optomechanical effect has a spatial symmetry partner. Thus, the description of these effects can be cast in terms of conceptually similar figures of merit. The scheme assumes a red-detuned optical probe; "gain" and "laser" should respectively be replaced by "loss" and "cooling" for a blue-detuned optical probe. The meaning of the figures of merit is discussed in the main text.

In this chapter, we derive the dynamics of optomechanical cavities from that of Brillouin-active waveguides (fig.3.1). The transition holds for both co- and counter-propagating pump and Stokes waves, i.e. for *forward* and *backward* scattering, and for opto-acoustic coupling between two different or two identical optical modal fields, i.e. for *inter-* [144, 148–151] or *intra-* [43, 64, 66, 93, 150, 152–155] modal scattering (fig.2.2). Hence, all flavours of photon-phonon interaction are treated on the same footing. Moreover, this spatially averaged cavity dynamics is found to be equivalent to the standard Hamiltonian of cavity optomechanics [35] – even in the case of low-finesse phonons. It turns out that this cavity dynamics can be mapped – by swapping space and time ($z \leftrightarrow t$) – on the steady-state spatial evolution of the opto-acoustic fields in the waveguide.

This implies that the plethora of optomechanical effects, such as stimulated Brillouin scattering [36, 37, 156], slow light [39, 41, 157], optomechani-cally induced transparency [39, 158], ground-state cooling [38, 138] etc., are different aspects of the same feedback loop (fig.2.1). The rigorous transition decisively indicates that both fields are a subset of a larger theory of photon-phonon interaction, which we build on the single Hamiltonian of equation (2.32). This is not to say that they are identical: a Brillouin-active waveguide supports complex spatiotemporal phenomena [159–161] and noise dynamics [95, 96] not present in a high-finesse optomechanical cavity. Nevertheless, in the resulting picture (fig.3.2), both traveling-wave and cavity-based photon-phonon interaction can be classified according to (1) the damping hierarchy of the photons and phonons and (2) the strength of the photon-phonon coupling with respect to the largest dissipation channel. For weak coupling, the long-lived particle species – either photons or phonons – triggers the photon-phonon conversion. The short-lived particle species cannot truly build up and is thus slaved to its long-lived partner; it is merely created in short segments (of space or time) and immediately decays afterwards.

All Brillouin-active waveguides so far exhibited far stronger phononic than photonic propagation losses; in addition, the coupling was always weak relative to this phononic damping. Hence, there are two to date unexplored regimes of guided-wave optomechanics: (1) photon-assisted amplification of traveling phonons and (2) strong coupling between traveling photons and phonons (fig.3.2). The strong coupling regime produces either traveling entangled photon-phonon pairs or state swapping between light and sound along the waveguide, depending on the details (e.g. probe detuning) of the experiment. Although currently unobserved, both effects may be an asset in future quantum phononic networks [54, 139, 162–164]. For instance, in the strong coupling regime the flying phonon – entangled to its photonic partner – could be detected piezo-electrically or optically and thereby enable Bell tests [165–167] between two different particle species. Our back-of-the-envelope estimates show that these regimes can be achieved in existing systems, such as dual-web fibers and silicon nanowaveguides.

The transition (fig.3.1) assumes that the photonic and phononic modes of the waveguide are not disturbed too strongly by looping it into a cavity. This is justified in many cases since cavity designs aim to minimize the losses (e.g. due to bending) induced by any modal perturbations. Within this approximation, it permits translations between circuit- and cavity-oriented figures of merit. For instance, we identify a connection between the Brillouin gain coefficient $\tilde{\mathcal{G}}$ and the zero-point coupling rate $g_0$. The former ($\tilde{\mathcal{G}}$) quantifies the pump power and waveguide length required to amplify a Stokes seed appreciably [36, 37]. The latter ($g_0$) captures the interaction strength between a single photon and a single phonon in an optomechanical cavity [35]. The transition proves that these figures of merit are inextricably linked by

$$v_{\mathrm{p}} v_{\mathrm{s}} \frac{(\hbar\omega_{\mathrm{p}})\,\Omega_{\mathrm{m}}}{4L} \left(\frac{\tilde{\mathcal{G}}}{Q_{\mathrm{m}}}\right) = g_0^2 \tag{3.1}$$

with $v_{\mathrm{p}}$ and $v_{\mathrm{s}}$ the group velocities of the pump and Stokes waves, $\hbar\omega_{\mathrm{p}}$ the pump photon energy, $\frac{\Omega_{\mathrm{m}}}{2\pi}$ the mechanical resonance frequency, $L$ the cavity roundtrip length and $Q_{\mathrm{m}}$ the waveguide's mechanical quality factor. This link is independent of the type of driving optical force and of the relative photon and phonon damping. Similarly, we derive connections between each of the circuit- and cavity-oriented figures of merit: between the vacuum coupling rates ($\tilde{g}_0$ and $g_0$, see (3.21)), the cooperativities ($\tilde{\mathcal{C}}$ and $\mathcal{C}$, see (3.34)) and the gain coefficients ($\tilde{\mathcal{G}}$ and $\mathcal{G}$, see (3.36)).

Notably, this treatment goes beyond cavity optomechanical systems that have a clear circuit equivalent (as in fig.3.1). Indeed, the standard cavity Hamiltonian $\hat{\mathcal{H}} = \hbar\omega_{\mathrm{c}}(\hat{x})\hat{a}^\dagger\hat{a} + \hbar\Omega_{\mathrm{m}}\hat{b}^\dagger\hat{b}$ [35] also captures the temporal dynamics of cavity optomechanics based on Bose-Einstein condensates [168, 169] or plasmonic Raman cavities [51]. The physics of all these diverse systems can be understood in the scheme of fig.3.2. On top of the similar dynamics, this means that the photon-phonon interaction efficiency of a larger class of systems can now be compared in a single framework. For instance, the gain coefficient of a silicon nanowire can be converted to the vacuum coupling rate of a hypothetical cavity (through (3.1)); which can next be compared to that of any other cavity optomechanical system. In reverse, the link enables the conversion of a vacuum coupling rate of an actual cavity optomechanical system into a hypothetical guided-wave coupling rate (through (3.21)); which can next be compared to that of any other waveguide. We give examples of such conversions, which can be tested empirically in many cases, in section 3.5.

The chapter is organized as follows: in section 3.2 we describe a minimal model of circuit optomechanics and frame it in terms of a guided-wave vacuum coupling rate $\tilde{g}_0$ and cooperativity $\tilde{\mathcal{C}}$. Next, we make the mean-field transition to a cavity in section 3.3. At that point, we also discuss the limitations of the analysis. The resulting dynamical effects are treated in section 3.4. The prospects for observing new effects are considered in section 3.5 and we conclude in section 3.6.

## 3.2 Circuit optomechanics

In particular, we study the interaction between a pump field with envelope $a_{\mathrm{p}}(z,t)$ and a red-detuned Stokes field with envelope $a_{\mathrm{s}}(z,t)$ mediated by an acoustic field with envelope $b(z,t)$. This nomenclature does not impose restrictions in the following, as the Stokes field may in fact be stronger than the pump. In that case we will refer to the pump as the anti-Stokes probe and to the Stokes as the pump. The envelopes contain only the slowly varying part of the photonic-phononic fields; rapidly oscillating factors $e^{i(kz-\omega t)}$ were removed in each case (chapter 2, section 2.2). The guided optical modes correspond to the points $(\omega_{\mathrm{p}}, k_{\mathrm{p}})$ and $(\omega_{\mathrm{s}}, k_{\mathrm{s}})$ in the optical dispersion relation (fig.2.2). By energy and wave-momentum [111] conservation, the excited phonon has an angular frequency $\Omega = \omega_{\mathrm{p}} - \omega_{\mathrm{s}}$ and wavevector $q = k_{\mathrm{p}} \mp k_{\mathrm{s}}$. The nature of the optical modes (co/counter and fast/slow) and the acoustic dispersion relation determine the wavevector $q$ and group velocity $v_{\mathrm{m}}$ of the excited phonons (fig.2.2&2.3). See chapter 2 for more background on the model.

Traveling-wave photon-phonon coupling is governed by the following dynamical evolution [95, 125, 130]

$$\begin{aligned} v_{\mathrm{p}}^{-1}\partial_t a_{\mathrm{p}} + \partial_z a_{\mathrm{p}} &= -i\tilde{g}_0 a_{\mathrm{s}} b - \tilde{\chi}_{\mathrm{p}}^{-1} a_{\mathrm{p}} \\ v_{\mathrm{s}}^{-1}\partial_t a_{\mathrm{s}} \pm \partial_z a_{\mathrm{s}} &= -i\tilde{g}_0 b^{\dagger} a_{\mathrm{p}} - \tilde{\chi}_{\mathrm{s}}^{-1} a_{\mathrm{s}} \\ v_{\mathrm{m}}^{-1}\partial_t b + \partial_z b &= -i\tilde{g}_0 a_{\mathrm{s}}^{\dagger} a_{\mathrm{p}} - \tilde{\chi}_{\mathrm{m}}^{-1} b \end{aligned} \tag{3.2}$$

These equations were derived in chapter 2, see equations (2.37) in section 2.2. This starting point and the following treatment holds quantum mechanically if one takes care to treat the envelopes in (3.2) as operators [95, 130] obeying the equal-time commutators (2.35), reproduced here:

$$\left[a_{\gamma}(z,t), a_{\gamma'}^{\dagger}(z',t)\right] = \sqrt{v_{\gamma} v_{\gamma'}}\,\delta_{\gamma\gamma'}\delta(z-z') \tag{3.3}$$

with $\gamma$ an index running over the pump $p$, Stokes $s$ and mechanical wave $m$, $v_{\gamma}$ the group velocities, $\delta_{jj'}$ the Kronecker delta, $\delta(z)$ the Dirac delta distribution and $a_{\mathrm{m}} = b$ for notational convenience. We flux-normalized the field operators $a_{\gamma}$ such that $\Phi_{\mathrm{p}} = a_{\mathrm{p}}^{\dagger} a_{\mathrm{p}}$, $\Phi_{\mathrm{s}} = a_{\mathrm{s}}^{\dagger} a_{\mathrm{s}}$ and $\Phi_{\mathrm{m}} = b^{\dagger} b$ correspond to the number of pump photons, Stokes photons and phonons passing through a cross-section of the waveguide per second. We will treat highly occupied (i.e. large mean flux $\langle \Phi_{\gamma} \rangle$) modes as classical in the remainder of the chapter, as is standard [35, 170–172]. Further, we denote $\tilde{g}_0$ the traveling-wave vacuum coupling rate (to be discussed further on), $\tilde{\chi}_{\gamma}^{-1} = \frac{\alpha_{\gamma}}{2} - i\tilde{\Delta}_{\gamma}$ the susceptibilities, $\alpha_{\gamma}$ the propagation losses and $\tilde{\Delta}_{\gamma}$ the wavevector offsets between externally applied fields and the intrinsic waveguide modes.

In some systems, e.g. for the Raman-like low-group-velocity phonons (fig.2.3) associated with forward intra-modal scattering [43, 64, 65, 150, 152], the phonon wavelength $\frac{2\pi}{K}$ can be substantially larger than its decay length $\alpha_{\mathrm{m}}^{-1}$ – so its slowly-varying amplitude treatment breaks down. Then the acoustic excitation is better treated as a localized series of mechanical oscillators [64, 125, 152], essentially dealing with each cross-sectional slice of the waveguide as an artificial Raman-active molecule. The above dynamics (3.2), however, contains these systems as well by letting the phonon decay length $\alpha_{\mathrm{m}}^{-1}$ vanish. Further, the sign ($\pm$) in the Stokes equation indicates the difference between forward ($+$) and backward ($-$) photon-phonon coupling. Cascaded scattering [94, 152] and noise [95, 96] can and should be added to this model in some instances. In fact, (3.2) can be regarded as the unique, minimal model for guided-wave Brillouin scattering [95, 125, 130]. We discuss potential extensions in section 3.5; in the following, we only need the minimal model (3.2), future extended versions can be dealt with similarly.

The Manley-Rowe relations [36, 173] guarantee that a single, unique figure of merit $\tilde{g}_0$ captures all conservative optical forces and scattering. Indeed, in the lossless case ($\alpha_\gamma = 0$), the rate of pump photon destruction must equal the rate of Stokes photon and phonon creation:

$$-\partial_z \Phi_{\mathrm{p}} = \pm \partial_z \Phi_{\mathrm{s}} = \partial_z \Phi_{\mathrm{m}} = -\tilde{g}_0 \left( i a_{\mathrm{s}}^\dagger b^\dagger a_{\mathrm{p}} + \mathrm{h.c.} \right) \tag{3.4}$$

Similar to $g_0$ in a cavity [35], $\tilde{g}_0$ quantifies the interaction strength between a single photon and a single phonon, but in this case flying along a waveguide instead of trapped in a cavity. We take $\tilde{g}_0$ real and positive without loss of generality. Briefly specializing to forward intra-modal scattering, the mean-field transition of section 3.3 will show that (see appendix A)

$$\tilde{g}_0 = -\tilde{x}_{\mathrm{ZPF}} \left. \frac{\partial k_{\mathrm{p}}}{\partial x} \right|_{\omega_{\mathrm{p}}} \tag{3.5}$$

with

$$\tilde{x}_{\mathrm{ZPF}} = x_{\mathrm{ZPF}} \sqrt{\frac{\delta L}{v_{\mathrm{m}}}} = \sqrt{\frac{\hbar}{2 m_{\mathrm{eff}} v_{\mathrm{m}} \Omega_{\mathrm{m}}}} \tag{3.6}$$

the guided-wave zero-point motion and $m_{\mathrm{eff}}$ the effective mass of the mechanical mode per unit length. Indeed, a short waveguide section of length $\delta L$ contains $\langle n_{\mathrm{m}} \rangle = \frac{\delta L}{v_{\mathrm{m}}} \langle \Phi_{\mathrm{m}} \rangle$ phonons with $\langle \Phi_{\mathrm{m}} \rangle$ the mean phonon flux. As particle fluxes – instead of numbers – are fundamental in the waveguide's Manley-Rowe relations (3.4), the zero-point motion is rescaled by precisely this factor $(\delta L / v_{\mathrm{m}})^{1/2}$ relative to the actual zero-point motion $x_{\mathrm{ZPF}}$ [35] of the $\delta L$-section

$$x_{\mathrm{ZPF}} = \sqrt{\frac{\hbar}{2 m_{\mathrm{eff}} \delta L \Omega_{\mathrm{m}}}} \tag{3.7}$$

Therefore, the traveling-wave vacuum coupling rate $\tilde{g}_0$ is determined by the wavevector shift induced by mechanical motion at fixed frequency, while the cavity vacuum coupling rate $g_0$ is determined by the frequency shift induced by mechanical motion at fixed wavevector [35]. Notably, the interpretation of $\tilde{g}_0$ as the coupling strength between a single traveling photon and phonon holds also for inter-modal and backward scattering (see appendix A).

In steady-state ($\partial_t \to 0$) and for a constant, strong pump ($\Phi_\mathrm{p}(z) = \Phi_\mathrm{p}(0)$), the evolution (3.2) reduces to

$$\begin{aligned} \partial_z a_\mathrm{s} &= \mp i\tilde{g}_0 b^\dagger a_\mathrm{p} \mp \tilde{\chi}_\mathrm{s}^{-1} a_\mathrm{s} \\ \partial_z b &= -i\tilde{g}_0 a_\mathrm{s}^\dagger a_\mathrm{p} - \tilde{\chi}_\mathrm{m}^{-1} b \end{aligned} \tag{3.8}$$

The phonon decay length $\alpha_\mathrm{m}^{-1}$ is generally largest for backward scattering. Even then, it typically does not exceed $\alpha_\mathrm{m}^{-1} \sim 100\,\mu\mathrm{m}$ [36, 98]. Therefore, the photon decay length massively exceeds the phonon decay length in Brillouin-active waveguides to date ($\alpha_\mathrm{s} \ll \alpha_\mathrm{m}$). A full solution of (3.8) exists but yields little intuitive insight (see appendix A). Therefore, we initially focus on two subcases: the conventional ($\alpha_\mathrm{s} \ll \alpha_\mathrm{m}$) and the reversed case ($\alpha_\mathrm{m} \ll \alpha_\mathrm{s}$), both in the weak coupling regime ($\tilde{g}_0\sqrt{\Phi_\mathrm{p}} \ll \alpha_\mathrm{s} + \alpha_\mathrm{m}$). These examples illustrate how one can formulate guided-wave optomechanics, including the classical stimulated Brillouin regime, in terms of the vacuum coupling rate $\tilde{g}_0$ and cooperativity $\tilde{\mathcal{C}}$.

First, strongly damped phonons ($\alpha_\mathrm{s} \ll \alpha_\mathrm{m}$) act as a localized slave wave ($\partial_z b \to 0$) given by $b = -i\tilde{\chi}_\mathrm{m}\tilde{g}_0 a_\mathrm{s}^\dagger a_\mathrm{p}$. On resonance ($\tilde{\Delta}_\gamma = 0$), we thus have

$$\partial_z a_\mathrm{s} = \mp(1 - \tilde{\mathcal{C}})\frac{\alpha_\mathrm{s}}{2} a_\mathrm{s} \tag{3.9}$$

with

$$\tilde{\mathcal{C}} = \frac{4\tilde{g}_0^2 \Phi_\mathrm{p}}{\alpha_\mathrm{s}\alpha_\mathrm{m}} = \frac{4\tilde{g}^2}{\alpha_\mathrm{s}\alpha_\mathrm{m}} \tag{3.10}$$

the guided-wave cooperativity and $\tilde{g} = \tilde{g}_0\sqrt{\Phi_\mathrm{p}}$ the pump-enhanced spatial coupling rate. Therefore, $\tilde{\mathcal{C}} = 1$ is the threshold for net phonon-assisted gain on flying photons. Since $P_\mathrm{p} = \hbar\omega_\mathrm{p}\Phi_\mathrm{p}$ is the pump power, we obtain $\tilde{\mathcal{C}} = \frac{\tilde{\mathcal{G}} P_\mathrm{p}}{\alpha_\mathrm{s}}$ and

$$\tilde{\mathcal{G}} = \frac{4\tilde{g}_0^2}{\hbar\omega_\mathrm{p}\alpha_\mathrm{m}} \tag{3.11}$$

the well-known Brillouin gain coefficient [36, 37], here framed in terms of a spatial coupling rate $\tilde{g}_0$ and cooperativity $\tilde{\mathcal{C}}$. It characterizes the spatial exponential build-up of a Stokes seed in case of highly damped phonons ($\alpha_\mathrm{s} \ll \alpha_\mathrm{m}$). Since $\langle \Phi_\mathrm{m} \rangle = \frac{\alpha_\mathrm{s}}{\alpha_\mathrm{m}}\tilde{\mathcal{C}}\langle \Phi_\mathrm{s} \rangle \ll \langle \Phi_\mathrm{s} \rangle$, there are on average far fewer phonons than photons flying along the waveguide in this case. The system enters the strong coupling regime as soon as $\tilde{\mathcal{C}} \sim \frac{\alpha_\mathrm{m}}{\alpha_\mathrm{s}}$ (see section 3.4).

Second, when the phononic damping is lowest ($\alpha_\mathrm{m} \ll \alpha_\mathrm{s}$), we similarly get a slaved Stokes wave ($\partial_z a_\mathrm{s} \to 0$) given by $a_\mathrm{s} = -i\tilde{\chi}_\mathrm{s}\tilde{g}_0 b^\dagger a_\mathrm{p}$ resulting in ($\tilde{\Delta}_\gamma = 0$)

$$\partial_z b = -(1 - \tilde{\mathcal{C}})\frac{\alpha_\mathrm{m}}{2} b \tag{3.12}$$

such that $\tilde{\mathcal{C}} = 1$ also yields the threshold for net photon-assisted gain on flying phonons. Since $\langle \Phi_\mathrm{s} \rangle = \frac{\alpha_\mathrm{m}}{\alpha_\mathrm{s}}\tilde{\mathcal{C}}\langle \Phi_\mathrm{m} \rangle \ll \langle \Phi_\mathrm{m} \rangle$, there are far fewer photons than phonons flying along the waveguide in this case. The system enters the strong coupling regime as soon as $\tilde{\mathcal{C}} \sim \frac{\alpha_s}{\alpha_m}$. By replacing the undepleted pump with an undepleted, strong Stokes mode ($\tilde{g} = \tilde{g}_0\sqrt{\Phi_s}$), it follows that an anti-Stokes seed sees larger loss by a factor $(1 + \tilde{\mathcal{C}})$ conventionally ($\alpha_s \ll \alpha_m$) and that a guided-wave phonon channel can be cooled by a factor $(1 + \tilde{\mathcal{C}})$ when it has the lowest propagation loss ($\alpha_m \ll \alpha_s$). An undepleted, strong phononic beam ($\tilde{g} = \tilde{g}_0\sqrt{\Phi_m}$) yields similar coupling between the pump and Stokes wave.

The coupling rate $\tilde{g}$ and the cooperativity $\tilde{\mathcal{C}}$ respect the symmetry between flying photons and phonons, whereas the gain coefficient $\tilde{\mathcal{G}}$ (3.11) is most relevant in case of stronger phonon damping. Therefore, we regard $\tilde{g}$ and $\tilde{\mathcal{C}}$ as more natural and fundamental figures of merit. It is straightforward to extend the above discussion for absorptive decay of the pump flux, i.e. $\Phi_p(z) = \Phi_p(0)e^{-\alpha_p z}$ and non-zero wavevector detunings $\tilde{\Delta}_\gamma \neq 0$.

So far, we discussed two subcases of guided-wave Brillouin scattering. We treat the strong coupling regime in section 3.4 and the full solution in the appendix A. Next, we move on to cavity optomechanics via the mean-field transition.

## 3.3 Bridge to cavity optomechanics

In this section, we transition to an optical cavity – made from a Brillouin-active waveguide – of roundtrip length $L$ (fig.3.1). To do so, we introduce the *mean-field* envelope operators

$$\bar{a}(t) = \frac{1}{L}\int_0^L a(z,t)\mathrm{d}z \tag{3.13}$$

for both the optical ($\bar{a}_{p/s}(t)$) and acoustic ($\bar{b}(t)$) fields. Such mean-field models have found early use in the treatment of fluorescence [174] and recently also in the context of frequency combs [175]. During roundtrip propagation, each field obeys dynamics of the form (see (3.2))

$$v^{-1}\partial_t a + \partial_z a = \zeta - \tilde{\chi}^{-1} a \tag{3.14}$$

with $\zeta$ the nonlinear interaction term. To describe the cavity feedback (fig.1), we add the boundary condition

$$a(0,t) = \sqrt{1-\alpha'}\sqrt{1-\mu}\, e^{i\varphi} a(L,t) + \sqrt{\mu}\, s(t) \tag{3.15}$$

with $\alpha'$ the additional loss fraction along a roundtrip (on top of $\alpha$, such as bending losses), $\mu$ the fraction of photons or phonons coupled to an in- or output channel, $\varphi$ the roundtrip phase shift and $s(t)$ the flux-normalized envelope of injected photons or phonons. By Taylor-expansion of (3.15), we get

$$a(L,t) - a(0,t) \approx \left(\frac{\alpha' + \mu}{2} - i\varphi\right)\bar{a}(t) - \sqrt{\mu}\, s(t) \tag{3.16}$$

with higher-order terms negligible and $a(L,t) \approx \bar{a}(t)$ in the high-finesse limit. Low-finesse situations, particularly relevant for phonons, are treated further on (see (3.24)). We operate close to the cavity resonance, such that $\varphi \ll 2\pi$. Next, we let (3.13) operate on (3.14) and use $\overline{\partial_t a} = \dot{\bar{a}}(t)$:

$$v^{-1}\dot{\bar{a}}(t) + L^{-1}\{a(L,t) - a(0,t)\} = \bar{\zeta}(t) - \tilde{\chi}^{-1}\bar{a}(t) \tag{3.17}$$

We insert (3.16) in (3.17) and find

$$\dot{\bar{a}} = v\bar{\zeta} - \chi^{-1}\bar{a} + \frac{\sqrt{\mu}}{T}s \tag{3.18}$$

with $\chi^{-1} = \frac{\kappa}{2} - i\Delta$ the cavity's photonic or phononic response function, $\kappa = \kappa_{\mathrm{i}} + \kappa_{\mathrm{c}}$ the total decay rate, $\kappa_{\mathrm{i}} = \frac{\alpha' + \alpha L}{T}$ the intrinsic decay rate, $\kappa_{\mathrm{c}} = \frac{\mu}{T}$ the coupling rate, $\Delta = \frac{\varphi + \tilde{\Delta} L}{T}$ the detuning and $T = \frac{L}{v}$ the roundtrip time.

Next, we multiply (3.18) by $\sqrt{T}$ and switch from flux- to number-normalized fields ($\bar{a} \mapsto \sqrt{T}\bar{a}$):

$$\dot{\bar{a}} = v\sqrt{T}\,\bar{\zeta} - \chi^{-1}\bar{a} + \sqrt{\kappa_{\mathrm{c}}}s \tag{3.19}$$

From here on, $n = \bar{a}^\dagger\bar{a}$ represents the number of quanta in the cavity, while $s^\dagger s$ still corresponds to the injected photon or phonon flux. The transition from (3.14) to (3.19) still holds when we replace $z \mapsto -z$ because condition (3.16) also reverses. Therefore, potential dynamical differences between forward and backward scattering disappear in a high-finesse traveling-wave cavity – at least in the minimal model (3.2) of guided-wave optomechanics.

Comparing (3.2) to (3.14), we see that $\zeta \propto fg$ with $f$ and $g$ equal to $a_{\mathrm{p/s}}$ or $b$. In the mean-field approximation, we assume these envelopes vary little over a roundtrip such that $\overline{fg} = \bar{f}\,\bar{g}$ holds (see appendix A). Finally, we apply the mean-field (3.14)-to-(3.19) transition to (3.2). Hence, an optomechanical cavity – constructed from a Brillouin-active waveguide – is governed by

$$\begin{aligned}
\dot{\bar{a}}_{\mathrm{p}} &= -ig_0\bar{a}_{\mathrm{s}}\bar{b} - \chi_{\mathrm{p}}^{-1}\bar{a}_{\mathrm{p}} + \sqrt{\kappa_{\mathrm{cp}}}s_{\mathrm{p}} \\
\dot{\bar{a}}_{\mathrm{s}} &= -ig_0\bar{b}^\dagger\bar{a}_{\mathrm{p}} - \chi_{\mathrm{s}}^{-1}\bar{a}_{\mathrm{s}} + \sqrt{\kappa_{\mathrm{cs}}}s_{\mathrm{s}} \\
\dot{\bar{b}} &= -ig_0\bar{a}_{\mathrm{s}}^\dagger\bar{a}_{\mathrm{p}} - \chi_{\mathrm{m}}^{-1}\bar{b} + \sqrt{\kappa_{\mathrm{cm}}}s_{\mathrm{m}}
\end{aligned} \tag{3.20}$$

with

$$\boxed{\sqrt{\frac{v_{\mathrm{p}}v_{\mathrm{s}}v_{\mathrm{m}}}{L}}\tilde{g}_0 = g_0} \tag{3.21}$$

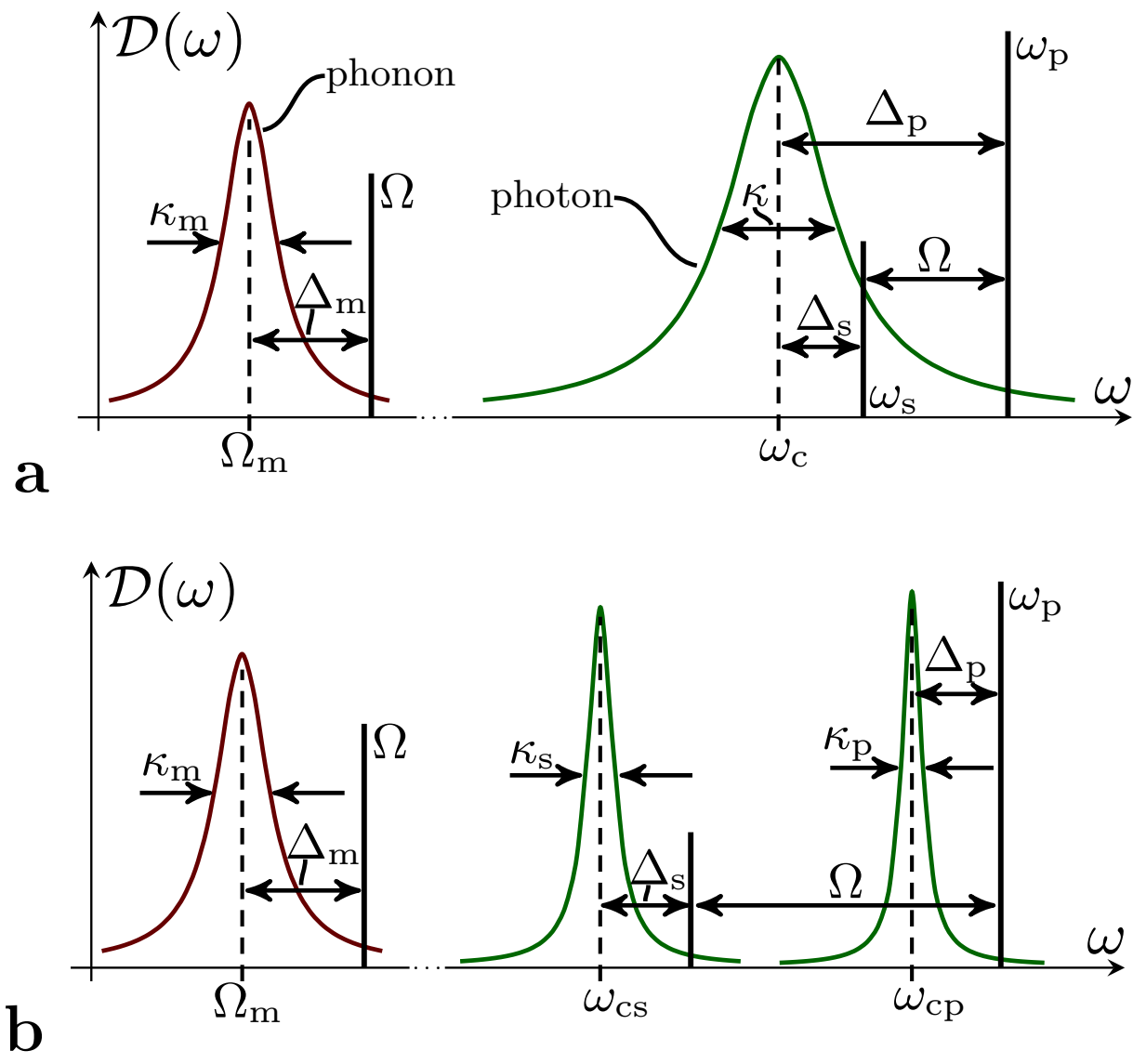

**Figure 3.3: Cavity description.** The photonic and phononic density of states $\mathcal{D}(\omega)$. The mean-field equations (3.20) describe coupling between one phononic and either one (**a**) or two (**b**) photonic resonances. The latter case (**b**) is most power-efficient, although hard to achieve in practice [136].

the well-known temporal zero-point coupling rate [35]. Indeed, equations (3.20) are equivalent (see appendix A) to the Heisenberg equations of motion resulting from the well-known Hamiltonian $\hat{\mathcal{H}} = \hbar\omega_c(\hat{x})\hat{a}^\dagger\hat{a} + \hbar\Omega_m\hat{b}^\dagger\hat{b}$ [35]. Remarkably, the equivalence holds even for inter-modal and backward scattering. The connection (3.21) between the traveling-wave and the cavity-based vacuum coupling rates $\tilde{g}_0$ and $g_0$ is at the heart of this chapter: other links such as (3.1) are based on this result. Further, the mean-field transition transforms the guided-wave commutator (3.3) into

$$\left[\bar{a}_\gamma, \bar{a}_{\gamma'}^\dagger\right] = \frac{\sqrt{v_\gamma v_{\gamma'}}}{L^2}\delta_{\gamma\gamma'}\int_0^L\int_0^L \mathrm{d}z\mathrm{d}z'\delta(z-z') = \frac{\sqrt{v_\gamma v_{\gamma'}}}{L}\delta_{\gamma\gamma'} \tag{3.22}$$

and through rescaling $\bar{a}_\gamma$ by $\sqrt{T_\gamma}$ into

$$\left[\bar{a}_\gamma, \bar{a}_{\gamma'}^\dagger\right] = \delta_{\gamma\gamma'} \tag{3.23}$$

thus correctly retrieving the standard harmonic oscillator commutators [35].

To derive (3.20), we made the same mean-field transition for photons and phonons. In particular, this supposes a large phonon finesse $\mathcal{F}_m = \frac{2\pi}{\kappa_m T_m} \gg 1$. Often there is only intrinsic phonon loss such that $\kappa_m = v_m\alpha_m$ and thus this requires $\frac{2\pi}{\alpha_m L} \gg 1$. In many systems, the phonon decay length $\alpha_m^{-1}$ is much shorter than the roundtrip length $L$. Then this phonon high-finesse limit does not hold. However, we can completely neglect phonon propagation ($\partial_z b \to 0$ in (3.2)) if $\alpha_{\mathrm{m}}$ is sufficiently large. The phonons' envelope operator $b$ then obeys

$$v_{\mathrm{m}}^{-1}\partial_t b = -i\tilde{g}_0 a_{\mathrm{s}}^{\dagger} a_{\mathrm{p}} - \tilde{\chi}_{\mathrm{m}}^{-1} b$$

Applying (3.13), multiplying by $\sqrt{T_{\mathrm{m}}}$ and switching from flux- to number-normalized envelopes results in

$$\dot{\bar{b}} = -i g_0 \bar{a}_{\mathrm{s}}^{\dagger} \bar{a}_{\mathrm{p}} - v_{\mathrm{m}} \tilde{\chi}_{\mathrm{m}}^{-1} \bar{b} \tag{3.24}$$

where we used (3.21). Hence, this localized low-phonon-finesse approach yields the same result as the previous high-finesse limit with $s_{\mathrm{m}} = 0$ (see (3.20)). Therefore, even low-finesse phonons produce the same dynamics as is commonly studied in cavity optomechanics [35].

Notably, the standard treatment of cavity optomechanics [35] does not consider an explicit space variable: the Hamiltonian $\hat{\mathcal{H}}$ performs an implicit spatial average by describing the entire object as single mechanical oscillator, in contrast to the explicit spatial average (3.13) performed in this chapter. However, even the implicit average in $\hat{\mathcal{H}}$ requires low-loss acoustic excitations to set up a global mechanical mode self-consistently, precisely as in the high-finesse approximation leading to (3.20). In the localized, low-finesse phonon approach that generates (3.24), the spatial averaging can still be performed and yields the same classical dynamics – but its meaning changes. Now ($\mathcal{F}_{\mathrm{m}} < 1$) the acoustic wave is too lossy to set up a global mechanical mode for the entire cavity. Instead, the cavity consists of an ensemble of independent Raman-like mechanical oscillators. It is no longer possible to address phonons circulating in the cavity.

Finally, we combine (3.21) and (3.11). Using $\alpha_{\mathrm{m}} v_{\mathrm{m}} = \Omega_{\mathrm{m}}/Q_{\mathrm{m}}$, we obtain result (3.1) immediately. Note that $Q_{\mathrm{m}}$ is defined here as the waveguide's intrinsic phonon quality factor, which could be different from the cavity's phonon quality factor if there were e.g. non-negligible phonon coupling or bending losses. In case of doubt, it is safe to alternatively write (3.1) as

$$v_{\mathrm{p}} v_{\mathrm{s}} v_{\mathrm{m}} \frac{(\hbar\omega_{\mathrm{p}})\,\alpha_{\mathrm{m}}}{4L} \tilde{\mathcal{G}} = g_0^2 \tag{3.25}$$

Both $\tilde{\mathcal{G}}$ and $g_0$ are well-established in the study of photon-phonon interaction, but they operate on different levels. The Planck constant $\hbar$ enters (3.1) because $\tilde{\mathcal{G}}$ is classical while $g_0$ is inherently quantum mechanical. In addition, $\tilde{\mathcal{G}}$ quantifies the combined action of forces and scattering and contains the phonon loss – while $g_0$ does not. Further, larger $L$ yields a smaller $g_0$ while $\tilde{\mathcal{G}}$ is length-independent. Therefore, $g_0^2 \propto \frac{\hbar}{L}\frac{\tilde{\mathcal{G}}}{Q_{\mathrm{m}}}$. This mean-field derivation is but one way to prove the $\tilde{\mathcal{G}} \leftrightarrow g_0$ conversion, other approaches yield the same result (see appendix A). This proof captures all reversible photon-phonon coupling mechanisms.

## 3.4 Symmetry between circuit and cavity optomechanical effects

In this section, we describe both guided-wave and cavity-based regimes of photon-phonon coupling. To begin with, we recover and briefly review the known cavity-based regimes of photon lasing, phonon lasing and strong coupling. Next, we map these regimes on the guided-wave spatial evolution of the opto-acoustic fields. The mapping unveils two unobserved regimes of guided-wave Brillouin scattering. We pay particular attention to the strong coupling regime ($\tilde{g} \gg \alpha_{\mathrm{s}} + \alpha_{\mathrm{m}}$).

Here, we assume zero photon and phonon input flux and an undepleted pump. Then (3.20) reduces to

$$\begin{aligned} \dot{\overline{a}}_{\mathrm{s}} &= -ig_0\overline{b}^{\dagger}\overline{a}_{\mathrm{p}} - \chi_{\mathrm{s}}^{-1}\overline{a}_{\mathrm{s}} \\ \dot{\overline{b}} &= -ig_0\overline{a}_{\mathrm{s}}^{\dagger}\overline{a}_{\mathrm{p}} - \chi_{\mathrm{m}}^{-1}\overline{b} \end{aligned} \tag{3.26}$$

These equations treat the photons and phonons identically. Therefore, every photonic phenomenon must have a phononic counterpart and vice versa. Even more, the *temporal* cavity dynamics (3.26) can be mapped ($t \mapsto z$) on the *spatial* steady-state waveguide evolution (3.8). Each effect known from cavities therefore has a waveguide counterpart (but not vice versa as we will see). This also implies that the spatial figures of merit have a temporal symmetry partner and vice versa; we prepared for this at the end of section 3.2 by defining a guided-wave vacuum coupling rate $\tilde{g}_0$ and cooperativity $\tilde{\mathcal{C}}$. To clearly expose these symmetries, we solve (3.26); keeping in mind that the very same discussion holds spatially for (3.8). First, we decouple (3.26) and get

$$\left(\frac{\mathrm{d}}{\mathrm{d}t} + \chi_{\mathrm{s}}^{-\star}\right)\left(\frac{\mathrm{d}}{\mathrm{d}t} + \chi_{\mathrm{m}}^{-1}\right)\overline{b}(t) = g^2\overline{b}(t) \tag{3.27}$$

Here, we introduced the pump-enhanced coupling rate $g = g_0\sqrt{n_{\mathrm{p}}}$. Next, we insert the ansatz $\overline{b} \propto e^{\gamma t}$ in (3.27) and find two roots $\gamma_{\pm}$

$$\gamma_{\pm} = \frac{1}{2}\left\{-\left(\chi_{\mathrm{s}}^{-\star} + \chi_{\mathrm{m}}^{-1}\right) \pm \sqrt{\left(\chi_{\mathrm{s}}^{-\star} - \chi_{\mathrm{m}}^{-1}\right)^2 + 4g^2}\right\} \tag{3.28}$$

In general, these roots strongly mix the photon and phonon response: the photon-phonon pair forms a polariton [56, 90, 169, 171, 176]. The guided-

wave analog of (3.27) is

$$\left(\partial_z \pm \tilde{\chi}_s^{-\star}\right)\left(\partial_z + \tilde{\chi}_m^{-1}\right) b(z) = \pm \tilde{g}^2 b(z) \tag{3.29}$$

and it can be treated identically. The full spatial and temporal dynamics is governed by the general solution (3.28) (see appendix A). However, it is more instructive to consider the limiting cases of weak and strong photon-phonon interaction relative to the system's damping.

First, if the photon-phonon interaction is sufficiently weak, i.e. $g \ll |\kappa_s - \kappa_m|$, the two roots in (3.28) disconnect. Usually, the photon and phonon decay rates differ significantly. Then there are two scenario's depending on the relative photonic and phononic decay rates. Essentially, the dynamics of the short-lived particle can be adiabatically eliminated, although it may still strongly modify the response of its long-lived partner.

In particular, when the phonons decay slowly ($\kappa_m \ll \kappa_s$), the photonic response is barely modified: $\dot{\bar{a}}_s \to 0$ and therefore $\bar{a}_s = -i\chi_s g_0 \bar{b}^\dagger \bar{a}_p$ to a good approximation. However, the phononic response can then dramatically change to $\chi_m^{-1} - g^2\chi_s^\star$. Hence, we recover the optical spring effect ($\delta\Omega_m = g^2\Im\chi_s^\star$) and phonon lasing ($\delta\kappa_m = -2g^2\Re\chi_s^\star$) [35]. At the photon resonance ($\Delta_s = 0$), the phonon linewidth equals $\kappa_m + \delta\kappa_m = (1 - \mathcal{C})\,\kappa_m$ with $\mathcal{C} = \frac{4g^2}{\kappa_s\kappa_m}$ the cooperativity. Therefore, the threshold for mechanical lasing is $\mathcal{C} = 1$. This instability was first contemplated by Braginsky [62] in the context of gravitational wave detection and received further study in systems ranging from gram-scale mirrors [177] to optomechanical crystals [178, 179]. Since $\langle n_s \rangle = \frac{\kappa_m}{\kappa_s}\mathcal{C}\langle n_m \rangle \ll \langle n_m \rangle$, there are far fewer Stokes photons than phonons in the cavity in this situation. The system enters the strong coupling regime as soon as $\mathcal{C} \sim \frac{\kappa_s}{\kappa_m}$.

Similarly, when the photons decay slowly ($\kappa_s \ll \kappa_m$), the phononic response is barely modified: $\dot{\bar{b}} \to 0$ and therefore $\bar{b} = -i\chi_m g_0 \bar{a}_s^\dagger \bar{a}_p$ to a good approximation. However, the photonic response can then dramatically change to $\chi_s^{-1} - g^2\chi_m^\star$. Hence, we recover the cavity frequency pull ($\delta\omega_{cs} = g^2\Im\chi_m^\star$) and photon lasing ($\delta\kappa_s = -2g^2\Re\chi_m^\star$) [37, 180, 181]. At the phonon resonance ($\Delta_m = 0$), the Stokes linewidth equals $\kappa_s + \delta\kappa_s = (1 - \mathcal{C})\,\kappa_s$ with $\mathcal{C}$ the same temporal cooperativity as above. Therefore, the threshold for Brillouin lasing is also $\mathcal{C} = 1$. First realized in fibers [182], this case was recently also studied in $CaF_2$ resonators [140], silica disks [136] and chalcogenide waveguides [141]. Such lasers are known for their excellent spectral purity [181, 183] and received attention for quantum-limited amplification [142]. Since $\langle n_m \rangle = \frac{\kappa_s}{\kappa_m}\mathcal{C}\langle n_s \rangle \ll \langle n_s \rangle$, there are far fewer phonons than Stokes photons in the cavity in this situation. The system enters the strong coupling regime as soon as $\mathcal{C} \sim \frac{\kappa_m}{\kappa_s}$.

Further, if the photon-phonon coupling rate is sufficiently strong, (3.27) simplifies to $\ddot{\bar{b}} = g^2\bar{b}$. An identical derivation yields $\ddot{\bar{b}} = -g^2\bar{b}$ if the Stokes wave is considered undepleted. Therefore, a red-detuned probe produces entangled photon-phonon pair generation ($\bar{b}(t) \propto e^{\pm g t}$), whereas a blue-detuned probe produces Rabi flopping between photons and phonons ($\bar{b}(t) \propto e^{\pm i g t}$) [35]. A situation of equally strong optical and mechanical damping ($\kappa_\mathrm{s} \approx \kappa_\mathrm{m}$) invalidates the above weak coupling treatment even for small $g$. However, this is not sufficient to see strong coupling behavior. From the general solution (see appendix A), this requires $g \gg \kappa_\mathrm{s} + \kappa_\mathrm{m}$. Indeed, in the strong coupling regime the hybridized photon-phonon polariton sees half the optical and half the mechanical damping [56]. Therefore, the state-swap frequency $\frac{g}{\pi}$ must be high compared to the average decay rate $\frac{\kappa_\mathrm{s}+\kappa_\mathrm{m}}{2}$ to observe an actual Rabi swap before the population decreases by $1/e$.

By comparing (3.26) and (3.8), we prove an analogy between spatial and temporal optomechanical effects (fig.3.2): the above cavity-based discussion still largely holds for guided-wave optomechanics with the mapping $g_0 \mapsto \tilde{g}_0$, $\mathcal{C} \mapsto \tilde{\mathcal{C}}$, $\kappa_\mathrm{s} \mapsto \alpha_\mathrm{s}$, $\kappa_\mathrm{m} \mapsto \alpha_\mathrm{m}$ and $n \mapsto \Phi$. In case of all co-propagating waves, and in the absence of cascading [94, 152] and noise [95, 96], the mapping of cavity optomechanics onto a Brillouin-active waveguide in steady-state is an exact equivalence. However, when for instance one of the particles counter-propagates, such as the Stokes photons in backward scattering, important differences arise that have no equivalent in cavity optomechanics. Indeed, as proven in section 3.3, information regarding the propagation direction of the waves disappears in the mean-field transition. Instead comparing (3.27) and (3.29), much can still be learned by instead mapping $g_0^2 \mapsto -\tilde{g}_0^2$ and $\kappa_\mathrm{s} \mapsto -\alpha_\mathrm{s}$. Note that this particular difference disappears if the counter-propagating particle species is undepleted: then it vanishes from the dynamics and the situation is identical to the co-propagating case.

Thus, guided-wave weak coupling requires $\tilde{g} \ll |\alpha_\mathrm{s} \mp \alpha_\mathrm{m}|$ with $\tilde{g} = \tilde{g}_0\sqrt{\Phi_\mathrm{p}}$ the spatial coupling rate (see appendix A). Under weak coupling, there are two cases depending on the relative photon and phonon propagation losses. We have touched upon these subcases at the end of section 3.2 and briefly consider them again here to show the similarity with cavity-optomechanical effects. First, when the phonons propagate far ($\alpha_\mathrm{m} \ll \alpha_\mathrm{s}$), the photonic loss $\alpha_\mathrm{s}$ barely changes. However, the phononic response can then drastically change to $\tilde{\chi}_\mathrm{m}^{-1} - \tilde{g}^2\tilde{\chi}_\mathrm{s}^\star$, which includes a shift in both the phononic propagation loss ($\delta\alpha_\mathrm{m} = -2\tilde{g}^2\Re\tilde{\chi}_\mathrm{s}^\star$) and group velocity ($\propto \Im\tilde{\chi}_\mathrm{s}^\star$); i.e. traveling-phonon amplification and light-induced slowing down of sound. In section 3.5, we show that this unobserved regime can be achieved in existing systems.

Second, when the Stokes photons propagate far ($\alpha_\mathrm{s} \ll \alpha_\mathrm{m}$), the phononic loss $\alpha_\mathrm{m}$ barely changes. However, the photonic response can then drastically change to $\tilde{\chi}_\mathrm{s}^{-1} - \tilde{g}^2\tilde{\chi}_\mathrm{m}^\star$. Hence, we are back in the conventional domain of phonon-assisted amplification of traveling photons ($\delta\alpha_\mathrm{s} = -2\tilde{g}^2\Re\tilde{\chi}_\mathrm{m}^\star$) and sound-induced slowing down of light ($\propto \Im\tilde{\chi}_\mathrm{m}^\star$) [41]. At resonance ($\tilde{\Delta}_\mathrm{m} = 0$), the Stokes propagation loss is $(1-\tilde{\mathcal{C}})\alpha_s$ as in (3.9).

If the coupling is sufficiently strong compared to the propagation losses ($\tilde{g} \gg \alpha_s + \alpha_m$), (3.29) simplifies to $\partial_z^2 b = \pm\tilde{g}^2 b$ (see appendix A). In the forward (+) case, and with boundary condition $b(0) = 0$, this yields

$$\begin{aligned} a_s(z) &= a_s(0)\cosh\tilde{g}z \\ b(z) &= -ia_s^\dagger(0)\sinh\tilde{g}z \end{aligned} \tag{3.30}$$

such that $\Phi_s(z) = \Phi_s(0)\cosh^2\tilde{g}z$ and $\Phi_m(z) = \Phi_s(0)\sinh^2\tilde{g}z$. Therefore, $\Phi_s(z) - \Phi_m(z) = \Phi_s(0)$ and $\partial_z\Phi_s = \partial_z\Phi_m$ as required by the Manley-Rowe relations (3.4). In the backward (−) case, with $L$ the waveguide length and boundary condition $b(0) = 0$, the evolution along the waveguide has no cavity-optomechanics counterpart. Specifically, we retrieve

$$\begin{aligned} a_s(z) &= \frac{a_s(L)}{\cos\tilde{g}L}\cos\tilde{g}z \\ b(z) &= -i\frac{a_s^\dagger(L)}{\cos\tilde{g}L}\sin\tilde{g}z \end{aligned} \tag{3.31}$$

such that $\Phi_s(z) = \frac{\Phi_s(L)}{\cos^2\tilde{g}L}\cos^2\tilde{g}z$ and $\Phi_m(z) = \frac{\Phi_s(L)}{\cos^2\tilde{g}L}\sin^2\tilde{g}z$. Therefore, $\Phi_s(z) + \Phi_m(z) = \frac{\Phi_s(L)}{\cos^2\tilde{g}L}$ and $-\partial_z\Phi_s = \partial_z\Phi_m$ as required by Manley-Rowe (3.4). The system has an instability at $\tilde{g}L = \frac{\pi}{2}$, which is reached before a full state swap between light and sound can be completed. This situation is called *contraflow Hermitian coupling* in classifications of coupled-mode interactions [184, 185]. In case of anti-Stokes (instead of Stokes) seeding in the strong coupling regime, an identical derivation leads to $\partial_z^2 b = -\tilde{g}^2 b$ – which produces the same Rabi oscillations for forward and backward scattering. Although familiar in resonators [35], these strong-coupling effects have not yet been observed in the field of guided-wave Brillouin scattering; see section 3.5 for the prospects.

We conclude this section by analyzing the relation between the guided-wave and cavity-based cooperativities ($\tilde{\mathcal{C}}$ and $\mathcal{C}$) and by introducing a gain coefficient ($\mathcal{G}$) for an optomechanical cavity. Note that the temporal cooperativity

$$\mathcal{C} = \frac{4g^2}{\kappa_s\kappa_m} \tag{3.32}$$

is the ratio between the roundtrip gain and loss: inserting $g^2 = g_0^2 n_p$, $n_p = \frac{P_p T_p}{\hbar\omega_p}$ and (3.21) in (3.32) indeed leads to

$$\mathcal{C} = \frac{\tilde{\mathcal{G}}P_p}{\frac{\kappa_s}{v_s}}\frac{v_m\alpha_m}{\kappa_m} = \frac{\tilde{\mathcal{G}}P_p L}{\kappa_s T_s}\frac{v_m\alpha_m}{\kappa_m} \tag{3.33}$$

with $P_\mathrm{p}$ the intracavity pump power and $\frac{v_\mathrm{m}\alpha_\mathrm{m}}{\kappa_\mathrm{m}}$ a naturally appearing correction factor that allows for higher phonon losses, so effectively lower $\mathcal{C}$, in the cavity than in the waveguide. This directly shows that

$$\tilde{\mathcal{C}} \geq \mathcal{C} \tag{3.34}$$

given (3.33), $\kappa_\gamma \geq v_\gamma \alpha_\gamma$ and $\tilde{\mathcal{C}} = \frac{\tilde{\mathcal{G}} P_\mathrm{p}}{\alpha_\mathrm{s}}$. Clearly, the guided-wave cooperativity exceeds the cavity-based cooperativity since the cavity has additional dissipation (e.g. coupling and bending losses). Finally, we define a gain coefficient $\mathcal{G}$ for a cavity in analogy to (3.11)

$$\mathcal{G} = \frac{4g_0^2}{\hbar\omega_\mathrm{p}\kappa_\mathrm{m}} \tag{3.35}$$

which characterizes the temporal exponential build-up of the Stokes when the phonons are heavily damped. The gain coefficients therefore obey

$$\tilde{\mathcal{G}} \geq \frac{L}{v_\mathrm{p} v_\mathrm{s}} \mathcal{G} \tag{3.36}$$

given $\kappa_\mathrm{m} \geq v_\mathrm{m}\alpha_\mathrm{m}$ and (3.21). Hence, the guided-wave and cavity-based optomechanical figures of merit are now conceptually similar and the relations between each of them were given in (3.1), (3.21), (3.34) and (3.36).

## 3.5 Prospects

In this section, we first give a couple of examples of how the $\tilde{\mathcal{G}} \leftrightarrow g_0$ connection (3.1) can be implemented – including several systems in which it can be tested empirically. Next, we move on to the prospects for observing new regimes of guided-wave optomechanics, simultaneously illustrating the application of our framework. Finally, we briefly discuss potential extensions to the minimal model (3.2) of traveling-wave Brillouin scattering.

Table 3.1 presents four implementations of the conversion from the gain coefficient $\tilde{\mathcal{G}}$ to the vacuum coupling rate $g_0$ ($\tilde{\mathcal{G}} \rightarrow g_0$) and four in reverse ($\tilde{\mathcal{G}} \leftarrow g_0$). The systems range from silicon nanowires and dual-web fibers to ultracold atom clouds and GaAs disks. In five cases, such as for silicon nanowires, the conversion can clearly be tested empirically by measuring $\tilde{\mathcal{G}}$ and $g_0$ through independent, established methods [35, 64]. In three cases, the conversion is hypothetical but still allows for comparison of the photon-phonon interaction strengths. For instance, an ultracold atom cloud in a Fabry-Pérot cavity [168] has no obvious traveling-wave equivalent. Nevertheless, its hypothetical waveguide partner would have a large gain coefficient of $\sim 10^8\,\mathrm{W^{-1}m^{-1}}$ – which compares favorably to optomechanical waveguides realized to date.

So far, all Brillouin-active waveguides had far lower phonon than photon propagation lengths ($\alpha_m^{-1} \ll \alpha_s^{-1}$). Cavity-optomechanical systems, by contrast, more often than not had far lower phonon than photon damping rates ($\kappa_m \ll \kappa_s$) [35]. Only uniquely high-optical-quality [136, 140, 141, 180, 186, 187] systems succeed at reversing the latter hierarchy ($\kappa_m \gg \kappa_s$). The common reversal of this damping hierarchy (going from waveguides to cavities) stems from the small phonon group velocities ($v_m \ll v_s$).

The question naturally arises if waveguides with larger phonon than photon propagation length ($\alpha_m^{-1} \gg \alpha_s^{-1}$) can be made, while still keeping high cooperativities $\tilde{\mathcal{C}} = \frac{4\tilde{g}^2}{\alpha_s \alpha_m} \sim 1$. Currently, the largest phonon decay lengths are of the order $\alpha_m^{-1} \sim 100\,\mu$m in backward Brillouin scattering [64, 98]. To realize larger phonon propagation lengths, one must look for waveguides with large acoustic group velocities $v_m$ and small linewidths $\kappa_m$. Thus, one promising approach uses low-frequency flexural modes ($\Omega \propto q^2$) in backward mode (large $q$) at low temperatures (large $Q_m$). Indeed, then we have both large $v_m \propto K$ and small $\kappa_m = \frac{\Omega_m}{Q_m} \sim \frac{10^7}{10^4}$ Hz $= 1$ kHz [56, 188, 189]. Since $\alpha_m^{-1} = \frac{v_m}{\kappa_m}$ in a waveguide (where there is only propagation loss), we find that decay lengths up to $\alpha_m^{-1} \sim 10$ m are feasible given $v_m \sim 10^4$ m/s and $\kappa_m \sim 1$ kHz. Such a small phonon propagation loss would strongly boost the cooperativity $\tilde{\mathcal{C}}$, which could compensate for a potentially lower coupling rate $\tilde{g}$ in backward mode. Clearly, nothing intrinsically forbids amplification of traveling phonons in systems such as the dual-web fiber [147] – where $\alpha_s^{-1} \sim 10$ cm. Besides its scientific interest, such a traveling-phonon amplifier may be useful in phonon networks [139, 162, 163, 190, 191].

Next, we look into achieving the strong coupling regime in the typical situation of high acoustic loss ($\alpha_m \gg \alpha_s$). To see traveling-wave Rabi flopping, entangled photon-phonon pair production or contraflow Hermitian coupling (see section 3.4), one must obtain $\tilde{g} > \alpha_m$ or equivalently $\tilde{\mathcal{C}} > \frac{\alpha_m}{\alpha_s}$. In optical fibers, in backward mode and given $\alpha_s^{-1} \sim 10$ km and $\alpha_m^{-1} \sim 100\,\mu$m, this requires $\tilde{\mathcal{C}} > 10^8$. This necessitates an unrealistic continuous-wave pump power of $P_p > 10^8 \frac{\alpha_s}{\tilde{\mathcal{G}}} = 10$ kW with $\tilde{\mathcal{G}} \sim 1\,\text{W}^{-1}\text{m}^{-1}$.

In contrast, silicon chips can produce significantly lower $\frac{\alpha_m}{\alpha_s}$ ratios and therefore ease the condition on $\tilde{\mathcal{C}}$ for strong coupling. One can expect phonon propagation distances up to $\alpha_m^{-1} \sim 1$ mm, as these are readily achieved in surface-acoustic-wave devices [192]. Together with $\alpha_s^{-1} \sim 1$ cm [64], this yields $\tilde{\mathcal{C}} > 10$ as the strong coupling condition, which requires a reasonable pump power of $P_p > 10 \frac{\alpha_s}{\tilde{\mathcal{G}}} = 100$ mW with $\tilde{\mathcal{G}} \sim 10^4\,\text{W}^{-1}\text{m}^{-1}$ [64, 65]. Indeed, current nanoscale silicon systems have already demonstrated $\tilde{\mathcal{C}} \approx 2$ [65, 94]. Hence, taking into account the rapid progress in state-of-the-art devices [64, 65, 94, 132], we expect demonstrations of traveling-phonon amplification and spatial strong coupling in the coming years. Such observations would open up entirely new realms of optomechanics.

Finally, this chapter can be extended on several fronts. First, the mean-field transition can be applied to the noise models of [95, 96]. Second, the regime of *nonlinear quantum optomechanics* [35, 193–195] should be transferred to waveguides. This requires that strong coupling is reached for merely one pump photon ($\Phi_{\mathrm{p}} \mapsto 1\,\mathrm{s}^{-1}$): $\tilde{g} = \tilde{g}_0\sqrt{\Phi_{\mathrm{p}}} = \tilde{g}_0 > \alpha_{\mathrm{s}} + \alpha_{\mathrm{m}}$. As $\alpha_{\mathrm{m}} \gg \alpha_{\mathrm{s}}$ usually, traveling-wave nonlinear quantum optomechanics is achieved when $\tilde{g}_0 > \alpha_{\mathrm{m}}$. Third, the coupling between the phononic mode and the thermal bath [95, 96] must be treated carefully to obtain truly quantum-coherent [56] coupling. And fourth, we focused mainly on the dynamics that optomechanical waveguides and cavities have in common, but wisdom may be found in the differences as well. We gave the example of contraflow strong coupling in section 3.4. In addition, the cavity has input fluxes that have no equivalent in a typical guided-wave set-up, while the waveguide can display spatiotemporal effects (both $\partial_z$ and $\partial_t$ in (3.2)) that are absent in a cavity (only $\partial_t$ in (3.20)). On top of this, the cavity breaks the symmetry between Stokes and anti-Stokes scattering, whereas this symmetry prevents exponential build-up of noise in low-dispersion forward intra-modal scattering [95]. It has also yet to be determined whether different cavity dynamics results in the medium-finesse case, as section 3.3 was limited to a low or high finesse.

Notably, with slight modifications, (3.2) also captures Raman scattering [37, 134, 196]. For instance, the phonon frequency is much larger so an optical phase-mismatch can arise. Still, equation (3.1) should hold with $\tilde{\mathcal{G}}$ the Raman gain coefficient. Therefore, this chapter also applies to guided-wave [37, 197–199], cavity-based [15, 200–202] and surface-enhanced Raman scattering [51].

| | $\tilde{\mathcal{G}}$ [W$^{-1}$m$^{-1}$] | $\overset{\longleftrightarrow}{(3.1)}$ | $\frac{g_0}{2\pi}$ [Hz] | $\frac{\Omega_\mathrm{m}}{2\pi}$ [Hz] | $Q_\mathrm{m}$ [−] | $L$ [$\mu$m] | $n_\mathrm{g}$ [−] | $\lambda$ [$\mu$m] |
|---|---|---|---|---|---|---|---|---|
| Silicon nanowire [64, 65] | $10^4$ | $\xrightarrow{\star}$ | $\frac{1.5\cdot10^6}{\sqrt{L\,[\mu\mathrm{m}]}}$ | $10^{10}$ | $10^3$ | – | 4.6 | 1.55 |
| Silica standard fiber [37] | 1 | $\xrightarrow{\star}$ | $\frac{70}{\sqrt{L\,[\mathrm{cm}]}}$ | $10^{10}$ | 500 | – | 1.45 | 1.55 |
| Silica dual-web fiber [147] | $4\cdot10^6$ | $\xrightarrow{\star}$ | $\frac{3\cdot10^3}{\sqrt{L\,[\mathrm{cm}]}}$ | $6\cdot10^6$ | $4\cdot10^4$ | – | 1.7 | 1.55 |
| Chalcogenide rib [97, 203] | $3\cdot10^2$ | $\xrightarrow{\star}$ | $\frac{7\cdot10^5}{\sqrt{L\,[\mu\mathrm{m}]}}$ | $8\cdot10^9$ | 230 | – | 2.6 | 1.55 |
| Silica microtoroid [56] | 600 | $\longleftarrow$ | $3\cdot10^3$ | $8\cdot10^7$ | $2\cdot10^4$ | 97 | 1.45 | 0.78 |
| Silicon optomechanical crystal [179] | $4\cdot10^4$ | $\xleftarrow{\star}$ | $6\cdot10^5$ | $6\cdot10^9$ | $2\cdot10^3$ | 5 | 5 | 1.55 |
| Rb ultracold atom cloud [168] | $10^8$ | $\longleftarrow$ | $6\cdot10^5$ | $4\cdot10^4$ | 42 | 400 | 1 | 0.78 |
| GaAs optomechanical disk [204] | $5\cdot10^4$ | $\longleftarrow$ | $3\cdot10^5$ | $10^9$ | $2\cdot10^3$ | 8 | 4 | 1.55 |

**Table 3.1: Translation between waveguides and cavities.** The table contains four conversions from a gain coefficient to a vacuum coupling rate and four in reverse. The right five columns contain the parameters necessary for the conversion through formula (3.1). These are order-of-magnitude estimates. In some cases, indicated by a $\star$, the conversion can be empirically tested. In other situations, the conversion is hypothetical: e.g. an ultracold atom cloud in a Fabry-Pérot cavity has no obvious guided-wave equivalent.

## 3.6 Conclusion

In conclusion, we unveiled a connection between Brillouin-active waveguides and optomechanical cavities. The link between the Brillouin gain coefficient $\tilde{\mathcal{G}}$ and the zero-point coupling rate $g_0$ was derived in a platform-independent way. As illustrated for silicon nanowires and ultracold atom clouds, it significantly expands the variety of systems whose photon-phonon interaction efficiency can be compared. Through the mean-field transition, we derived the dynamics of optomechanical cavities from that of Brillouin-active waveguides. We framed the behavior of both systems in terms of cooperativities and vacuum coupling rates. Next, we showed that phenomena familiar from cavity optomechanics all have guided-wave partners, but not the other way around. The broader theory predicts that several novel regimes, such as spatial strong coupling, will be accessible in state-of-the-art systems in the coming years. Hence, we showed that Brillouin scattering and cavity optomechanics are subsets of a larger realm of photon-phonon interaction. In the following chapters, we experimentally demonstrate silicon waveguides in the weak-coupling regime. In particular, these waveguides currently feature orders-of-magnitude stronger phononic than photonic propagation loss, so they lead to sound-assisted amplification of light (equation (3.9)).

# 4

# Partially suspended silicon waveguide

*We are all in the gutter, but some of us are looking at the stars.*
Oscar Wilde

This chapter is based on [64].

## Contents

IN THE PAST DECADE, *there has been a surge in research at the boundary between photonics and phononics. Most efforts centered on coupling light to motion in a high-quality optical cavity, typically geared towards manipulating the quantum state of a mechanical oscillator. It was recently predicted that the strength of the light-sound interaction would increase drastically in nanoscale silicon wires. In this chapter, we demonstrate such a giant overlap between near-infrared light and gigahertz sound co-localized in a small-core silicon wire. The wire is supported by a tiny pillar to block the path for external phonon leakage, trapping* 10 *GHz phonons in* *an area below* 0.1 $\mu m^2$*. Since the geometry can also be studied in microrings, it paves the way for complete fusion between the fields of cavity optomechanics and Brillouin scattering. The result bodes well for the realization of optically-driven lasers/sasers, isolators and comb generators on a densely integrated silicon chip.*

## 4.1 Introduction

The diffraction of light by sound was first studied by Léon Brillouin in the early 1920s. Therefore, such inelastic scattering has long been called *Brillouin scattering* [36]. The effect is known as *stimulated* Brillouin scattering when the sound is generated by a strong intensity-modulated light field. This sets the stage for a feedback loop: First, the beat note between two optical waves (called the *pump* and the *Stokes seed*) generates sound. Next, this sound creates a travelling index grating that scatters light. On the quantum level, the process annihilates pump photons while creating acoustic phonons and Doppler red-shifted Stokes photons.

In a seminal experimental study [133], Brillouin scattering was viewed as a source of intense coherent sound. Later, the effect became better known as a noise source in quantum optics [150] and for applications such as phononic band mapping [42, 205], slow and stored light [41, 159], spectrally pure lasing [136, 140, 206] and microwave processing [34, 137].

Traditionally [34, 36, 41, 93, 132, 133, 136, 137, 140, 143, 144, 149, 150, 152, 159, 205–207], the photon-phonon interaction was mediated by the material nonlinearity. Electrostriction drove the phonon creation and phonon-induced permittivity changes scattered photons. This conventional image of SBS as a bulk effect, without reference to geometry, breaks down in nanoscale waveguides. In such waveguides, boundary effects can no longer be neglected [104]. Thus, both electrostrictive forces and radiation pressure create acoustic waves: the former generate density variations and the latter transfers momentum to the waveguide boundaries. Equivalently, the new theory [104] takes into account not only bulk permittivity changes but also the shifting material boundaries. The impressive progress in engineering radiation pressure in micro- and nanoscale systems [27–30, 105, 138, 208, 209] recently inspired the prediction of enormously enhanced SBS [66, 104, 210, 211] in silicon nanowires. The strong optical confinement offered by these wires boosts both bulk and boundary forces. However, destructive interference between the two contributions may still completely cancel the photon-phonon coupling. The giant light-sound overlap arises particularly when both types of optical forces align with the acoustic field [104, 211].

Unfortunately, typical silicon-on-insulator (SOI) wires provide only weak acoustic confinement because there is little elastic mismatch between the silicon core and the silicon dioxide substrate. The large SBS strength was thus thought to be accessible only in silicon waveguides that are fully suspended in air [43, 66, 104, 210, 211]. This requirement severely compromises the ability to make centimeter-scale interaction lengths, which are paramount to reduce the required pump power. Hence, Brillouin scattering remained elusive in silicon photonic nanowires.

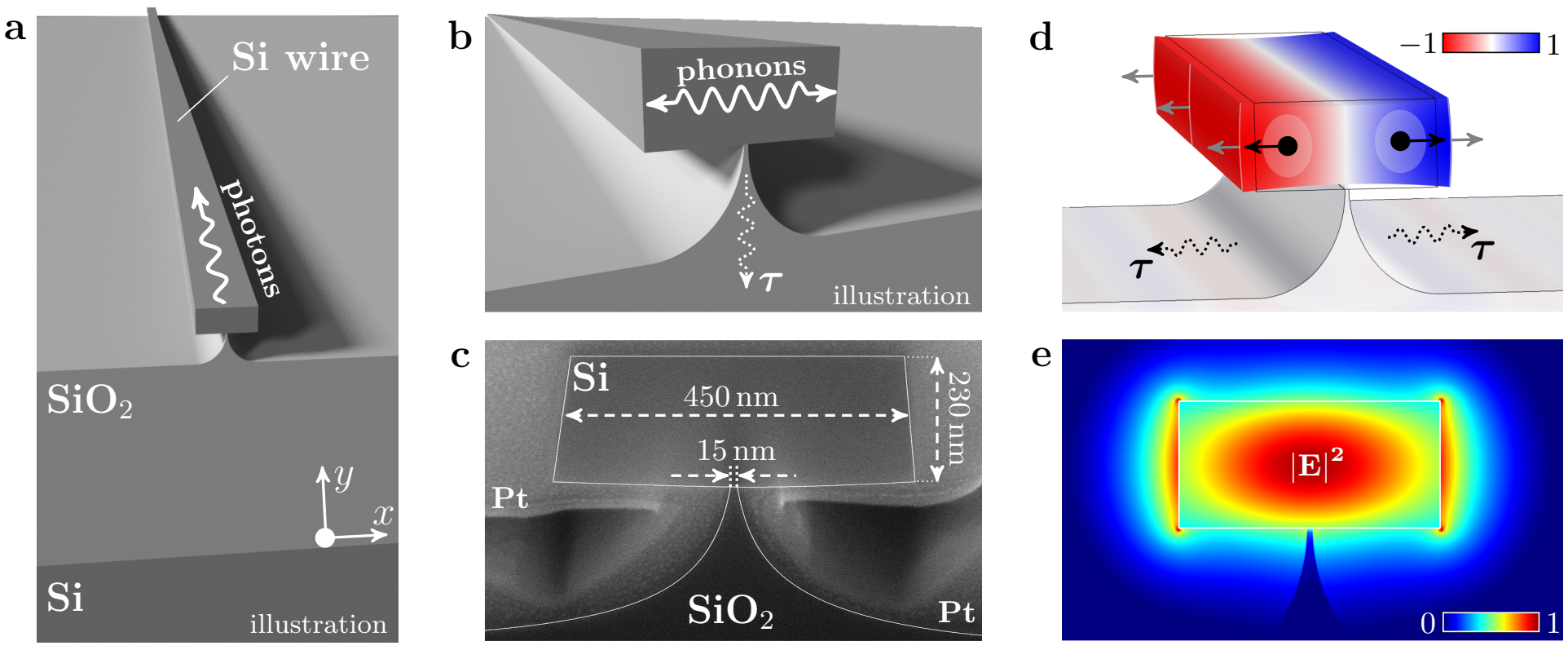

**Figure 4.1: A silicon wire on a pillar as an acoustic phonon cavity.** **a**, Top view of the silicon wire. Light propagates along the wire. It confines photons owing to the high optical contrast with the silicon dioxide substrate and the air. **b**, Unlike the photons, the phonons are trapped transversally. The leakage of phonons through the pillar determines their lifetime $\tau \approx 5\,\text{ns}$. **c**, A scanning electron micrograph of the $450 \times 230\,\text{nm}$ cross-section. Before ion milling, we deposit platina (Pt) on the wire for better visualization (section 4.7). We fabricate pillars as narrow as 15 nm reliably. **d**, The horizontal component of the observed acoustic mode **u** (red: $-$, blue: $+$) aligns with the bulk electrostrictive forces (black arrows) and the boundary radiation pressure (grey arrows). **e**, Electric field norm of the quasi-TE optical mode.

## 4.2 Summary of results

Here, we take the middle ground between these conflicting demands. By partially releasing a silicon wire from its substrate, we drastically improve acoustic confinement (fig.4.1a-c). There is still some leakage through the pillar, but it is sufficiently limited to tap the large overlap between the optical forces and the hypersonic mode (fig.4.1d). Moreover, this way it is straightforward to increase the interaction length. Building on this compromise, we demonstrate an order-of-magnitude performance leap in the light-sound coupling strength.

The observed mechanical mode strongly interacts with the fundamental quasi-TE optical mode (fig.4.1e). The main contribution to the coupling stems from the good overlap between the horizontal optical forces and displacement profile. In particular, the bulk electrostrictive forces $\mathbf{f}_{\text{es}}$ and the bound-ary radiation pressure $\mathbf{f}_{\rm rp}$ both point in the same direction as the acoustic field $\mathbf{u}$ (fig.4.1d) The negative $p_{11}$ photoelastic coefficient of silicon plays a crucial role here. It flips the horizontal electrostrictive force such that it aligns with the boundary force [124, 211]. Therefore they interfere constructively, leading to a total overlap $\langle \mathbf{f}, \mathbf{u} \rangle = \langle \mathbf{f}_{\rm es}, \mathbf{u} \rangle + \langle \mathbf{f}_{\rm rp}, \mathbf{u} \rangle$ up to twice as large as each individual component. Since the SBS gain $\tilde{\mathcal{G}}$ at the phonon resonance frequency $\Omega_{\rm m}$ scales as $|\langle \mathbf{f}, \mathbf{u} \rangle|^2$, the total scattering from pump to Stokes photons may be up to four times as efficient as by electrostriction or radiation pressure individually.

Such force interference [104, 211] was previously studied in hybrid silicon nitride/silicon waveguides [43]. In that case, the light was confined to the silicon core but the sound to the silicon nitride membrane. In this chapter, both light and sound are compressed to the same silicon core. The elastic mode (fig.4.1d) can be understood as the fundamental mode of a Fabry-Pérot cavity for hypersonic waves (fig.4.1b), formed by the silicon/air boundaries. Therefore, its frequency can be estimated as $\frac{\Omega_{\rm m}}{2\pi} = \frac{v}{2w} = 10.1\,\text{GHz}$ with $v = 9130\,\text{m/s}$ the longitudinal speed of sound along $\langle 110 \rangle$ silicon [212] and $w = 450\,\text{nm}$ the waveguide width.

### 4.2.1 Fabrication

To create the pillar structure, we start from an SOI wire fabricated by deep UV lithography [213] through the silicon photonics platform ePIXfab (www.ePIXfab.eu). Next, we perform an additional oxide etch with hydrofluoric acid. By carefully controlling the etching speed, a narrow pillar is left underneath the wire (fig.4.1a-c). Through this simple fabrication method, we obtain wires up to 4 cm long. To retain compactness, wires longer than 3 mm are coiled up into a low-footprint spiral (fig.4.2). Despite the additional etch, the wires still exhibit optical propagation losses $\alpha$ of only 2.6 dB/cm.

### 4.2.2 Experiments

In our experiments (fig.4.3), we investigate straight and spiral waveguides (fig.4.2) with lengths $L$ ranging from 1.4 mm to 4 cm. We couple 1550 nm TE-light to the waveguides through focusing grating couplers [214] and perform both gain (fig.4.3a-b) and cross-phase modulation (fig.4.3c-d) experiments. The resonances (fig.4.3a and c) observed in these experiments allow for a characterization of the photon-phonon coupling in two independent ways.

#### Gain

First, we monitor the power in a Stokes seed as a function of frequency spacing $\frac{\Omega}{2\pi}$ with a strong co-propagating pump wave (fig.4.3a-b). We observe a

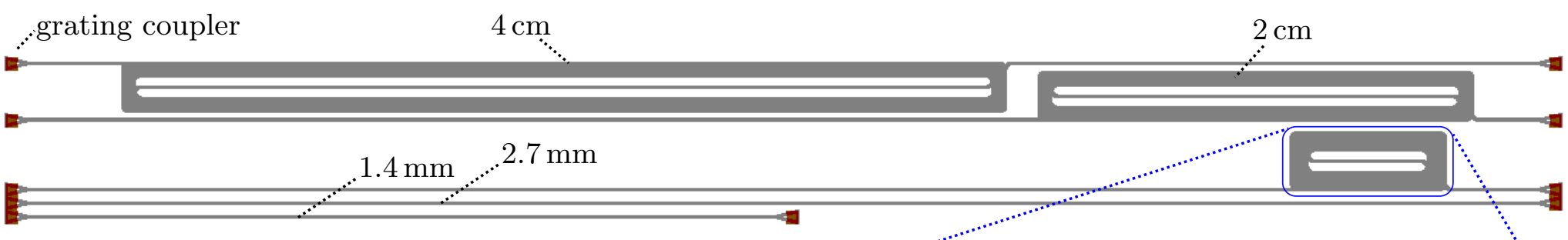

**a**

| $L$ [mm] | 1.4 | 2.7 | 10 | 20 | 40 |
|---|---|---|---|---|---|
| $L_{\mathrm{eff}}$ [mm] | 1.3 | 2.5 | 7.5 | 12 | 15 |
| $L_x/L_z$ [%] | 0 | 0 | 24 | 5.9 | 2.8 |
| $L_{\mathrm{bend}}/L_z$ [%] | 0 | 0 | 12 | 3.6 | 1.7 |
| $L_{\mathrm{single}}/L_{\mathrm{rest}}$ [%] | 100 | 100 | 42 | 22 | 16 |

**b**

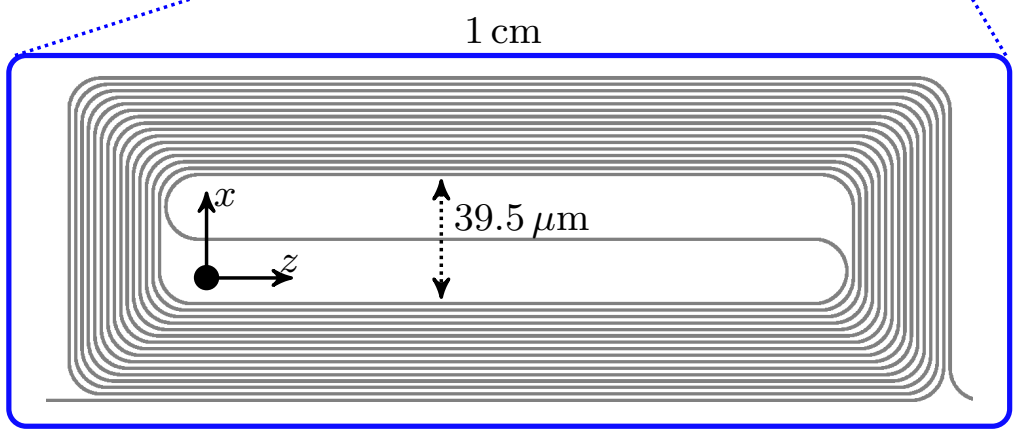

**c**

**Figure 4.2: Overview of the straight and spiral waveguides. a**, Top view of the silicon wires on a pillar. The 1.4 mm- and 2.7 mm-long wires are straight, the other waveguides are coiled up into a spiral. **b**, Properties of the waveguides: $L$ is the total waveguide length and $L_{\mathrm{eff}} = \frac{1-\exp(-\alpha L)}{\alpha}$ the effective length with $\alpha = 2.6\,\mathrm{dB/cm}$. The total wire length can be decomposed as $L = L_x + L_z + L_{\mathrm{bend}}$ into the length $L_x$ along the $x$-axis, the length $L_z$ along the $z$-axis and the length of the bends $L_{\mathrm{bend}}$. The fractions $L_x/L_z$ and $L_{\mathrm{bend}}/L_z$ drop rapidly as the spirals become longer. We also define the ratio $L_{\mathrm{single}}/L_{\mathrm{rest}}$ between the effective length $L_{\mathrm{single}}$ of isolated wires (i.e. more than $10\,\mu$m from another wire) and the effective length $L_{\mathrm{rest}}$ of the wires in close proximity ($< 2\,\mu$m). **c**, The 1 cm-long spiral has the highest fraction of wires along the $x$-axis, the highest fraction of bends and the smallest fraction of wires in proximity of another wire. The spacing between adjacent wires is $1.55\,\mu$m.

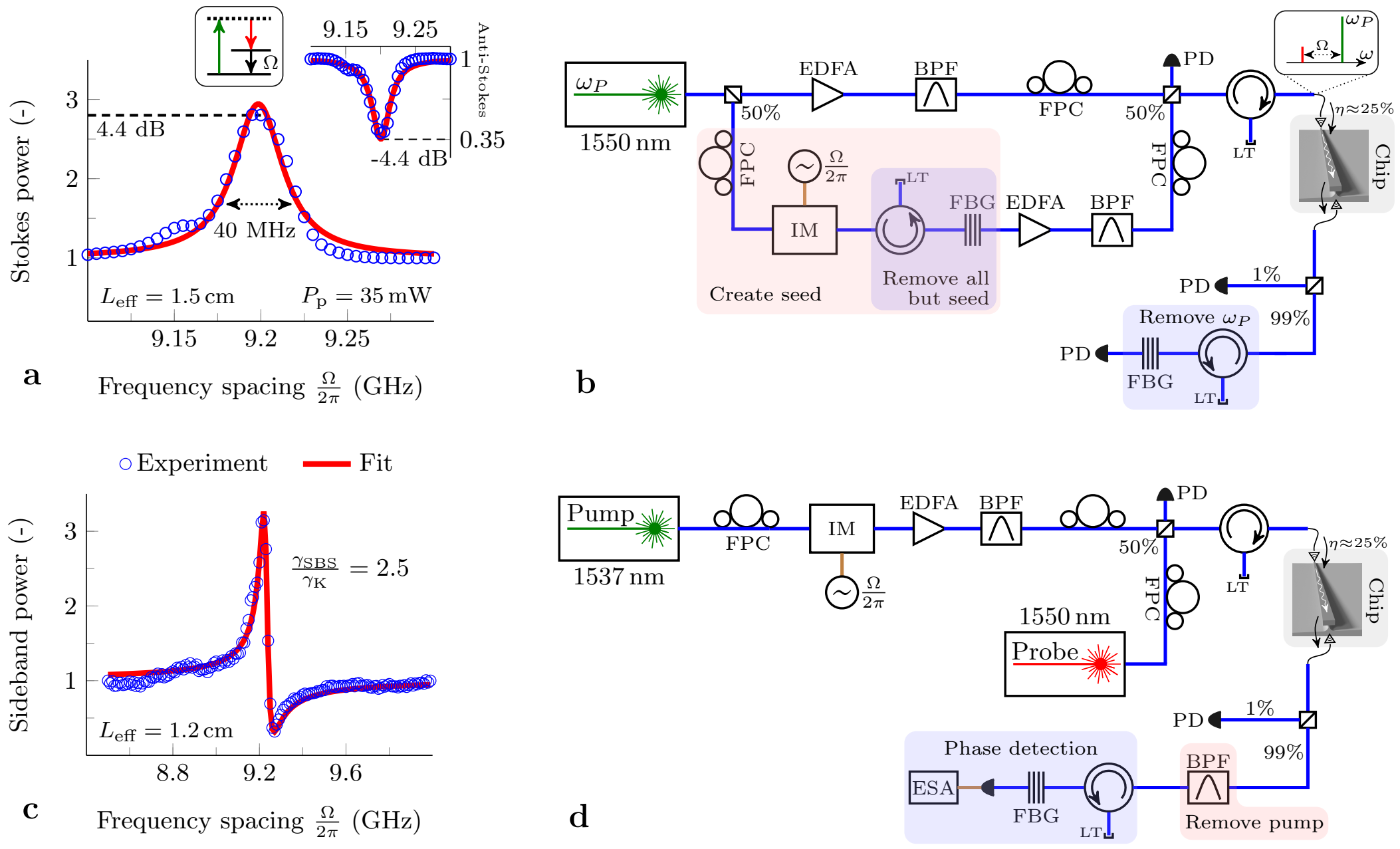

**Figure 4.3: Experimental characterization of the photon-phonon coupling.** **a**, A typical Lorentzian gain profile on a Stokes seed and (inset) a depletion profile on an anti-Stokes seed. In both cases, the interaction generates acoustic phonons and red-shifted photons (energy diagram). We observe such resonances in wires as short as 2.7 mm (section 4.7) and obtain the highest on/off gain of 175% (4.4 dB) in the 4 cm spiral ($L_{\text{eff}} = 1.5$ cm) with 35 mW on-chip pump power. There is a remnant of a second resonance at 9.15 GHz (section 4.7). We fit the normalized Stokes power to the exponential of a Lorentzian function (equation (B.6), appendix B). **b**, The fiber-based set-up used to study forward SBS. A tunable laser is amplified in the upper arm by an erbium-doped fiber amplifier (EDFA) to serve as a pump. In the lower arm, the laser light is intensity modulated (IM) to generate a blue- and red-shifted sideband. Next, a fiber Bragg grating (FBG) rejects all but the Stokes seed (section 4.7). The pump and Stokes seed are coupled to the chip through curved grating couplers. Finally, the power in the pump and Stokes wave is monitored separately. Light traps (LTs) prevent backscattered light from entering the circuit. With minor modifications, this set-up can be reconfigured to observe the loss on an anti-Stokes seed or the backward SBS. The latter is weak in our wires (section 4.7), so we focus on the forward SBS. **c**, A typical Fano signature obtained from the XPM-experiment, which we use to calibrate the Brillouin with respect to the Kerr nonlinearity ($\gamma_{\text{SBS}}/\gamma_{\text{K}} = 2.5$). **d**, A pump is intensity modulated, amplified, combined with a probe wave and sent to the chip. The pump is removed at the output by a band-pass filter (BPF). The phase modulation on the probe wave is transducted to intensity modulation by filtering out the red-shifted sideband. Finally, we use an electrical spectrum analyzer (ESA) to observe the beat between the probe and the imprinted blue sideband.

Lorentzian gain profile at $\frac{\Omega_m}{2\pi} = 9.2\,\text{GHz}$, as expected in the low-cascading regime (see B). Similarly, we observe an identical depletion profile on an anti-Stokes seed (fig.4.3a). The Stokes seed experiences amplification as long as the pump remains undepleted. Exactly on resonance, the on/off gain is given by $2\gamma_{\text{SBS}} P_p L_{\text{eff}}$ – with $2\gamma_{\text{SBS}} = \tilde{\mathcal{G}}$ the Brillouin gain coefficient, $P_p$ the input pump power and $L_{\text{eff}} = \frac{1-\exp(-\alpha L)}{\alpha}$ the effective interaction length. The effective length has an upper limit of $\frac{1}{\alpha} = 1.7\,\text{cm}$ in our wires. To extract the Brillouin parameter $\gamma_{\text{SBS}}$, we sweep the pump power in a 2.7 mm-long wire (fig.4.4a). Above 25 mW on-chip power, nonlinear absorption saturates the on/off gain. Then free carriers, created by two-photon absorption (TPA), result in a power-dependent optical loss $\alpha(P_p)$. We extract $2\gamma_{\text{SBS}} = 3218\,\text{W}^{-1}\text{m}^{-1}$ below this threshold. The Lorentzian fit yields an acoustic linewidth of $\frac{\kappa_m}{2\pi} = 30\,\text{MHz}$ and thus a quality factor of $Q_m = \frac{\Omega_m}{\kappa_m} = 306$ and a phonon lifetime of $\tau = \frac{1}{\kappa_m} = 5.3\,\text{ns}$ in the same short wire. The largest on/off gain of 0.6 dB below the TPA-threshold falls narrowly short of the linear loss $\alpha L = 0.7\,\text{dB}$ (fig.4.4a). Thus the wire is close to net optical amplification, which is necessary for Brillouin lasing. The on/off gain reaches 175% in the longest 4 cm-wires (fig.4.3a), improving by a factor 19 on previous results in a silicon/silicon nitride waveguide [43]. In the next chapter, we demonstrate net gain in a cascade of fully suspended nanowires.

### Forward vs. backward.

Similarly, we observe backward SBS (section 4.7): for counter-propagating pump and Stokes waves that generate elastic waves with a large wavevector $K \approx 2k_0$ and $k_0$ the pump wavevector. However, we achieve the giant light-sound overlap only for forward SBS: for co-propagating pump and Stokes waves that generate low group velocity acoustic phonons with small wavevector $K \approx \frac{\Omega}{v_g}$ and $v_g$ the optical group velocity. Therefore, we focus on forward SBS here.

### Cross-phase modulation

Second, we measure the strength of the cross-phase modulation (XPM) imprinted on a weak probe by a strong intensity-modulated pump (fig.4.3c-d). The experiment yields a distinct asymmetric Fano signature at $\frac{\Omega_m}{2\pi} = 9.2$ GHz caused by interference between the resonant Brillouin and the non-resonant Kerr response (appendix B). The lineshape follows $|\frac{\gamma_{\text{XPM}}(\Omega)}{2\gamma_K}|^2$, with $\gamma_K$ the

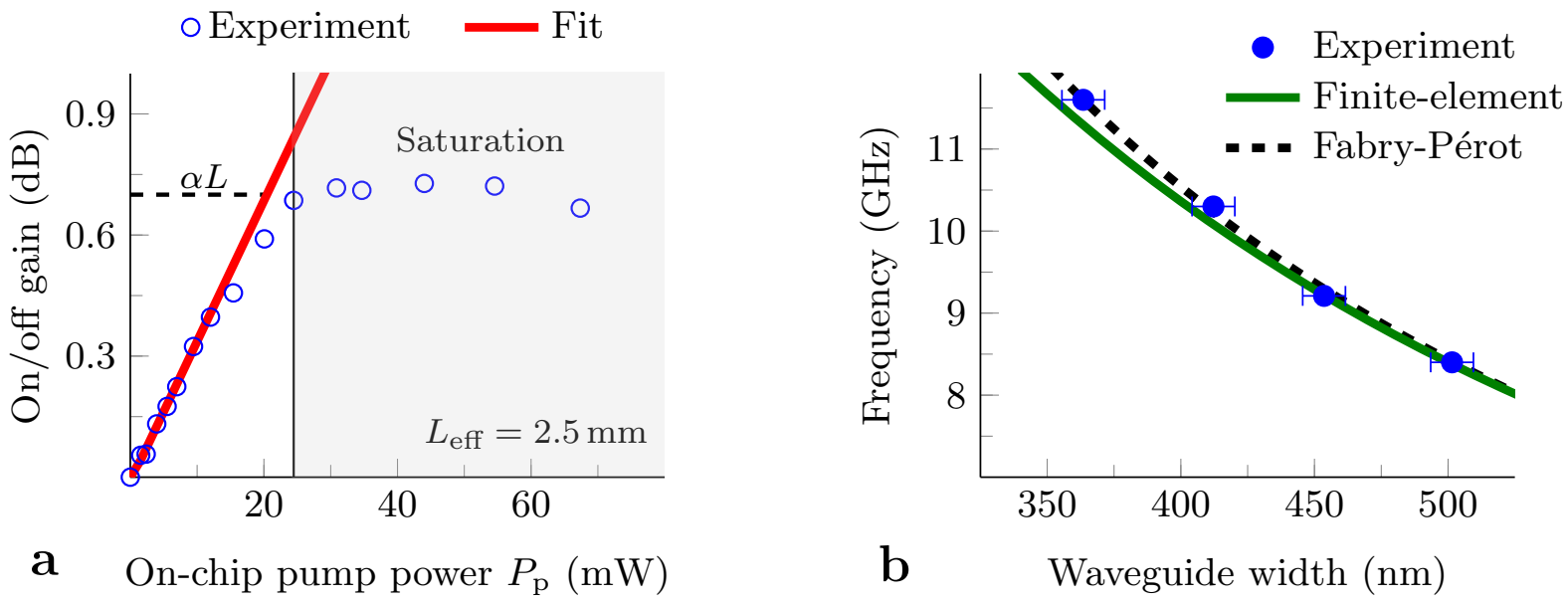

**Figure 4.4: Analysis of the Brillouin gain and phononic resonance frequency. a**, Scaling of the on/off Brillouin gain with input pump power. Above a power threshold of 25 mW, the on/off gain saturates because of nonlinear absorption. We perform a fit to obtain the Brillouin nonlinearity below that threshold. **b**, The phononic resonance frequency for different waveguide widths. Both a simple Fabry-Pérot and a rigorous finite-element model agree with the data. The black Fabry-Pérot curve took 8433 m/s as the longitudinal speed of sound in silicon, whereas 9130 m/s is the speed of sound along ⟨110⟩ silicon [212]. This would shift the black curve upwards by 8%, still in remarkable agreement with the data given the simplicity of this picture.

Kerr parameter and

$$\gamma_{\mathrm{XPM}}(\Omega) = 2\gamma_{\mathrm{K}} + \gamma_{\mathrm{SBS}}\mathcal{L}(\Omega)$$
$$\mathcal{L}(\Omega) = \frac{1}{-2\Delta_{\mathrm{r}} + i} \qquad \Delta_{\mathrm{r}} = \frac{\Omega - \Omega_{\mathrm{m}}}{\kappa_{\mathrm{m}}}$$

We deduce the ratio $\gamma_{\mathrm{SBS}}/\gamma_{\mathrm{K}} = 2.5$ and $Q_{\mathrm{m}} = 249$ from the fit. The Kerr parameter $\gamma_{\mathrm{K}}$ of similar silicon wires has been studied extensively, with values reported at $\gamma_{\mathrm{K}} = 566\,\mathrm{W}^{-1}\mathrm{m}^{-1}$ for our cross-section [215]. Because of the pillar etch, the light is more confined to the high-index silicon core. We simulate that this yields a slight increase of the Kerr effect by 8% to $\gamma_{\mathrm{K}} = 611\,\mathrm{W}^{-1}\mathrm{m}^{-1}$. Thus we have $2\gamma_{\mathrm{SBS}} = 3055\,\mathrm{W}^{-1}\mathrm{m}^{-1}$, within 5% of the value obtained from the gain experiments. This nonlinearity is at least a factor $10^3$ stronger than in photonic crystal and highly nonlinear fibres [152, 155]. Further, the resonance frequency, quality factor and interaction strength are in good agreement with the models.

## 4.3 Resonance frequency

To study the frequency, we perform the XPM-experiment for waveguide widths from 350 nm to 500 nm (fig.4.4b). Both a simple Fabry-Pérot ($\frac{\Omega_{\mathrm{m}}}{2\pi} = \frac{v}{2w}$) and a sophisticated finite-element model match the observed resonances. The finite-element model takes into account the exact geometry of the wires as obtained from a scanning electron micrograph (fig.4.1c). This includes the waveguide height, pillar size, sidewall angle and the $\langle 110 \rangle$ crystal orientation of our wires (section 4.7). We find that the waveguide width alone pins down the resonance frequency, with other geometrical parameters inducing minor shifts. For a 450 nm-wide waveguide, the frequency sensitivity to width changes is 19.2 MHz/nm (fig.4.4b). In contrast, the calculated sensitivity to height changes is only 2.3 MHz/nm. This supports the intuitive Fabry-Pérot view, in which the height does not appear at all.

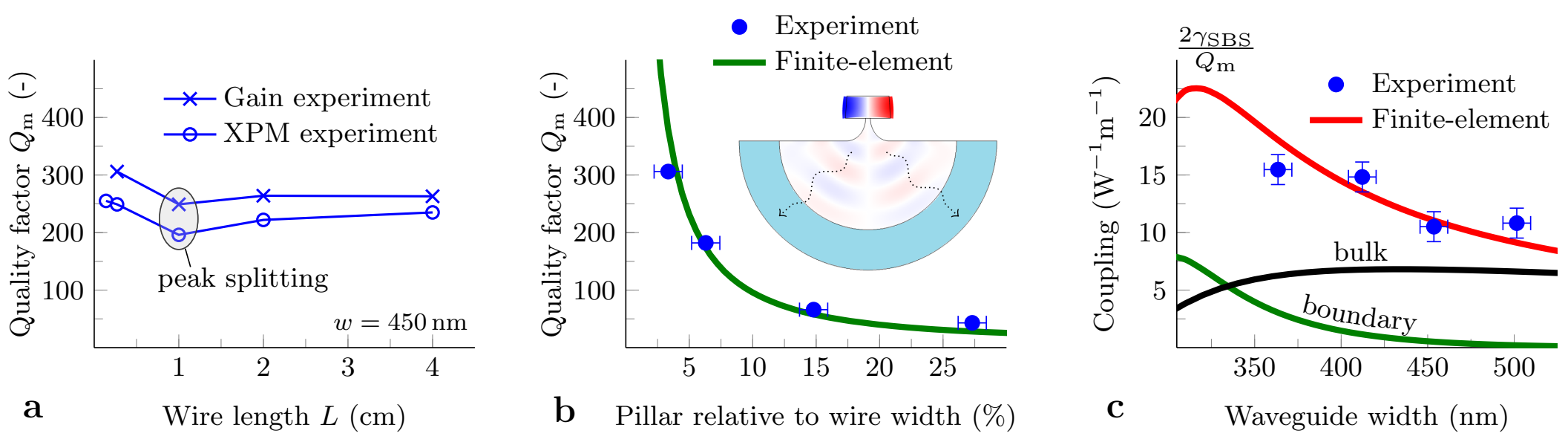

**Figure 4.5: Study of the mechanical quality factor and the intrinsic photon-phonon overlap.** **a**, The quality factor stays above 250 even in 4 cm-long spirals, showing no length-dependent line broadening. Besides, there is peak splitting in the 1 cm spiral (section 4.7). Neither the longer spirals nor the straight wires exhibit such splitting. **b**, A finite-element model of phonon leakage through the pillar accurately predicts the observed quality factors ($L = 2.7$ mm). **c**, The non-resonant nonlinearity $\frac{2\gamma_{SBS}}{Q_m}$ is a measure of the intrinsic photon-phonon coupling. The electrostriction (black) and radiation pressure (green) interfere constructively, bringing about the total overlap (red). As the width increases, the boundary contribution vanishes rapidly.

The large sensitivity to waveguide width implies that a 2 nm width fluctuation shifts the resonance by more than a linewidth. Therefore inhomogeneous broadening may affect both the lineshape and -width, similar to Doppler-broadening in gain media. Surprisingly, we achieve acoustic quality factors above 250 even in the longest 4 cm-wires (fig.4.5a). This suggests that there is, if at all, only limited length-dependent line broadening (section 4.7). In the next chapter, however, we observe strong indications of inhomogeneous broadening in a cascade of suspended nanowires. Besides, this large sensitivity can be exploited to tailor the resonance frequency.

## 4.4 Quality factor

By sweeping the pillar size in a short 450 nm-wide waveguide, we establish leakage through the pillar as the dominant phononic loss mechanism (fig.4.5b). The pillar acts as a channel for elastic waves that propagate down into the substrate. We rigorously model this mechanism by adding an artificial absorbing layer at the boundary of the simulation domain (section 4.7). As predicted by such a model, the observed quality factors diminish rapidly with increasing pillar size. The pillar should be seen as a moving acoustic membrane [216], not as a fixed point. Therefore, it affects neither the acoustic field profile nor its associated stiffness $k_{\text{eff}}$ considerably. We regard a short phonon lifetime as the prime reason that SBS was not observed so far in regular, unsuspended SOI nanowires.

## 4.5 Interaction strength

Finally, we show that the photon-phonon coupling is a constructive combination of bulk (electrostriction) and boundary (radiation pressure) effects (fig.4.5c). The resonant Brillouin gain coefficient is given by $\tilde{\mathcal{G}} = 2\gamma_{\text{SBS}} = \omega_0 Q_{\text{m}} |\langle \mathbf{f}, \mathbf{u} \rangle|^2 / (2k_{\text{eff}})$ (appendix B), so the non-resonant part $\frac{2\gamma_{\text{SBS}}}{Q_{\text{m}}}$ is a direct measure of the photon-phonon overlap [104, 211]. In our finite-element simulations of $\langle \mathbf{f}, \mathbf{u} \rangle$ and $k_{\text{eff}}$, we take into account the mechanical anisotropy of silicon but not the pillar. We also approximate the cross-section as rectangular, neglecting the small sidewall angle. Still, the simulations match the experimentally determined coupling strength. Neither electrostriction nor radiation pressure separately explain the experimental values of $\frac{2\gamma_{\text{SBS}}}{Q_{\text{m}}} \approx 12\,\text{W}^{-1}\text{m}^{-1}$. These values are a factor 10 larger than in on-chip chalcogenide [132] and silicon nitride/silicon waveguides [43], showing the benefit of compressing light and sound to the same nanoscale core.

## 4.6 Conclusion

These observations provide robust evidence for incredibly strong photon-phonon coupling in silicon nanowires. The simulations of the interaction strength match the experiments, indicating that the new nanoscale SBS theory [104] is on the right track. Moreover, simple finite-element models accurately capture both the phononic resonance frequency and lifetime.

Building on the good light-sound overlap, some typical SBS applications are now squarely in reach. For lasing [141], the gain must exceed the loss over an optical cavity roundtrip. Currently, we achieve 0.6 dB on/off gain in a wire with 0.7 dB propagation loss (fig.4.4a). In the next chapter, we demonstrate gain exceeding the propagation losses. We note that lasing – as opposed to sasing – also requires a resonator that is less optically than acoustically damped [63, 136, 141, 142, 186]. Another example is the microwave filter [34], since such filters can be driven by even sub-1 dB gain [217, 218]. A 50 dB-notch microwave filter was recently demonstrated [218] using a silicon pedestal waveguide.

For other devices, such as isolators [149, 219] and slow light [41], the performance in terms of optical losses and SBS strength must be improved more substantially. Optical losses below 1 dB/cm have already been demonstrated in comparable silicon wires; by moving from a 200 mm- to a 300 mm-wafer CMOS pilot line with more advanced lithography tools [220]. Significant net gain should be accessible in such low-loss wires, in which effective interaction lengths up to $L_{\text{eff}} = 5$ cm may be obtained.

Further, free-carrier absorption saturates the SBS gain above a pump power of 25 mW (fig.4.4a). Therefore, the SBS gain has to be improved by other means: either the phonon lifetime $\tau$ or the photon-phonon overlap $\langle \mathbf{f}, \mathbf{u} \rangle$ should be enhanced. Currently limited by the phonon leakage, the lifetime could be increased – at the cost of smaller bandwidth – by exciting asymmetric elastic modes [221]. It has been predicted that such modes can be generated efficiently by cross-coupling between the quasi-TE and quasi-TM optical mode [211]. Alternatively, the overlap $\langle \mathbf{f}, \mathbf{u} \rangle$ may be increased by trapping light in a narrow horizontal air-slot between two specially designed silicon wires [66]. Such ideas may further boost the Brillouin nonlinearity to a level sufficient for milliwatt-threshold lasing [141, 186], frequency comb generation [147, 152] and fully non-resonant optomechanics [32].

In fact, each application comes with a specific figure of merit. For comb generation with a dual-pump [152], the forward SBS gain is critical. Then it is equivalent to increase the lifetime $\tau$ or the overlap $\langle \mathbf{f}, \mathbf{u} \rangle$. In other cases, such as for slow light [41], the bandwidth is equally important. Further, it is often desirable to have the acoustic resonance in the gigahertz range [34] – which implicitly sets the stiffness $k_{\text{eff}}$ given a certain mass. Even so, a large light-sound overlap $\langle \mathbf{f}, \mathbf{u} \rangle$ – clearly manifested in this chapter – is greatly beneficial in all cases.

In conclusion, we demonstrated efficient interaction between near-infrared light and gigahertz sound trapped in a small-core silicon photonic wire. The structure exhibits an extraordinary light-sound overlap, at the same time allowing for a centimeter-scale Brillouin-active interaction length. The combination of both opens the door to practical Brillouin devices integrated on a CMOS-compatible silicon chip.

## 4.7 Additional information

### Device fabrication and characterization

The SOI wires are fabricated by 193 nm deep UV lithography on a 200 mm wafer CMOS pilot line at imec. We underetch the wires with 2% diluted hydrofluoric acid at a vertical etching rate of 10 nm/min. The horizontal etch rate was about 17% slower. We did not succeed in fabricating narrow pillars on some chips taken from different pilot runs, likely because of an inhomogeneity in the oxide substrate. We deposit platina (Pt) on the wire and then mill the cross-section (fig.4.1c) by a focused ion beam. The platina deposition ensures a straight cross-section and prevents charging effects when the cross-section is viewed by a scanning electron beam. The straight wires have lengths 1.4 and 2.7 mm, while the spirals are 1, 2 and 4 cm long (fig.4.2). The 1, 2 and 4 cm spirals have a footprint of 275 $\mu$m × 100 $\mu$m, 775 $\mu$m × 90 $\mu$m and 1570 $\mu$m × 90 $\mu$m respectively. Adjacent wires are spaced by 1.55 $\mu$m inside the spiral. We find the propagation loss $\alpha$ of 2.6 dB/cm and the coupling loss of 6 dB per grating coupler by the cut-back method. Thus we assumed in- and output grating couplers to be identical. Since both coupling and propagation losses were known, we derived the on-chip pump power from a continuous power measurement of some tapped light both before and after the chip. The error on these estimates is about 10-20% and is larger in longer waveguides with increased spectral transmission fluctuations (caused by distributed backscatter).

### Experimental set-up

We use the following abbreviations (fig.4.3): erbium-doped fibre amplifier (EDFA), band-pass filter (BPF), fibre polarization controller (FPC), intensity modulator (IM), electrical spectrum analyser (ESA), light trap (LT), fibre Bragg grating (FBG) and photodetector (PD). The FBGs were a crucial part of our set-up. Produced by TeraXion Inc., these filters were custom-designed to have a flat response within the passband and drop to −30 dB within 2.5 GHz. We use the steep flanks to filter out either the red- or blue-shifted sidebands. Their bandwidth is 60 GHz. In addition, we employ a pair of well aligned FBGs for the gain experiment (fig.4.3b). We measured the short-term linewidth of our external cavity tunable lasers through the self-heterodyne method based on a 20 km fiber delay and a 200 MHz acousto-optic frequency shifter. All sources had linewidths below 1 MHz if care was taken to turn off the laser's coherence control options – which are sometimes used to broaden the laser's linewidth through phase modulation to e.g. avoid SBS in fibers. Therefore, the laser noise does not affect the measured ≈ 35 MHz acoustic linewidths.

### Finite-element modelling

We obtain the photonic and phononic modes from the finite-element solver COMSOL. They were exported to MATLAB to calculate the photon-phonon coupling [104, 211]. Since our wires are aligned along a $\langle 110 \rangle$ axis, we rotated both the elasticity $(c_{11}, c_{12}, c_{44}) = (166, 64, 79)$ GPa and the photoelasticity $(p_{11}, p_{12}, p_{44}) = (-0.09, 0.017, -0.051)$ matrix by $\pi/4$. To simulate the clamping loss, we add an artificial silica matching layer with Young's modulus $\frac{i}{\zeta}E$ and density $-i\zeta\rho$. The layer absorbs incoming elastic waves without reflection. In a frequency-domain simulation, we then find the quality factor from $Q_\text{m} = \frac{\Re\Omega_\text{m}}{2\Im\Omega_\text{m}}$. We optimize $\zeta$ for minimal $Q_\text{m}$. A typical value is $\zeta = 2$ for a 420 nm-thick matching layer.

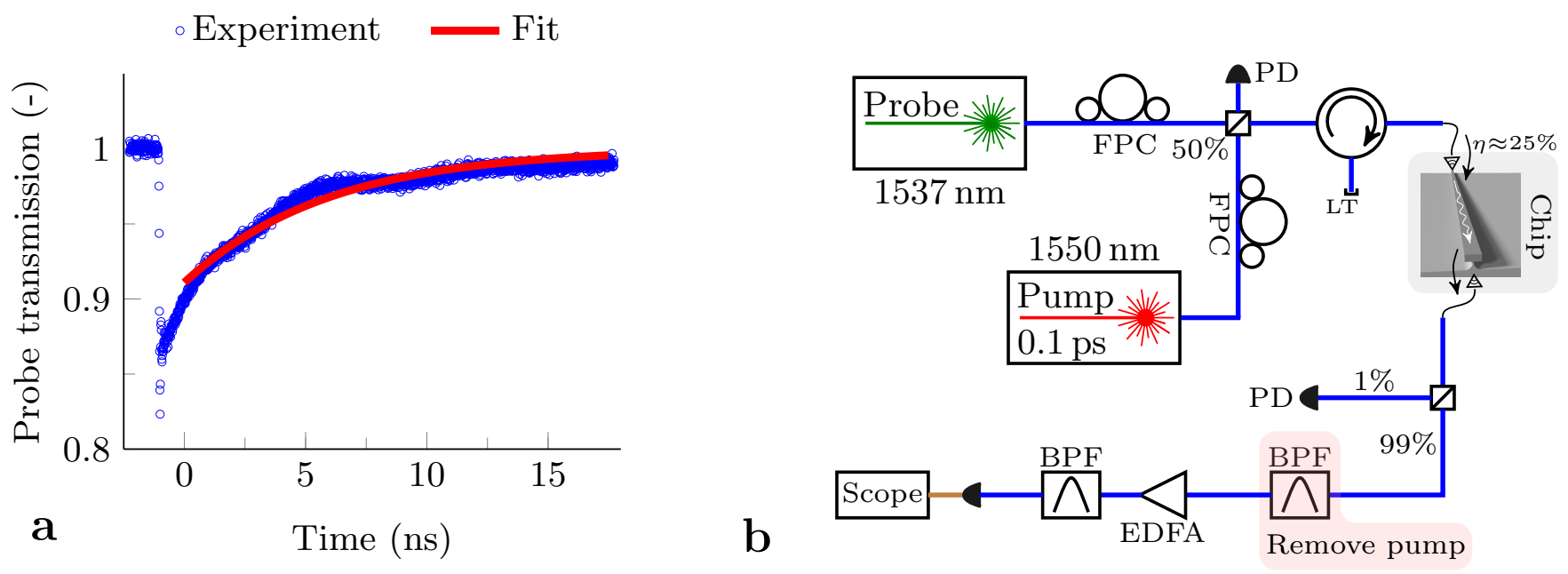

**Figure 4.6: Measurement of the free-carrier lifetime.** **a**, Oscilloscope trace of the probe power. The pump pulse arrives at $t = -1$ ns. We start the fit a nanosecond later to avoid fitting to photodiode ringing artifacts. **b**, Pump-probe set-up used to obtain the trace. The band-pass filter (BPF) has more than 50 dB extinction at 1550 nm.

## Measurement of the free-carrier lifetime

Free electrons and holes, created by two-photon absorption (TPA) in our experiments, induce significant free-carrier absorption (FCA) and free-carrier index changes (FCI) above a certain power threshold. As reflected in the saturation of the SBS gain (fig.4.4a), this threshold is about 25 mW in our $450 \times 230$ nm silicon wires. From the observations

- that our finite-element and coupled-mode modelling of the Brillouin effect matches the experiments
- and that the off-resonance background is flat in the XPM experiment

we have evidence that the free carriers are not noticeably influencing our results below the threshold. Nevertheless, we performed a cross-FCA experiment (fig.4.6) to exclude the possibility of a significant drop in free-carrier lifetime $\tau_\text{c}$ caused by the underetch of our wires.

The pump was a $\approx$ 100 fs-pulse with a repetition rate of $\frac{1}{50\,\text{ns}}$ and peak power of $\approx$ 1 kW. When a pump pulse arrives, it creates many free carriers by TPA. The free carriers recombine before the next pump pulse arrives. Their presence is read out by monitoring the power of a c.w. probe wave on a high-

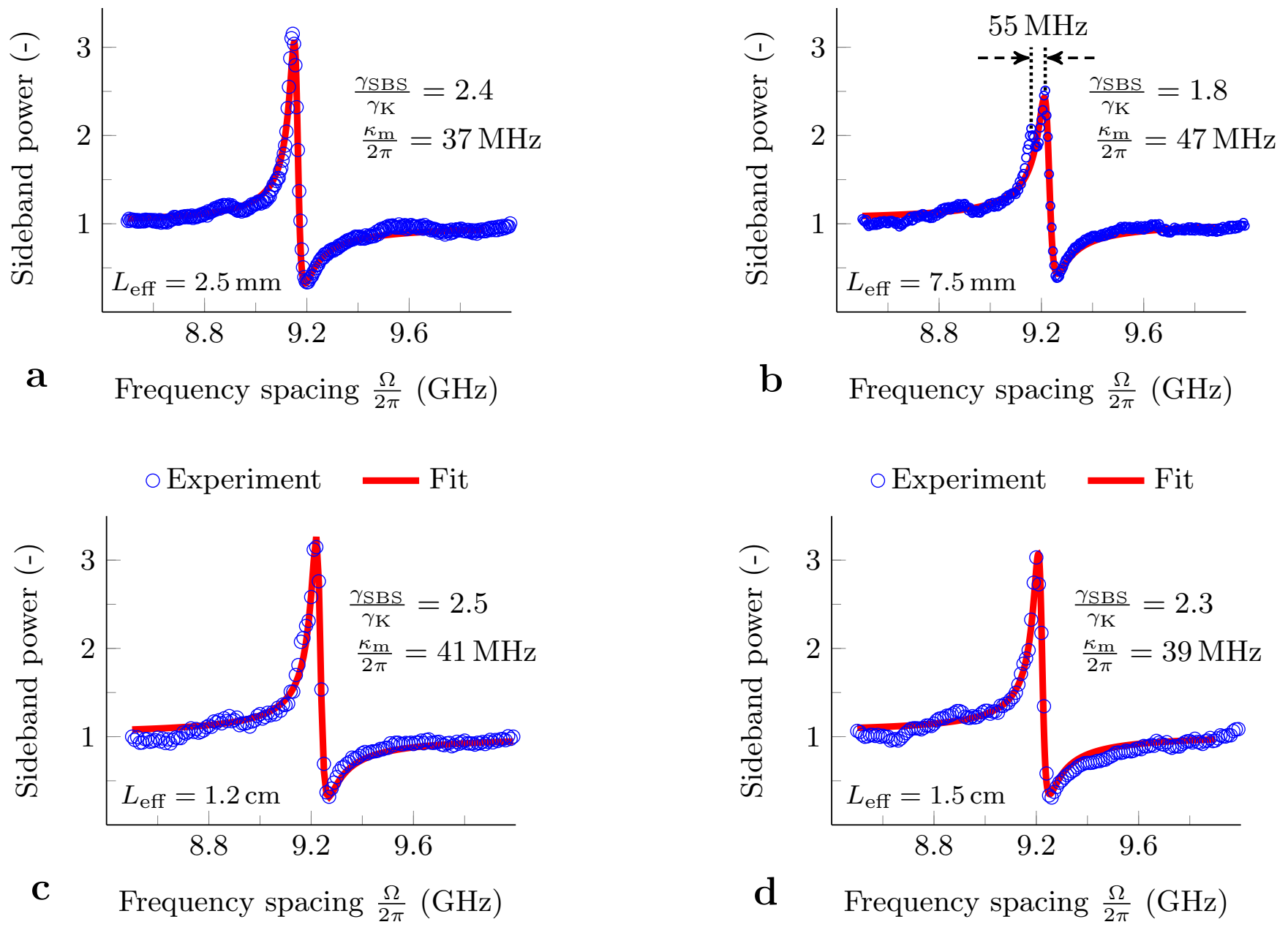

**Figure 4.7: Fano resonances in wires of increasing length.** We conduct the XPM-experiment (fig.4.3d) in wires of lengths 1.4 mm (not shown), 2.7 mm (**a**), 1 cm (**b**), 2 cm (**c**) and 4 cm (**d**). The resonance is identical in the 1.4 mm and the 2.7 mm wires. The extracted Brillouin nonlinearity $\gamma_{SBS}$ and linewidth $\frac{\kappa_m}{2\pi}$ are nearly identical in the 2.7 mm straight wire, the 2 cm spiral and the 4 cm spiral. Only the 1 cm spiral (**b**) exhibits peak splitting and a reduced $\gamma_{SBS}$. The peak is split by 55 MHz.

speed oscilloscope. Thus the transmission $T$ of the probe is

$$T = \exp\left(-\alpha_{FCA}(t)\right) = \exp\left(-\alpha_{FCA}(t_0)\exp\left(-\frac{t}{\tau_c}\right)\right)$$

where we normalized the transmission to the case without FCA. Here we exploited the relation $\alpha_{FCA}(t) \propto N(t)$ with $N(t)$ the free-carrier concentration.

The experiments yield typical values of $\tau_c = 6.2$ ns before the etch and $\tau_c = 5.7$ ns after the etch in identical waveguides. Hence there is, if at all, only a minor decrease of $\tau_c$ due to the underetch. The associated bandwidth of $f_{3\,dB} = \frac{1}{2\pi\tau_c} = 28$ MHz suggests a negligible FCI-effect at 10 GHz. As a precaution, we work when possible – in the longer wires – with low power (below 15 mW on-chip) in the XPM-experiments. The free-carrier nonlinearity $\gamma_{FCI}$ can, in principle, always be reduced below $\gamma_K$ because $\gamma_{FCI} \propto P_{pump}$ while $\gamma_K$ does not depend on $P_{pump}$.

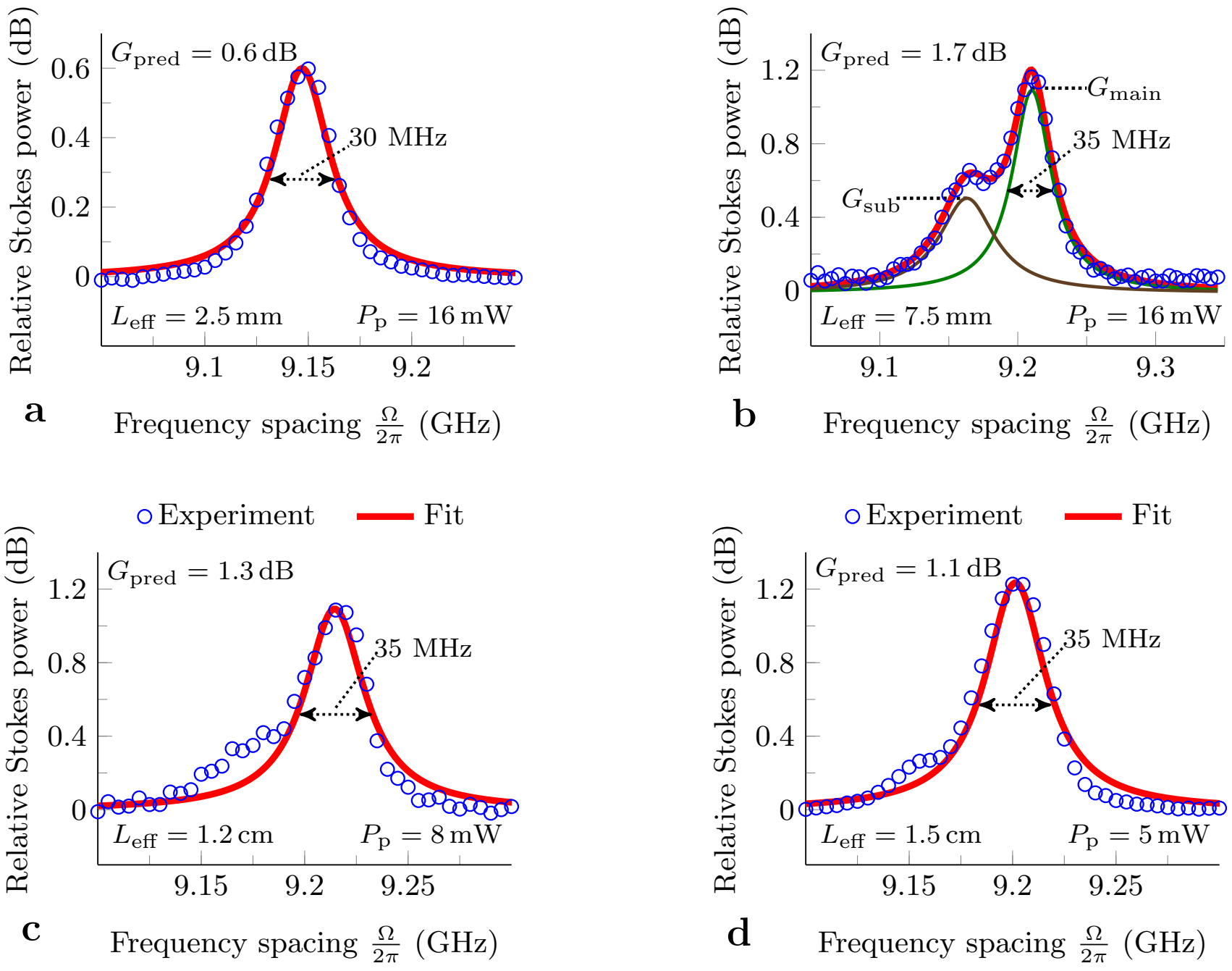

**Figure 4.8: Lorentzian gain resonances in wires of increasing length.** We conduct the gain experiment (fig.4.3b) in wires of lengths 2.7 mm (**a**), 1 cm (**b**), 2 cm (**c**) and 4 cm (**d**). The 1 cm spiral shows two resonances: the minor peak (brown) and the main peak (green) add up to the entire resonance (red). We also show the predicted gain $G_{\mathrm{pred}} = 2\gamma_{\mathrm{SBS}} P_{\mathrm{p}} L_{\mathrm{eff}}$ given $2\gamma_{\mathrm{SBS}} = 3218\,\mathrm{W}^{-1}\mathrm{m}^{-1}$. The predicted and observed Brillouin gain match closely in the 2.7 mm straight wire, the 2 cm spiral and the 4 cm spiral. Only the 1 cm spiral (**b**) exhibits a reduced gain coefficient, caused by the peak splitting. Similar to the XPM-experiment, the peak is split by 55 MHz. There is a remnant of the minor peak in the 2 cm and 4 cm spirals. The linewidth is about 35 MHz in all resonances, where we show the width of only the main resonance in the 1 cm spiral.

### Inhomogeneous broadening

The sensitivity of the resonance frequency to width changes is 19.2 MHz/nm (fig.4.4b). Therefore, a width change of 2 nm shifts the resonance by as much as its 35 MHz linewidth. The phononic resonance is more than an order of magnitude less sensitive to height variations (2.3 MHz/nm) and pillar size variations (0.5 MHz/nm). Hence geometrical non-uniformities, particularly in the wire width, may broaden the mechanical resonance. This potential broadening cannot account for most of the mechanical linewidth. Otherwise, the measured $Q_m$-factors would not agree – within the measurement error – with the simulated leakage-limited $Q_m$-factors (fig.4.5b). Thus other sources of acoustic linewidth must be small relative to the clamping loss.

Nevertheless, we estimate an upper limit on the broadening caused by width non-uniformities. These width fluctuations can be measured indirectly from variations in the optical mode index, and thus the center wavelength of ring resonators, Mach-Zehnder interferometers and arrayed waveguide gratings. The standard deviation of the long-range variations in center wavelength of such devices is less than 0.6 nm for our deep UV process [222]. In our silicon wires, a 1 nm shift in waveguide width (height) yields a 1 nm (2 nm) shift in center wavelength [222]. Thus, width (height) variations typically ($1\sigma$) do not exceed 0.6 nm (0.3 nm). Therefore, the width (height) fluctuations contribute less than 11 MHz (0.7 MHz) or 31% (2%) of the measured linewidth. Besides, it was recently shown that 300 mm-diameter (instead of 200 mm-diameter) wafer processing results in a factor 2 better width and height uniformity [220].

Furthermore, the pillar size broadening is smaller than the width-induced broadening. By repeated pillar size measurements (as in fig.4.1c) in different sections of the spirals, we find that the pillar size variation falls within the $\pm 5$ nm error of the scanning electron microscope. This yields a linewidth contribution of less than 4 MHz.

In conclusion, there is no evidence for inhomogeneous broadening caused by geometrical non-uniformities. The measured and simulated leakage-limited $Q_m$-factors are – within the measurement error – in good agreement (fig.4.5b). The upper limits on width-, height- and pillar-induced broadening are 11 MHz, 0.7 MHz and 4 MHz. In the next chapter, however, we observe strong indications of inhomogeneous broadening in a cascade of suspended nanowires.

### Peak splitting in spirals

As discussed in the previous section, there is no evidence for inhomogeneous broadening of the resonances – in the sense that a geometric parameter varies continuously along the wire. However, our experiments do show peak split-ting in the 1 cm spiral (fig.4.2c). This splitting occurs in both the XPM (fig.4.7) and gain experiment (fig.4.8).

We find that the Fano resonances (fig.4.7) are nearly identical in the 2.7 mm straight wire and the 2 cm and 4 cm spirals. Thus neither the interaction strength $\gamma_{\text{SBS}}$ nor the linewidth $\frac{\kappa_{\text{m}}}{2\pi}$ is affected by the wire length. The Fano resonance shows two sharp peaks in the 1 cm spiral (fig.4.7b). The peaks are separated by 55 MHz. This particular spiral also has a reduced $\gamma_{\text{SBS}}$.

Similarly, the gain coefficient $2\gamma_{\text{SBS}}$ and the linewidth $\frac{\kappa_{\text{m}}}{2\pi}$ are nearly identical in the 2.7 mm straight wire and the 2 cm and 4 cm spirals (fig.4.8). The 1 cm spiral again consists of two resonances separated by 55 MHz (fig.4.8b). There is also a small remnant of the minor peak in the 2 cm and 4 cm spirals, which was not visible in the XPM-experiments because of the larger frequency step. Apart from the 1 cm spiral, the observed and predicted ($G_{\text{pred}} = 2\gamma_{\text{SBS}} P_{\text{p}} L_{\text{eff}}$) gain factors are in good agreement.

The 55 MHz splitting may be explained by a 2.8 nm shift in wire width in a sub-section of the waveguide. We suspect that this is caused by the lithographical proximity effect, in which closely adjacent wires are slightly less wide than an isolated wire. Such proximity effects are known to play an important role in deep UV lithography [213]. A drop in wire width would cause an upshift in resonance frequency, as we indeed observe going from the 2.7 mm straight wire to the spirals (fig.4.7-4.8).

For the relative height of the sub-peak and the main resonance (fig.4.8b), we find $G_{\text{sub}}/G_{\text{main}} = 45\%$. This fraction corresponds to the ratio $L_{\text{single}}/L_{\text{rest}} = 42\%$ (fig.4.2b), with $L_{\text{single}}$ the effective length of the waveguide where there is only a single wire (i.e. separated from other wires by more than 10 $\mu$m) and $L_{\text{rest}}$ the effective length of the spiral section where wires are in close proximity ($< 2\,\mu$m). The spirals are not located in the middle between the gratings (fig.4.2). Therefore, a detailed comparison also depends on the order of the wires.

In conclusion, the 1 cm spiral shows a reduced Brillouin nonlinearity. We suspect that this is caused by the lithographical proximity effect, which reduces the width of closely adjacent wires. The 2 cm and 4 cm spiral exhibit a Brillouin nonlinearity at full strength; as strong as in the 2.7 mm straight wire (fig.4.7-4.8).

### Backward scattering

So far we focused on forward SBS, in which the excited phonons have a very short wavevector $K = \frac{\Omega}{v_{\text{g}}}$ because the pump and Stokes have nearly equal wavevectors. However, the phononic mode demonstrated in this chapter (fig.4.1d) can also be operated at another point in its dispersion diagram. When the Stokes and pump counterpropagate through the wire, they generate the class of phonons that obey $K = 2k_0$. The propagating version (fig.4.9a) of the Fabry-Pérot mechanical mode (fig.4.1d) may then induce gain as well. These modes have an acoustic group velocity that nearly equals the bulk acoustic velocity. Therefore, they have a significantly larger acoustic decay length ($\sim 10\,\mu$m) – making them more attractive for phononic circuits. We reconfigured our gain experiment (fig.4.9b) to study such modes.

We also find a Lorentzian gain profile (fig.4.9c-d), but this time at 13.66 GHz. We do not observe this peak in unsuspended waveguides. For instance, we observe on/off gain of 0.22 dB with $P_\mathrm{p} = 11.8\,\mathrm{mW}$ and $L = 2\,\mathrm{cm}$ (fig.4.9d). We extract $2\gamma_\mathrm{SBS} = 359\ \mathrm{W}^{-1}\mathrm{m}^{-1}$ and $Q_\mathrm{m} = 971$. Thus we have $\frac{2\gamma_\mathrm{SBS}}{Q_\mathrm{m}} = 0.37\,\mathrm{W}^{-1}\mathrm{m}^{-1}$: a factor 30 lower than in the forward case (fig.4c). We attribute this reduction to destructive interference between electrostriction and radiation pressure, as predicted before [211] for fully suspended wires. Because of this low overlap, we observe these resonances only in the long spirals (fig.4.2). Based on our finite-element models, we expect this propagating mode (fig.4.9a) at 14.4 GHz with a coupling of $\frac{2\gamma_\mathrm{SBS}}{Q_\mathrm{m}} = 0.41\,\mathrm{W}^{-1}\mathrm{m}^{-1}$. Therefore, we suspect that this is indeed the observed mode. Further investigations should resolve this issue, as the simulations predict that there are propagating modes with slightly better coupling at higher frequencies. However, even the highest simulated coupling strength reaches only $\frac{2\gamma_\mathrm{SBS}}{Q_\mathrm{m}} = 1.04\,\mathrm{W}^{-1}\mathrm{m}^{-1}$ for an elastic mode at 27.3 GHz. The sweep of the frequency spacing $\frac{\Omega}{2\pi}$ between the pump and Stokes seed was limited to 16 GHz in the current set-up.

Finally, it has been predicted that backward SBS is suppressed in waveguides consisting of a cascade of fully suspended regions [98]. The issue is that the acoustic wave cannot build up to its full strength in a suspended waveguide shorter than the acoustic decay length. In the limit of long suspended regions, the length of each region is effectively reduced by one acoustic decay length ($\sim 10\,\mu$m). However, our wires are partially suspended all along their length. Therefore the acoustic build-up has little effect on the backward SBS gain.

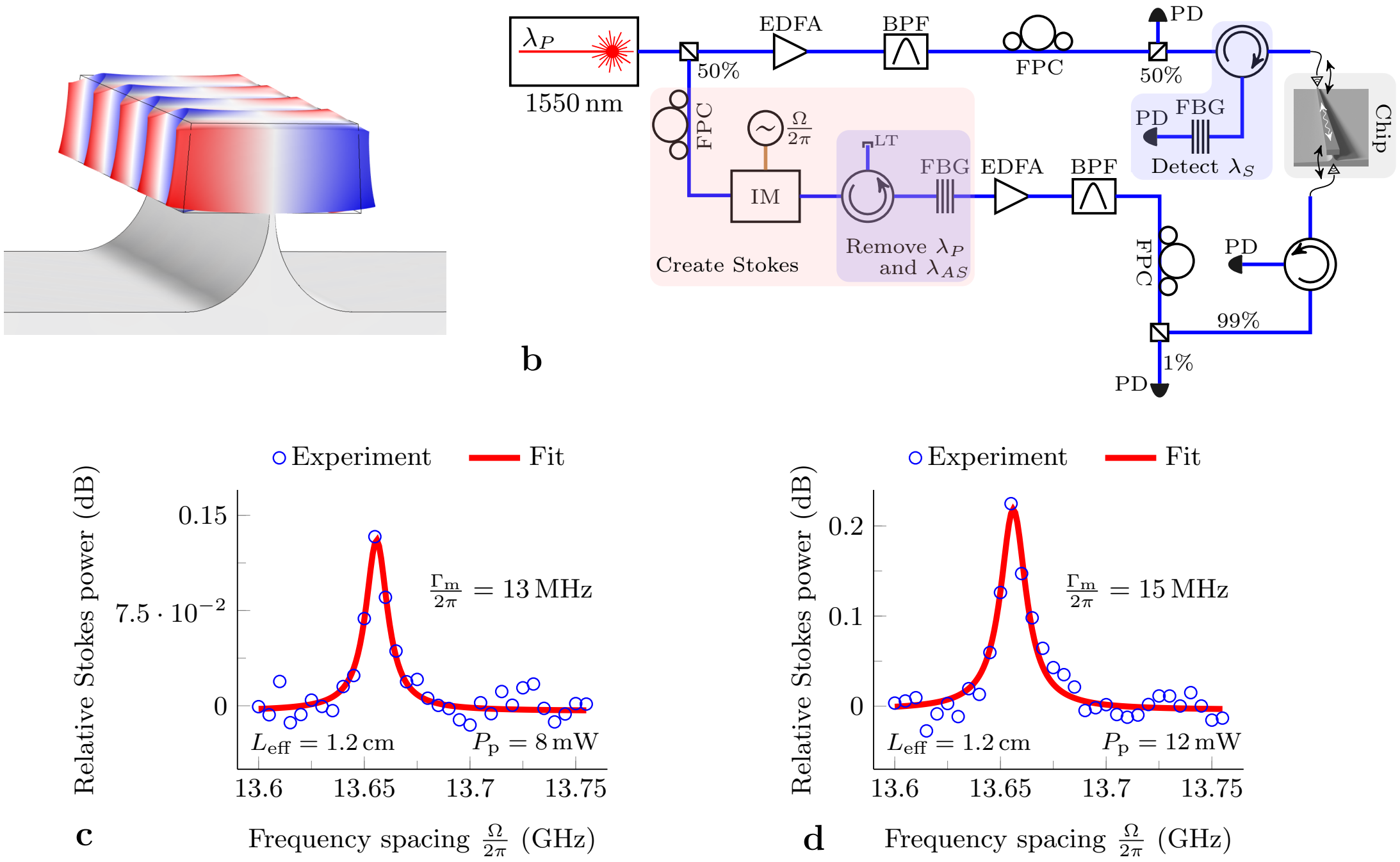

**Figure 4.9: Characterization of the backward Brillouin scattering.** **a**, Propagating version of the Fabry-Pérot phononic mode (see fig.4.1d for comparison). **b**, Experimental set-up used to observe the backward SBS gain. This time the Stokes and pump wave counterpropagate through the chip, exciting phonons that satisfy $K = 2k_0$. These acoustic waves have a significantly longer decay length on the order of $10\,\mu$m. **c**-**d**, Lorentzian gain profile on the Stokes seed in the 2 cm spiral ($L_{\text{eff}} = 1.2$ cm) for on-chip pump powers $P_{\text{p}}$ of 8 mW (**c**) and 12 mW (**d**).

# 5

# Cascade of suspended silicon nanowires

*Go to the edge of the cliff and jump off. Build your wings on the way down.*
Ray Bradbury

This chapter is based on [65].

**Contents**

THE CENTURY-OLD *study of photon-phonon coupling has seen a remarkable revival in the past decade. Driven by early observations of dynamical back-action, the field progressed to ground-state cooling and the counting of individual phonons. A recent branch investigates the potential of traveling-wave, optically broadband photon-phonon interaction in silicon circuits. In this chapter, we report continuous-wave Brillouin gain exceeding the optical losses in a series of suspended silicon beams, a step towards selective on-chip amplifiers. We obtain efficiencies up to* $10^4$ $W^{-1}m^{-1}$*, the highest to date in the phononic gigahertz range. We also find indications that geometric disorder poses a significant challenge towards nanoscale phonon-based technologies.*

## 5.1 Introduction

The interaction between photons and acoustic phonons has been investigated in bulk materials since the 1920s [36, 37]. In case the phonons are generated by optical forces, such as radiation pressure and electrostriction [43, 64, 104, 125, 211], the interaction is often called stimulated Brillouin scattering – a feedback loop in which energy flows from the optical waves to the mechanical oscillator. Its signature is the narrowband amplification of an optical probe that is red-detuned by the phonon resonance frequency from a strong optical pump.

Although the mechanical linewidth does not exceed 100 MHz typically, there is no such inherent optical bandwidth restriction in SBS: the pump wavelength can be freely tuned in a wide span ($\sim$ 100 nm) and multiple pumps can be combined to tailor the SBS response [41]. Compared to cavity-based optomechanics [35, 138, 140, 179, 206, 223, 224], such a circuit-oriented approach – exploiting wideband waveguides – is intrinsically less power-efficient as the optical field is not resonantly enhanced. Nevertheless, the removal of the optical bandwidth restriction and the accompanying optical versatility has motivated a great deal of SBS work in small-core waveguides, from photonic crystal [93, 149, 152], dual-web [147] and subwavelength [154] fibres to chalcogenide [132, 141, 203, 225] and silicon waveguides [43, 64, 161]. It may provide new integrated signal processing capabilities such as tunable RF notch filters [161, 218] and true time delays [159]. The prospect is especially appealing in silicon photonic wires, whose strong confinement enhances the light-matter coupling. Mass-manufacturable silicon-on-insulator chips are therefore an exciting platform for high-density optomechanical circuitry, perhaps even at the quantum level [139, 162, 163, 226].

Recent work on this front has demonstrated promising photon-phonon coupling efficiencies in all-silicon waveguides [64]. The coupling was sufficiently strong to bring the waveguides into transparency (i.e. on/off gain equal to the propagation loss), but phonon leakage and free-carrier absorption precluded actual amplification above the optical propagation loss. Here, we eliminate the phonon clamping loss – observing an increase of the phonon quality factor from 300 up to 1000 at room temperature – by fully suspending the silicon nanowires. Thus, we achieve a modest amount of gain exceeding the optical losses. The waveguides consist of a series of suspended beams, supported by silicon dioxide anchors (fig.5.1).

Finally, we observed a strong dependence of the phonon quality factor on the number of and distance between the suspensions. This indicates the presence of geometric disorder that broadens and splits the phonon dispersion relation in some cases, similar to Doppler broadening in gas lasers [227]. From a wider perspective and not limited to our system, such geometric disorder may hinder development of nanoscale phonon circuits quite generally

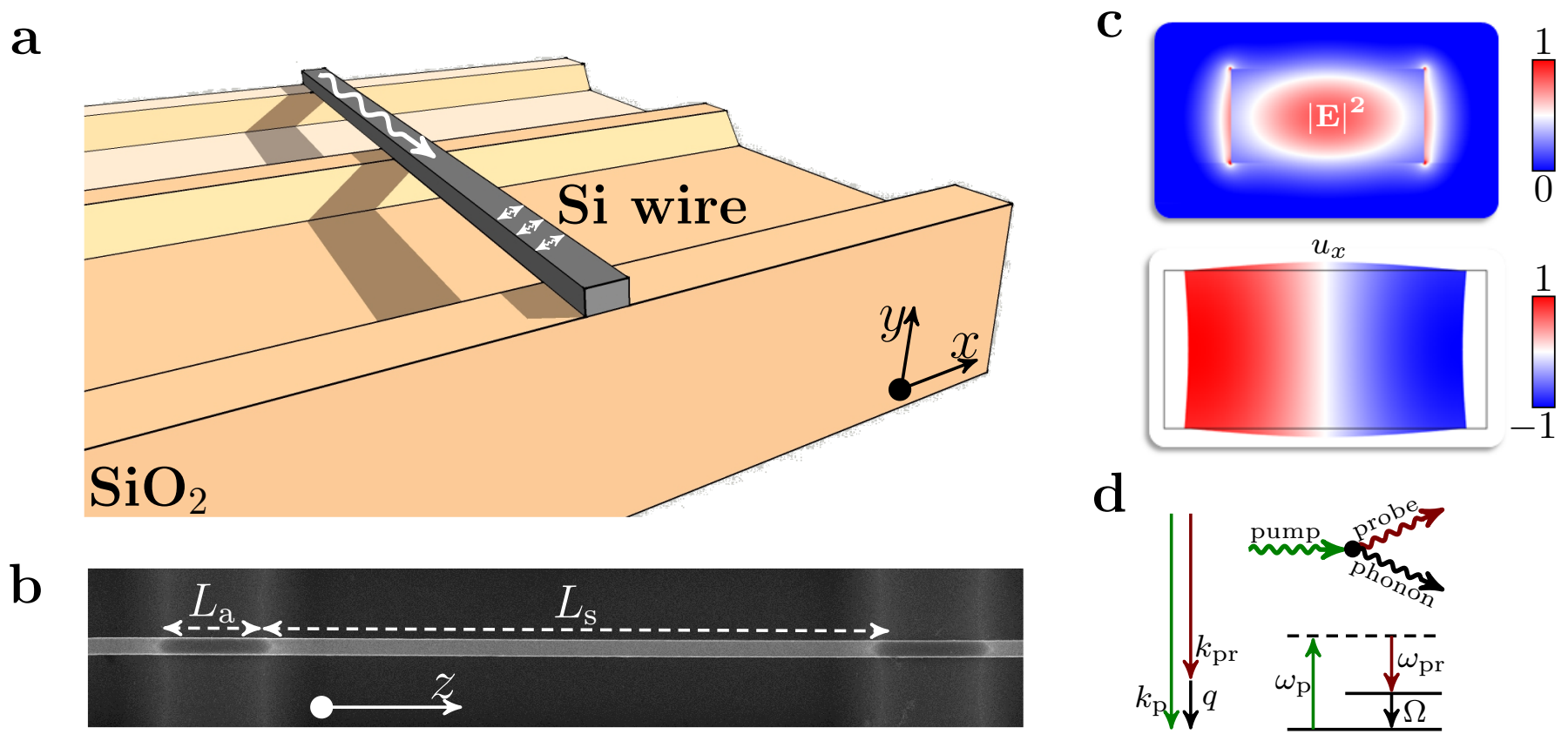

**Figure 5.1: A series of suspended silicon nanowires.** **a**, Impression of a silicon-on-insulator waveguide that consists of a series of suspensions and anchors. The photons propagate along the wire while the phonons are localized at their $z$-point of creation. **b**, Scanning electron micrograph of an actual suspension of length $L_s = 25.4\,\mu$m held by $L_a = 4.6\,\mu$m long anchors. **c**, Photonic (top) and phononic (bottom) traveling modes. **d**, The Brillouin process converts incoming pump photons with energy-momentum $(\hbar\omega_p, \hbar k_p)$ into redshifted probe (Stokes) photons $(\hbar\omega_{pr}, \hbar k_{pr})$ and phonons $(\hbar\Omega, \hbar q)$.

[139, 162–164].

## 5.2 Results and discussion

### 5.2.1 Theoretical background

The following discussion is concerned with forward intra-modal scattering, in which co-propagating pump and probe waves generate low-wavevector, low-group-velocity acoustic phonons (fig.5.1d).

#### Brillouin gain

First, we briefly treat the small-signal Brillouin gain in a waveguide consisting of suspensions of length $L_s$ and anchors of length $L_a$. The section length is $L_{sec} = L_s + L_a$ and there are $N = \frac{L}{L_{sec}}$ such sections with $L$ the total waveguide length (fig.5.1). We denote the input pump power $P_p$ and the red-detuned probe (Stokes) power $P_{pr}$. As previously shown (appendix B and equation (3.9)), the power $P_{pr}$ of the probe obeys

$$\frac{dP_{pr}}{dz} = -\left(\tilde{\mathcal{G}} P_p e^{-\alpha z} \Im \mathcal{L} + \alpha\right) P_{pr} \qquad \text{suspensions} \qquad (5.1)$$

$$\frac{dP_{pr}}{dz} = -\alpha P_{pr} \qquad \text{anchors}$$

in the low-cascading regime and with $\tilde{\mathcal{G}}$ the Brillouin gain coefficient, $\mathcal{L}(\Delta) = \frac{1}{-2\Delta+i}$ the complex Lorentzian, $\Delta = \frac{\Omega-\Omega_m}{\kappa_m}$ the normalized detuning, $\kappa_m$ the phonon linewidth and $\Im\mathcal{L} = -\frac{1}{4\Delta^2+1}$. To derive (5.1), we assumed that the phonon propagation loss far exceeds the photon propagation loss and that the photon-phonon coupling is weak relative to the spatial phonon decay [63]. In particular, in this work the photon decay length $\alpha^{-1}$ is about a centimeter, while the phonons spatially decay over a couple of nanometers in the $z$-direction [64]. Indeed, the flat dispersion (fig.2.3) of these Raman-like [104, 152] phonons yields an exceedingly low group velocity [64]. Therefore, each suspension consists of a series of independent mechanical oscillators – whose frequency depends on the local width [64]. This phonon locality also implies that the anchor does not contribute to the Brillouin gain.

We treat the optical loss $\alpha$ as distributed, although it may in fact be partially localized at the interfaces or be unequal in the suspensions and anchors. This is a good approximation as the SBS strength in the next section only depends on the remaining pump power, not on how some of it was lost in the previous sections. The ansatz $P_{pr} = g(z)e^{-\alpha z}$ and piecewise integration of (5.1) results in

$$\ln \frac{g(L)}{g(0)} = -\tilde{\mathcal{G}} P_p L_{s,eff} \Im\mathcal{L} \sum_{k=0}^{N-1} e^{-\alpha k L_{sec}} = -\tilde{\mathcal{G}} P_p L_{s,eff} \Im\mathcal{L} \frac{1-e^{-\alpha L}}{1-e^{-\alpha L_{sec}}} \approx -\tilde{\mathcal{G}} P_p f_s L_{eff} \Im\mathcal{L}$$

with the effective suspension length $L_{s,eff} = \frac{1-e^{-\alpha L_s}}{\alpha} \approx L_s$ since the sections are much smaller than the optical decay length ($L_{sec} \ll \alpha^{-1}$) and $f_s = \frac{L_s}{L_{sec}}$. Therefore we obtain

$$\ln \frac{P_{pr}(L)}{P_{pr}(0)e^{-\alpha L}} = \frac{\tilde{\mathcal{G}} P_p f_s L_{eff}}{4\Delta^2+1} \tag{5.2}$$

These experiments are the circuit analog of cavity-based optomechanically induced transparency [158, 228]. However, our system features spatially stronger mechanical than optical damping, such that it is the optical response that is modified here [63].

### Cross-phase modulation

Gain measurements provide access to all relevant optomechanical parameters, but require careful calibration of the on-chip pump power $P_p$. In contrast, a cross-phase modulation (XPM) measurement [43, 64] is, in absence of free-carriers, intrinsically calibrated: it provides access to the ratio of the photon-phonon coupling and the electronic Kerr effect independent of pump power. These experiments are the circuit analog of cavity-based coherent wavelength conversion [229], although the conversion need not take place between two optical resonances in our case.

We assume weak XPM and denote the envelopes of the pump and its red- and blue-detuned sidebands $a_{\text{p}}$, $a_{\text{p}-}$ and $a_{\text{p}+}$ and similarly for the injected probe $a_{\text{pr}}$ and the XPM-imprinted blue-shifted sideband $a_{\text{pr}+}$. The imprinted sideband grows as (appendix B)

$$\frac{\mathrm{d}a_{\text{pr}+}}{\mathrm{d}z} = -\frac{i}{2}\left(4\gamma_{\text{Ks}} + \tilde{\mathcal{G}}\mathcal{L}\right)\left(a_{\text{p}}a_{\text{p}-}^{\star} + a_{\text{p}+}a_{\text{p}}^{\star}\right)a_{\text{pr}} - \frac{\alpha}{2}a_{\text{pr}+} \qquad \text{suspensions}$$
$$\frac{\mathrm{d}a_{\text{pr}+}}{\mathrm{d}z} = -i2\gamma_{\text{Ka}}\left(a_{\text{p}}a_{\text{p}-}^{\star} + a_{\text{p}+}a_{\text{p}}^{\star}\right)a_{\text{pr}} - \frac{\alpha}{2}a_{\text{pr}+} \qquad \text{anchors}$$

with $\gamma_{\text{K}}$ the Kerr parameter. Note that the XPM can also be seen as a four-wave mixing process with photon creations ($a^{\star}$) and annihilations ($a$) given by $a_{\text{pr}+}^{\star}\left(a_{\text{p}}a_{\text{p}-}^{\star} + a_{\text{p}+}a_{\text{p}}^{\star}\right)a_{\text{pr}}$. We assume that the pump and probe remain undepleted by the XPM, but include their absorptive decay. Then we get

$$\frac{\mathrm{d}a_{\text{pr}+}}{\mathrm{d}z} = C_{\text{s}}e^{-\frac{3\alpha}{2}z} - \frac{\alpha}{2}a_{\text{pr}+} \qquad \text{suspensions}$$
$$\frac{\mathrm{d}a_{\text{pr}+}}{\mathrm{d}z} = C_{\text{a}}e^{-\frac{3\alpha}{2}z} - \frac{\alpha}{2}a_{\text{pr}+} \qquad \text{anchors}$$

with $C_{\text{s}} = \frac{1}{2}\left(4\gamma_{\text{Ks}} + \tilde{\mathcal{G}}\mathcal{L}\right)C$, $C_{\text{a}} = 2\gamma_{\text{Ka}}C$ and $C = -i\left(a_{\text{p}}a_{\text{p}-}^{\star} + a_{\text{p}+}a_{\text{p}}^{\star}\right)a_{\text{pr}}|_{z=0}$. Inserting the ansatz $a_{\text{pr}+}(z) = g(z)e^{-\frac{\alpha}{2}z}$ and piecewise integrating ($g(0) = 0$) yields

$$g(L) \approx \left(C_{\text{s}}L_{\text{s}} + C_{\text{a}}L_{\text{a}}\right)\sum_{k=0}^{N-1} e^{-\alpha k L_{\text{sec}}} = \left(C_{\text{s}}L_{\text{s}} + C_{\text{a}}L_{\text{a}}\right)\frac{1-e^{-\alpha L}}{1-e^{-\alpha L_{\text{sec}}}} \approx \left(C_{\text{s}}f_{\text{s}} + C_{\text{a}}f_{\text{a}}\right)L_{\text{eff}}$$

where we used $L_{\text{a,eff}} \approx L_{\text{a}}$, $L_{\text{eff}} = \frac{1-e^{-\alpha L}}{\alpha}$ and $f_{\text{a}} = 1 - f_{\text{s}}$. Therefore,

$$\left|a_{\text{pr}+}(L)\right|^2 = 4\overline{\gamma}_{\text{K}}^2\mathcal{F}|C|^2L_{\text{tot,eff}}^2e^{-\alpha L} \propto \mathcal{F}$$

with the averaged Kerr parameter $\overline{\gamma}_{\text{K}} = \gamma_{\text{Ks}}f_{\text{s}} + \gamma_{\text{Ka}}f_{a}$, the normalized Fano function

$$\mathcal{F}(\Delta) = \left|1 + f_{\text{s}}\frac{\tilde{\mathcal{G}}}{4\overline{\gamma}_{\text{K}}}\mathcal{L}(\Delta)\right|^2 = |1 + r\mathcal{L}(\Delta)|^2 \tag{5.3}$$

and $r = f_{\text{s}}\frac{\tilde{\mathcal{G}}}{4\overline{\gamma}_{\text{K}}}$ the ratio between the mechanically- and Kerr-driven XPM.

Including two-photon absorption, free-carrier index changes and free-carrier absorption, (5.3) no longer holds. For instance, in case of two-photon absorption the Kerr parameter should be replaced by $\overline{\gamma}_{\text{K}} \to \overline{\gamma}_{\text{K}} - i\overline{\gamma}_{\text{TPA}}$ and (5.3) becomes

$$\mathcal{F}(\Delta) = \left|1 + e^{i\phi}f_{\text{s}}\frac{\tilde{\mathcal{G}}}{4\overline{\gamma}_{\text{tot}}}\mathcal{L}(\Delta)\right|^2 = \left|1 + e^{i\phi}r\mathcal{L}(\Delta)\right|^2 \tag{5.4}$$

with $\overline{\gamma}_{\text{tot}} = \sqrt{\overline{\gamma}_{\text{K}}^2 + \overline{\gamma}_{\text{TPA}}^2}$ and $\tan\phi = \frac{\overline{\gamma}_{\text{TPA}}}{\overline{\gamma}_{\text{K}}}$. In our case, the two-photon $\phi$ is small and positive as $\frac{\overline{\gamma}_{\text{TPA}}}{\overline{\gamma}_{\text{K}}} \approx 0.1$. The free-carrier nonlinearity $\overline{\gamma}_{\text{FC}}$, however, can give rise to negative $\phi$ (section 5.4).

### 5.2.2 Fabrication and passive characterization

We started from air-cladded 220 nm thick, 450 nm wide silicon-on-insulator wires fabricated by 193 nm UV lithography (`www.ePIXfab.eu`) at imec. Next, we patterned an array of apertures in a resist spinned atop the wires. Then we immersed the chip in buffered hydrofluoric acid, which selectively etches the silicon dioxide substrate, until the wires were released. The end result was a series of suspended beams, each typically 25 $\mu$m long and held by 5 $\mu$m silicon dioxide anchors (fig.5.1). Simulations and measurements show that the reflections caused by these anchors are negligibly small (section 5.4). We found optical losses $\alpha \approx 5.5$ dB/cm by the cut-back method, which are a factor 2 larger than before the etch. This is likely related to a deterioration of the wires' surface state and consistent with both (1) the measured drop in free-carrier lifetime (section 5.4) and (2) the decrease in free-carrier absorption found in the gain experiment (fig.5.2b).

### 5.2.3 Optomechanical experiments

In this section, we discuss the guided-wave optomechanical characterization of a series of suspended silicon nanobeams. We used the experimental setups presented in [64] and reproduced in the Appendix. Our device is characterized by the suspension length $L_{\text{s}}$, the anchor length $L_{\text{a}}$, the section length $L_{\text{sec}} = L_{\text{s}} + L_{\text{a}}$, the number of suspensions $N$, the total length $L = NL_{\text{sec}}$, the total effective length $L_{\text{eff}} = \frac{1-e^{-\alpha L}}{\alpha}$, the suspended fraction $f_{\text{s}} = \frac{L_{\text{s}}}{L_{\text{sec}}}$ and the waveguide width $w$. Unless stated otherwise, these parameters have values $L_{\text{s}} = 25.4\,\mu$m, $L_{\text{a}} = 4.6\,\mu$m, $L_{\text{sec}} = 30\,\mu$m, $N = 85$, $L = 2535\,\mu$m, $L_{\text{eff}} = 2168\,\mu$m, $f_{\text{s}} = 0.85$ and $w = 450$ nm. In some cases, our waveguides have a non-suspended input/output section before/after the cascade of suspended nanobeams. We take this into account when calculating input pump powers (gain experiment) or suspended fractions (XPM experiment). We focus on this sub-centimeter wire as longer waveguides suffer from a decreased mechanical quality factor (fig.5.3a).

#### Brillouin gain

First, we measured the amplitude response of our system. We injected a weak probe red-detuned from a strong pump and retrieved the probe power as a function of detuning (fig.5.2a). As before [64], we find gain resonances around 9.1 GHz. The on/off gain increases with pump power (fig.5.2b) and reaches the transparency point $\tilde{\mathcal{G}} P_p = \alpha$ around $P_p = 30$ mW. Beyond this pump power, the ouput exceeds the input probe photon flux. At the maximum pump power of 39 mW, we obtain guided-wave cooperativities [63] of $\tilde{\mathcal{C}} = \frac{(f_s \tilde{\mathcal{G}}) P_p}{\alpha} = 1.7$. This modest net gain of 0.5 dB (fig.5.2b) is a step towards selective on-chip amplifiers that could be used for homodyne detection, in order to eliminate the requirement of a phase-stabilized local oscillator [230].

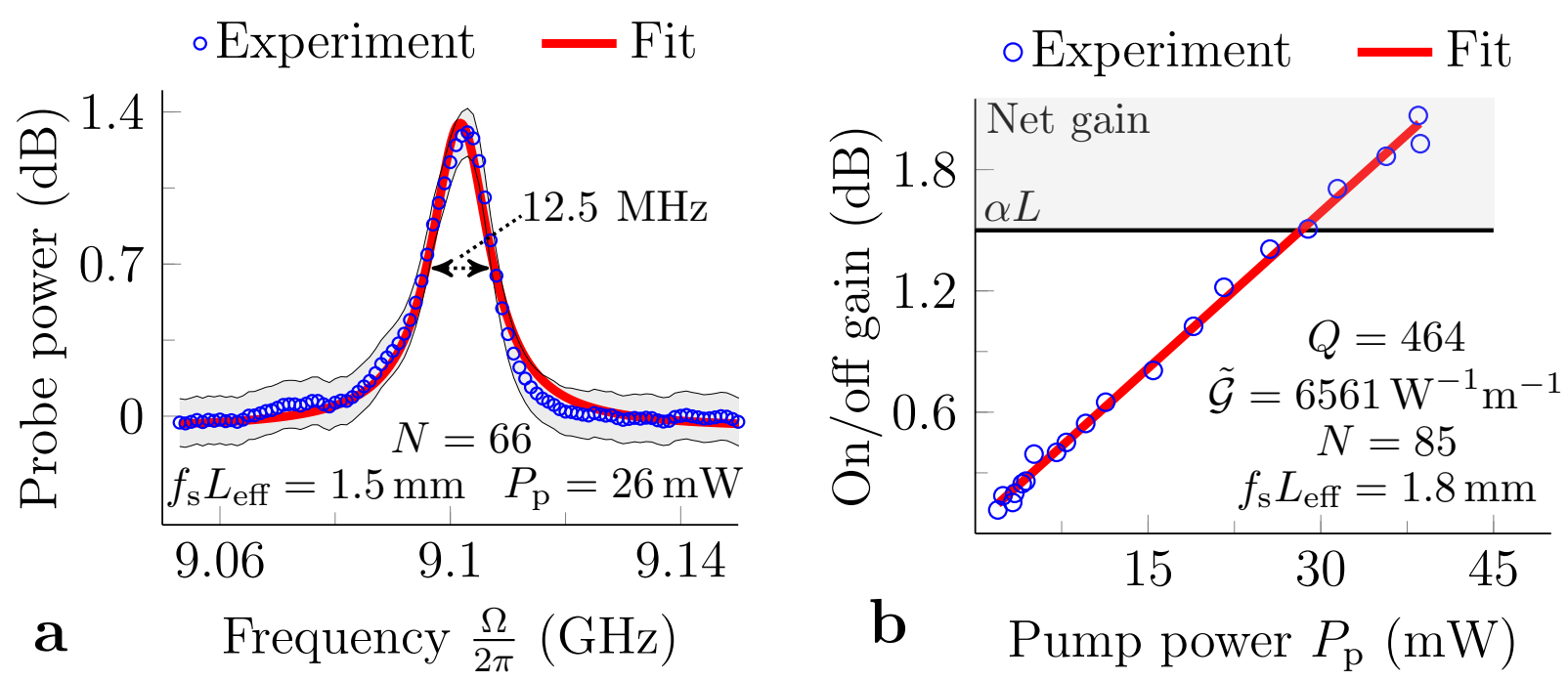

**Figure 5.2: Brillouin gain exceeding the optical losses. a**, An example of a Brillouin gain resonance, in this case with an on/off gain of 1.4 dB, quality factor of $Q_m = 728$ and an on-chip input pump power of 26 mW. The shaded grey area indicates uncertainty in the probe power. **b**, Scan of the on/off gain with pump power. At a pump power of 30 mW the transparency point is reached. For $P_p > 30$ mW, more probe photons leave than enter the waveguide. The slope yields the Brillouin gain coefficient $\tilde{\mathcal{G}} = 6561\,\mathrm{W^{-1}m^{-1}}$ with a quality factor of 464 in this particular waveguide. Notably, the on/off gain scales linearly with pump power across the entire sweep – indicating the absence of free-carrier absorption in this range.

Notably, the linear scaling between on/off gain and pump power (fig.5.2b) indicates the absence of free-carrier absorption up to 40 mW [98, 231]. In contrast, we previously measured increased nonlinear absorption already at 25 mW in silicon wires on a pillar [64]. Both this finding and the higher propagation losses (5.5 dB/cm instead of 2.6 dB/cm [64]) likely originate in a deterioration of the wires' surface state during the fabrication of the suspensions. In agreement with this hypothesis, we measured a drop in the free-carrier lifetime (section 5.4).

In case this structure were to be placed in a cavity, such as a silicon microring, it would also have to overcome coupling losses to achieve the photon/phonon lasing threshold ($\tilde{\mathcal{C}} > \mathcal{C} = 1$ [63]). We note that the acoustic linewidth ($\sim 10$ MHz) is a factor $10^2$ smaller than typical optical linewidths of silicon microcavities ($\sim 1$ GHz). Therefore, this would produce stimulated emission of phonons, not photons [35, 63, 134, 142, 179, 209]. Such a device would not benefit from the spectral purification associated with Brillouin lasers [55, 183]. The origin of this reversal of the damping hierarchy (going from waveguides to cavities) [63, 142] lies in the exceedingly low group velocity of these Raman-like [152] acoustic phonons; indeed, despite enormously higher propagation losses they usually still have lower linewidths than photons [35, 63, 142]. Only uniquely high-quality optical cavities, to date realized only using silica [136, 141, 180, 186, 187] or crystalline [140] materials, can produce lower photonic than phononic damping rates.

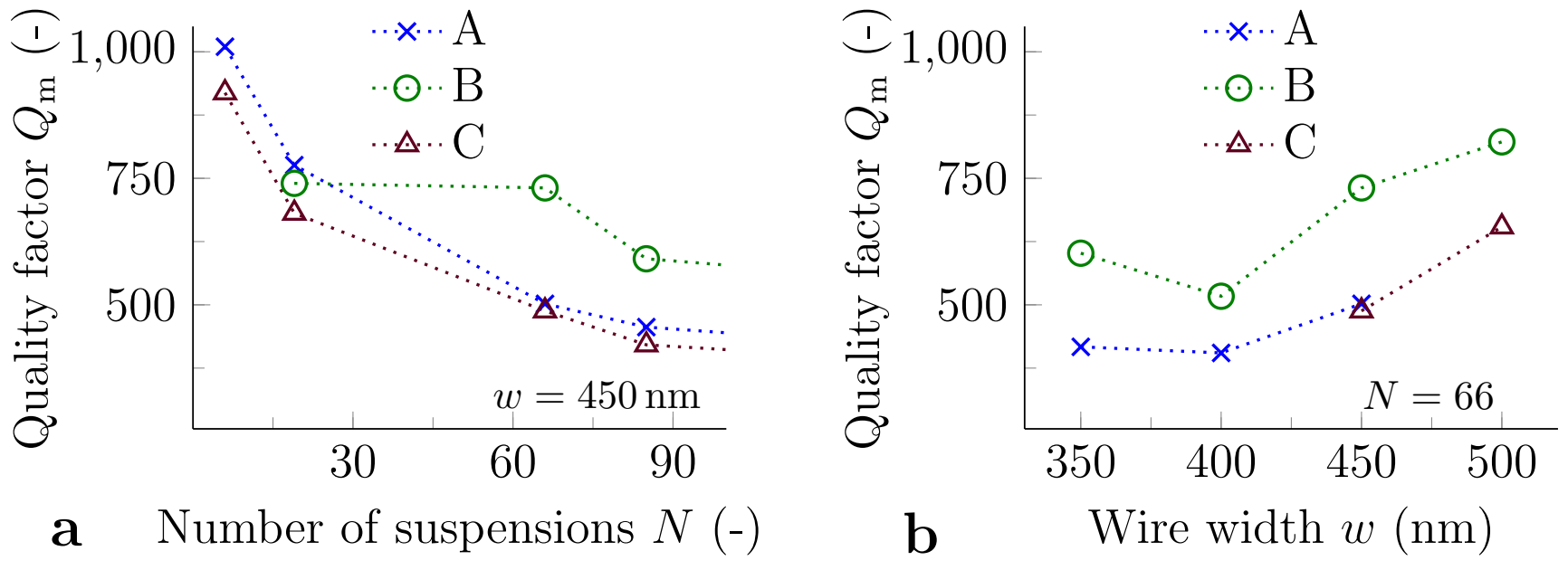

**Figure 5.3: The quality factor decreases with the number of suspensions.** We study the phonon quality factor for three samples (A, B and C) from the same wafer. The samples were designed to be identical. **a**, The quality factor increases up to 1010 when there are only 6 suspensions. For larger $N$, the quality factors approach $\approx 400$. Unless stated otherwise, all resonances are still well-fit by a Lorentzian function (fig.5.2a). **b**, In general, wider waveguides exhibit slightly larger $Q_m$. However, this pattern is neither linear (samples A/C) nor monotonic (sample B). Some waveguides were defective, possibly because of a collapsed beam, and were excluded from the study.

### Geometric disorder

Next, we study the quality factor extracted from the gain resonances (fig.5.2a). We find that it strongly decreases with the number of suspensions $N$ (fig.5.3a); from $Q_m \approx 10^3$ at $N = 6$ to $Q_m \approx 400$ at $N = 85$. For larger numbers of suspensions in a spiral configuration ($N = 1332$, not shown), the quality factor levels off around $Q_m \approx 340$. Notably, this relation changes from sample to sample – even if they originate from the same wafer (fig.5.3a). We attribute such variations to inhomogeneous broadening by geometric disorder, presumably in the width of the nanowires [64]. Indeed, the sensitivity of the resonance frequency $\frac{\Omega_m}{2\pi}$ to width variations is $\frac{1}{2\pi}\frac{d\Omega_m}{dw} \approx 20\,\text{MHz/nm}$ [64]. Therefore, realistic width variations $\delta w$ of about 0.5 nm [220, 222] yield inhomogeneous linewidths of about $\frac{\Gamma_{inh}}{2\pi} \approx 10\,\text{MHz}$ – comparable to those measured (fig.5.2a). Similar disorder has been studied in snowflake crystals [139].

Further, we investigated the influence of the width $w$ on the quality factor

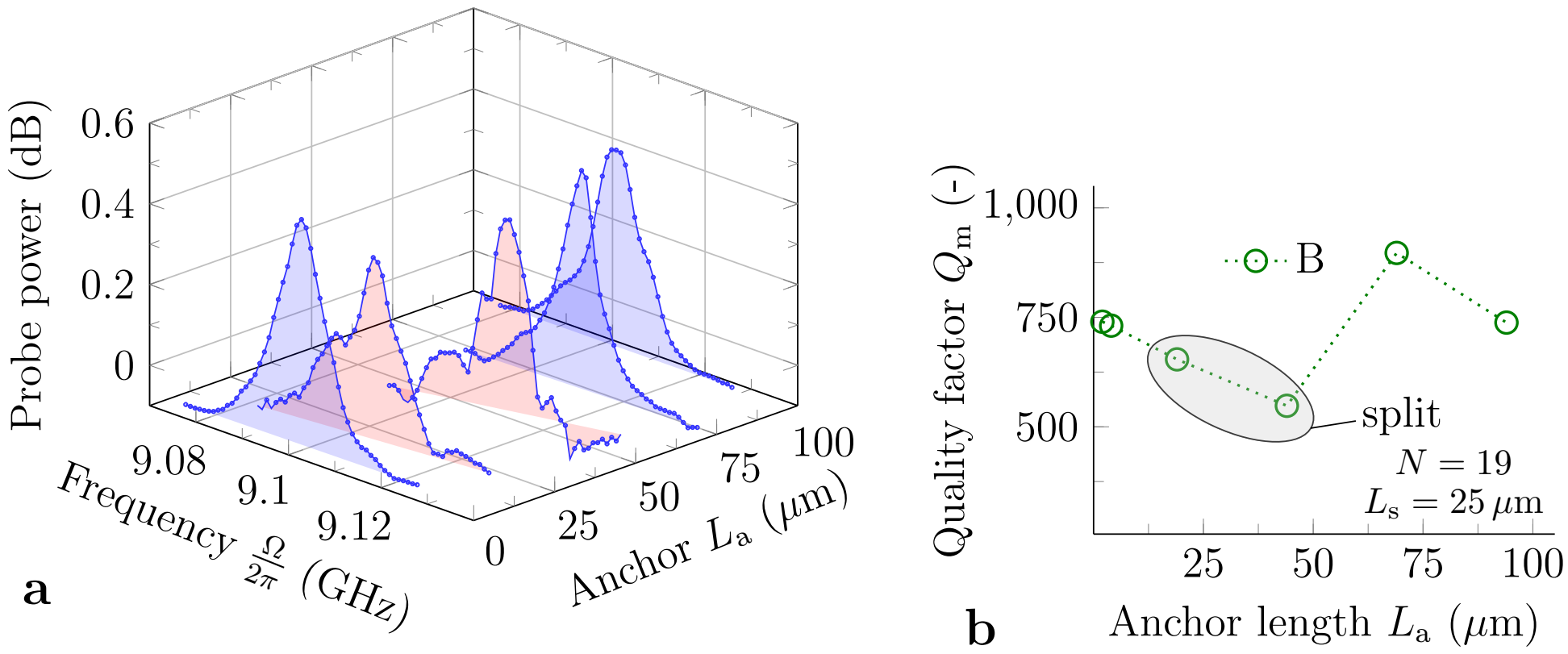

**Figure 5.4: The phonon resonance splits at certain anchor lengths. a**, As we sweep the anchor length, the initially clean curve splits at $L_a = 19\,\mu$m and $44\,\mu$m but recombines at $L_a = 69\,\mu$m. The pump power was $P_p = 26\,$mW and the position of the first suspension was fixed in this sweep. **b**, A Lorentzian fit to the gain curves of (**a**) yields high quality factors at short and long anchors. We suspect that (**a**) and (**b**) arise from a nanometer-scale ($\delta w$) width fluctuation in this straight silicon wire.

$Q_m$. Since the resonance frequency scales inversely with width ($\frac{\Omega_m}{2\pi} \propto w^{-1}$) [64], its sensitivity scales inverse quadratically with width ($\frac{d\Omega_m}{dw} \propto w^{-2}$). Subsequently, the inhomogeneously broadened linewidth scales similarly ($\Gamma_{inh} \propto w^{-2}$) in case the size of the width variations $\delta w$ does not depend on $w$. Then the quality factor scales linearly with width ($Q_m = \frac{\Omega_m}{\kappa_m} \propto w$). We indeed observe overall larger quality factors for wider wires, although this pattern is neither linear nor monotonic (fig.5.3b). We note that, unless stated otherwise, all resonances were still well-fit by a Lorentzian function (fig.5.2a). In case of sufficiently sampled geometric disorder, the gain curves would become convolutions of a Lorentzian and a probability function describing the geometric disorder (e.g. distribution of the width $w$). The largest deviations of such Voigt curves with respect to a Lorentzian occur in the tails (large $\Delta = \frac{\Omega - \Omega_m}{\kappa_m}$), precisely where the relative uncertainty in the measured probe power is highest (fig.5.2a). Given this uncertainty, both Lorentzian and e.g. Gaussian-shaped curves produce good fits to the gain resonances. A low-temperature characterization would yield more information regarding the nature of the acoustic broadening mechanisms.

There are two types of potential width fluctuations: (1) fast sidewall roughness with a small coherence length $L_{coh}$ (on the order of 50 nm [232]) and (2) slow variations in the average waveguide width $w$ with a much larger coherence length (on the order of 100 $\mu$m). We suspect that mechanism (2) is at play here, since even an individual section ($L_{sec} \approx 30\,\mu$m) is much larger than the coherence length ($\approx$ 50 nm [232]) of the surface roughness. Therefore, sidewall roughness cannot explain the significant changes of $Q_m$ with $N$ (fig.5.3a): even a single section samples it fully ($L_{\text{sec}}/L_{\text{coh}} \approx 600$). In contrast, slow excursions of the waveguide width are consistent with such behavior. We confirm this by scanning the anchor length $L_a$ while keeping the number of suspensions $N$ and the suspension length $L_s$ constant (fig.5.4). In this sweep, the position of the first suspension is fixed. As $L_a$ increases, the initially clean resonance first splits at $L_a = 19\,\mu$m and then recombines at $L_a = 69\,\mu$m (fig.5.4a). Remarkably, the $L_a = 69\,\mu$m wire even produces the highest quality factor (fig.5.4b). In light of the above discussion, this behavior likely stems from a nanometer-scale ($\delta w$) width excursion: short and long anchor waveguides avoid the width fluctuations and thus yield clean profiles. Both fig.5.3&5.4 are fingerprints of geometric disorder that hinders the development of integrated Brillouin-based technologies.

The disorder-induced broadening may be overcome by moving away from the forward intra-modal configuration. In general, the observed resonance broadens through two mechanisms: First, the direct acoustic broadening as observed in this chapter. Second, indirect broadening of the acoustic resonance via the phase-matching condition. For an slight increase in waveguide width, the direct mechanism decreases the acoustic frequency. However, it increases the optical wavevector – which indirectly increases the acoustic frequency in the backward configuration. The direct and indirect shifts can thus have opposite sign and similar magnitude. So they could cancel out. This has been studied in simulation [68], but it has yet to be demonstrated. In contrast, the indirect broadening disappears in case of forward intra-modal scattering.

### Photon-phonon overlap

Finally, we measure the XPM resonances (fig.5.5a) for a subset of waveguides to obtain an independent estimate of the photon-phonon interaction efficiency. Combined with the gain data, we obtain the $(\tilde{\mathcal{G}}, Q_m)$-pairs for a large set of waveguides (fig.5.5b) with fixed waveguide width $w = 450\,$nm. A fit (fig.5.5b) to this dataset yields a non-resonant nonlinearity of $\frac{\tilde{\mathcal{G}}}{Q_m} = 10.3\,\text{W}^{-1}\text{m}^{-1}$ – in good agreement with earlier experiments [64] and predictions [104, 211]. The efficiencies reach up to $\tilde{\mathcal{G}} = 10360\,\text{W}^{-1}\text{m}^{-1}$, the highest value obtained thus far in the gigahertz range. Flexible megahertz-class systems in vacuum can produce up to $\tilde{\mathcal{G}} = 10^6\,\text{W}^{-1}\text{m}^{-1}$ at 6 MHz [147]. Previous gigahertz systems achieved $\tilde{\mathcal{G}} = 3100\,\text{W}^{-1}\text{m}^{-1}$ in silicon pedestal waveguides at 9.2 GHz [64], $\tilde{\mathcal{G}} = 2328\,\text{W}^{-1}\text{m}^{-1}$ in hybrid silicon/silicon nitride waveguides at 1.3 GHz [43] and $\tilde{\mathcal{G}} = 304\,\text{W}^{-1}\text{m}^{-1}$ in chalcogenide rib waveguides at 7.7 GHz [34].

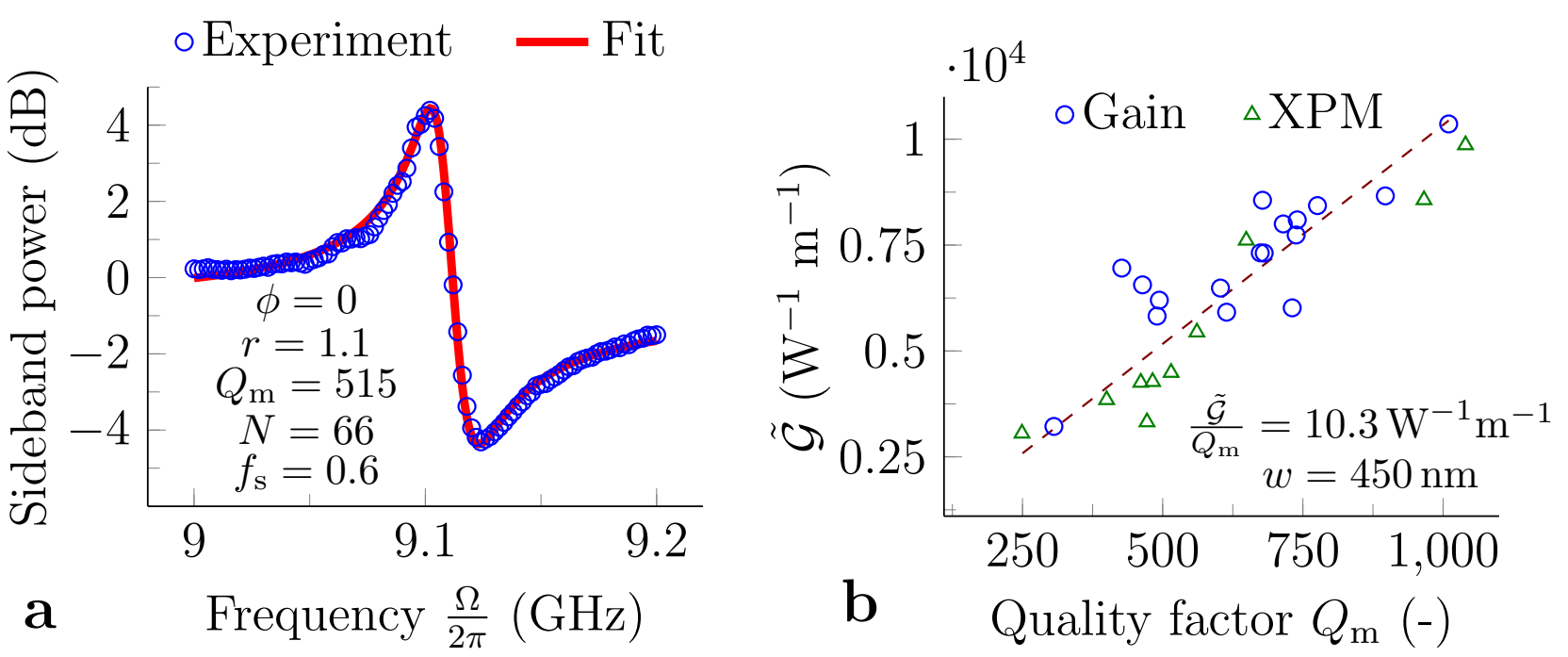

**Figure 5.5: The efficiency $\tilde{\mathcal{G}}$ reaches up to $10^4\,\mathbf{W}^{-1}\mathbf{m}^{-1}$.** **a**, Example of a Fano resonance obtained from the XPM experiment, used to determine the quality factor $Q_m$ and gain coefficient $\tilde{\mathcal{G}}$ given $\overline{\gamma}_{tot} = 610\,W^{-1}m^{-1}$ independently from the gain resonances. **b**, Plot of $(\tilde{\mathcal{G}}, Q_m)$-pairs for a large set of waveguides obtained from both the gain (fig.5.2a) and the XPM experiment (fig.5.5a). A linear fit without offset yields $\tilde{\mathcal{G}}/Q_m = 10.3\,W^{-1}m^{-1}$. Most variation results from uncertainty in the coupling efficiency ($\approx$ 25%). The two points at $Q_m < 375$ concern a silicon wire on a pillar [64].

## 5.3 Conclusion

Through a novel opto-acoustic nanodevice, a series of suspended silicon wires, we demonstrate modest (0.5 dB) net Brillouin gain with high efficiencies (up to $10^4\,W^{-1}m^{-1}$). This device is a step towards integrated selective amplifiers. We find that fabrication disorder, likely in the waveguide width, broadens and splits the phonon resonances in some cases. In particular, the phonon quality factor strongly decreases as the number of suspended silicon beams increases. Such disorder is expected to hinder development of nanoscale phonon-based technologies quite generally – new cancellation techniques [68] or better fabrication tools must be developed to address this issue.

## 5.4 Additional information

### Influence of two-photon absorption and free-carriers on XPM

In this section, we describe the influence of two-photon absorption (TPA) and free-carriers (FCs) on the Fano resonances. First, we recall that (see (5.3)), in absence of TPA and FCs, the sideband power is proportional to $\mathcal{F}$ with

$$\mathcal{F}(\Delta) = \left|1 + r\frac{1}{-2\Delta + i}\right|^2 \tag{5.5}$$

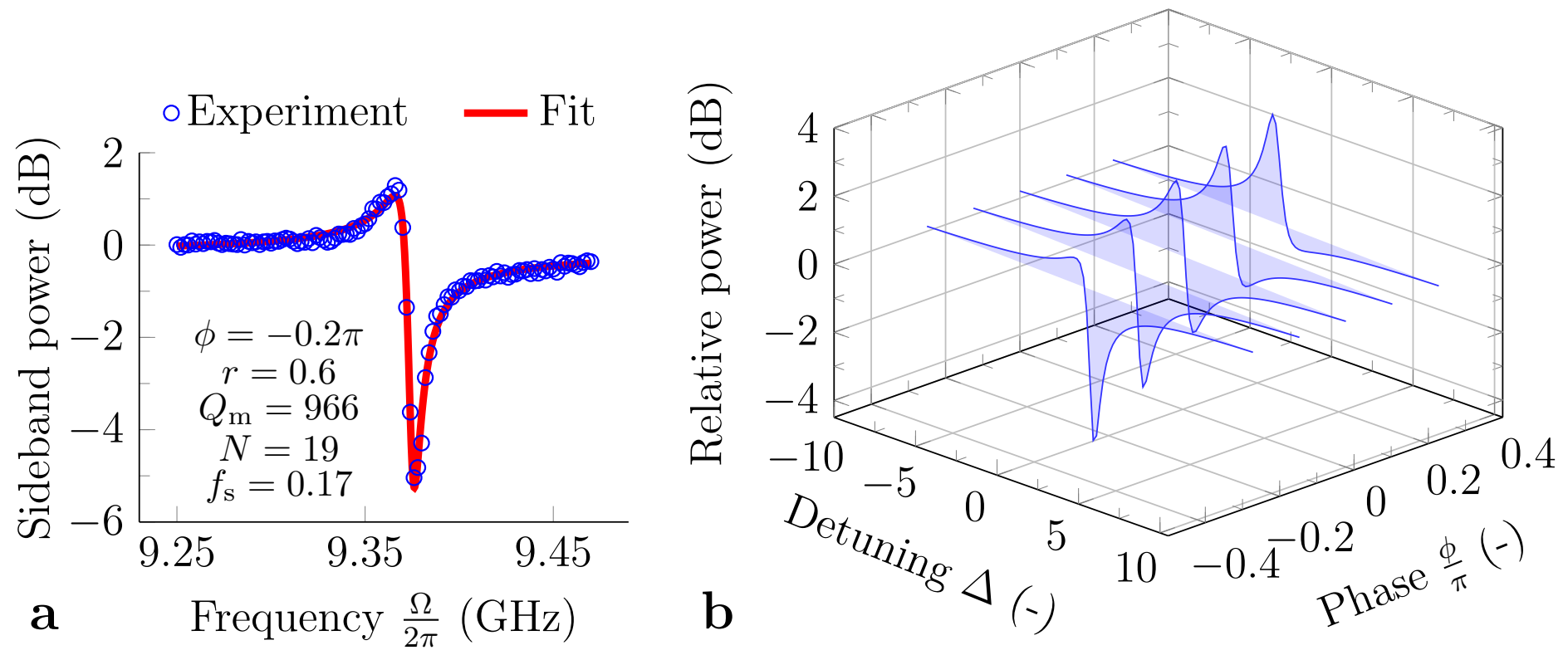

**Figure 5.6: Influence of phase $\phi$ on the Fano resonance.** **a**, In some cases, particularly for small $N$, we observe asymmetric ($\mathcal{F}_{\max}\,[\mathrm{dB}] \ll |\mathcal{F}_{\min}\,[\mathrm{dB}]|$) Fano resonances. The data is well-fit by including a phase shift $\phi < 0$ – physically linked to free-carrier generation (see (5.8)). **b**, Plot of the Fano function $\mathcal{F}(\Delta)$ as the phase shift $\phi$ is scanned ($r = 0.5$). The Fano resonance is symmetric ($\mathcal{F}_{\max}\,[\mathrm{dB}] = |\mathcal{F}_{\min}\,[\mathrm{dB}]|$) at $\phi = 0$; while for $\phi < 0$, the resonance becomes significantly deeper as in (**a**).

with $r = f_s \frac{\tilde{\mathcal{G}}}{4\overline{\gamma}_K}$. One can show that this Fano function $\mathcal{F}$ has one maximum $\mathcal{F}_{\max}$ and one minimum $\mathcal{F}_{\min}$ given by

$$\mathcal{F}_{\max} = \frac{r^2 + 4 + r\sqrt{r^2+4}}{r^2 + 4 - r\sqrt{r^2+4}} = \frac{1}{\mathcal{F}_{\min}} \tag{5.6}$$

that are fully determined by $r$. This implies $\mathcal{F}_{\max}\,[\mathrm{dB}] = -\mathcal{F}_{\min}\,[\mathrm{dB}]$, as evident in fig.5.5a. Inverting this for $r$ yields

$$r = \frac{\mathcal{F}_{\max} - 1}{\sqrt{\mathcal{F}_{\max}}} \tag{5.7}$$

Applied to fig.5.5a, we have $\mathcal{F}_{\max} = 4.4\,\mathrm{dB} = 2.75$ and thus $r = 1.1$ through (5.7) – in agreement with the $r$ obtained from a least-square fit (fig.5.5a). The extrema are reached at a detuning of $\Delta_{\max/\min} = \frac{1}{4}\left(r \mp \sqrt{r^2+4}\right)$. In the large $r$ limit, we get $\Delta_{\max} \to -\frac{1}{2r}$ and $\Delta_{\min} \to \frac{r}{2}$. In the small $r$ limit, we have $\Delta_{\max/\min} \to \frac{r \mp 2}{4}$. Therefore, the maximum XPM is always reached at a negative detuning between $-\frac{1}{2}$ and 0, typically ($r > 1$) close to the phonon resonance ($\Delta = 0$).

In some cases, we observe $\mathcal{F}_{\max}\,[\mathrm{dB}] \ll |\mathcal{F}_{\min}\,[\mathrm{dB}]|$ – a clear indication that (5.5) is too simplistic. It turns out that the Fano function (5.5) must be replaced by [43]

$$\mathcal{F}(\Delta) = \left|1 + e^{i\phi} r \frac{1}{-2\Delta + i}\right|^2 \tag{5.8}$$

with $r = f_s \frac{\tilde{\mathcal{G}}}{4\overline{\gamma}_{\mathrm{tot}}}$ and $\overline{\gamma}_{\mathrm{tot}} = |\overline{\gamma}_K - i\overline{\gamma}_{\mathrm{TPA}} + \overline{\gamma}_{\mathrm{FC}}\overline{P}|$ the *total* nonlinearity, including two-photon absorption and free-carrier effects, $\phi = -\angle\left(\overline{\gamma}_K - i\overline{\gamma}_{\mathrm{TPA}} + \overline{\gamma}_{\mathrm{FC}}\overline{P}\right)$

and $\overline{P}$ the average power in the waveguide. The free-carrier nonlinearity $\overline{\gamma}_{FC}$ is complex as free carriers modulate both the index and the absorption – both effects create an imprinted sideband on the probe; in addition, $\overline{\gamma}_{FC}(\Omega)$ depends on the modulation frequency since free carriers do not respond instantaneously. This is a slow dependency, so we take $\overline{\gamma}_{FC}$ constant in the range of our sweep [43]. This can be shown by solving for the carrier dynamics [233]

$$\partial_t N_c = \frac{\beta_{TPA}}{2\hbar\omega} P^2 - \kappa_c N_c \tag{5.9}$$

in frequency-domain and using the proportionality $\Delta n \propto -N_c$ and $\Delta\alpha \propto N_c$ between both the index and absorption and the carrier concentration [233]. Here we denote $\beta_{TPA}$ the two-photon absorption coefficient and $\kappa_c$ the free-carrier recombination rate.

Notably, $\phi > 0$ in absence of free carriers ($\overline{\gamma}_{FC} = 0$). The observed $\phi < 0$ (fig.5.6a) is thus linked to the presence of free carriers; in addition, we still use $\overline{\gamma}_{tot} \approx \overline{\gamma}_K \approx 610\,W^{-1}m^{-1}$ on the assumption that the Kerr effect remains the dominant background nonlinearity. This is consistent with the observations that (1) in most cases $\mathcal{F}_{max}\,[dB] \approx -\mathcal{F}_{min}\,[dB]$ and thus $\phi \approx 0$, (2) the background is flat (fig.5.5&5.6a) and (3) the Brillouin efficiencies deduced from the XPM experiment are in reasonable agreement with those inferred from the gain experiment (fig.5.5b).

### Drop in free-carrier lifetime

Using the set-up presented in [64], we measured a significant drop in the free-carrier lifetime in the suspended beams with respect to the regular non-suspended waveguide. Both this finding and the higher propagation losses likely originate in a deterioration of the silicon wires' surface state during the fabrication of the suspended beams.

### The interface reflections are negligible

Our device has discontinuities between the suspended nanobeams and the beams fixed at the anchors (fig.5.1). Since the beams are spaced periodically, optical reflections may build up. However, we simulated an upper bound for the Fresnel reflection at the discontinuity of less than $10^{-4}$ – indicating that reflections are negligibly small. Empirically, there are indeed no notable differences in the transmission spectrum of a regular waveguide versus that of a suspended waveguide (fig.5.8). Therefore, our device can be treated as a single-pass structure.

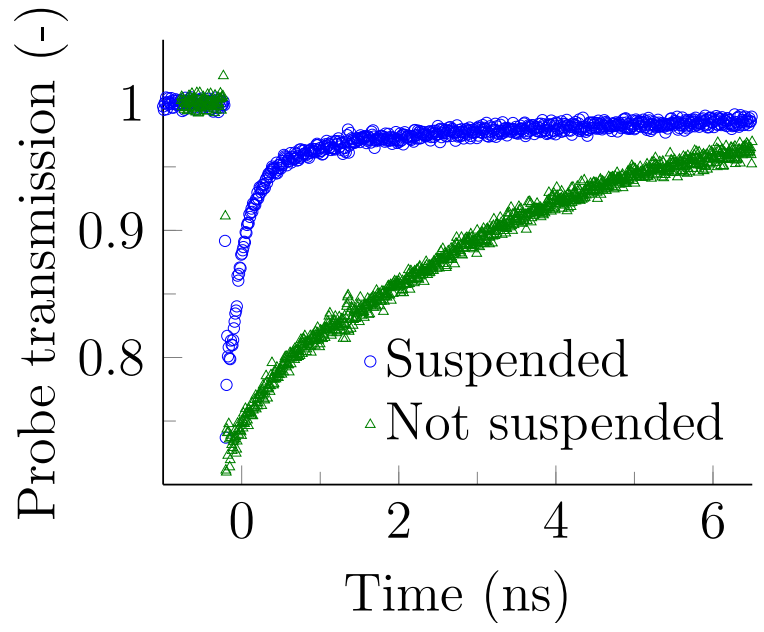

**Figure 5.7: Drop in free-carrier lifetime.** We measured an increase in the free-carrier recombination rate after the suspension of the silicon beams. Both this finding and the higher propagation losses likely originate in a deterioration of the silicon wires' surface state during the fabrication of the suspended beams.

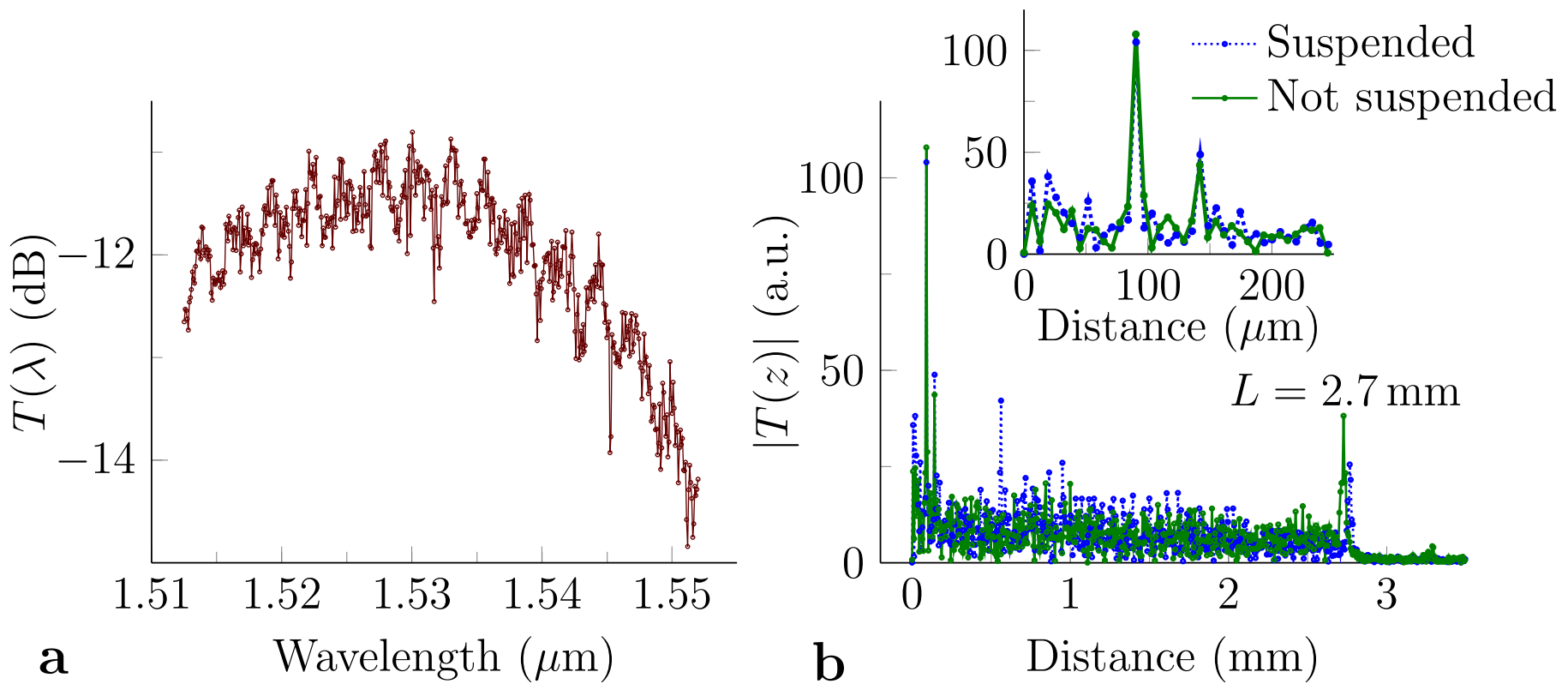

**Figure 5.8: The interface reflections are small.** **a**, The fiber-to-fiber transmission spectrum shows the grating bandwidth. It exhibits a slight (1 dB) variation at a free spectral range of about 2.9 nm, both in standard and suspended silicon-on-insulator waveguides. **b**, The Fourier transform of this spectrum reveals potential cavities present in the wire. We compare the regular wire (green) to a wire with 85 suspensions ($L_s = 25.4\,\mu$m) and anchors ($L_a = 4.6\,\mu$m) (blue). The distance-axis was calibrated with a group index of $n_g = 4.7$ and shows the waveguide length of 2.7 mm because of grating coupler reflections. These are weaker in the suspended case owing to the higher propagation loss. The 2.9 nm free spectral range (**a**) shows up at 90 $\mu$m (**b**, inset). Generally, the two Fourier spectra are nearly identical and do not exhibit notable peaks related to the suspension.

### The measurement set-ups

The measurement set-ups for the gain and XPM experiment were identical to those presented in the previous chapter (fig.4.3).

### Fabrication recipe

We started from air-cladded 220 nm thick, 450 nm wide silicon-on-insulator wires fabricated by 193 nm UV lithography at imec. Next, we patterned an array of apertures in a resist spinned atop the wires. Then we immersed the chip in buffered hydrofluoric acid, which selectively etches the silicon dioxide substrate, until the wires were released. The vertical oxide etch rate was 70 nm/min, whereas the horizontal rate was up to a factor 4 faster – depending on which pilot run the chip was taken from. This is likely related to inhomogeneities in the oxide substrate. The detailed post-imec steps are:

1. Acetone and IPA clean
2. Spin coat Ti35e reversal resist at 7000 rpm for 40 seconds, typically 2 $\mu$m thick
3. Soft bake at 100 °C for 3 minutes
4. UV exposure in MA6 mask aligner to the I-line (365 nm) of a mercury gas-discharge lamp using a chrome-on-quartz photomask
5. Resist rest for 20 minutes
6. Soft bake at 125 °C for 2 minutes
7. UV flood exposure for 185 seconds without photomask
8. Immerse in solution containing three volumes of $H_2O$ and one volume of AZ-400K developer for 1 minute and 45 seconds
9. Hard bake at 150 °C for 10 minutes
10. Wet etch in AF 875-125 buffered HF for 7 to 17 minutes
11. Immerse slowly in deionized water
12. Immerse slowly in acetone for 5 minutes
13. Immerse slowly in IPA
14. Soft dry on hotplate at 60 °C for 2 minutes

# 6

# Silicon slot waveguides

*Never trust an experimental result until it has been confirmed by theory.*
Arthur Eddington

This chapter is based on [66].

## Contents

STIMULATED BRILLOUIN SCATTERING *has attracted renewed interest with the promise of highly tailorable integration into the silicon photonics platform. In an effort to engineer a structure with large photon-phonon coupling, we analyze both forward and backward Brillouin scattering in high-index-contrast silicon slot waveguides. The calculations predict that boundary forces enhance the Brillouin gain in narrow slots. We estimate a gain coefficient of about* $10^5\,\mathrm{W}^{-1}m^{-1}$ *in* 5 *nm-gap horizontal slot waveguides, which is an order of magnitude larger than those observed in stand-alone silicon wires. Such efficient coupling could enable a host of phonon-based technologies on a mass-producible silicon chip.*

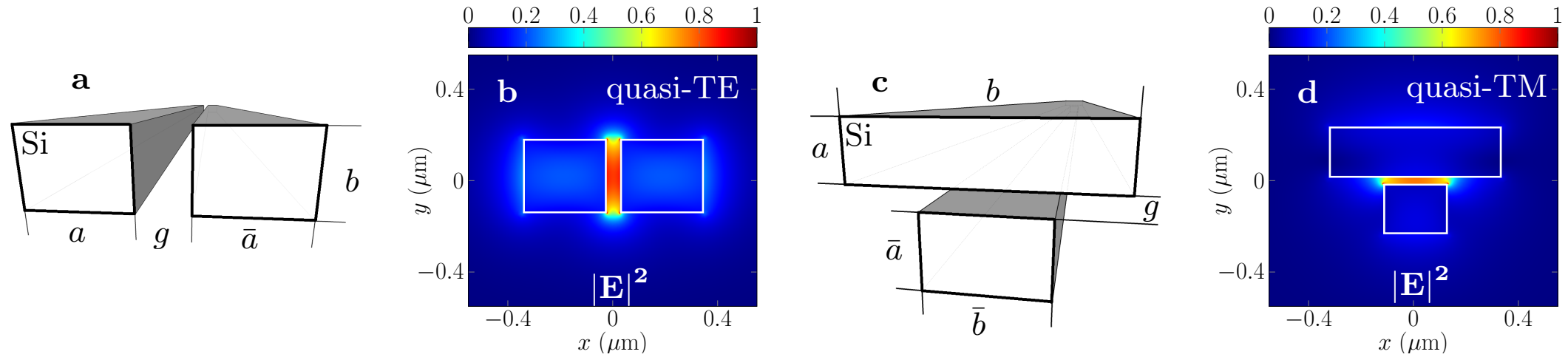

**Figure 6.1:** Vertical (**a**) and horizontal (**c**) silicon slot waveguides suspended in air, with the corresponding optical mode (**b**,**d**).

## 6.1 Introduction

Stimulated Brillouin scattering is a powerful means to control light, with applications ranging from lasing [187], comb generation [152, 234, 235] and isolation [149] to RF-waveform synthesis [137], slow/stored light [157, 236] and reconfigurable filtering [237]. Silicon nanowires are known for their large Kerr and Raman nonlinearity [238]. However, Brillouin scattering long lagged behind in silicon. The culprit was the silica substrate on which the silicon wires are typically made. It severely decreases the phonons' lifetime. Unlike in chalcogenide rib waveguides [203, 239], elastic waves in silicon cannot easily be guided by internal reflection because sound is faster in silicon than in silica (fig.4.5 and section 4.7). In chapters 4 and 5, we demonstrated efficient Brillouin scattering in partially and fully suspended silicon nanowires. Nonetheless, the quest for better photon-phonon overlaps remains open.

In this chapter, we take the study of Brillouin scattering to silicon slot waveguides, to exploit their strong optical mode confinement [240, 241] and large boundary forces [242–244]. We perform full-vectorial coupled optical and mechanical simulations of the Brillouin gain coefficient using the finite-element solver COMSOL. We use isotropic elasticity coefficients $(c_{11}, c_{12}, c_{44}) = (217, 85, 66)$ GPa for easy comparison with [104, 211]. Silicon is mechanically anisotropic, so in a more accurate calculation the coefficients $(c_{11}, c_{12}, c_{44}) = (166, 64, 80)$ GPa should be used for a guide along a $\langle 100 \rangle$ crystal axis [245]. Further, we use the photoelastic coefficients $(p_{11}, p_{12}, p_{44}) = (-0.094, 0.017, -0.051)$ [246], which is also valid in case the guide is aligned along a $\langle 100 \rangle$ axis. We perform our calculations using the weak-form [247] COMSOL module with the MATLAB Livelink.

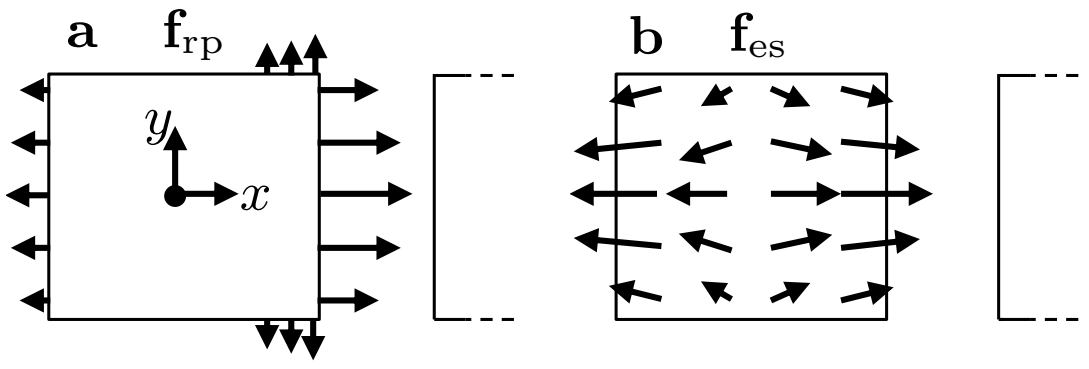

**Figure 6.2:** Typical optical force profile on left beam of vertical slot waveguide: radiation pressure (**a**) and electrostrictive body force (**b**). The radiation pressure is large close to the slot.

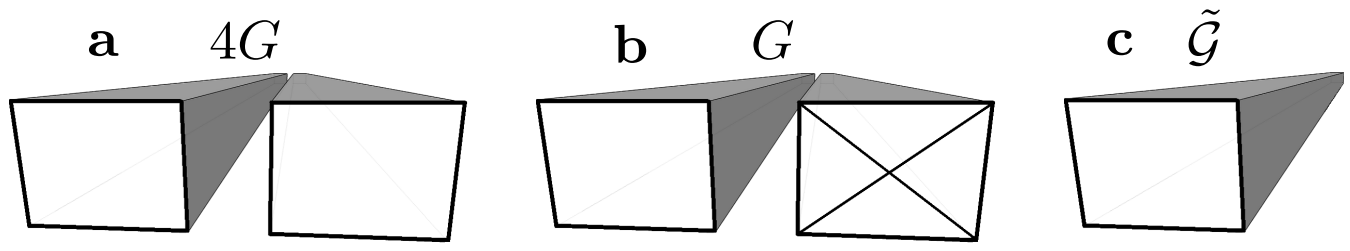

**Figure 6.3:** **a**, A slot with two mechanically excited beams. **b**, Slot with just one excited beam. **c**, Stand-alone wire. We work in scenario **b**.

## 6.2 Background and assumptions

We consider vertical (fig.6.1a-b) and horizontal (fig.6.1c-d) slot waveguides suspended in air. Both waveguides strongly confine light, creating large radiation pressure close to the slot. This gives rise to the possibility of (1) improving the photon-phonon coupling, (2) verifying SBS theory in a regime dominated by radiation pressure instead of electrostriction and (3) exciting new types of phononic modes.

A particular mechanical mode with displacement field $\mathbf{u}$, wavevector $q$, stiffness $k_{\text{eff}}$ and quality factor $Q_{\text{m}}$ has a peak SBS gain $G$ of $\omega Q_{\text{m}}|\langle \mathbf{f}, \mathbf{u}\rangle|^2/(2k_{\text{eff}})$ (appendix B), with $\omega$ the optical frequency, $\mathbf{f} = \mathbf{f}_{\text{rp}} + \mathbf{f}_{\text{es}}$ the power-normalized force distribution and $\langle \mathbf{f}, \mathbf{u}\rangle = \int \mathbf{f}^* \cdot \mathbf{u}\, dA$ the photon-phonon overlap [104, 211]. The radiation pressure $\mathbf{f}_{\text{rp}}$ is located on the waveguide boundaries (fig.6.2a), while the electrostrictive force $\mathbf{f}_{\text{es}}$ has both a body (fig.6.2b) and a boundary (not shown) component. The boundary component of $\mathbf{f}_{\text{es}}$ is an order of magnitude smaller than $\mathbf{f}_{\text{rp}}$. Furthermore, we define $G_{\text{rp}}$ and $G_{\text{es}}$ as the SBS gain when only $\mathbf{f}_{\text{rp}}$ or $\mathbf{f}_{\text{es}}$ is present. The total gain $G$ is determined by interference between $\mathbf{f}_{\text{rp}}$ and $\mathbf{f}_{\text{es}}$. Additionally, the pump and Stokes seed co- (counter-) propagate in forward (backward) SBS. Phase-matching then requires that $q \approx 0$ ($q \approx 2k_{\text{p}}$), with $k_{\text{p}}$ the pump wavevector. Finally, we work at $\lambda = 1.55\,\mu$m, use a flat $Q_{\text{m}}$ of $10^3$ as in [104, 211] and we launch the pump and Stokes seed into the same optical mode (fig.6.1b-d).

If the two beams have identical width and height (fig.6.3a), their mechanical resonances can be addressed simultaneously. Then the gain is $4G$, since the total overlap $\langle \mathbf{f}, \mathbf{u} \rangle$ is twice the overlap over a single beam. However, these dimensions cannot differ by more than a fraction $1/Q_m$ to align the resonances within one mechanical linewidth. This is technologically challenging with a $Q_m$ of about $10^3$. So we assume just one beam of dimensions $(a, b)$ contributes to SBS (fig.6.3b), even though the unexcited beam may also be suspended. Moreover, in wide slots the optical mode evolves into the symmetric supermode of two weakly coupled silicon wires. So $4G \to \tilde{\mathcal{G}}$ as $g \to \infty$, with $\tilde{\mathcal{G}}$ the gain coefficient associated with an acoustic mode of a stand-alone wire (fig.6.3c). In other words, the slot waveguide (fig.6.3b) has to overcome a factor 4 to reach a $G/\tilde{\mathcal{G}} > 1$ gain enhancement situation.

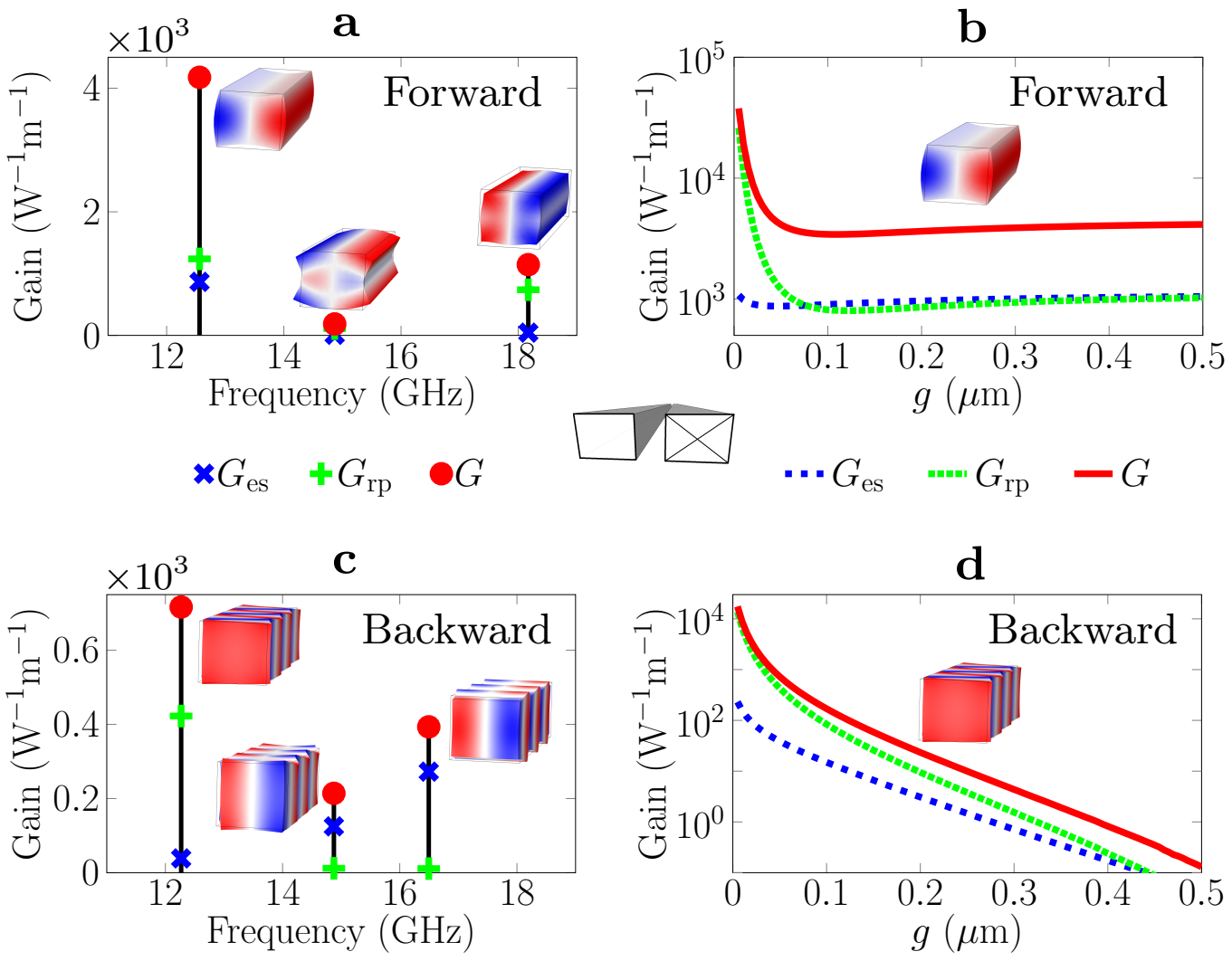

**Figure 6.4:** (**a-c**) Brillouin spectrum of a vertical slot waveguide and (**b-d**) the gain of the most promising mode increases rapidly in narrow slots. The color of the modes indicates the sign of $u_x$ (red: $+$, blue: $-$).

## 6.3 SBS in vertical slot waveguides

Figures 6.4a-c show the forward and backward SBS spectrum for a vertical slot waveguide with dimensions $(a, b, \bar{a}, g) = (315\,\text{nm}, 0.9a, a, 50\,\text{nm})$, including only the three modes with largest gain.

In the forward case (fig.6.4a), the mechanical modes are identical to those of a stand-alone wire. The maximum gain among all modes is $4.2 \times 10^3\,\text{W}^{-1}\text{m}^{-1}$. This is smaller than $\tilde{\mathcal{G}}/4 = 4.3 \times 10^3\,\text{W}^{-1}\text{m}^{-1}$ [211], despite the increase in radiation pressure close to the slot. The cause is a decrease in the pressure on the far side from the slot (fig.6.2a). This decrease cancels out the enhanced forces close to the slot, even in slots as narrow as 50 nm. This also explains why $G$ does not change significantly for $g > 50\,\text{nm}$ (fig.6.4b). Hence, smaller gaps are necessary to boost $G$ substantially. Indeed, for the most promising mode we numerically find that $G \propto 1/g$ as $g$ falls below 50 nm (fig.6.4b). Eventually $G$ approaches a maximum of $\approx 1.1 \times 10^5\,\mathrm{W}^{-1}\mathrm{m}^{-1}$ as $g \to 0$.

In the backward case, the mechanical modes are different from those of a stand-alone wire since the phonon wavevector $q \approx 2k_\mathrm{p}$ depends on the effective index $n_p$ of the optical mode. From the point of view of a single beam, horizontal symmetry is broken by the slot waveguide. So modes that were previously forbidden by symmetry can have non-zero gain in the slot waveguide. Such a previously forbidden phonon has the largest backward SBS gain in the slot waveguide (fig.6.4c). For $g = 50$ nm, this phononic mode yields a gain of $7.2 \times 10^2\,\mathrm{W}^{-1}\mathrm{m}^{-1}$. The optical forces are symmetric again in wide slots. Then this mode is forbidden, which means that $G \to 0$ as $g \to \infty$ (fig.6.4d). Going from wide to narrow slots, $G$ first increases exponentially, then its growth accelerates like $G \propto 1/g^{1.6}$ and ultimately converges to a maximum of $\approx 4.5 \times 10^4\,\mathrm{W}^{-1}\mathrm{m}^{-1}$ as $g \to 0$.

In general, boundary forces dominate the SBS gain in narrow slots (fig.6.4b-d). The slot enhances these forces despite the reduced dispersion in such waveguides. As $g \to 0$, the group and effective indices $n_g$ and $n_p$ approach those of a single wire of width $a + \bar{a}$. Thus the waveguide dispersion decreases (fig.6.5a), contrary to the prediction that very dispersive waveguides are optimal for large boundary forces [248]. Writing the boundary force density as $\mathbf{p}\delta\left(\mathbf{r} - \mathbf{r}_{\partial\mathrm{wg}}\right)$, it was shown that $c\int \mathbf{p}\cdot\mathbf{r}dl = n_\mathrm{g} - n_\mathrm{p}$ from the scale-invariance of Maxwell's equations [248]. For a stand-alone wire the integral becomes $\int \mathbf{p}\cdot\mathbf{r}dl = A_\mathrm{wg}\left(\bar{p}_x + \bar{p}_y\right)$ with $A_\mathrm{wg} = ab$ and $\bar{p}$ the magnitude of the spatially averaged radiation pressure. However, this no longer holds for a slot waveguide. Then the integral yields $\int \mathbf{p}\cdot\mathbf{r}dl = A_\mathrm{g}\left(\bar{p}_{x,L} - \bar{p}_{x,R}\right) + 2A_\mathrm{wg}\left(\bar{p}_{x,L} + \bar{p}_y\right)$, with $\bar{p}_{x,L/R}$ the pressure on the left/right boundary, $A_g = gb$ and $a = \bar{a}$. Since $A_\mathrm{g} \to 0$ as $g \to 0$, $\bar{p}_{x,R}$ and thus $\bar{p}_x + \bar{p}_y$ can increase drastically in narrow slots (fig.6.5a).

Next, we investigate the effect of $\bar{a}$ (fig.6.5b-d). As $\bar{a} \to 0$, there is no slot-enhancement. Then $G \to \tilde{\mathcal{G}}$, regardless of all other parameters. Furthermore, the optical mode increasingly retreats into the widest beam. This implies $G \to 0$ when $\bar{a} \to \infty$, although this effect is more pronounced in wider slots.

In the forward case (fig.6.5b), $\bar{a}$ affects only the force distribution. The gain $G(\bar{a})$ has a maximum in narrow slots, but decreases monotonically otherwise. This confirms that small gaps are required for substantial gain enhancement in vertical slot waveguides.

In the backward case (fig.6.5d), $G(\bar{a})$ always has a maximum because this phonon is forbidden in a stand-alone wire. However, the maximum increases by a factor 26 when the slot is narrowed from 50 nm to 5 nm. The gain is dominated by boundary forces regardless of $(\bar{a}, g)$.

Last, we scan $(a, b)$ with $\bar{a} = a$ and $g$ fixed at 5 nm. These parameters influence both the optical and mechanical mode. The $(a, b)$-optimum depends heavily on the slot size and on the mechanical mode. Nonetheless, fig.6.5c shows that there actually exists such an optimum. We find a maximum gain of $7.0 \times 10^4\,\mathrm{W^{-1}m^{-1}}$ for $(a, b) = (260, 150)$ nm.

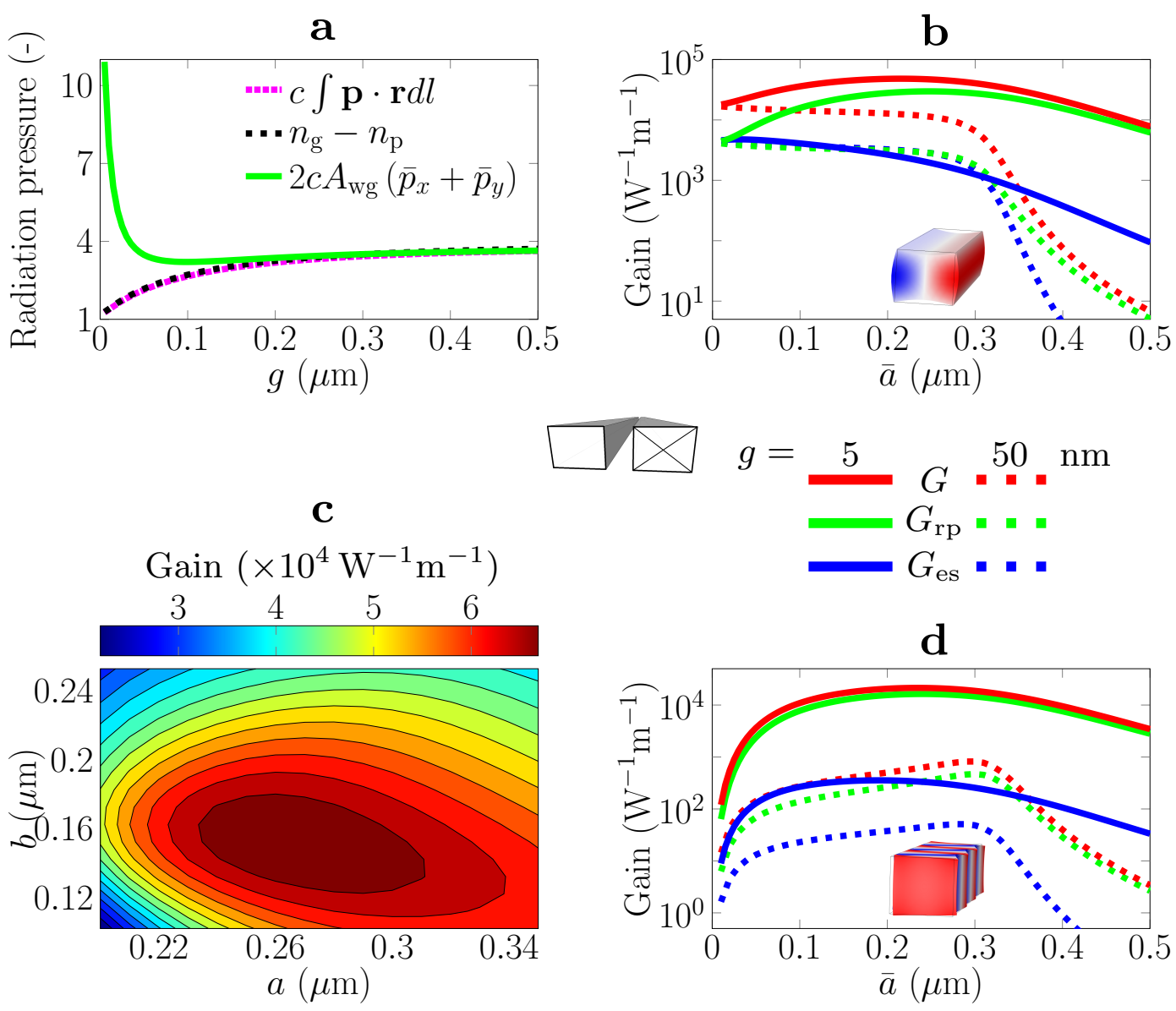

**Figure 6.5:** Gradient forces can be large despite low dispersion (**a**), Narrow slots perform better than a stand-alone wire for a range of $\bar{a}$-values (**b**-**d**) and $G$ has a clear optimum in the $(a, b)$-plane for the same mode as in **b** with $g = 5$ nm (**c**).

## 6.4 SBS in horizontal slot waveguides

The horizontal slot (fig.6.1c-d) has the potential advantage of (1) the extra degree of freedom $\bar{b}$ and (2) smaller gaps. In such a slot, the gap $g$ is not limited by the resolution of lithography techniques. As a result, SBS enhancement may be within reach of current technology. As long as $\bar{b} = b$, the horizontal slot waveguide is but a rotated version of the vertical one. Therefore we immediately explore the case $\bar{b} \neq b$. We calculate the forward and backward Brillouin spectrum for a horizontal slot waveguide with dimensions $(a, b, \bar{a}, \bar{b}, g) = (160, 620, a, 240, 5)$ nm.

In the forward case (fig.6.6a), the fundamental flexural mode couples most efficiently. This mode has negligible SBS gain in a stand-alone wire because of cancellations in the photon-phonon overlap. Indeed, the $u_y$ component has two nodes, while the $y$-component of the boundary force does not change sign. Owing to $b > \bar{b}$, the cancellations can be avoided by confining the optical mode between the nodes of $u_y$.

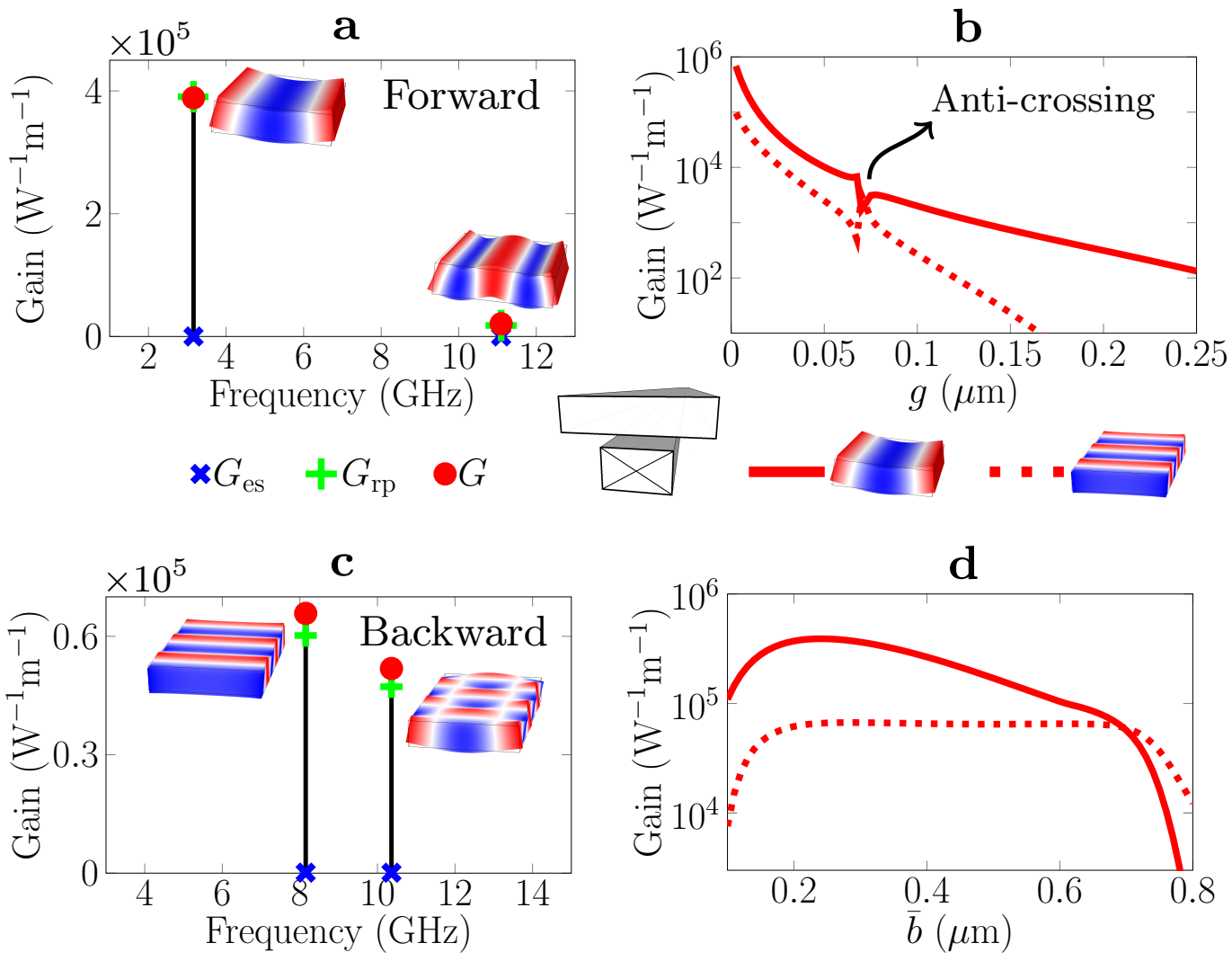

**Figure 6.6:** Both forward and backward SBS is very efficient in narrow horizontal slots (**a**-**b**-**c**) and the flexural mode is sensitive to $\bar{b}$ (**d**). The color of the modes indicates the sign of $u_y$ (red: $+$, blue: $-$).

In the backward case (fig.6.6c), there are two modes with enhanced coupling. The first mode has a nearly uniform $u_y$ component. It is a rotated version of the mode we previously studied in fig.6.4-6.5d. The second mode is the fundamental flexural mode, but at the operating point $q \approx 2k_\text{p}$ in its dispersion diagram.

The gain increases by four orders of magnitude when $g$ drops from 250 to 5 nm (fig.6.6b). This radical enhancement is superexponential in $g$ for gaps below 50 nm. The forward (backward) gain approaches $\approx 1.3 \times 10^6\,\text{W}^{-1}\text{m}^{-1}$ ($1.5 \times 10^5\,\text{W}^{-1}\text{m}^{-1}$) as $g \to 0$. At $g = 70\,\text{nm}$, an optical mode anti-crossing causes a dip in the SBS gain. However, $G(g)$ quickly recovers its original path as $g$ leaves the anti-cross region. We only show the total gain $G$ because $G_\text{es}$ is at least a factor $10^5$ ($10^2$) smaller than $G_\text{rp}$ across the entire sweep range in the forward (backward) case. Thus SBS by these modes is driven by boundary forces only, with a vanishing electrostrictive contribution.

Finally, we sweep $\bar{b}$ (fig.6.6d). In the forward case, $k_\text{eff}$ and $\mathbf{u}$ do not depend on $\bar{b}$. Then we explore purely the effect of the boundary force density $\mathbf{f}_\text{rp}(\bar{b})$ on the photon-phonon overlap $\langle \mathbf{f}_\text{rp}(\bar{b}), \mathbf{u} \rangle$. The coupling is optimal for $\bar{b} = 240\,\text{nm}$. For smaller $\bar{b}$, $G$ decreases because the slot-enhancement occurs only in a small region. For larger $\bar{b}$, $G$ decreases because the optical mode is no longer confined between the nodes of $u_y$. In the backward case, the operating point $q \approx 2k_\text{p}$ changes as $n_p$ depends on $\bar{b}$. This propagating phononic mode is less sensitive to $\bar{b}$ because of its nearly uniform $u_y$ component.

## 6.5 Conclusion

To conclude, we found that strong boundary forces improve the efficiency of Brillouin scattering in narrow silicon slot waveguides. However, appreciable enhancement compared to a stand-alone wire is currently only accessible in horizontal slots. In such slots, we expect very efficient SBS because (1) small gaps should be technologically accessible and (2) the fundamental mechanical flexural mode can be excited. Notably, 5 nm gaps filled with oxide have already been fabricated [249]. The suspension of long silicon beams remains one important hurdle towards testing these predictions. A practical device may consist of a partially suspended waveguide or a cascade of full suspensions (chapters 4 and 5). With an efficiency in excess of $10^5\,\mathrm{W}^{-1}\mathrm{m}^{-1}$, the simulations predict that 20 dB gain is feasible with 50 mW on-chip pump power over 1 mm propagation length. This could enable vacuum coupling rates above $\frac{g_0}{2\pi} \approx 3\,\mathrm{MHz}$ if the structure were to be placed in a cavity (equation (3.1)). In addition, it may allow for observations of the elastic [67] gain coefficient $\frac{G}{Q_\mathrm{m}} \approx 400\,\mathrm{W}^{-1}\mathrm{m}^{-1}$ (fig.6.6a) – which could be comparable to the Kerr effect of regular silicon waveguides [250]. Finally, other materials could be deposited in the slot. This may enable acoustic confinement without suspension and stronger photoelasticity [251].

# 7
# Conclusion and prospect

*Somewhere, something incredible is waiting to be known.*
Carl Sagan

THIS WORK aimed to grasp the physics of photon-phonon interactions across the widest range of structures and materials, to create strong, tailorable photon-photon control mediated by phonons in nanoscale silicon waveguides, to scale up the resonance frequency of these phonons to the gigahertz range and to perform optical signal processing tasks using the newly developed structures.

We addressed these goals as follows. Chapter 2 presented a rigorous quantum theory of photon-phonon interactions in nanoscale waveguides. Next, chapter 3 built on this theory, exploring the landscape of physical effects in both waveguides and cavities. The study unveiled the connection between the Brillouin gain coefficient and the zero-point coupling rate. It united the language of Brillouin and Raman scattering, framed in gain coefficients, with that of cavity QED and optomechanics, framed in vacuum coupling rates and cooperativities. In addition, it turned out that each cavity-based effect had a waveguide-based counterpart – but not vice versa. This led to the prediction of several unexplored effects, including (1) the photon-assisted amplification or cooling of traveling phonons and (2) the spatial strong-coupling regime. The former allows for non-reciprocal, optical control over sound and heat. The latter may lead to observations of entanglement and population oscillations between traveling light and sound. The formalism brings Brillouin interactions in line with major themes in cavity QED and quantum information processing.

Chapters 4 and 5 experimentally demonstrated interaction between 193 THz photons and 9.3 GHz acoustic phonons. The photons travel about a million times farther than the phonons. In addition, their coupling strength is swamped by the propagation losses. Therefore, the experiments gave rise to phonon-assisted amplification of photons – the traditional Brillouin gain. Specifically, we made the first observation of forward and backward Brillouin scattering in silicon nanowires (chapter 4). The nanowires were partially suspended to limit phonon leakage. Compressing both light and sound to the nanoscale silicon core tremendously enhanced the nonlinearity, as it did a decade earlier for the Kerr and Raman effect. We demonstrated efficiencies and on/off gain up to 3218 $W^{-1}m^{-1}$ and 4.4 dB in centimeter-long waveguides. The shorter wires were transparent in a 35 MHz-band 9.2 GHz red-detuned from the pump.

However, the partially suspended wires fell narrowly short of net amplification. Chapter 5 overcame this hurdle by fully suspending a cascade of silicon beams. The full, not partial, suspension eliminated the acoustic leakage, enhancing the phonon lifetime from 5.3 ns to 17.5 ns. This enabled the first observation of Brillouin gain above the propagation losses in a silicon waveguide. We obtained efficiencies up to $10^4$ $W^{-1}m^{-1}$, the highest so far in the gigahertz range. Alas, the net amplification amounted to a meager 0.5 dB. It was limited by (1) the available on-chip pump power (no free-carrier effects were seen), (2) the higher propagation losses after suspension and (3) inhomogeneous broadening of the phononic resonance. In particular, we observed line broadening from 9.2 MHz for 6 suspensions to 20 MHz for 66 suspensions. The amount of broadening differed even between nominally identical samples from the same wafer. Better fabrication aside, we proposed to cancel the broadening via the indirect acoustic sensitivity to the optical dispersion.

So how do our devices compare to others? The answer depends on the goal: is it to demonstrate large amplification, to observe novel effects, to make microwave notch filters, to ease fabrication or simply to achieve large efficiency? Regardless on the goal, however, there are a few generally acknowledged figures of merit. In table 7.1, we list these figures of merit for ours and a set of other waveguides.

| | $\frac{\Omega_\mathrm{m}}{2\pi}$ [GHz] | $\tilde{\mathcal{G}}$[W$^{-1}$m$^{-1}$] | $Q_\mathrm{m}$[$-$] | $\frac{\tilde{\mathcal{G}}}{Q_\mathrm{m}}$[W$^{-1}$m$^{-1}$] | $\alpha^{-1}$ [cm] | $\tilde{\mathcal{C}}$ [$-$] | net gain [dB] | $P_\mathrm{p}$ [mW] |
|---|---|---|---|---|---|---|---|---|
| Silicon pedestal wires [64] | 9.2 | 3100 | 300 | 10.3 | 1.7 | 1 | 0 | 23 |
| Suspended silicon beams [65] | 9.2 | $10^4$ | 1010 | 10.3 | 0.8 | 3 | 0.5 | 38 |
| Silicon/nitride structures [43] | 1.3 | 2328 | 1750 | 1.3 | 0.6 | 0.3 | -2.6 | 20 |
| Suspended silicon ridges [94] | 4.3 | 1840 | 1019 | 1.8 | 22 | 18 | 5 | 62 |
| Chalcogenide ribs [203] | 7.7 | 304 | 226 | 1.3 | 5.5 | 5 | 10.3 | 300 |
| Silica fibers [207] | 10.8 | 0.5 | 540 | $10^{-3}$ | $10^6$ | $10^4$ | >50 | 100 |
| Photonic crystal fibers [152] | 1.8 | 1.5 | 122 | $10^{-2}$ | $10^4$ | 46 | >50$^\star$ | 500 |
| Chalcogenide fibers [252] | 7.9 | 65 | 159 | 0.4 | 500 | 22 | 38 | 68 |
| Slotted fibers [147] | $6 \cdot 10^{-3}$ | $4 \cdot 10^6$ | 6000$^\dagger$ | 670 | 7.5 | $3 \cdot 10^3$ | >50$^\star$ | 10 |
| Silicon ridges$^*$ [197, 202] | $1.5 \cdot 10^4$ | 65 | 149 | 0.4 | 22 | 13 | 8 | $10^3$ |

**Table 7.1: Comparison of waveguides in terms of optomechanical figures of merit.** The $^\star$ marks hypothetical gain: in fact combs were generated. The † marks a system operated in vacuum. The $^*$ marks a structure that harnesses optical phonons and a p-i-n diode to extract free carriers. We estimated some values. Overall, the silicon devices perform best in terms of efficiency and overlap and worst in terms of losses and power handling. See table 3.1 for comparisons of these waveguides with yet to be studied waveguide-equivalents of optomechanical cavities.

Nanoscale silicon waveguides perform best in terms of efficiency and worst in terms of losses. In this sense, these waveguides are both a blessing and a curse: they essentially lose in absorption and power-handling what they gain in photon-phonon overlap. It is extremely challenging to realize large efficiencies and low losses simultaneously. This requires exceptionally accurate fabrication techniques: even atomic-scale disorder strongly degrades the optical and acoustic performance. However, the prospect of efficiently processing light with sound in mass-producible integrated circuits drives hope that novel techniques will circumvent the downsides. In only a few years, the field has witnessed tremendous improvements. In 2013, silicon/nitride waveguides gave 9% on/off gain [43]. A year later, silicon pedestal wires generated 175% on/off gain and a factor 10 higher overlaps (chapter 4). The pedestal wires were soon used to demonstrate a microwave filter [218]. Cascades of suspended wires enabled net gain in 2015 (chapter 5). A couple of months ago, 5 dB net gain was observed in suspended silicon ridges [94]. Not only their quantitative performance, but also the variety of the structures is remarkable. There is clearly more to come.

So far, phonons were mostly exploited as mediators and interfaces. But could they not take center stage? Intriguingly, microsound circuits were once seen as a promising way to make microwave circuits fifteen orders of magnitude smaller [253]. Phonons may not be ideal messengers, but they do have a unique set of properties. This may very well be our most exciting prospect of all.

# A
# Mean-field transition

**Contents**

## A.1 Link to cavity Hamiltonian

With the mean-field transition derived in the main text, we take a step beyond the $\tilde{\mathcal{G}} \leftrightarrow g_0$ link. As we show in this section, the mean-field equations are equivalent to the cavity Langevin equations in the resolved-sideband limit ($\kappa \ll \Omega_m$). In the case of coupling between one mechanical and one optical resonance (fig.3.3), the standard theory [35] starts from the Hamiltonian

$$\hat{\mathcal{H}} = \hbar\omega_c \hat{a}^\dagger \hat{a} + \hbar\Omega_m \hat{b}^\dagger \hat{b} + \hat{\mathcal{H}}_{\text{int}}$$

with

$$\hat{\mathcal{H}}_{\text{int}} = \hbar g_0 \hat{a}^\dagger \hat{a} \left(\hat{b} + \hat{b}^\dagger\right)$$

the interaction Hamiltonian, $\hat{x} = x_{\text{ZPF}} \left(\hat{b} + \hat{b}^\dagger\right)$ the mechanical oscillator's position, $x_{\text{ZPF}}$ the zero-point motion, $\hat{a}$ and $\hat{b}$ ladder operators for the optical and mechanical oscillator and $g_0 = x_{\text{ZPF}} \frac{\partial \omega_c}{\partial x}$ the zero-point coupling rate.

When the pump is undepleted, the interaction Hamiltonian can be linearized: $\hat{a} = \overline{\alpha} + \delta\hat{a}$ with $\delta\hat{a}$ a small fluctuation. Then we have

$$\hat{\mathcal{H}}_{\mathrm{int}}^{(\mathrm{lin})} = \hbar g_0 \overline{\alpha} \left( \delta\hat{a} + \delta\hat{a}^\dagger \right) \left( \hat{b} + \hat{b}^\dagger \right)$$

Using the equation of motion $\dot{\hat{a}} = -\frac{i}{\hbar}[\hat{a}, \hat{\mathcal{H}}]$ and the commutator $[\hat{a}, \hat{a}^\dagger] = 1$ (the same for $\hat{b}$), this linearized Hamiltonian leads straight to the coupled equations [35]

$$\begin{aligned} \delta\dot{\hat{a}} &= -\left( \frac{\kappa}{2} - i\Delta \right) \delta\hat{a} - i g_0 \overline{\alpha} \left( \hat{b} + \hat{b}^\dagger \right) \\ \dot{\hat{b}} &= -\left( \frac{\kappa_{\mathrm{m}}}{2} - i\Omega_{\mathrm{m}} \right) \hat{b} - i g_0 \overline{\alpha} \left( \delta\hat{a} + \delta\hat{a}^\dagger \right) \end{aligned}$$

with $\Delta = \omega_{\mathrm{p}} - \omega_{\mathrm{c}}$. Next, we consider a blue-detuned pump in the resolved-sideband regime ($\kappa \ll \Omega_{\mathrm{m}}$). Then we can write the ladder operators as $\delta\hat{a} \to \hat{a}_{\mathrm{s}} e^{i\Omega t}$ and $\hat{b} \to \hat{b} e^{-i\Omega t}$, with $\hat{a}_{\mathrm{s}}$ and $\hat{b}$ now slowly-varying. We neglect the $\hat{b}$-term in the optical equation and the $\delta\hat{a}$-term in the mechanical equation because they are off-resonant. This is the rotating-wave approximation, which corresponds to the classical slowly-varying envelope approximation [36, 37]. Hence, the above equations reduce to

$$\begin{aligned} \dot{\hat{a}}_{\mathrm{s}} &= -i g_0 \overline{\alpha} \hat{b}^\dagger - \chi_{\mathrm{s}}^{-1} \hat{a}_{\mathrm{s}} \\ \dot{\hat{b}} &= -i g_0 \overline{\alpha} \hat{a}_{\mathrm{s}}^\dagger - \chi_{\mathrm{m}}^{-1} \hat{b} \end{aligned} \tag{A.1}$$

and we find that equations (A.1) are identical to equations (16) given $\hat{a}_{\mathrm{s}} \mapsto \overline{a}_{\mathrm{s}}$ and $\hat{b} \mapsto \overline{b}$. Remarkably, the equivalence holds even though the pump and Stokes could be counter-propagating or in different optical modes. In the unresolved-sideband limit ($\Omega_{\mathrm{m}} \ll \kappa$), anti-Stokes generation and cascading must be added for forward intra-modal, but not necessarily for backward or inter-modal Brillouin scattering. Indeed, comb generation is usually not accessible by backward or inter-modal coupling because of the phase-mismatch (fig.2.2). This assumption can be violated in Fabry-Pérot cavities [254] or when the first-order Stokes becomes sufficiently strong to pump a second-order Stokes wave [137].

## A.2 Manley-Rowe relations

In this section, we prove that the Manley-Rowe relations guarantee the existence of a single real, positive photon-phonon coupling coefficient in waveguides ($\tilde{g}_0$) and in cavities ($g_0$). In waveguides, the Manley-Rowe relations are formulated at the level of photon and phonon fluxes $\Phi$. In cavities, they are written down in terms of the total photon and phonon numbers $n$.

**Manley-Rowe in waveguides**

A Brillouin-active waveguide in steady-state ($\partial_t \to 0$) obeys (see (2))

$$\begin{aligned} \partial_z a_\mathrm{p} &= -i\tilde{\kappa}_\mathrm{mop} a_\mathrm{s} b - \tilde{\chi}_\mathrm{p}^{-1} a_\mathrm{p} \\ \pm\partial_z a_\mathrm{s} &= -i\tilde{\kappa}_\mathrm{mos} b^\dagger a_\mathrm{p} - \tilde{\chi}_\mathrm{s}^{-1} a_\mathrm{s} \\ \partial_z b &= -i\tilde{\kappa}_\mathrm{om} a_\mathrm{s}^\dagger a_\mathrm{p} - \tilde{\chi}_\mathrm{m}^{-1} b \end{aligned} \tag{A.2}$$

with arbitrary normalizations of the pump, Stokes and acoustic envelope such that generally $\tilde{\kappa}_\mathrm{mop} \neq \tilde{\kappa}_\mathrm{mos} \neq \tilde{\kappa}_\mathrm{om}$ are different complex numbers. Using $\partial_z \left(a^\dagger a\right) = (\partial_z a^\dagger) a + a^\dagger (\partial_z a)$, we find

$$\begin{aligned} \partial_z \Phi_\mathrm{p} &= -\alpha_\mathrm{p}\Phi_\mathrm{p} + \left(i\tilde{\kappa}_\mathrm{mop}^\star a_\mathrm{s}^\dagger b^\dagger a_\mathrm{p} + \mathrm{h.c.}\right) \\ \pm\partial_z \Phi_\mathrm{s} &= -\alpha_\mathrm{s}\Phi_\mathrm{s} - \left(i\tilde{\kappa}_\mathrm{mos} a_\mathrm{s}^\dagger b^\dagger a_\mathrm{p} + \mathrm{h.c.}\right) \\ \partial_z \Phi_\mathrm{m} &= -\alpha_\mathrm{m}\Phi_\mathrm{m} - \left(i\tilde{\kappa}_\mathrm{om} a_\mathrm{s}^\dagger b^\dagger a_\mathrm{p} + \mathrm{h.c.}\right) \end{aligned} \tag{A.3}$$

Suppose now that the envelopes are flux-normalized such that $\Phi_\mathrm{p} = a_\mathrm{p}^\dagger a_\mathrm{p}$, $\Phi_\mathrm{s} = a_\mathrm{s}^\dagger a_\mathrm{s}$ and $\Phi_\mathrm{m} = b^\dagger b$ correspond to the number of pump photons, Stokes photons and phonons passing through a cross-section of the waveguide per second. Then we demand that, in the lossless case ($\alpha_\mathrm{j} = 0$), the rate of pump photon destruction equals the rate of Stokes photon and phonon creation

$$-\partial_z \Phi_\mathrm{p} = \pm\partial_z \Phi_\mathrm{s} = \partial_z \Phi_\mathrm{m} \tag{A.4}$$

These are the Manley-Rowe relations [36, 185] for a Brillouin waveguide. We deduce from (A.3) and (A.4) that

$$\tilde{\kappa}_\mathrm{mop}^\star = \tilde{\kappa}_\mathrm{mos} = \tilde{\kappa}_\mathrm{om} \tag{A.5}$$

This proves the existence of a single coupling coefficient that captures all reversible optical forces and scattering. Note that (A.5) also guarantees power-conservation since

$$\partial_z \left(\hbar\omega_\mathrm{p}\Phi_\mathrm{p} \pm \hbar\omega_\mathrm{s}\Phi_\mathrm{s} + \hbar\Omega\Phi_\mathrm{m}\right) = 0$$

leads with (A.3) in the lossless case to

$$-\omega_\mathrm{p}\tilde{\kappa}_\mathrm{mop}^\star + \omega_\mathrm{s}\tilde{\kappa}_\mathrm{mos} + \Omega\tilde{\kappa}_\mathrm{om} = 0 \tag{A.6}$$

which is true given (A.5) and $\omega_\mathrm{p} = \omega_\mathrm{s} + \Omega$. Next, we show that this coefficient (A.5) can be taken real and positive without loss of generality. Renormalizing the envelopes to $c_\mathrm{p} a_\mathrm{p}$, $c_\mathrm{s} a_\mathrm{s}$ and $c_\mathrm{m} b$ yields new coupling coefficients

$$\frac{c_\mathrm{p}}{c_\mathrm{s} c_\mathrm{m}}\tilde{\kappa}_\mathrm{mop} \qquad \frac{c_\mathrm{s}}{c_\mathrm{p} c_\mathrm{m}^\star}\tilde{\kappa}_\mathrm{mos} \qquad \frac{c_\mathrm{m}}{c_\mathrm{p} c_\mathrm{s}^\star}\tilde{\kappa}_\mathrm{om} \tag{A.7}$$

as can be seen from (A.2). Suppose that $\tilde{\kappa}_{\mathrm{om}} = \tilde{g}_0 e^{i\varphi}$ is complex with $\tilde{g}_0$ real and positive. Then we take $c_{\mathrm{p}} = c_{\mathrm{s}} = c_{\mathrm{m}} = e^{-i\varphi}$. Using (A.5) and (A.7), it follows that the renormalized coupling coefficients are real and positive:

$$\tilde{\kappa}_{\mathrm{mop}} = \tilde{\kappa}_{\mathrm{mos}} = \tilde{\kappa}_{\mathrm{om}} = \tilde{g}_0 \tag{A.8}$$

This unique coupling coefficient quantifies the coupling strength between a single photon and a single phonon propagating along a waveguide. Indeed, suppose that $a_{\mathrm{p}} = a_{\mathrm{s}} = b \mapsto 1\,\mathrm{s}^{-1/2}$ such that $\Phi_{\mathrm{p}} = \Phi_{\mathrm{s}} = \Phi_{\mathrm{m}} \mapsto 1\,\mathrm{s}^{-1}$ at a certain point along the waveguide. In the lossless case, (A.3) then becomes

$$\begin{aligned} \partial_z \Phi_{\mathrm{p}} &= -2\tilde{g}_0 \\ \pm\partial_z \Phi_{\mathrm{s}} &= 2\tilde{g}_0 \\ \partial_z \Phi_{\mathrm{m}} &= 2\tilde{g}_0 \end{aligned} \tag{A.9}$$

So $2\tilde{g}_0$ gives the rate (per meter) at which the pump flux decreases and the Stokes and phonon flux increase at a point along waveguide through which one pump photon, one Stokes photon and one phonon are passing.

The waveguide coupling coefficient $\tilde{g}_0$ can also be interpreted in terms of a zero-point motion. As shown in (14), we have

$$\tilde{g}_0 = \sqrt{\frac{L}{v_{\mathrm{p}} v_{\mathrm{s}} v_{\mathrm{m}}}}\, g_0 \tag{A.10}$$

For forward intra-modal scattering ($v_{\mathrm{p}} = v_{\mathrm{s}} = v_{\mathrm{g}}$)

$$g_0 = x_{\mathrm{ZPF}} \left.\frac{\partial \omega_{\mathrm{p}}}{\partial x}\right|_{k_{\mathrm{p}}} \tag{A.11}$$

is defined in terms of the zero-point motion and the cavity frequency pull at fixed wavevector [35]. Combining (A.10), (A.11) and (A.30), we obtain

$$\tilde{g}_0 = -\frac{\omega_{\mathrm{p}}}{c} \tilde{x}_{\mathrm{ZPF}} \left.\frac{\partial n_{\mathrm{eff}}}{\partial x}\right|_{\omega_{\mathrm{p}}} = -\tilde{x}_{\mathrm{ZPF}} \left.\frac{\partial k_{\mathrm{p}}}{\partial x}\right|_{\omega_{\mathrm{p}}} \tag{A.12}$$

with

$$\tilde{x}_{\mathrm{ZPF}} = x_{\mathrm{ZPF}} \sqrt{\frac{L}{v_{\mathrm{m}}}} = \sqrt{\frac{\hbar}{2 m_{\mathrm{eff}} v_{\mathrm{m}} \Omega_{\mathrm{m}}}} \tag{A.13}$$

the waveguide "zero-point motion" and $m_{\mathrm{eff}}$ the effective mass per unit length. Indeed, a waveguide section of length $L$ contains on average $\langle n_{\mathrm{m}} \rangle = \frac{L}{v_{\mathrm{m}}} \langle \Phi_{\mathrm{m}} \rangle$ phonons with $\langle \Phi_{\mathrm{m}} \rangle$ the mean phonon flux. As fluxes – instead of numbers – are the fundamental quantities in waveguides, the zero-point motion is corrected by precisely a factor $\sqrt{\frac{L}{v_{\mathrm{m}}}}$ in (A.13).

Often the optical envelopes are power-normalized and the acoustic envelope displacement-normalized. Starting from flux-normalized envelopes, one can switch to such normalizations through

$$c_{\mathrm{p}} = \sqrt{\hbar\omega_{\mathrm{p}}} \qquad c_{\mathrm{s}} = \sqrt{\hbar\omega_{\mathrm{s}}} \qquad c_{\mathrm{m}} = \sqrt{\frac{2\hbar\Omega_{\mathrm{m}}}{k_{\mathrm{eff}}v_{\mathrm{m}}}} = 2\tilde{x}_{\mathrm{ZPF}} \tag{A.14}$$

with $k_{\mathrm{eff}}$ the effective stiffness per unit length and by applying (A.7).

**Manley-Rowe in cavities**

Here, we apply the discussion of the previous section to the mean-field cavity equations. With arbitrary envelope normalizations and without input, equations (13) are

$$\begin{aligned}\dot{\overline{a}}_{\mathrm{p}} &= -i\kappa_{\mathrm{mop}}\overline{a}_{\mathrm{s}}\overline{b} - \chi_{\mathrm{p}}^{-1}\overline{a}_{\mathrm{p}} \\ \dot{\overline{a}}_{\mathrm{s}} &= -i\kappa_{\mathrm{mos}}\overline{b}^{\dagger}\overline{a}_{\mathrm{p}} - \chi_{\mathrm{s}}^{-1}\overline{a}_{\mathrm{s}} \\ \dot{\overline{b}} &= -i\kappa_{\mathrm{om}}\overline{a}_{\mathrm{s}}^{\dagger}\overline{a}_{\mathrm{p}} - \chi_{\mathrm{m}}^{-1}\overline{b}\end{aligned} \tag{A.15}$$

with generally $\kappa_{\mathrm{mop}} \neq \kappa_{\mathrm{mos}} \neq \kappa_{\mathrm{om}}$. Applying $\frac{\mathrm{d}}{\mathrm{d}t}\left(\overline{a}^{\dagger}\overline{a}\right) = (\frac{\mathrm{d}}{\mathrm{d}t}\overline{a}^{\dagger})\overline{a} + \overline{a}^{\dagger}(\frac{\mathrm{d}}{\mathrm{d}t}\overline{a})$ to (A.15), we find

$$\begin{aligned}\frac{\mathrm{d}}{\mathrm{d}t}n_{\mathrm{p}} &= -\kappa_{\mathrm{p}}n_{\mathrm{p}} + \left(i\kappa_{\mathrm{mop}}^{\star}\overline{a}_{\mathrm{s}}^{\dagger}\overline{b}^{\dagger}\overline{a}_{\mathrm{p}} + \mathrm{h.c.}\right) \\ \frac{\mathrm{d}}{\mathrm{d}t}n_{\mathrm{s}} &= -\kappa_{\mathrm{s}}n_{\mathrm{s}} - \left(i\kappa_{\mathrm{mos}}\overline{a}_{\mathrm{s}}^{\dagger}\overline{b}^{\dagger}\overline{a}_{\mathrm{p}} + \mathrm{h.c.}\right) \\ \frac{\mathrm{d}}{\mathrm{d}t}n_{\mathrm{m}} &= -\kappa_{\mathrm{m}}n_{\mathrm{m}} - \left(i\kappa_{\mathrm{om}}\overline{a}_{\mathrm{s}}^{\dagger}\overline{b}^{\dagger}\overline{a}_{\mathrm{p}} + \mathrm{h.c.}\right)\end{aligned} \tag{A.16}$$

Suppose now that the envelopes are number-normalized such that $n_{\mathrm{p}} = \overline{a}_{\mathrm{p}}^{\dagger}\overline{a}_{\mathrm{p}}$, $n_{\mathrm{s}} = \overline{a}_{\mathrm{s}}^{\dagger}\overline{a}_{\mathrm{s}}$ and $n_{\mathrm{m}} = \overline{b}^{\dagger}\overline{b}$ correspond to the number of pump photons, Stokes photons and phonons in the cavity. We demand that, in the lossless case ($\kappa_{\mathrm{j}} = 0$), the rate of pump photon destruction equals the rate of Stokes photon and phonon creation

$$-\dot{n}_{\mathrm{p}} = \dot{n}_{\mathrm{s}} = \dot{n}_{\mathrm{m}} \tag{A.17}$$

These are the Manley-Rowe equations for an optomechanical cavity. We deduce from (A.16) and (A.17) that

$$\kappa_{\mathrm{mop}}^{\star} = \kappa_{\mathrm{mos}} = \kappa_{\mathrm{om}} \tag{A.18}$$

This proves the existence of a single coupling coefficient that captures all conservative optical forces and scattering. Note that (A.18) also guarantees energy-conservation since

$$\frac{\mathrm{d}}{\mathrm{d}t}\left(\hbar\omega_{\mathrm{p}}n_{\mathrm{p}} + \hbar\omega_{\mathrm{s}}n_{\mathrm{s}} + \hbar\Omega n_{\mathrm{m}}\right) = 0$$

leads with (A.16) in the lossless case to

$$-\omega_{\mathrm{p}}\kappa_{\mathrm{mop}}^{\star}+\omega_{\mathrm{s}}\kappa_{\mathrm{mos}}+\Omega\kappa_{\mathrm{om}}=0 \tag{A.19}$$

which holds given (A.18) and $\omega_{\mathrm{p}}=\omega_{\mathrm{s}}+\Omega$. As in the previous section, one can show that this coupling coefficient can be chosen real and positive. This unique coupling coefficient must then be the well-known $g_0$. It quantifies the interaction strength between a single photon and a single phonon trapped in a cavity. Indeed, suppose that $\bar{a}_{\mathrm{p}}=\bar{a}_{\mathrm{s}}=\bar{b}\mapsto 1$ such that $n_{\mathrm{p}}=n_{\mathrm{s}}=n_{\mathrm{m}}\mapsto 1$ at a certain point in time. In the lossless case, (A.16) then becomes

$$\begin{aligned}\dot{n}_{\mathrm{p}}&=-2g_0\\ \dot{n}_{\mathrm{s}}&=2g_0\\ \dot{n}_{\mathrm{m}}&=2g_0\end{aligned} \tag{A.20}$$

So $2g_0$ gives the rate (per second) at which the number of pump photons decreases and the number of Stokes photons and phonons increases when there is one pump photon, one Stokes photon and one phonon in the cavity.

Often the optical envelopes are energy-normalized and the acoustic envelope displacement-normalized. Starting from number-normalized envelopes, one can switch to such normalizations through

$$c_{\mathrm{p}}=\sqrt{\hbar\omega_{\mathrm{p}}}\qquad c_{\mathrm{s}}=\sqrt{\hbar\omega_{\mathrm{s}}}\qquad c_{\mathrm{m}}=\sqrt{\frac{2\hbar\Omega_{\mathrm{m}}}{k_{\mathrm{eff}}L}}=2x_{\mathrm{ZPF}} \tag{A.21}$$

with $x_{\mathrm{ZPF}}$ the zero-point motion and by applying (A.7).

## A.3 Mean-field approximation

### Justification of $\overline{fg}=\bar{f}\bar{g}$

We denote $f(z,t)$ and $g(z,t)$ two operators that vary slowly on a lengthscale $L$. The mean-field operators are defined as $\bar{f}(t)=\frac{1}{L}\int_0^L f(z,t)\mathrm{d}z$. Clearly, when $f(z,t)=f(0,t)$ and $g(z,t)=g(0,t)$ are constants then $\overline{fg}(t)=f(0,t)g(0,t)=\bar{f}(t)\bar{g}(t)$. Let us assume now that the amplitudes vary slowly enough such that they can be Taylor-expanded as $f(z,t)=f(0,t)+f'z$ with $f'=\partial_z f(0,t)$ and the same for $g$. Then we have

$$\begin{aligned}\bar{f}&=\frac{1}{L}\left(f(0)L+f'(0)\frac{L^2}{2}\right)\\ \bar{g}&=\frac{1}{L}\left(g(0)L+g'(0)\frac{L^2}{2}\right)\end{aligned}$$

where we dropped the time-dependence. Thus, we have

$$\overline{f}\,\overline{g} = f(0)g(0) + \left(f'g(0) + f(0)g'\right)\frac{L}{2} + f'g'\frac{L^2}{4}$$

Similarly,

$$\begin{aligned}\overline{fg} &= \frac{1}{L}\int_0^L \left(f(0)g(0) + \left(f'g(0) + f(0)g'\right)z + f'g'z^2\right)\mathrm{d}z \\ &= f(0)g(0) + \left(f'g(0) + f(0)g'\right)\frac{L}{2} + f'g'\frac{L^2}{3}\end{aligned}$$

Therefore $\overline{fg} - \overline{f}\,\overline{g} = f'g'\frac{L^2}{12}$ which can be neglected if $L$ is sufficiently small compared to the lengthscale on which $f(z,t)$ and $g(z,t)$ vary.

## A.4 Alternative proofs of the $\tilde{\mathcal{G}} \leftrightarrow g_0$ link

In this section, we describe two other approaches to derive the link

$$g_0^2 = v_{\mathrm{g}}^2 \frac{(\hbar\omega_p)\,\Omega_{\mathrm{m}}}{4L}\left(\frac{\tilde{\mathcal{G}}}{Q_{\mathrm{m}}}\right) \tag{A.22}$$

### From independent full-vectorial definitions

Here, we derive equation (A.22) from the full-vectorial definitions of $\tilde{\mathcal{G}}$ and $g_0$ – specializing to intra-modal forward scattering. We focus on the moving boundary contribution. From the perturbation theory of Maxwell's equations with respect to moving boundaries [255], the cavity frequency shift $\frac{\partial\omega_{\mathrm{c}}}{\partial x}$ can be expressed as

$$\frac{\partial\omega_{\mathrm{c}}}{\partial x} = \frac{\omega_{\mathrm{p}}}{2}\frac{\oint dA\,(\mathbf{u}\cdot\hat{\mathbf{n}})\left(\Delta\epsilon|\mathbf{E}_{\parallel}|^2 - \Delta\epsilon^{-1}|\mathbf{D}_{\perp}|^2\right)}{\int dV\epsilon|\mathbf{E}|^2}$$

with $\mathbf{u}$ the normalized ($\max(|\mathbf{u}|) = 1$) acoustic field, $\hat{\mathbf{n}}$ the unit normal pointing from material 1 to material 2, $\Delta\epsilon = \epsilon_1 - \epsilon_2$ and $\Delta\epsilon^{-1} = \epsilon_1^{-1} - \epsilon_2^{-1}$. The upper integral is over the entire surface area of the cavity, the lower integral across the cavity volume. Further, $\mathbf{E}_{\parallel}$ is the electric field parallel to the boundary and $\mathbf{D}_{\perp}$ the displacement field perpendicular to the boundary. For a longitudinally invariant cavity, the surface integral can be reduced to a line integral and the volume integral to a surface integral:

$$\frac{\partial\omega_{\mathrm{c}}}{\partial x} = \frac{\omega_{\mathrm{p}}}{2}\frac{\oint dl\,(\mathbf{u}\cdot\hat{\mathbf{n}})\left(\Delta\epsilon|\mathbf{E}_{\parallel}|^2 - \Delta\epsilon^{-1}|\mathbf{D}_{\perp}|^2\right)}{\int dA\epsilon|\mathbf{E}|^2} \tag{A.23}$$

Further, the gain coefficient $\tilde{\mathcal{G}}$ is given by [64, 104, 211]

$$\tilde{\mathcal{G}} = \omega_{\mathrm{p}} \frac{Q_{\mathrm{m}}}{2k_{\mathrm{eff}}} |\langle \mathbf{f}, \mathbf{u} \rangle|^2 \tag{A.24}$$

with $\mathbf{f}$ the power-normalized optical force density and $\langle \mathbf{f}, \mathbf{u} \rangle = \int \mathbf{f}^* \cdot \mathbf{u} \, dA$. Note that $k_{\mathrm{eff}}$ is the effective stiffness *per unit length*. In the case of radiation pressure forces $\mathbf{f}_{\mathrm{rp}}$ we have [211]

$$\mathbf{f}_{\mathrm{rp}} = \frac{1}{2} \left( \Delta\epsilon |\mathbf{e}_{\parallel}|^2 - \Delta\epsilon^{-1} |\mathbf{d}_{\perp}|^2 \right) \hat{\mathbf{n}} \delta(\mathbf{r} - \mathbf{r}_{\mathrm{boundary}})$$

with $\delta(\mathbf{r} - \mathbf{r}_{\mathrm{boundary}})$ a spatial delta-distribution at the waveguide boundaries. The fields $\mathbf{e}$ and $\mathbf{d}$ are power-normalized. Here we already assumed that the Stokes and pump field profiles are nearly identical, which holds for intra-modal SBS given the small frequency shifts. Hence, we get

$$\langle \mathbf{f}_{\mathrm{rp}}, \mathbf{u} \rangle = \frac{1}{2} \oint dl \, (\mathbf{u} \cdot \hat{\mathbf{n}}) \left( \Delta\epsilon |\mathbf{e}_{\parallel}|^2 - \Delta\epsilon^{-1} |\mathbf{d}_{\perp}|^2 \right) \tag{A.25}$$

Additionally, the guided optical power $P$ is given by

$$P = \frac{v_{\mathrm{g}}}{2} \langle \mathbf{E}, \epsilon \mathbf{E} \rangle = \frac{v_{\mathrm{g}}}{2} \int dA \epsilon |\mathbf{E}|^2 \tag{A.26}$$

Combining equations (A.23), (A.25) and (A.26), we find

$$\frac{\partial \omega_{\mathrm{c}}}{\partial x} = \frac{v_{\mathrm{g}} \omega_{\mathrm{p}}}{2} \langle \mathbf{f}_{\mathrm{rp}}, \mathbf{u} \rangle$$

A similar derivation can be done for the strained bulk, so we have

$$\begin{aligned} \frac{\partial \omega_{\mathrm{c}}}{\partial x} &= \frac{v_{\mathrm{g}} \omega_{\mathrm{p}}}{2} \langle \mathbf{f}, \mathbf{u} \rangle \\ \implies \langle \mathbf{f}, \mathbf{u} \rangle &= \frac{2}{v_{\mathrm{g}} \omega_{\mathrm{p}}} \frac{\partial \omega_{\mathrm{c}}}{\partial x} \end{aligned} \tag{A.27}$$

with $\mathbf{f} = \mathbf{f}_{\mathrm{rp}} + \mathbf{f}_{\mathrm{es}}$ and $\mathbf{f}_{\mathrm{es}}$ the electrostrictive force density. Substituting equation (A.27) in (A.24) yields

$$\tilde{\mathcal{G}} = \frac{2Q_{\mathrm{m}}}{\omega_{\mathrm{p}} v_{\mathrm{g}}^2 k_{\mathrm{eff}}} \left( \frac{\partial \omega_{\mathrm{c}}}{\partial x} \right)^2 \tag{A.28}$$

Finally, we use the definition of the zero-point coupling rate $g_0 = x_{\mathrm{ZPF}} \frac{\partial \omega_{\mathrm{c}}}{\partial x}$ and the zero-point motion $x_{\mathrm{ZPF}} = \sqrt{\frac{\hbar}{2 m_{\mathrm{eff}} L \Omega_{\mathrm{m}}}}$ with $m_{\mathrm{eff}}$ the effective mass *per unit length*. Inserting these in (A.28) yields

$$\begin{aligned} \tilde{\mathcal{G}} &= \frac{2Q_{\mathrm{m}}}{\omega_{\mathrm{p}} v_{\mathrm{g}}^2 k_{\mathrm{eff}}} \frac{2 m_{\mathrm{eff}} L \Omega_{\mathrm{m}}}{\hbar} g_0^2 \\ &= Q_{\mathrm{m}} \frac{4L}{(\hbar \omega_{\mathrm{p}}) \, \Omega_{\mathrm{m}}} \frac{g_0^2}{v_{\mathrm{g}}^2} \end{aligned} \tag{A.29}$$

and (A.29) is identical to (A.22). In this derivation, we started from full-vectorial definitions that are only valid for intra-modal forward scattering. In contrast, the mean-field transition shows that this result remains true with $v_g \to \sqrt{v_p v_s}$ for inter-modal coupling.

**From independent derivative definitions**

The cavity resonance condition is $k_p L = 2\pi m$ with $m$ an integer. Given $k_p = \frac{\omega_p n_{eff}}{c}$ and $c$ the speed of light, this implies that

$$\left.\frac{\partial \omega_p}{\partial x}\right|_{k_p} = -\frac{\omega_p}{n_{eff}} \left.\frac{\partial n_{eff}}{\partial x}\right|_{k_p}$$

This can be recast in terms of the index sensitivity at fixed frequency by

$$\left.\frac{\partial n_{eff}}{\partial x}\right|_{k_p} = \frac{n_{eff}}{n_g} \left.\frac{\partial n_{eff}}{\partial x}\right|_{\omega_p}$$

with $v_{ph} = \frac{c}{n_{eff}}$ the phase velocity and $n_g = \frac{c}{v_g}$ the group index. Thus we have

$$\left.\frac{\partial \omega_p}{\partial x}\right|_{k_p} = -\frac{\omega_p}{n_g} \left.\frac{\partial n_{eff}}{\partial x}\right|_{\omega_p} \tag{A.30}$$

The cavity frequency pull must be calculated at fixed wavevector ($g_0 = x_{ZPF} \left.\frac{\partial \omega_p}{\partial x}\right|_{k_p}$), so this yields

$$\left( \left.\frac{\partial n_{eff}}{\partial x}\right|_{\omega_p} \right)^2 = g_0^2 \left( x_{ZPF} \frac{\omega_p}{n_g} \right)^{-2} \tag{A.31}$$

Previously [64], we showed that

$$\tilde{\mathcal{G}} = 2\omega_p \frac{Q_m}{k_{eff}} \left( \frac{1}{c} \left.\frac{\partial n_{eff}}{\partial x}\right|_{\omega_p} \right)^2 \tag{A.32}$$

Substitution of (A.31) in (A.32) with $x_{ZPF} = \sqrt{\frac{\hbar}{2 m_{eff} L \Omega_m}}$ results in

$$\tilde{\mathcal{G}} = \frac{4 L Q_m}{\hbar \omega_p v_g^2 \Omega_m} g_0^2$$

or the other way around

$$g_0^2 = v_g^2 \frac{(\hbar \omega_p)\, \Omega_m}{4L} \left( \frac{\tilde{\mathcal{G}}}{Q_m} \right) \tag{A.33}$$

This proof only holds for forward intra-modal scattering – whereas the mean-field transition applies to backward and inter-modal scattering as well.

## A.5 Full solution of guided-wave evolution

In this section, we give the full solution of the traveling-wave spatial dynamics (3.8) in the constant pump approximation ($\Phi_\mathrm{p}(z) = \Phi_\mathrm{p}(0)$) where the pump is strong enough to be treated classically. From this solution, one can derive the regimes treated in section 3.4 as limiting cases. In addition, this solution can directly be mapped on the cavity-based temporal dynamics (3.26). Thus, we start from

$$\begin{aligned} \partial_z a_\mathrm{s} &= \mp i \tilde{g}_0 b^\dagger a_\mathrm{p} \mp \tilde{\chi}_\mathrm{s}^{-1} a_\mathrm{s} \\ \partial_z b &= -i \tilde{g}_0 a_\mathrm{s}^\dagger a_\mathrm{p} - \tilde{\chi}_\mathrm{m}^{-1} b \end{aligned} \tag{A.34}$$

which immediately yields

$$\left(\partial_z + \tilde{\chi}_\mathrm{m}^{-\star}\right)\left(\partial_z \pm \tilde{\chi}_\mathrm{s}^{-1}\right) a_\mathrm{s}(z) = \pm \tilde{g}^2 a_\mathrm{s}(z) \tag{A.35}$$

where $\pm$ stands for forward ($+$) and backward ($-$) scattering. Inserting the ansatz $a_\mathrm{s}(z) \propto e^{\gamma z}$ leads to

$$\gamma^2 + \left(\tilde{\chi}_\mathrm{m}^{-\star} \pm \tilde{\chi}_\mathrm{s}^{-1}\right)\gamma \pm \left(\tilde{\chi}_\mathrm{m}^{-\star}\tilde{\chi}_\mathrm{s}^{-1} - \tilde{g}^2\right) = 0 \tag{A.36}$$

Its solution is

$$\begin{aligned} \gamma_1 &= \frac{1}{2}\left\{-\left(\tilde{\chi}_\mathrm{m}^{-\star} \pm \tilde{\chi}_\mathrm{s}^{-1}\right) + \sqrt{\left(\tilde{\chi}_\mathrm{m}^{-\star} \mp \tilde{\chi}_\mathrm{s}^{-1}\right)^2 + 4\tilde{g}^2}\right\} \\ \gamma_2 &= \frac{1}{2}\left\{-\left(\tilde{\chi}_\mathrm{m}^{-\star} \pm \tilde{\chi}_\mathrm{s}^{-1}\right) - \sqrt{\left(\tilde{\chi}_\mathrm{m}^{-\star} \mp \tilde{\chi}_\mathrm{s}^{-1}\right)^2 + 4\tilde{g}^2}\right\} \end{aligned} \tag{A.37}$$

Given these two roots, one can determine the exact evolution of the photon-phonon fields along the waveguide if the correct boundary conditions are known. The boundary condition $b(0) = 0$ and fixed probe flux $\Phi_\mathrm{s}(0) = a_\mathrm{s}^\dagger(0)a_\mathrm{s}(0)$ are appropriate for forward scattering such that

$$\begin{aligned} a_\mathrm{s}(z) &= \frac{a_\mathrm{s}(0)}{\gamma_2 - \gamma_1}\left\{\left(\gamma_2 + \tilde{\chi}_\mathrm{s}^{-1}\right)e^{\gamma_1 z} - \left(\gamma_1 + \tilde{\chi}_\mathrm{s}^{-1}\right)e^{\gamma_2 z}\right\} \\ b^\dagger(z) &= i\frac{a_\mathrm{s}(0)}{\tilde{g}}\frac{\left(\gamma_2 + \tilde{\chi}_\mathrm{s}^{-1}\right)\left(\gamma_1 + \tilde{\chi}_\mathrm{s}^{-1}\right)}{\gamma_2 - \gamma_1}\left\{e^{\gamma_1 z} - e^{\gamma_2 z}\right\} \end{aligned} \tag{A.38}$$

The backward case (fixed $\Phi_\mathrm{s}(L)$ with $L$ the waveguide length) can be solved similarly. This full solution contains the important regimes discussed in section 3.4. For instance, in the strong coupling regime ($\tilde{g} \gg \alpha_\mathrm{s} + \alpha_\mathrm{m}$) and at resonance ($\tilde{\Delta}_\mathrm{j} = 0$) one can show that

$$\begin{aligned} \gamma_1 &\approx -\frac{\alpha_\mathrm{m} + \alpha_\mathrm{s}}{4} + \tilde{g} \overset{\text{large } \tilde{g}}{\longrightarrow} \tilde{g} \\ \gamma_2 &\approx -\frac{\alpha_\mathrm{m} + \alpha_\mathrm{s}}{4} - \tilde{g} \longrightarrow -\tilde{g} \end{aligned} \tag{A.39}$$

Therefore, the spatial coupling rate $\tilde{g}$ must overcome the average photonic and phononic propagation loss before actual photon-phonon pair generation can be seen. The photons and phonons indeed equally share the total propagation loss $\alpha_{\mathrm{m}} + \alpha_{\mathrm{s}}$ in this regime, as in cavity settings [56]. The spatial evolution (A.38) then becomes identical to (3.30). The weak coupling regimes of stimulated photon ($\alpha_{\mathrm{s}} \ll \alpha_{\mathrm{m}}$) and phonon ($\alpha_{\mathrm{m}} \ll \alpha_{\mathrm{s}}$) emission can equally well be obtained from the full solution (A.38). This solution also contains acoustic ($\alpha_{\mathrm{s}} \ll \alpha_{\mathrm{m}}$) [98] and optical ($\alpha_{\mathrm{m}} \ll \alpha_{\mathrm{s}}$) build-up effects.

# B

# Model of forward intra-modal Brillouin scattering in the presence of a Kerr effect

## Contents

In this appendix we present a model that captures the essential dynamics of forward intra-modal stimulated Brillouin scattering in the presence of a background Kerr effect. The analysis is very similar to earlier discussions of forward SBS [152, 155] and Raman scattering [36]. Specifically, we describe under which circumstances forward SBS can still be seen as a pure gain process. Thus the model includes

- cascading into higher-order Stokes and anti-Stokes waves,
- four-wave mixing contributions from both the Brillouin and the Kerr effect and
- the effect of these contributions on the SBS gain.

## B.1 Derivation of spatial evolution

We assume that the electromagnetic field is composed of discrete lines, with $a_n(z)$ the complex amplitude of component $n$ with angular frequency $\omega_n = \omega_0 + n\Omega$ at position $z$ along the guide. By definition, $\omega_0$ is the frequency of the pump. In the presence of weak nonlinear coupling between the waves, the evolution of the slowly-varying amplitudes is [36]

$$\frac{\mathrm{d}a_n}{\mathrm{d}z} = -i\frac{\omega_0}{2cn_{\mathrm{eff}}\epsilon_0}p_n^{\mathrm{NL}}$$

with $p_n^{\mathrm{NL}}$ the complex amplitude at frequency $\omega_n$ of the nonlinear polarization $P^{\mathrm{NL}}(z,t) = \epsilon_0\chi^{\mathrm{NL}}(z,t)E(z,t)$ with $E(z,t) = \frac{1}{2}\sum_n a_n(z)\exp\left(i(\omega_n t - k_n z)\right) +$ c.c. the electric field. Here we assumed that the cascading is limited to tens of higher-order sidebands, such that $\omega_n \approx \omega_0$ and that all components experience the same effective mode index $n_{\mathrm{eff}}$. Further, the nonlinear susceptibility is given by

$$\chi^{\mathrm{NL}}(z,t) = 2n_{\mathrm{eff}}\Delta n_{\mathrm{eff}}(z,t)$$

in case the index changes $\Delta n_{\mathrm{eff}}(z,t)$ are small. These index changes are composed of an instantaneous Kerr component and a delayed Brillouin component:

$$\begin{aligned}\Delta n_{\mathrm{eff}}(z,t) &= \Delta n_{\mathrm{eff,Kerr}}(z,t) + \Delta n_{\mathrm{eff,Brillouin}}(z,t) \\ &= \frac{\bar{n}_2}{A_{\mathrm{eff}}}P(z,t) + \left.\frac{\partial n_{\mathrm{eff}}}{\partial q}\right|_{q_{\mathrm{avg}}} q(z,t)\end{aligned}$$

with $\bar{n}_2$ the nonlinear Kerr index averaged over the waveguide cross-section, $A_{\mathrm{eff}}$ the effective mode area, $P(z,t)$ the total optical power and $q$ a coordinate describing the mechanical motion. There is a small shift in the average value of $q$ due to the constant component of the power $P(z,t)$. However, we can always redefine $q$ such that $q_{\mathrm{avg}} \equiv 0$. In addition, $\frac{\partial n_{\mathrm{eff}}}{\partial q}$ is the sensitivity of the effective index with respect to motion. This factor contains contributions from both the moving boundary (radiation pressure) and the bulk (electrostriction). It should be calculated at fixed optical frequency, since this frequency is externally applied. We characterize the mechanical mode as a harmonic oscillator in each cross-section $z$:

$$\ddot{q}(z,t) + \kappa_{\mathrm{m}}\dot{q}(z,t) + \Omega_{\mathrm{m}}^2 q(z,t) = \frac{F(z,t)}{m_{\mathrm{eff}}}$$

with $\frac{\kappa_{\mathrm{m}}}{2\pi}$ the Brillouin linewidth, $\Omega_{\mathrm{m}}^2 = \frac{k_{\mathrm{eff}}}{m_{\mathrm{eff}}}$ the angular frequency, $m_{\mathrm{eff}}$ the effective mass of the mechanical mode per unit length and $F(z,t)$ the total force acting on that mode per unit length. Since this equation does not explicitly depend on $z$, $q(z,t)$ directly inherits its position-dependency from $F(z,t)$. Note that any propagation of phonons along the waveguide is neglected in this step. Each cross-section oscillates independently, reminiscent of the molecular vibration in Raman scattering [36, 152]. This assumption is justified by the very low group velocity of the acoustic phonons. From phase-matching, the acoustic wavevector along the $z$-axis is $K = k_0 - k_{-1} = \frac{\Omega}{v_g}$. Therefore, the acoustic phase velocity $v_a = \frac{\Omega}{K}$ equals the optical group velocity $v_g$. At the same time, we must have $v_a v_{a,g} = c_a^2$ with $v_{a,g}$ the acoustic group velocity and $c_a$ the bulk acoustic velocity. Here we treat the silicon wire acoustically as a slab waveguide close to cut-off. This yields a low acoustic group velocity of $v_{a,g} \approx 1\,\text{m/s}$ for our silicon wires, which is confirmed by the finite-element model. Therefore we treat the acoustic wave as a localized oscillator, following the success of this description in other systems [152, 155].

From power-conservation [256], the optical force $F(z,t)$ per unit length can be related to $\frac{\partial n_{\text{eff}}}{\partial q}$ as

$$F(z,t) = \frac{1}{c} \left. \frac{\partial n_{\text{eff}}}{\partial q} \right|_{q_{\text{avg}}} P(z,t)$$

The power $P(z,t) = 2E^2(z,t)$ contains frequencies $n\Omega\ \forall n$ up to the total number of lines. However, we assume that only the component at $\Omega$ excites the mechanical motion. So we take $F(z,t) = \frac{1}{2} f_\Omega \exp\left(i(\Omega t - Kz)\right) + \text{c.c.}$ with $f_\Omega = \frac{1}{c}\frac{\partial n_{\text{eff}}}{\partial q} p_\Omega$ and $p_\Omega = 2\sum_n a_n a_{n-1}^\star$.

We normalized the amplitudes $a_n$ such that the power of wave $\omega_n$ is $|a_n|^2$. Close to resonance ($\Omega \approx \Omega_m$), the steady-state response of the harmonic oscillator is $q_\Omega = Q_m \frac{f_\Omega}{k_{\text{eff}}} \mathcal{L}(\Omega)$ with the Lorentzian function $\mathcal{L}(\Omega) = \frac{1}{-2\Delta_r + i}$, the relative detuning $\Delta_r = \frac{\Omega - \Omega_m}{\kappa_m}$ and the quality factor $Q_m = \frac{\Omega_m}{\kappa_m}$. Therefore, we can write the nonlinear index change in terms of the nonlinear Kerr and Brillouin parameters $\gamma_K$ and $\gamma_{\text{SBS}}$:

$$\begin{aligned}
\Delta n_\Omega &= \Delta n_{\Omega,\text{Kerr}} + \Delta n_{\Omega,\text{Brillouin}} \\
&= \frac{\gamma_K}{k_0} p_\Omega + \frac{\gamma_{\text{SBS}}}{k_0} p_\Omega \mathcal{L}(\Omega) \\
&= \frac{p_\Omega}{k_0} \gamma(\Omega)
\end{aligned}$$

where we defined the total nonlinearity parameter $\gamma(\Omega) = \gamma_K + \gamma_{\text{SBS}}\mathcal{L}(\Omega)$, using $\gamma_K \equiv k_0 \frac{\bar{n}_2}{A_{\text{eff}}}$ and $\gamma_{\text{SBS}} \equiv \omega_0 \frac{Q_m}{k_{\text{eff}}} \left(\frac{1}{c}\frac{\partial n_{\text{eff}}}{\partial q}\right)^2$. We note that this formula for the Brillouin nonlinearity is identical to the rigorous [104, 211] $\gamma_{\text{SBS}} = \omega_0 Q_m |\langle \mathbf{f}, \mathbf{u}\rangle|^2/(4k_{\text{eff}})$ if we identify $\frac{1}{c}|\frac{\partial n_{\text{eff}}}{\partial q}| \equiv \frac{|\langle \mathbf{f}, \mathbf{u}\rangle|}{2}$. Hence the evolution of

the amplitudes is

$$\frac{\mathrm{d}a_n}{\mathrm{d}z} = -i\frac{k_0}{2}\left(\Delta n_\Omega a_{n-1} + \Delta n_\Omega^\star a_{n+1}\right) \tag{B.1}$$

$$\Delta n_\Omega = \frac{p_\Omega}{k_0}\gamma(\Omega) = \frac{2\gamma(\Omega)}{k_0}\sum_n a_n a_{n-1}^\star$$

## B.2 Solving for the evolution

These equations can be solved analytically since $\Delta n_\Omega$ turns out to be a constant of motion. Indeed, derivation yields

$$\begin{aligned}\frac{\mathrm{d}\Delta n_\Omega}{\mathrm{d}z} &\propto \sum_n \left(a_n \frac{\mathrm{d}a_{n-1}^\star}{\mathrm{d}z} + \frac{\mathrm{d}a_n}{\mathrm{d}z} a_{n-1}^\star\right)\\ &\propto \sum_n \Delta n_\Omega \left(|a_n|^2 - |a_{n-1}|^2\right)\\ &+ \Delta n_\Omega^\star \left(a_n a_{n-2}^\star - a_{n+1} a_{n-1}^\star\right)\\ &= 0\end{aligned}$$

Consequently, equation (B.1) can be solved either directly by using properties of the Bessel functions or indirectly by noting that $\Delta n_\Omega(z) = \Delta n_\Omega(0)$ such that the nonlinear interaction is equivalent to spatiotemporal phase-modulation. Specifically,

$$\begin{aligned}E(z,t) &= \frac{1}{2}\sum_n a_n(z)\exp\left(i(\omega_n t - k_n z)\right) + \text{c.c.}\\ &= \frac{1}{2}\exp\left(-ik_0 z \Delta n_{\text{eff}}(z,t)\right)\times\\ &\qquad \sum_n a_n(0)\exp\left(i(\omega_n t - k_n z)\right) + \text{c.c.}\end{aligned} \tag{B.2}$$

Moreover, we have

$$\begin{aligned}\Delta n_{\text{eff}}(z,t) &= |\Delta n_\Omega(0)|\sin\left(\Omega t - Kz + \varphi_0\right)\\ &= \frac{2|\gamma(\Omega)|}{k_0}\left|\sum_n a_n(0)a_{n-1}^\star(0)\right|\sin\left(\Omega t - Kz + \varphi_0\right)\end{aligned} \tag{B.3}$$

with $\varphi_0 = \angle\left\{\Delta n_\Omega(0)\exp\left(i\frac{\pi}{2}\right)\right\}$. As previously noted in the context of photonic crystal fibres [152], this is equivalent to phase-modulation with a depth $\xi$ determined by the strength of the input fields, the interaction length and the nonlinear parameter $|\gamma(\Omega)|$. The amplitudes of the individual components can finally be found by inserting $\exp\left(i\xi\sin\Phi\right) = \sum_n \mathcal{J}_n(\xi)\exp\left(in\Phi\right)$ with $\mathcal{J}_n$ the $n$th-order Bessel function of the first kind. To arrive at this phase-modulation picture, we assumed that all index changes originate from the beating at frequency $\Omega$. This is correct for the mechanical effect since it is weak off resonance. However, the Kerr response is non-resonant at telecom wavelengths. Thus its strength is the same at $\omega_0 + n\Omega$ for all $n$. We include the $n\Omega$ ($n \neq 1$) Kerr-mediated coupling in the next paragraph, keeping in mind that equations (B.2)-(B.3) are only entirely correct when $\gamma_K = 0$.

To see how the modulation picture (B.2)-(B.3) relates to the traditional view of SBS as a pure gain process, we simplify equation (B.1) to the case of an undepleted pump, a Stokes and an anti-Stokes signal. Neglecting higher-order cascading, this yields

$$\begin{aligned} \frac{\mathrm{d}a_\mathrm{s}}{\mathrm{d}z} &= -i\gamma^\star(\Omega)\left(|a_\mathrm{p}|^2 a_\mathrm{s} + a_\mathrm{p}^2 a_\mathrm{a}^\star\right) \\ \frac{\mathrm{d}a_\mathrm{a}}{\mathrm{d}z} &= -i\gamma(\Omega)\left(|a_\mathrm{p}|^2 a_\mathrm{a} + a_\mathrm{p}^2 a_\mathrm{s}^\star\right) \end{aligned} \tag{B.4}$$

In case $a_\mathrm{a}(0) = 0$, the initial evolution of the Stokes power is

$$\frac{\mathrm{d}P_\mathrm{s}}{\mathrm{d}z} = -2\Im\{\gamma(\Omega)\}P_\mathrm{p}P_\mathrm{s}$$

Since $\Im\{\gamma(\Omega)\} = -\frac{\gamma_\mathrm{SBS}}{4\Delta_\mathrm{r}^2+1}$, we recover a Lorentzian Brillouin gain profile in this approximation:

$$\begin{aligned} \frac{\mathrm{d}P_\mathrm{s}}{\mathrm{d}z} &= G_\mathrm{SBS}(\Omega)P_\mathrm{p}P_\mathrm{s} \\ G_\mathrm{SBS}(\Omega) &= \frac{2\gamma_\mathrm{SBS}}{4\Delta_\mathrm{r}^2+1} \end{aligned} \tag{B.5}$$

Similarly, the anti-Stokes experiences a Lorentzian loss profile if $a_\mathrm{s}(0) = 0$. Thus the Kerr effect has no impact on the initial evolution of the Stokes wave. Therefore, forward SBS is a pure gain process as long as the anti-Stokes build-up is negligible. By numerically integrating equations (B.4), including linear losses, we confirm that this is the case in our experiments. The $n\Omega$ ($n \neq 1$) Kerr-mediated coupling does not change this conclusion. We can see this as follows. In the Lorentz-model for the permittivity, the Kerr response can be treated as a second-order nonlinear spring [36]

$$\ddot{x} + \Gamma_\mathrm{e}\dot{x} + \Omega_e^2(x)x = -\frac{e}{m_\mathrm{e}}E$$

with $x$ the displacement of the electron cloud, $m_\mathrm{e}$ the electron mass, $\Omega_\mathrm{e}^2(x) = \frac{k_\mathrm{e}(x)}{m_\mathrm{e}}$ and $k_\mathrm{e}(x) = k_\mathrm{e}(0) + \frac{\partial^2 k_\mathrm{e}}{\partial x^2}x^2$ the nonlinear spring constant. Since $\omega_n \ll \Omega_\mathrm{e}$, the oscillator responds instantaneously to the Lorentz-force $-eE$:

$$\Omega_e^2(x)x = -\frac{e}{m_\mathrm{e}}E$$

Thus the linear solution is $x_{\mathrm{L}}(z,t) = \frac{-e}{k_{\mathrm{e}}(0)} E(z,t)$. In the first Born approximation, the nonlinear displacement is

$$x_{\mathrm{NL}} = -\frac{1}{k_{\mathrm{e}}(0)} \frac{\partial^2 k_{\mathrm{e}}}{\partial x^2} x_{\mathrm{L}}^3$$

And the nonlinear polarization is $P^{\mathrm{NL}} = \epsilon_0 \chi^{\mathrm{NL}} E = -Nex_{\mathrm{NL}}$ with $N$ the atomic number density. This implies that the nonlinear polarization is proportional to $E^3(z,t)$. Unlike in the Brillouin case, the Lorentz oscillator does not filter out $0\Omega$, $2\Omega$, $3\Omega$, etc. terms. Selecting the right components of $P^{\mathrm{NL}}$, we find that equations (B.4) are modified to

$$\begin{aligned} \frac{\mathrm{d}a_{\mathrm{s}}}{\mathrm{d}z} &= -i\gamma^\star(\Omega)\left(|a_{\mathrm{p}}|^2 a_{\mathrm{s}} + a_{\mathrm{p}}^2 a_{\mathrm{a}}^\star\right) - i\gamma_{\mathrm{K}}|a_{\mathrm{p}}|^2 a_{\mathrm{s}} \\ \frac{\mathrm{d}a_{\mathrm{a}}}{\mathrm{d}z} &= -i\gamma(\Omega)\left(|a_{\mathrm{p}}|^2 a_{\mathrm{a}} + a_{\mathrm{p}}^2 a_{\mathrm{s}}^\star\right) - i\gamma_{\mathrm{K}}|a_{\mathrm{p}}|^2 a_{\mathrm{a}} \end{aligned}$$

for a strong, undepleted pump. The added terms on the right generate a constant phase shift and do, therefore, not alter the conclusion that these equations yield Brillouin gain when $a_{\mathrm{a}}(0) = 0$. However, such added terms do invalidate the phase-modulation solution (B.2)-(B.3).

Back to that solution (B.2)-(B.3), at first sight we expect a Fano-like resonance for the Stokes power because the modulation depth depends on $|\gamma(\Omega)|$ and not on $\Im\{\gamma(\Omega)\}$. However, the input phase $\varphi_0$ also contains phase information on $\gamma(\Omega)$. We analytically check that the phase-modulation picture is equivalent to a pure gain process in the low-gain regime. Combining equations (B.2) and (B.3) with only an initial pump and Stokes wave, we find

$$a_{\mathrm{s}}(z) = a_{\mathrm{s}}(0) - \mathcal{J}_1(\xi)\, a_{\mathrm{p}}(0) \exp\left(-i(\varphi_0 + \pi)\right)$$

with $\xi = 2|\gamma(\Omega)|\sqrt{P_{\mathrm{s}}(0) P_{\mathrm{p}}} z$ the unitless cascading parameter. The power of the Stokes wave then becomes

$$P_{\mathrm{s}}(z) = P_{\mathrm{s}}(0)\left(1 - 2\Im\{\gamma(\Omega)\} P_{\mathrm{p}} z\right) + \frac{\xi^2}{4} P_{\mathrm{p}}$$

Here we approximated the Bessel function as $\mathcal{J}_1(\xi) \approx \frac{\xi}{2}$, which is valid in the low-$\xi$ regime. The last term, containing $\xi^2$, gives rise to a Fano resonance but is smaller than the other terms in this regime. Taking the derivative and letting $z \to 0$, we indeed recover the gain equation (B.5). In our experiments we reach values of $\xi \approx 0.4$ in the longest waveguides and at maximum pump power. To conclude, we can safely neglect higher-order cascading and treat forward SBS as a pure gain process driven exclusively by the Brillouin nonlinearity. In the presence of linear optical losses, the modified evolution

of the Stokes wave is

$$\frac{\mathrm{d}P_\mathrm{s}}{\mathrm{d}z} = \left(G_\mathrm{SBS}(\Omega)P_\mathrm{p}\exp\left(-\alpha z\right) - \alpha\right) P_\mathrm{s}$$
$$G_\mathrm{SBS}(\Omega) = \frac{2\gamma_\mathrm{SBS}}{4\Delta_\mathrm{r}^2 + 1}$$

with $\alpha$ the linear optical loss and $P_\mathrm{p}$ the input pump power. The analytical solution of this equation is

$$P_s(L) = P_s(0)\exp\left(G_\mathrm{SBS}(\Omega)P_\mathrm{p}L_\mathrm{eff} - \alpha L\right) \tag{B.6}$$

with $L_\mathrm{eff} = \frac{1-\exp(-\alpha L)}{\alpha}$ the effective interaction length. In the case of nonlinear losses $\alpha(P_\mathrm{p})$ the equations can be integrated numerically.

## B.3 The Brillouin gain coefficient

The Brillouin gain coefficient $\tilde{\mathcal{G}} = 2\gamma_\mathrm{SBS}$ at the mechanical resonance ($\Omega = \Omega_\mathrm{m}$) is given by

$$\tilde{\mathcal{G}} = 2\omega_0 \frac{Q_\mathrm{m}}{k_\mathrm{eff}}\left(\frac{1}{c}\frac{\partial n_\mathrm{eff}}{\partial q}\right)^2$$
$$= \omega_0\frac{Q_\mathrm{m}}{2k_\mathrm{eff}}|\langle \mathbf{f}, \mathbf{u}\rangle|^2 \tag{B.7}$$

with $\omega_0$ the optical angular frequency [Hz], $Q_\mathrm{m}$ the mechanical quality factor [-], $k_\mathrm{eff}$ the effective stiffness per unit length [N/m$^2$], $c$ the speed of light [m/s], $\frac{\partial n_\mathrm{eff}}{\partial q}$ the derivative of the optical mode index with respect to mechanical motion [1/m], $\mathbf{f} = \mathbf{f}_\mathrm{rp} + \mathbf{f}_\mathrm{es}$ the power-normalized force density [N/(m$^3$W)] and $\langle \mathbf{f}, \mathbf{u}\rangle = \int \mathbf{f}^\star \cdot \mathbf{u}\, dA$ the overlap integral between the optical forces and the mechanical mode $\mathbf{u}$ [-]. Note that we chose the mechanical mode profile $\mathbf{u}$ to be dimensionless, so the mechanical coordinate $q$ is expressed in [m]. Therefore the overlap integral $\langle \mathbf{f}, \mathbf{u}\rangle$ has dimension [s/m$^2$], as does $\frac{1}{c}\frac{\partial n_\mathrm{eff}}{\partial q}$. The effective stiffness is defined as $k_\mathrm{eff} \equiv \Omega_\mathrm{m}^2 m_\mathrm{eff}$, where $m_\mathrm{eff} \equiv \langle \mathbf{u}, \rho\mathbf{u}\rangle$ is the effective mass per unit length [kg/m] with $\rho$ the mass density. Typically, the elastic mode profile $\mathbf{u}$ is normalized such that $\max(|\mathbf{u}|) = 1$. Then we have $m_\mathrm{eff} \approx m$ (and $m_\mathrm{eff} \leq m$) with $m$ the true mass of the waveguide per unit length. With these definitions, the gain coefficient indeed has dimensions

$[\text{W}^{-1}\text{m}^{-1}]$:

$$\begin{aligned}[\tilde{\mathcal{G}}] &= [\omega_0][\frac{Q_\text{m}}{k_\text{eff}}][\langle \mathbf{f}, \mathbf{u}\rangle]^2 \\ &= \frac{1}{\text{s}}\frac{\text{m}^2}{\text{N}}\frac{\text{s}^2}{\text{m}^4} \\ &= \frac{\text{s}}{\text{Nm}^2} \\ &= \frac{1}{\text{Wm}}\end{aligned}$$

Furthermore, the Brillouin gain (B.7) is identical to formulas presented in earlier theoretical work [104, 211]. Specifically, the gain coefficient (formula (10) of Qiu (2013) [211]) is

$$\tilde{\mathcal{G}} = \frac{2\omega_0 Q_\text{m}}{\Omega_\text{m}^2 v_\text{gp} v_\text{gs}} \frac{|\langle \tilde{\mathbf{f}}, \mathbf{u}\rangle|^2}{\langle \mathbf{E}_\text{p}, \epsilon \mathbf{E}_\text{p}\rangle \langle \mathbf{E}_\text{s}, \epsilon \mathbf{E}_\text{s}\rangle \langle \mathbf{u}, \rho \mathbf{u}\rangle}$$

with $v_\text{gp}$ and $v_\text{gs}$ the pump and Stokes optical group velocity, $\mathbf{E}_\text{p}$ and $\mathbf{E}_\text{s}$ the pump and Stokes electric field distribution, $\epsilon$ the dielectric permittivity and $\tilde{\mathbf{f}}$ the force distribution. The total power in a guided wave is $P = \frac{v_\text{g}}{2}\langle \mathbf{E}, \epsilon \mathbf{E}\rangle$, so we get

$$\tilde{\mathcal{G}} = \frac{\omega_0 Q_\text{m}}{2\Omega_\text{m}^2} \frac{|\langle \tilde{\mathbf{f}}, \mathbf{u}\rangle|^2}{P_\text{p} P_\text{s} m_\text{eff}}$$

where we used $m_\text{eff} = \langle \mathbf{u}, \rho \mathbf{u}\rangle$. Defining the power-normalized force distribution $\mathbf{f}$ as $\mathbf{f} \equiv \frac{\tilde{\mathbf{f}}}{\sqrt{P_\text{p} P_\text{s}}}$, we arrive at formula (B.7):

$$\tilde{\mathcal{G}} = \omega_0 \frac{Q_\text{m}}{2k_\text{eff}} |\langle \mathbf{f}, \mathbf{u}\rangle|^2$$

Our finite-element calculations of the SBS coefficient are based on this formula. This theory completely reproduces the conventional backward SBS coefficients in the limit of transverse waveguide dimensions much larger than the free-space wavelength [104, 211]. It predicts strongly enhanced photon-phonon coupling in sub-wavelength waveguides – as in our silicon nanowires.

## B.4 Model of the XPM experiments

In the cross-phase modulation (XPM) experiments, we study the phase modulation imprinted on a probe wave by a strong intensity-modulated pump. The pump and its sidebands are located at frequencies $\omega_0$, $\omega_1 = \omega_0 + \Omega$ and $\omega_{-1} = \omega_0 - \Omega$. The probe has frequency $\omega_\text{pr}$. The four-wave mixing interaction between these waves imprints sidebands $\omega_{\text{pr}} \pm \Omega$ on the probe. We monitor the power $P_{\text{imprint}}$ in the $\omega_{\text{imprint}} = \omega_{\text{pr}} + \Omega$ sideband at the end of the waveguide as a function of $\Omega$.

If there were only Brillouin coupling between the waves, the effective index would be modulated exclusively at frequency $\Omega$. However, the Kerr effect responds equally well to the beat notes $\Delta_0 = \omega_0 - \omega_{\text{pr}}$ and $\Delta_{-1} = \omega_{-1} - \omega_{\text{pr}}$. So there are four pathways to $\omega_{\text{imprint}}$:

$$\begin{aligned}
\omega_{\text{imprint}} &= \omega_{\text{pr}} + (\omega_1 - \omega_0) \\
\omega_{\text{imprint}} &= \omega_{\text{pr}} + (\omega_0 - \omega_{-1}) \\
\omega_{\text{imprint}} &= \omega_1 - \Delta_0 \\
\omega_{\text{imprint}} &= \omega_0 - \Delta_{-1}
\end{aligned}$$

Both the Kerr and the Brillouin effect take the first two, but only the Kerr effect takes the latter two pathways. Therefore the Kerr effect manifests itself with double strength in these experiments. Building on the formalism of section B.1, we calculate the imprinted sideband power $P_{\text{imprint}}$. The index modulation is

$$\begin{aligned}
\Delta n_{\text{eff}}(z,t) = {} & |\Delta n_\Omega| \sin\left(\Omega t - Kz + \varphi_\Omega\right) \\
& + |\Delta n_{\Delta_0}| \sin\left(\Delta_0 t - \left(k_0 - k_{\text{pr}}\right) z + \varphi_{\Delta_0}\right) \\
& + |\Delta n_{\Delta_{-1}}| \sin\left(\Delta_{-1} t - \left(k_{-1} - k_{\text{pr}}\right) z + \varphi_{\Delta_{-1}}\right)
\end{aligned} \tag{B.8}$$

with the following definitions

$$\begin{aligned}
\Delta n_\Omega &= \frac{p_\Omega}{k_0} \left\{ \gamma_{\text{K}} + \gamma_{\text{SBS}} \mathcal{L}(\Omega) \right\} \\
\Delta n_{\Delta_0} &= \frac{p_{\Delta_0}}{k_0} \gamma_{\text{K}} \\
\Delta n_{\Delta_{-1}} &= \frac{p_{\Delta_{-1}}}{k_0} \gamma_{\text{K}}
\end{aligned}$$

As before, we denote the angles $\varphi_{\text{j}} = \angle \left\{ \Delta n_{\text{j}} \exp\left(i\frac{\pi}{2}\right) \right\}$. We also define a modulation depth $\xi_{\text{j}} = k_0 z |\Delta n_{\text{j}}|$ for each beat note. Next, we insert equation (B.8) in equation (B.2) and apply the Bessel expansion $\exp\left(i\xi \sin\Phi\right) = \sum_n \mathcal{J}_n(\xi) \exp\left(in\Phi\right)$ to each of the beat notes. This results in

$$\begin{aligned}
E(z,t) = {} & \frac{1}{2} \sum_{klm} \mathcal{J}_k(\xi_\Omega) \mathcal{J}_l(\xi_{\Delta_0}) \mathcal{J}_m(\xi_{\Delta_{-1}}) \times \\
& \exp\left(-i(k\Phi_\Omega + l\Phi_{\Delta_0} + m\Phi_{\Delta_{-1}})\right) \times \\
& \sum_n a_n(0) \exp\left(i(\omega_n t - k_n z)\right) + \text{c.c.}
\end{aligned} \tag{B.9}$$

Only three terms in the Bessel expansion influence $P_{\text{imprint}}$ when $\xi$ is small. In particular, for $(klm) = (-100)$, $(010)$ and $(001)$ the frequencies $\omega_{\text{pr}}$, $\omega_1$ and $\omega_0$ are shifted to $\omega_{\text{imprint}}$ respectively. Working out equation (B.9) for these terms, we obtain

$$a_{\text{imprint}}(z) = -\frac{\xi_\Omega}{2}\exp\left(i\varphi_\Omega\right)a_{\text{pr}} + \frac{\xi_{\Delta_0}}{2}\exp\left(-i\varphi_{\Delta_0}\right)a_1 + \frac{\xi_{\Delta_{-1}}}{2}\exp\left(-i\varphi_{\Delta_{-1}}\right)a_0$$

for the amplitude $a_{\text{imprint}}(z)$ of the imprinted tone. Here we used $\mathcal{J}_1\left(\xi\right) \approx \frac{\xi}{2}$ for small $\xi$. Since the beat note amplitudes are $p_\Omega = 2\left(a_1 a_0^\star + a_0 a_{-1}^\star\right)$, $p_{\Delta_0} = 2a_0 a_{\text{pr}}^\star$ and $p_{\Delta_{-1}} = 2a_{-1}a_{\text{pr}}^\star$, we finally obtain

$$P_{\text{imprint}}(z) = |\gamma_{\text{XPM}}(\Omega)|^2 |p_\Omega|^2 P_{\text{pr}} \frac{z^2}{4}$$

with $\gamma_{\text{XPM}}(\Omega) = 2\gamma_{\text{K}} + \gamma_{\text{SBS}}\mathcal{L}(\Omega)$. Therefore we use the Fano lineshape $|\frac{\gamma_{\text{XPM}}(\Omega)}{2\gamma_{\text{K}}}|^2$ as a fitting function for the normalized probe sideband power.

# List of Figures

# List of Tables